\documentclass[a4paper,11pt]{article}
\usepackage{header}
\usepackage{local}

\begin{document}

\pagestyle{empty}
\hfil{\LARGE\bf 
The ground state construction\par
\medskip
\hfil of bilayer graphene}
\par
\seqskip

{\bf\hfil Alessandro Giuliani}\par
{\small\hfil Dipartimento di Matematica e Fisica, Universit\`a degli Studi Roma Tre,\par
\hfil L.go S.L. Murialdo, 1, 00146 Roma, Italy.\par}
\subseqskip

\hfil{\bf Ian Jauslin}\par
{\small\hfil University of Rome ``Sapienza'', Dipartimento di Fisica\par
\hfil P.le Aldo Moro, 2, 00185 Rome, Italy.\par}
\seqskip

\begin{abstract}
We consider a model of half-filled bilayer graphene, in which the three dominant Slonczewski-Weiss-McClure hopping parameters are retained, 
in the presence of short range interactions. Under a smallness assumption on the interaction strength $U$ as well as on the inter-layer hopping $\epsilon$, 
we construct the ground state in the thermodynamic limit, and prove its analyticity 
in $U$, uniformly in $\epsilon$. The interacting Fermi surface is degenerate, and consists of eight Fermi points, two of which are protected by symmetries, while the locations of the other six 
are renormalized by the interaction, and the effective dispersion relation at the Fermi points is conical.
The construction reveals the presence of different energy regimes, where the effective behavior of correlation functions 
changes qualitatively. The analysis of the crossover between regimes plays an important role in the proof of analyticity and in the uniform control of the radius of convergence. 
The proof is based on a rigorous implementation of fermionic renormalization group methods, including determinant estimates for the renormalized expansion.
\end{abstract}

\pagebreak

\tableofcontents
\pagebreak

\pagestyle{plain}
\setcounter{page}1

\section{Introduction}
\label{introduction}
\indent Graphene, a one-atom thick layer of graphite, has captivated a large part of the scientific community for the past decade. With good reason: as was shown by A.~Geim's team, graphene is a stable two-dimensional crystal with very peculiar electronic properties~\cite{ngeZF}. The mere fact that a two-dimensional crystal can be synthesized, and manipulated, at room temperature without working inside a vacuum~\cite{geOZ} is, in and of itself, quite surprising. But the most interesting features of graphene lay within its electronic properties. Indeed, electrons in graphene were found to have an extremely high mobility~\cite{ngeZF}, which could make it a good candidate to replace silicon in microelectronics; and they were later found to behave like massless Dirac Fermions~\cite{ngeZFi,zteZFi}, which is of great interest for the study of fundamental Quantum Electro-Dynamics. These are but a few of the intriguing features~\cite{geiZSe} that have prompted a lively response from the scientific community.\par
\indent These peculiar electronic properties stem from the particular energy structure of graphene. It consists of two energy bands, that meet at exactly two points, called the {\it Fermi points}~\cite{walFSe}. Graphene is thus classified as a {\it semi-metal}: it is not a {\it semi-conductor} because there is no gap between its energy bands, nor is it a {\it metal} either since the bands do not overlap, so that the density of charge carriers vanishes at the Fermi points. Furthermore, the bands around the Fermi points are approximately conical~\cite{walFSe}, which explains the masslessness of the electrons in graphene, and in turn their high mobility.\par
\bigskip

\indent Graphene is also interesting for the mathematical physics community: its free energy and correlation functions, in particular its conductivity, can be computed non-perturbatively using constructive Renormalization Group (RG) techniques~\cite{giuOZ,gmpOO,gmpOT}, at least if it is at {\it half-filling}, the interaction is {\it short-range} and its strength is {\it small enough}. This is made possible, again, by the special energy structure of graphene. Indeed, since the {\it propagator} (in the quantum field theory formalism) diverges at the Fermi points, the fact that there are only two such singularities in graphene instead of a whole line of them (which is what one usually finds in two-dimensional theories), greatly simplifies the RG analysis. Furthermore, the fact that the bands are approximately conical around the Fermi points, implies that a short-range interaction between electrons is {\it irrelevant} in the RG sense, which means that one need only worry about the renormalization of the propagator, which can be controlled.\par
\indent Using these facts, the formalism developed in~\cite{benNZ} has been applied in~\cite{giuOZ, gmpOT} to express the free energy and correlation functions as convergent series.\par
\indent Let us mention that the  
case of Coulomb interactions is more difficult, in that the effective interaction is marginal in an RG sense. In this case, the theory has been constructed at all orders in renormalized perturbation theory
\cite{gmpOZ, gmpOOt}, but a non-perturbative construction is still lacking. 
\par
\bigskip

\indent In the present work, we shall extend the results of~\cite{giuOZ} by performing an RG analysis of half-filled {\it bilayer} graphene with short-range interactions. Bilayer graphene consists of two layers of graphene in so-called {\it Bernal} or {\it AB} 
stacking (see below). Before the works of A.~Geim et al.~\cite{ngeZF}, graphene was mostly studied in order to understand the properties of graphite, so it was natural to investigate the properties of multiple layers of 
graphene, starting with the bilayer~\cite{walFSe, sloFiE, mccFiSe}. A common model for hopping electrons on graphene bilayers is the so-called {\it Slonczewski-Weiss-McClure} model,
which is usually studied by retaining only certain hopping terms, depending on the energy regime one is interested in: including more hopping terms corresponds to probing the system at lower energies.
The fine structure of the Fermi surface and the behavior of the 
dispersion relation around it depends on which hoppings are considered or, equivalently, on the range of energies under inspection.

\indent In a first approximation, the energy structure of bilayer graphene is similar to that of the monolayer: there are only two Fermi points, 
and the dispersion relation is approximately conical around them. This picture is valid for energy scales larger than the transverse hopping between the two layers, referred to in the following as the {\it first regime}.
At lower energies, the effective dispersion relation around the two Fermi points appears to be approximately {\it parabolic}, instead of conical. This implies that the effective mass of the electrons in bilayer graphene does not vanish, unlike those in the monolayer, which has been observed 
experimentally~\cite{novZS}.\par
\indent From an RG point of view, the parabolicity implies that the electron interactions are {\it marginal} in bilayer graphene, thus making the RG analysis non-trivial. The flow of the effective couplings
has been studied by O.~Vafek~\cite{vafOZ, vayOZ}, who has found that it diverges logarithmically, and has identified the most divergent channels, 
thus singling out which of the possible quantum instabilities are dominant (see also~\cite{tvOT}). 
However, as was mentioned earlier, the assumption of parabolic dispersion relation is only an approximation,
valid in a range of energies between the scale of the transverse hopping and a second threshold, proportional to the cube of the transverse hopping (asymptotically, as this hopping goes to zero).
This range will be called the {\it second regime}.\par
\indent By studying the smaller energies in more detail, one finds~\cite{mccZS} that around each of the Fermi points, there are three extra Fermi points, forming a tiny equilateral triangle around the original ones. This is referred to in the literature as {\it trigonal warping}. Furthermore, around each of the now eight Fermi points, the energy bands are approximately conical. This means that, from an RG perspective, the
logarithmic divergence studied in \cite{vafOZ} is cut off at the energy scale where the conical nature of the eight Fermi points becomes observable (i.e. at the end of the second regime). At lower energies
the electron interaction is irrelevant in the RG sense, which implies that the flow of the effective interactions remains bounded at low energies. Therefore, the analysis of \cite{vafOZ} is meaningful 
only if the flow of the effective constants has grown significantly in the second regime. 
\par
\indent However, our analysis shows that the flow of the effective couplings in this regime does not grow at all, due to their smallness after integration over the first regime, 
which we quantify in terms both of the bare couplings and of the transverse hopping. This puts into question the physical relevance of the ``instabilities'' coming from the logarithmic divergence in the second regime, 
at least in the case we are treating, namely small interaction strength and small interlayer hopping.\par

\indent The transition from a normal phase to one with broken symmetry as the interaction strength is increased from small to intermediate values was studied in~\cite{ctvOT} at second order in perturbation theory. Therein, it was found that while at small bare couplings the infrared flow is convergent, at larger couplings it tends to increase, indicating a transition towards an {\it electronic nematic state}.\par

\indent Let us also mention that the third regime is not believed to give an adequate description of the system at arbitrarily small energies: at energies smaller than a third threshold (proportional to the fourth power of the transverse hopping) one finds~\cite{parZS} that the six extra Fermi points around the two original ones, are actually microscopic ellipses. The analysis of the thermodynamic properties of the system in this regime (to be called the fourth regime) requires new ideas and techniques, due to the extended nature of the singularity, and goes beyond the scope of this paper. It may be possible to adapt the ideas of \cite{benZS} to this regime, and we hope to come back to this issue in a future publication. 
\par
\bigskip

\indent To summarize, at weak coupling and small transverse hopping, we can distinguish four energy regimes: in the first, the system behaves like two uncoupled monolayers, in the second, the energy bands are approximately parabolic, in the third, the trigonal warping is taken into account and the bands are approximately conical, and in the fourth, six of the Fermi points become small curves. We shall treat the first, second and third regimes,
which corresponds to retaining only the three dominant Slonczewski-Weiss-McClure hopping parameters. 
Informally, we will prove that {\it the interacting half-filled system is analytically close to the non-interacting one} in these regimes, and that the effect of the interaction is merely to renormalize the 
hopping parameters. The proof depends on a sharp multiscale control of the crossover regions separating one regime from the next. 
\par
\bigskip

\indent We will now give a quick description of the model, and a precise statement of the main result of the present work, followed by a brief outline of its proof.\par
\subseqskip

\subsection{Definition of the model}
\label{intromodelsec}
\begin{figure}
\hfil\hskip-1.5cm\includegraphics{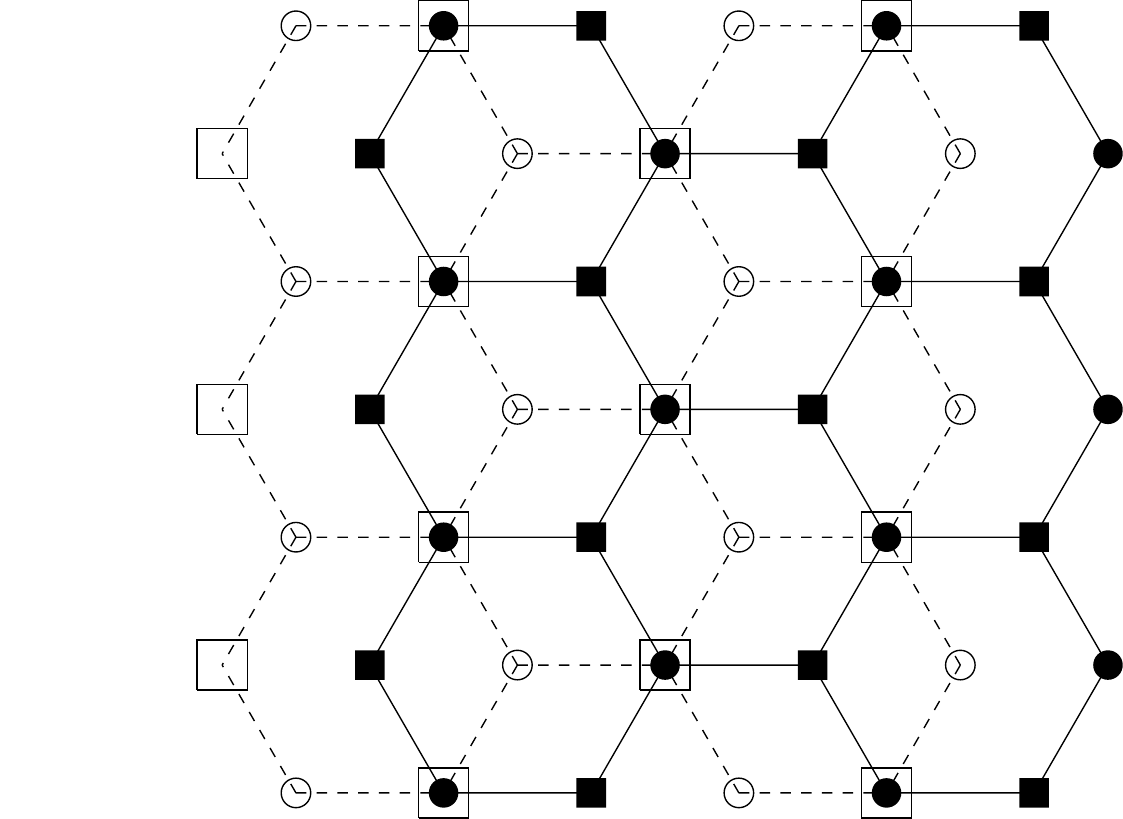}\par
\caption{\ta\ and \tb\ represent atoms of type $a$ and $b$ on the lower layer and \tat\ and \tbt\ represent atoms of type $\tilde a$ and $\tilde b$ on the upper layer. Full lines join nearest neighbors within the lower layer and dashed lines join nearest neighbors within the upper layer.}
\label{baregraph}\end{figure}

\indent We shall consider a crystal of bilayer graphene, which is made of two honeycomb lattices in {\it Bernal} or {\it AB} stacking, as shown in figure~\ref{baregraph}. We can identify four inequivalent types of sites in the lattice, which we denote by $a$, $\tilde b$, $\tilde a$ and $b$ (we choose this peculiar order for practical reasons which will become apparent in the following).\par
\bigskip

\indent We consider a Hamiltonian of the form
\begin{equation}\mathcal H=\mathcal H_0+\mathcal H_I\label{fullham}\end{equation}
where the {\it free Hamiltonian} $\mathcal H_0$ plays the role of a kinetic energy for the electrons, and the {\it interaction Hamiltonian} $\mathcal H_I$ describes the interaction between electrons.\par
\bigskip

\begin{figure}
\hfil\includegraphics{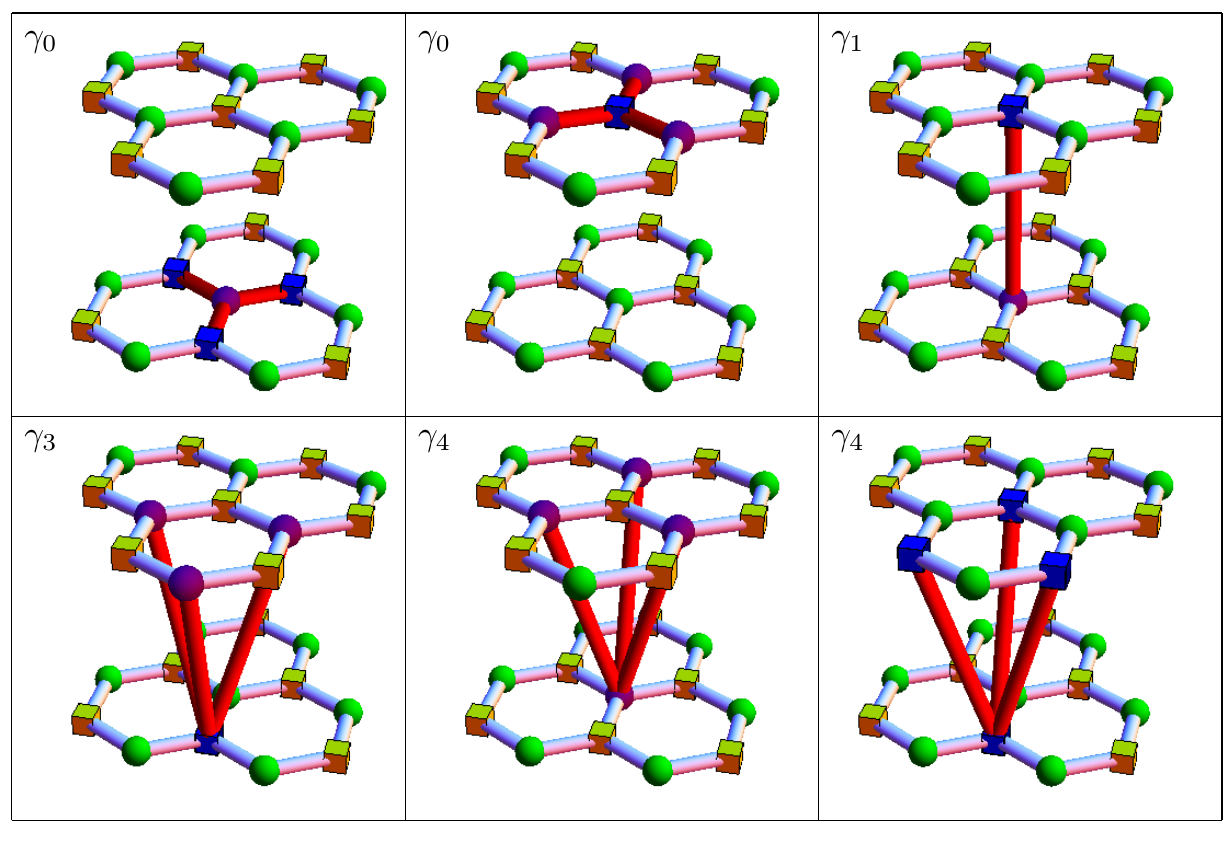}\par
\caption{The different types of hopping. From top-left to bottom-right: $a\leftrightarrow b$, $\tilde a\leftrightarrow\tilde b$, $a\leftrightarrow\tilde b$, $b\leftrightarrow\tilde a$, $a\leftrightarrow \tilde a$, $b\leftrightarrow\tilde b$. Atoms of type $a$ and $\tilde a$ are represented by spheres and those of type $b$ and $\tilde b$ by cubes; the interaction is represented by solid red (color online) cylinders; the interacting atoms are displayed either in purple or in blue.}\label{intergraph}\end{figure}

\indent $\mathcal H_0$ is given by a {\it tight-binding} approximation, which models the movement of electrons in terms of {\it hoppings} from one atom to the next. There are four inequivalent types of hoppings which we shall consider here, each of which will be associated a different {\it hopping strength} $\gamma_i$. Namely, the hoppings between neighbors of type $a$ and $b$, as well as $\tilde a$ and $\tilde b$ will be associated a hopping strength $\gamma_0$; $a$ and $\tilde b$ a strength $\gamma_1$; $\tilde a$ and $b$ a strength $\gamma_3$; $\tilde a$ and $a$, and $\tilde b$ and $b$ a strength $\gamma_4$ (see figure~\ref{intergraph}). We can thus express $H_0$ in {\it second quantized} form in {\it momentum space} at {\it zero chemical potential} as~\cite{walFSe, sloFiE, mccFiSe}
\begin{equation}
\mathcal H_0=\frac{1}{|\hat\Lambda|}\sum_{ k\in\hat \Lambda}\hat A_{ k}^\dagger H_0( k)\hat A_{ k}\label{hamkintro}\end{equation}
\begin{equation}\hat A_{ k}:=\left(\begin{array}{c}\hat a_{ k}\\\hat{\tilde b}_{ k}\\\hat{\tilde a}_{ k}\\\hat b_{ k}\end
{array}\right)\mathrm{\ and\ }H_0( k):=
-\left(\begin{array}{*{4}{c}}
\Delta&\gamma_1&\gamma_4\Omega(k)&\gamma_0\Omega^*(k)\\[0.2cm]
\gamma_1&\Delta&\gamma_0\Omega(k)&\gamma_4\Omega^*(k)\\[0.2cm]
\gamma_4\Omega^*(k)&\gamma_0\Omega^*(k)&0&\gamma_3\Omega(k)e^{3ik_x}\\[0.2cm]
\gamma_0\Omega(k)&\gamma_4\Omega(k)&\gamma_3\Omega^*(k)e^{-3ik_x}&0
\end{array}\right)\label{hmatintro}\end{equation}
in which $\hat a_k$, $\hat{\tilde b}_k$, $\hat{\tilde a}_k$ and $\hat b_k$ are {\it annihilation operators} associated to atoms of type $a$, $\tilde b$, $\tilde a$ and $b$, $k\equiv(k_x,k_y)$, $\hat\Lambda$ is the {\it first Brillouin zone}, and $\Omega(k):=1+2e^{-i\frac32k_x}\cos\left(\frac{\sqrt3}2k_y\right)$.
These objects will be properly defined in section~\ref{modelsec}. The $\Delta$ parameter in $H_0$ models a shift in the chemical potential around atoms of type $a$ and $\tilde b$~\cite{sloFiE, mccFiSe}. We choose the energy unit in such a way that $\gamma_0=1$. The hopping strengths have been measured experimentally in graphite~\cite{dreZT, toySeSe, misSeN, doeSeN} and in bilayer graphene samples~\cite{zhaZE, malZSe}; their values are given in the following table:
\begin{equation}
\begin{array}{|c|c|c|}
\cline{2-3}
\multicolumn1{c|}{}&\mathrm{bilayer\ graphene~\cite{malZSe}}&\mathrm{graphite~\cite{dreZT}}\\
\hline
\gamma_1&0.10&0.12\\
\gamma_3&0.034&0.10\\
\gamma_4&0.041&0.014\\
\Delta&0.006\ \mathrm{\cite{zhaZE}}&-0.003\\
\hline
\end{array}\label{tabgamma}\end{equation}
We notice that the relative order of magnitude of $\gamma_3$ and $\gamma_4$ is quite different in graphite and in bilayer graphene. In the latter, $\gamma_1$ is somewhat small, and $\gamma_3$ and $\gamma_4$ are of the same order, whereas $\Delta$ is of the order of $\gamma_1^2$. We will take advantage of the smallness of the hopping strengths and treat $\gamma_1=:\epsilon$ as a small parameter: we fix
\begin{equation} \frac{\gamma_1}{\epsilon}=1,\ \frac{\gamma_3}{\epsilon}=0.33,\ \frac{\gamma_4}{\epsilon}=0.40,\ \frac{\Delta}{\epsilon^2}=0.58\label{relge}\end{equation}
and assume that $\epsilon$ is as small as needed.
\par
\bigskip

{\bf Remark:} The symbols used for the hopping parameters are standard. The reason why $\gamma_2$ was omitted is that it refers to next-to-nearest layer hopping in graphite. 
In addition, for simplicity, we have neglected the intra-layer next-to-nearest neighbor hopping $\gamma_0'\approx 0.1\gamma_1$, which is known to play an analogous role to $\gamma_4$ and $\Delta$ \cite{zhaZE}.
\par\bigskip

\indent The interactions between electrons will be taken to be of {\it extended Hubbard} form, i.e. 
\begin{equation}
\mathcal H_I:=U\sum_{( x, y)}v( x- y)\left( n_{ x}-\frac{1}{2}\right)\left( n_{ y}-\frac{1}{2}\right)
\label{hamintxintro}\end{equation}
where $n_x:=\alpha_x^\dagger\alpha_x$ in which $\alpha_x$ is one of the annihilation operators $a_x$, $\tilde b_x$, $\tilde a_x$ or $b_x$; the sum over $(x,y)$ runs over all pairs of atoms in the lattice; $v$ is a short range interaction potential (exponentially decaying); $U$ is the {\it interaction strength} which we will assume to be small.\par
\bigskip

\indent We then define the {\it Gibbs average} as
$$\left<\cdot\right>:=\frac{1}{Z}\mathrm{Tr}\left( e^{-\beta\mathcal H}\cdot\right)$$
where
$$Z:=\mathrm{Tr}\left( e^{-\beta\mathcal H}\right)=:e^{-\beta|\Lambda| f}.$$
The physical quantities we will study here are the {\it free energy} $f$, and the {\it two-point Schwinger function} defined as the $4\times4$ matrix
\begin{equation}\check s_2(\mathbf x_1-\mathbf x_2):=\left(\left<\mathbf T(\alpha_{\mathbf x_1}'\alpha^\dagger_{\mathbf x_2})\right>\right)_{(\alpha',\alpha)\in\{a,\tilde b,\tilde a,b\}^2},
\label{schwindefintro}\end{equation}
where $\mathbf x_1:=(t_1,x_1)$ and $\mathbf x_2:=(t_2,x_2)$ includes an extra {\it imaginary time} component, $t_{1,2}\in[0,\beta)$, which is introduced in order to compute $Z$ and Gibbs averages,
$$\alpha_{t,x}:=e^{\mathcal H_0t}\alpha_{x}e^{-\mathcal H_0t}\quad\mathrm{for\ }\alpha\in\{a,\tilde b,\tilde a,b\}$$
and $\mathbf T$ is the {\it Fermionic time ordering operator}:
\begin{equation}
\mathbf T(\alpha'_{t_1,x_1}\alpha^\dagger_{t_2,x_2})=\left\{\begin{array}l\alpha_{t_1,x_1}'\alpha^\dagger_{t_2,x_2}\mathrm{\ if\ }t_1>t_2\\
-\alpha^\dagger_{t_2,x_2}\alpha'_{t_1,x_1}\mathrm{\ if\ }t_1\leq t_2\end{array}\right..
\label{timeorderdef}\end{equation}
We denote the Fourier transform of $\check s_2(\mathbf x)$ (or rather of its anti-periodic extension in imaginary time for $t$'s beyond $[0,\beta)$) by $s_2(\mathbf k)$ where $\mathbf k:=(k_0,k)$, and $k_0\in\frac{2\pi}{\beta}(\mathbb Z+\frac12)$.\par
\subseqskip

\subsection{Non-interacting system}
\indent In order to state our main results on the interacting two-point Schwinger function,
it is useful to first review the scaling properties of the non-interacting one,
$$s_2^{(0)}(\mathbf k)=-(ik_0\mathds1+H_0(k))^{-1},$$
including a 
discussion of the structure of its singularities in momentum space. 
\bigskip

\begin{figure}
\hfil\includegraphics{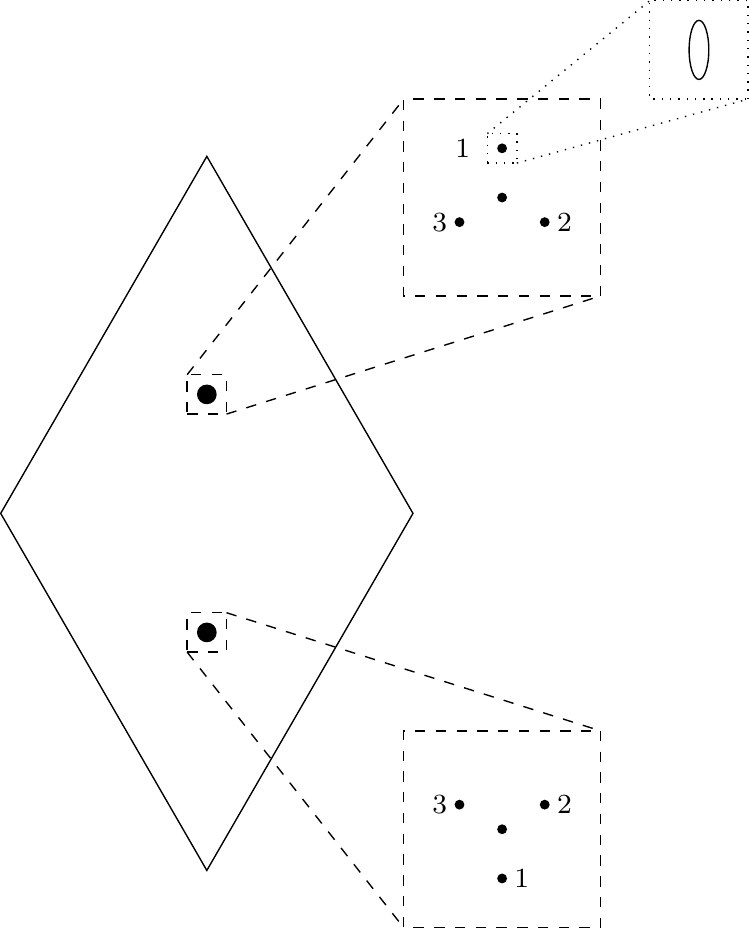}\par
\caption{Schematic representation of the Fermi points. Each dotted square represents a zoom into the finer structure of the Fermi points. The secondary Fermi points are labeled as indicated in the figure. In order not to clutter the drawing, only one of the zooms around the secondary Fermi point was represented.
}\label{figferm}\end{figure}

\point{Non-interacting Fermi surface} If $H_0(k)$ is not invertible, then $s_2^{(0)}(0,k)$ is divergent. The set of quasi-momenta $\mathcal F_0:=\{k,\ \det H_0(k)=0\}$ is called the non-interacting {\it Fermi surface}
at zero chemical potential, which has the following structure: it contains two isolated points located at
\begin{equation}p_{F,0}^{\omega}:=\left(\frac{2\pi}{3},\omega\frac{2\pi}{3\sqrt3}\right),\ \omega\in\{-1,+1\}\label{fermdefz}\end{equation}
around each of which there are three very small curves that are approximately elliptic (see figure~\ref{figferm}). The whole singular set $\mathcal F_0$ is contained within 
two small circles (of radius $O(\epsilon^2)$), so that on scales larger than $\epsilon^2$, $\mathcal F_0$ can be approximated by just two points, $\{p_{F,0}^\pm\}$, see figure~\ref{figferm}. As we zoom in, looking at smaller 
scales, we realize that each small circle contains four Fermi points: the central one, and three secondary points around it, called $\{p_{F,j}^\pm,\ j\in\{1,2,3\}\}$. 
A finer zoom around the secondary points reveals that they are actually curves of size $\epsilon^3$. \par
\bigskip

\par

\point{Non-interacting Schwinger function} We will now make the statements about approximating the Fermi surface more precise, and discuss the behavior of $s_2^{(0)}$ around its singularities. We will identify four regimes in which the Schwinger function behaves differently.\par
\bigskip

\begin{figure}
\hfil\includegraphics{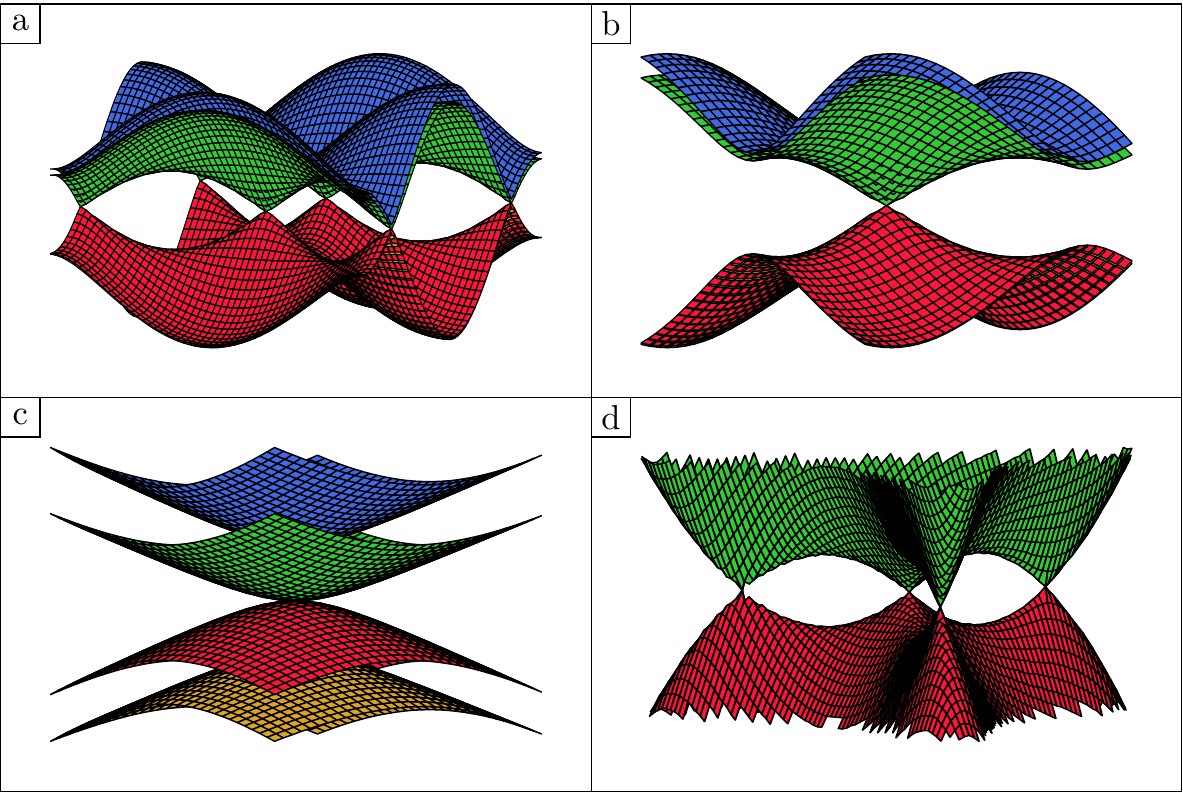}
\caption{Eigenvalues of $H_0(k)$. The sub-figures b,c,d are finer and finer zooms around one of the Fermi points.}
\label{figbands}
\end{figure}

\subpoint{First regime} One can show that, if $\mathbf p_{F,0}^\pm:=(0,p_{F,0}^\pm)$, and
$$\|(k_0,k_x',k_y') \|_{\mathrm I}:=\sqrt{k_0^2+(k_x')^2+(k_y')^2}$$
then 
\begin{equation}
s_2^{(0)}(\mathbf p_{F,0}^\pm+\mathbf k' )=\left(\mathfrak L_{\mathrm{I}}\hat A(\mathbf p_{F,0}^\pm+\mathbf k' )\right)^{-1}\left(\mathds1+O(\|\mathbf k' \|_{\mathrm{I}},\epsilon\|\mathbf k' \|_{\mathrm{I}}^{-1})\right)
\label{freeschwino}\end{equation}
in which 
$\mathfrak L_{\mathrm{I}}\hat A$ is a matrix, independent of $\gamma_1$, $\gamma_3$, $\gamma_4$ and $\Delta$, whose eigenvalues vanish {\it linearly} around $\mathbf p_{F,0}^\pm$ (see figure~\ref{figbands}b). We thus identify a {\it first regime}:
$$\epsilon\ll\|\mathbf k' \|_{\mathrm{I}}\ll1$$
in which the error term in~(\ref{freeschwino}) is {\it small}. In this first regime, $\gamma_1$, $\gamma_3$, $\gamma_4$ and $\Delta$ are negligible, and the Fermi surface is approximated by $\{p_{F,0}^\pm\}$, around which the Schwinger function diverges {\it linearly}.\par
\bigskip

\subpoint{Second regime} Now, if
$$\|(k_0,k_x',k_y') \|_{\mathrm{II}}:=\sqrt{k_0^2+\frac{(k_x')^4}{\gamma_1^2}+\frac{(k_y')^4}{\gamma_1^2}}$$
then
\begin{equation}
s_2^{(0)}(\mathbf p_{F,0}^\pm+\mathbf k' )=\left(\mathfrak L_{\mathrm{II}}\hat A(\mathbf p_{F,0}^\pm+\mathbf k' )\right)^{-1}\left(\mathds1+O(\epsilon^{-1}\|\mathbf k' \|_{\mathrm{II}},\epsilon^{3/2}\|\mathbf k' \|_{\mathrm{II}}^{-1/2})\right)
\label{freeschwint}\end{equation}
in which $\mathfrak L_{\mathrm{II}}\hat A$ is a matrix, independent of $\gamma_3$, $\gamma_4$ and $\Delta$. Two of its eigenvalues vanish {\it quadratically} around $\mathbf p_{F,0}^\pm$ (see figure~\ref{figbands}c) and two are bounded away from 0. The latter correspond to {\it massive} modes, whereas the former to {\it massless} modes. We thus identify a {\it second regime}:
$$\epsilon^3\ll\|\mathbf k' \|_{\mathrm{II}}\ll\epsilon$$
in which $\gamma_3$, $\gamma_4$ and $\Delta$ are negligible, and the Fermi surface is approximated by $\{p_{F,0}^\pm\}$, around which the Schwinger function diverges {\it quadratically}.\par
\bigskip

\subpoint{Third regime} If $\mathbf p_{F,j}^\pm:=(0,p_{F,j}^\pm)$, $j=0,1,2,3$, and 
$$\|(k_0,k'_{j,x},k'_{j,y})\|_{\mathrm{III}}:=\sqrt{k_0^2+\gamma_3^2(k'_{j,x})^2+\gamma_3^2(k'_{j,y})^2}$$
then
\begin{equation}
s_2^{(0)}(\mathbf p_{F,j}^\pm+\mathbf k'_{j})=\left(\mathfrak L_{\mathrm{III},j}\hat A(\mathbf p_{F,j}^\pm+\mathbf k'_{j})\right)^{-1}\left(\mathds1+O(\epsilon^{-3}\|\mathbf k'_{j}\|_{\mathrm{III}},\epsilon^{4}\|\mathbf k'_{j}\|_{\mathrm{III}}^{-1})\right)
\label{freeschwinth}\end{equation}
in which $\mathfrak L_{\mathrm{III},j}\hat A$ is a matrix, independent of $\gamma_4$ and $\Delta$, two of whose eigenvalues vanish {\it linearly} around $\mathbf p_{F,j}^\pm:=(0,p_{F,j}^\pm)$ (see figure~\ref{figbands}d) and two are bounded away from 0. We thus identify a {\it third regime}:
$$\epsilon^4\ll\|\mathbf k'_{j}\|_{\mathrm{III}}\ll\epsilon^3$$
in which $\gamma_4$ and $\Delta$ are negligible, and the Fermi surface is approximated by $\{p_{F,j}^\pm\}_{j\in\{0,1,2,3\}}$, around which the Schwinger function diverges {\it linearly}.
\par
\bigskip

{\bf Remark:} If $\gamma_4=\Delta0$, then the error term $O(\epsilon^{4}\|\mathbf k'_{j}\|_{\mathrm{III}}^{-1})$ in \eqref{freeschwinth} vanishes identically, which allows us to extend the third regime to all momenta satisfying 
$$\|\mathbf k'_{j}\|_{\mathrm{III}}\ll\epsilon^3.$$

\subseqskip

\subsection{Main Theorem}\label{maintheosec}
\indent We now state the Main Theorem, whose proof will occupy the rest of the paper. Roughly, our result is that as long as $|U|$ and $\epsilon$ are small enough and $\gamma_4=\Delta=0$ (see the remarks following the statement for an explanation of why this is assumed), the free energy and the two-point Schwinger function are well defined in the thermodynamic and zero-temperature limit $|\Lambda|,\beta\to\infty$, and that the two-point Schwinger function is analytically close to that with $U=0$. The effect of the interaction is shown to merely {\it renormalize} the constants of the non-interacting Schwinger function.\par
\bigskip

\indent We define
$$
\mathcal B_\infty:=\mathbb{R}\times\left(\mathbb{R}^2/(\mathbb ZG_1+\mathbb ZG_2)\right),\quad
G_1:=\left(\frac{2\pi}{3},\frac{2\pi}{\sqrt{3}}\right),\mathrm\  G_2:=\left(\frac{2\pi}{3},-\frac{2\pi}{\sqrt{3}}\right),
$$
where the physical meaning of  $\mathbb{R}^2/(\mathbb ZG_1+\mathbb ZG_2)$ is that of the {\it first Brillouin zone}, and 
$G_{1,2}$ are the generators of the dual lattice. \bigskip

\Theos{Main Theorem}
If $\gamma_4=\Delta=0$, then there exists $U_0>0$ and $\epsilon_0>0$ such that for all $|U|<U_0$ and $\epsilon<\epsilon_0$, the specific ground state energy 
$$e_0:=-\lim_{\beta\to\infty}\lim_{|\Lambda|\to\infty}\frac{1}{\beta|\Lambda|}\log(\mathrm{Tr}(e^{-\beta\mathcal H}))$$
exists and is analytic in $U$. In addition, there exist eight Fermi points $\{\tilde{\mathbf p}_{F,j}^\omega\}_{\omega=\pm,j=0,1,2,3}$ such that:
\begin{equation} \tilde{\mathbf p}_{F,0}^\omega={\mathbf p}_{F,0}^\omega, \qquad |\tilde{\mathbf p}_{F,j}^\omega-{\mathbf p}_{F,j}^\omega|\leqslant (\mathrm{const}.)\  |U|\epsilon^2, \ j=1,2,3,\end{equation}
and, $\forall\mathbf k\in\mathcal B_\infty\setminus\{\tilde{\mathbf p}_{F,j}^\omega\}_{\omega=\pm,j=0,1,2,3}$, the thermodynamic and zero-temperature limit of the two-point Schwinger function, $\lim_{\beta\to\infty}\lim_{|\Lambda|\to\infty}s_2(\mathbf k)$, exists and is analytic in $U$.
\endtheo
\bigskip

{\bf Remarks}:
\begin{itemizepp}
\item The theorem requires $\gamma_4=\Delta=0$. As we saw above, those quantities play a negligible role in the non-interacting theory as long as we do not move beyond the third regime. This suggests that the theorem should hold with $\gamma_4,\Delta\neq0$ under the condition that $\beta$ is not too large,
i.e., smaller than $(\mathrm{const}.)\ \epsilon^{-4}$. However, that case presents a number of extra technical complications, which we will spare the reader.
\item The conditions that $|U|<U_0$ and $\epsilon<\epsilon_0$ are independent, in that we do not require any condition on the relative values of $|U|$ and $\epsilon$. Such a result calls for tight bounds on the integration over the first regime. If we were to assume that $|U|\ll\epsilon$, then the discussion would be greatly simplified, but such a condition would be artificial, and we will not require it be satisfied. L.~Lu~\cite{luOTh} sketched 
the proof of a result similar to our Main Theorem, without discussing the first two regimes, which requires such an artificial condition on $U/\epsilon$. 
The renormalization of the secondary Fermi points is also ignored in that reference.
\end{itemizepp}
\bigskip

\indent In addition to the Main Theorem, we will prove that the dominating part of the two point Schwinger function is qualitatively the same as the non-interacting one, with renormalized constants. This result is detailed in Theorems~\ref{theoo}, \ref{theot} and~\ref{theoth} below, each of which refers to one of the three regimes.\par
\bigskip

\point{First regime} Theorem~\ref{theoo} states that in the first regime, the two-point Schwinger function behaves at dominant order like the non-interacting one with renormalized factors.\par
\bigskip

\Theo{Theorem}\label{theoo}
Under the assumptions of the Main Theorem, if $C\epsilon\leqslant \|\mathbf k-\mathbf p_{F,0}^\omega \|_{\mathrm I}\leqslant C^{-1}$ for a suitable $C>0$, then, in the thermodynamic and zero-temperature limit,
\begin{equation}
s_2(\mathbf k)=-\frac{1}{\tilde k_0\bar k_0+|\bar\xi|^2}\left(\begin{array}{*{4}{c}}
-i\bar k_0&0&0&\bar\xi^*\\
0&-i\bar k_0&\bar\xi&0\\
0&\bar\xi^*&-i\tilde k_0&0\\
\bar\xi&0&0&-i\tilde k_0\end{array}\right)
(\mathds1+r(\mathbf k))
\label{schwino}\end{equation}
where 
\begin{equation} r(\mathbf p^\omega_{F,0}+\mathbf k')=O\big(\big(1+|U||\log\|\mathbf k' \|_{\mathrm{I}}|\big)\|\mathbf k' \|_{\mathrm{I}},\epsilon\|\mathbf k' \|_{\mathrm{I}}\big),\end{equation}
and, for $(k_0,k_{x}',k_{y}'):=\mathbf k-\mathbf p^\omega_{F,0}$,
\begin{equation}
\bar k_0:=z_1 k_0,\quad
\tilde k_0:=\tilde z_1 k_0,\quad
\bar \xi:=\frac{3}{2}v_1(ik'_x+\omega k'_y)
\label{rcco}\end{equation}
in which $(\tilde z_1,z_1,v_1)\in\mathbb{R}^3$ satisfy
\begin{equation}
|1-\tilde z_1|\leqslant C_1|U|,\quad
|1-z_1|\leqslant C_1|U|,\quad
|1-v_1|\leqslant C_1|U|
\label{ineqrcco}\end{equation}
for some constant $C_1>0$ (independent of $U$ and $\epsilon$).
\endtheo
\bigskip

{\bf Remarks}:
\begin{itemizepp}
\item The singularities of $s_2$ are approached linearly in this regime.
\item By comparing (\ref{schwino}) with its non-interacting counterpart~(\ref{freepropzo}), we see that the effect of the interaction is to {\it renormalize} the constants in front of $k_0$ and $\xi$ in~(\ref{freepropzo}).
\item The {\it inter-layer correlations}, that is the $\{a,b\}\times\{\tilde a,\tilde b\}$ components of the dominating part of $s_2(\mathbf k)$ vanish. In this regime, the Schwinger function of bilayer graphene behave like that of two independent graphene layers.
\end{itemizepp}
\bigskip

\point{Second regime} Theorem~\ref{theot} states a similar result for the second regime. As was mentioned earlier, two of the components are {\it massive} in the second (and third) regime, and we first perform a change of variables to isolate them, and state the result on the massive and massless components, which are denoted below by $\bar s_2^{(M)}$ and $\bar s_2^{(m)}$ respectively.\par
\bigskip

\Theo{Theorem}
\label{theot}
Under the assumptions of the Main Theorem, if $C\epsilon^3\leqslant \|\mathbf k-\mathbf p_{F,0}^\omega \|_{\mathrm{II}}\leqslant C^{-1}\epsilon$ for a suitable $C>0$, then, in the thermodynamic and zero-temperature limit,
\begin{equation}
s_2(\mathbf k)
=\left(\begin{array}{*{2}{c}}\mathds1&M(\mathbf k)^\dagger\\0&\mathds1\end{array}\right)\left(\begin{array}{*{2}{c}}\bar s_2^{(M)}&0\\
0&\bar s_2^{(m)}(\mathbf k)
\end{array}\right)\left(\begin{array}{*{2}{c}}\mathds1&0\\ M(\mathbf k)&\mathds1\end{array}\right)
(\mathds1+r(\mathbf k))
\label{schwint}\end{equation}
where:
\begin{equation}
r(\mathbf p_{F,0}^\omega+\mathbf k')=
O(\epsilon^{-1/2}\|\mathbf k' \|_{\mathrm{II}}^{1/2},
\epsilon^{3/2}\|\mathbf k' \|_{\mathrm{II}}^{-1/2},
|U|\epsilon\, |\log\epsilon|),
\label{errort}\end{equation}
\begin{equation}
\bar s_2^{(m)}(\mathbf k)
=\frac{1}{\bar\gamma_1^2\bar k_0^2+|\bar\xi|^4}
\left(\begin{array}{*{2}{c}}
i\bar\gamma_1^2\bar k_0&
\bar\gamma_1(\bar\xi^*)^2\\
\bar\gamma_1\bar\xi^2&
i\bar\gamma_1^2\bar k_0
\end{array}\right),\quad
\bar s_2^{(M)}=-\left(\begin{array}{*{2}{c}}0&\bar\gamma_1^{-1}\\\bar\gamma_1^{-1}&0\end{array}\right),
\label{schwintcomps}\end{equation}
\begin{equation}
M(\mathbf k):=-\frac1{\bar\gamma_1}\left(\begin{array}{*{2}{c}}\bar\xi^*&0\\0&\bar\xi\end{array}\right)
\label{theoMt}\end{equation}
and, for $(k_0,k_{x}',k_{y}'):=\mathbf k-\mathbf p^\omega_{F,0}$,
\begin{equation}
\bar\gamma_1:=\tilde m_2 \gamma_1,\quad
\bar k_0:=z_2 k_0,\quad
\bar \xi:=\frac{3}{2}v_2(ik'_x+\omega k'_y)
\label{rcct}\end{equation}
in which $(\tilde m_2,z_2,v_2)\in\mathbb{R}^3$ satisfy
\begin{equation}
|1-\tilde m_2|\leqslant C_2|U|,\quad
|1-z_2|\leqslant C_2|U|,\quad
|1-v_2|\leqslant C_2|U|
\label{ineqrcct}\end{equation}
for some constant $C_2>0$ (independent of $U$ and $\epsilon$).
\endtheo
\bigskip

{\bf Remarks}:
\begin{itemizepp}
\item The {\it massless} components $\{\tilde a,b\}$ are left invariant under the change of basis that block-diagonalizes $s_2$. Furthermore, $M$ is {\it small} in the second regime, which implies that the {\it massive} components are {\it approximately} $\{a,\tilde b\}$.
\item As can be seen from~(\ref{schwintcomps}), the {\it massive} part $\bar s_2^{(M)}$ of $s_2$ is not singular in the neighborhood of the Fermi points, whereas the {\it massless} one, i.e. $\bar s_2^{(m)}$, is.
\item The massless components of $s_2$ approach the singularity {\it quadratically} in the spatial components of $\mathbf k$.
\item Similarly to the first regime, by comparing~(\ref{schwintcomps}) with~(\ref{freepropzt}), we find that the effect of the interaction is to {\it renormalize} constant factors.
\end{itemizepp}
\bigskip

\point{Third regime} Theorem~\ref{theoth} states a similar result as Theorem~\ref{theot} for the third regime, though the discussion is made more involved by the presence of the extra Fermi points.\par
\bigskip

\Theo{Theorem}
\label{theoth}
For $j=0,1$, under the assumptions of the Main Theorem, if $\|\mathbf k-\tilde{\mathbf p}_{F,j}^\omega\|_{\mathrm{III}}\leqslant C^{-1}\epsilon^3$ for a suitable $C>0$, then
\begin{equation}
s_2(\mathbf k)
=\left(\begin{array}{*{2}{c}}\mathds1&M(\mathbf k)^\dagger\\0&\mathds1\end{array}\right)\left(\begin{array}{*{2}{c}}\bar s_2^{(M)}&0\\
0&\bar s_2^{(m)}(\mathbf k)
\end{array}\right)\left(\begin{array}{*{2}{c}}\mathds1&0\\ M(\mathbf k)&\mathds1\end{array}\right)
(\mathds1+r(\mathbf k))
\label{schwinth}\end{equation}
where
\begin{equation}
r(\tilde{\mathbf p}_{F,j}^\omega+\mathbf k'_{j})=O(
\epsilon^{-3}\|\mathbf k'_{j}\|_{\mathrm{III}}(1+\epsilon|\log\|\mathbf k'_{j}\|_{\mathrm{III}}||U|),
\epsilon(1+|\log\epsilon||U|)
)
\label{errorth}\end{equation}
\begin{equation}
\bar s_2^{(m)}(\mathbf k)
=\frac{1}{\bar k_{0,j}^2+\gamma_3^2|\bar x_j|^2}
\left(\begin{array}{*{2}{c}}
i\bar k_{0,j}&
\gamma_3\bar x_j^*\\
\gamma_3\bar x_j&
i\bar k_{0,j}
\end{array}\right),\quad
\bar s_2^{(M)}=-\left(\begin{array}{*{2}{c}}0&\bar\gamma_{1,j}^{-1}\\\bar\gamma_{1,j}^{-1}&0\end{array}\right),
\label{schwinthcomps}\end{equation}
\begin{equation}
M(\mathbf k):=-\frac1{\bar\gamma_{1,j}}\left(\begin{array}{*{2}{c}}\bar\Xi_j^*&0\\0&\bar\Xi_j\end{array}\right)
\label{theoMth}\end{equation}
and, for $(k_0,k_{x}',k_{y}'):=\mathbf k-\mathbf p^\omega_{F,j}$,
\begin{equation}\begin{array}c
\bar k_{0,j}:=z_{3,j}k_0,\quad
\bar\gamma_{1,j}:=\tilde m_{3,j}\gamma_1,\quad
\bar x_0:=\tilde v_{3,0}\frac{3}{2}(ik'_x-\omega k'_y)=:-\bar\Xi_0^*\\[0.5cm]
\bar x_{1}:=\frac{3}{2}\left(3\tilde v_{3,1}ik'_x+\tilde w_{3,1}\omega k'_x\right),\quad
\bar\Xi_{1}:=m_{3,1}\gamma_1\gamma_3+\bar v_{3,1}ik'_x+\bar w_{3,1}k'_y
\end{array}\label{rccthj}\end{equation}
in which $(\tilde m_{3,j},m_{3,j},z_{3,j},\bar v_{3,j},\tilde v_{3,j},\bar w_{3,j},\tilde w_{3,j})\in\mathbb{R}^7$ satisfy
\begin{equation}\begin{array}c
|m_{3,j}-1|+
|\tilde m_{3,j}-1|
\leqslant C_{3}|U|,\quad
|z_{3,j}-1|\leqslant C_{3}|U|,\\[0.5cm]
|\bar v_{3,j}-1|+
|\tilde v_{3,j}-1|\leqslant C_{3}|U|,\quad
|\bar w_{3,j}-1|+
|\tilde w_{3,j}-1|\leqslant C_{3}|U|
\end{array}\label{ineqrccthj}\end{equation}
for some constant $C_{3}>0$ (independent of $U$ and $\epsilon$).
\endtheo
\bigskip

Theorem~\ref{theoth} can be extended to the neighborhoods of $\tilde{\mathbf p}_{F,j}^\omega$ with $j=2,3$, 
by taking advantage of the symmetry of the system under rotations of angle $2\pi/3$:\par
\medskip

\Theos{Extension to $j=2,3$}
For $j=2,3$, under the assumptions of the Main Theorem, if $\|\mathbf k-\tilde{\mathbf p}_{F,j}^\omega\|_{\mathrm{III}}\leqslant C^{-1}\epsilon^3$ for a suitable $C>0$, then
\begin{equation}
s_2(\mathbf k'_{j}+\tilde{\mathbf p}_{F,j}^\omega)=
\left(\begin{array}{*{2}{c}}\mathds1&0\\0&\mathcal T_{T\mathbf k'_{j}+\tilde{\mathbf p}_{F,j-\omega}^\omega}\end{array}\right) s_2(T\mathbf k'_{j}+\tilde{\mathbf p}_{F,j-\omega}^{\omega})\left(\begin{array}{*{2}{c}}\mathds1&0\\0&\mathcal T^\dagger_{T\mathbf k'_{j}+\tilde{\mathbf p}_{F,j-\omega}^\omega}\end{array}\right)
\label{schwinthj}\end{equation}
where $T(k_0,k_x,k_y)$ denotes the rotation of the $k_x$ and $k_y$ components by an angle $2\pi/3$, $\mathcal T_{(k_0,k_x,k_y)}:=e^{-i(\frac32k_x-\frac{\sqrt3}2k_y)\sigma_3}$, and $\tilde{\mathbf p}_{F,4}^{-}\equiv \tilde{\mathbf p}_{F,1}^{-}$.
\endtheo
\bigskip

{\bf Remarks}:
\begin{itemizepp}
\item The remarks below Theorem~\ref{theot} regarding the massive and massless fields hold here as well.
\item The massless components of $s_2$ approach the singularities {\it linearly}.
\item By comparing~(\ref{schwinth}) with (\ref{freepropzth}) and~(\ref{freepropoth}), we find that the effect of the interaction is to {\it renormalize} the constant factors.
\end{itemizepp}
\subseqskip

\subsection{Sketch of the proof}
\indent In this section, we give a short account of the main ideas behind the proof of the Main Theorem.\par
\bigskip

\point{Multiscale decomposition} The proof relies on a {\it multiscale} analysis of the model, in which the free energy and Schwinger function are expressed as successive integrations over individual scales. 
Each scale is defined as a set of $\mathbf k$'s contained inside an annulus at a distance of $2^h$ for $h\in\mathbb Z$ around the singularities located at $\mathbf p_{F,j}^\omega$.
The positive scales correspond to the ultraviolet regime, which we analyze in a multiscale fashion because of the (very mild) singularity of the free propagator at equal imaginary times. 
It may be possible to avoid the decomposition by employing ideas in the spirit of \cite{psZE}. The negative scales are treated differently, depending on the regimes they belong to (see below),
and they contain the essential difficulties of the problem, whose nature is intrinsically infrared. 
\par
\bigskip

\point{First regime} In the first regime, i.e. for $-1\gg h\gg h_\epsilon:=\log_2\epsilon$, the system behaves like two uncoupled graphene layers, so the analysis carried out in~\cite{giuOZ} holds. From a renormalization group perspective, this regime is {\it super-renormalizable}: the scaling dimension of diagrams with $2l$ external legs is $3-2l$, so that only the two-legged diagrams are relevant whereas all of the others are irrelevant (see section~\ref{powercountingsec} for precise definitions of scaling dimensions, relevance and irrelevance). This allows us to compute a strong bound on four-legged contributions:
$$|\hat W_4^{(h)}(\mathbf k)|\leqslant (\mathrm{const.})\ |U|2^{2h}$$
whereas a naive power counting argument would give $|U|2^{h}$ (recall that with our conventions $h$ is negative).\par
\bigskip

\indent The super-renormalizability in the first regime stems from the fact that the Fermi surface is 0-dimensional and that $H_0$ is linear around the Fermi points. While performing the multiscale integration, we deal with the two-legged terms by incorporating them into $H_0$, and one must therefore prove that by doing so, the Fermi surface remains 0-dimensional and that the singularity remains linear. This is guaranteed by a symmetry argument, which in particular shows the invariance of the Fermi surface.\par
\bigskip

\point{Second regime} In the second regime, i.e. for $3h_\epsilon\ll h\ll h_\epsilon$, the singularities of $H_0$ are quadratic around the Fermi points, which changes the {\it power counting} of the renormalization group analysis: the scaling dimension of $2l$-legged diagrams becomes $2-l$ so that the two-legged diagrams are still relevant, but the four-legged ones become marginal. One can then check~\cite{vafOZ} that they are actually marginally relevant, which means that their contribution increases proportionally to $|h|$. This turns out not to matter: since the second regime is only valid for $h\gg 3h_\epsilon$, $|\hat W_4^{(h)}|$ may only increase by $3|h_\epsilon|$, and since the theory is super-renormalizable in the first regime, there is an extra factor $2^{h_\epsilon}$ in $\hat W_4^{(h_\epsilon)}$, so that $\hat W_4^{(h)}$ actually increases from $2^{h_\epsilon}$ to $3|h_\epsilon|2^{h_\epsilon}$, that is to say it barely increases at all if $\epsilon$ is small enough.\par
\bigskip

\indent Once this essential fact has been taken into account, the renormalization group analysis can be carried out without major difficulties. As in the first regime, the invariance of the Fermi surface is guaranteed by a symmetry argument.\par
\bigskip

\point{Third regime} In the third regime, i.e. for $h\ll 3h_\epsilon$, the theory is again super-renormalizable (the scaling dimension is $3-2l$). There is however an extra difficulty with respect to the first regime, in that the Fermi surface is no longer invariant under the renormalization group flow, but one can show that it does remain 0-dimensional, and that the only effect of the multiscale integration is to move $p_{F,j}^\omega$ along the line between itself and $p_{F,0}^\omega$.\par
\subseqskip

\subsection{Outline}
\indent The rest of this paper is devoted to the proof of the Main Theorem and of Theorems~\ref{theoo}, \ref{theot} and~\ref{theoth}. The sections are organized as follows.

\begin{itemize}
\item In section~\ref{themodelsec}, we define the model in a more explicit way than what has been done so far; then we show how to compute the free energy and Schwinger function using a Fermionic path integral formulation and a {\it determinant expansion}, due to Battle, Brydges and Federbush~\cite{brySeE,batEF}, see also \cite{bkESe, arNE}; and finally we discuss the symmetries of the system.
\item In section~\ref{proppropsec}, we discuss the non-interacting system. In particular, we derive detailed formulae for the Fermi points and for the asymptotic behavior of the propagator around its singularities. 
\item In section~\ref{schemesec}, we describe the multiscale decomposition used to compute the free energy and Schwinger function.
\item In section~\ref{treeexpsec}, we state and prove a {\it power counting} lemma, which will allow us to compute bounds for the effective potential in each regime. The lemma is based on the Gallavotti-Nicol\`o tree expansion~\cite{galEFi}, and follows~\cite{benNZ, genZO, giuOZh}. We conclude this section by showing how to compute the two-point Schwinger function from the effective potentials.
\item In section~\ref{uvsec}, we discuss the integration over the {\it ultraviolet regime}, i.e. scales $h>0$.
\item In sections~\ref{osec}, \ref{tsec} and~\ref{thsec}, we discuss the multiscale integration in the first, second and third regimes,
and complete the proofs of the Main Theorem, as well as of Theorems \ref{theoo}, \ref{theot}, \ref{theoth}.
\end{itemize}
\seqskip

\section{The model}
\label{themodelsec}
\hfil\framebox{\bf From this point on, we set $\gamma_4=\Delta0$.}
\bigskip

\indent In this section, we define the model in precise terms, re-express the free energy and two-point Schwinger function in terms of Grassmann integrals and truncated expectations, which we will subsequently explain how to compute, and discuss the symmetries of the model and their representation in this formalism.\par
\subseqskip

\subsection{Precise definition of the model}
\label{modelsec}
\indent In the following, some of the formulae are repetitions of earlier ones, which are recalled for ease of reference. This section complements section~\ref{intromodelsec},
where the same definitions were anticipated in a less verbose form. The main novelty lies in the momentum-real space correspondence, which is 
made explicit.
\par
\bigskip

\begin{figure}
\hfil\hskip-1.5cm\includegraphics{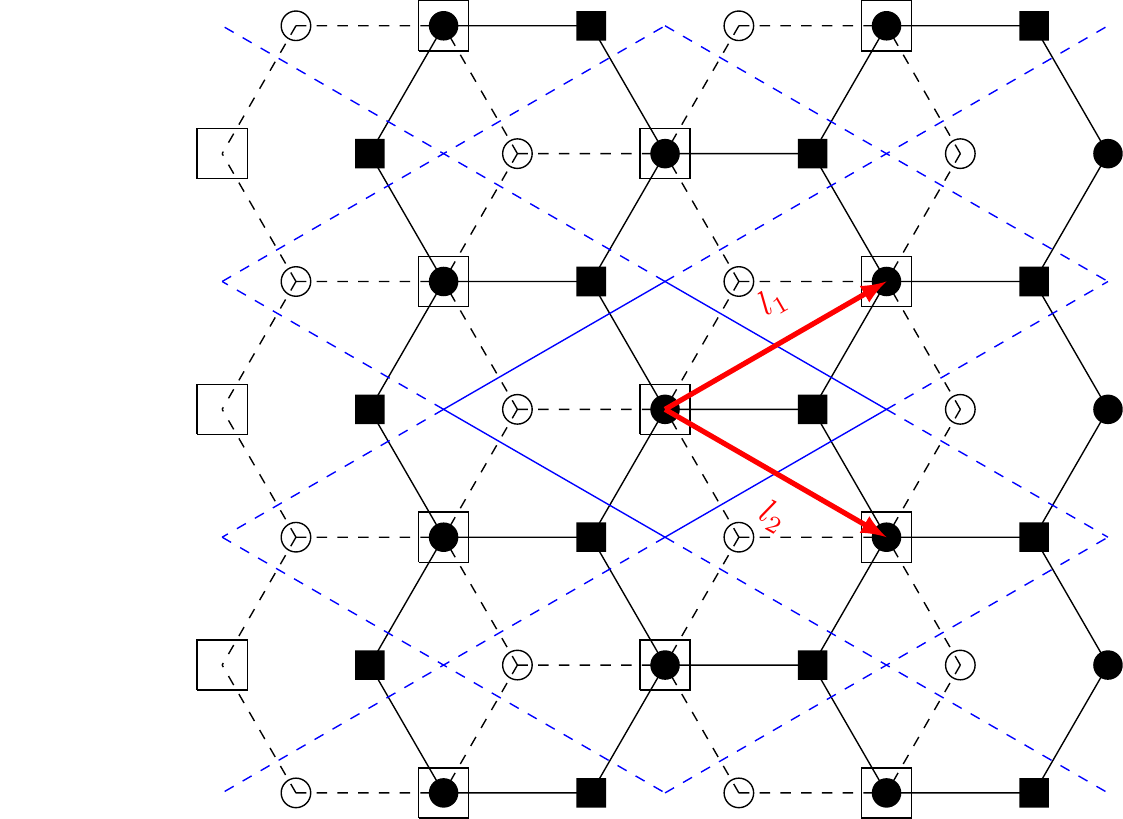}\par
\label{cellgraph}
\caption{decomposition of the crystal into elementary cells, represented by the blue (color online) rhombi. There are four atoms in each elementary cell: \ta of type $a$ at $(0,0,0)$, \tbt of type $\tilde b$ at $(0,0,c)$, \tat of type $\tilde a$ at $(-1,0,c)$ and \tb of type $b$ at $(0,0,c)$.}
\end{figure}

\point{Lattice} As mentioned in section~\ref{introduction}, the atomic structure of bilayer graphene consists in two honeycomb lattices in so-called {\it Bernal} or {\it AB} stacking, as was shown in figure~\ref{baregraph}. It can be constructed by copying an elementary cell at every integer combination of
\begin{equation}
 l_1:=\left(\frac{3}{2},\frac{\sqrt{3}}{2},0\right),\mathrm{\ } l_2:=\left(\frac{3}{2},-\frac{\sqrt{3}}{2},0\right)
\label{laeo}\end{equation}
where we have chosen the unit length to be equal to the distance between two nearest neighbors in a layer (see figure~\ref{cellgraph}). The elementary cell consists of four atoms at the following coordinates
$$(0,0,0);\mathrm{\ }(0,0,c);\mathrm{\ }(-1,0,c);\mathrm{\ }(1,0,0)$$
given relatively to the center of the cell. $c$ is the spacing between layers; it can be measured experimentally, and has a value of approximately 2.4~\cite{triNT}.\par
\bigskip

\par\medskip

\indent We define the lattice 
\begin{equation}\Lambda:=\left\{n_1 l_1+n_2 l_2,\ (n_1,n_2)\in\{0,\cdots,L-1\}^2\right\}\label{laet}\end{equation}
where $L$ is a positive integer that determines the size of the crystal, that we will eventually send to infinity, with periodic boundary conditions. We introduce the intra-layer  nearest neighbor vectors:
\begin{equation}
\delta_1:=(1,0,0),\quad\delta_2:=\left(-\frac{1}{2},\frac{\sqrt{3}}{2},0\right),\quad\delta_3:=\left(-\frac{1}{2},-\frac{\sqrt{3}}{2},0\right).
\label{lea}\end{equation}\par
\bigskip

\indent The {\it dual} of $\Lambda$ is
\begin{equation}\hat\Lambda:=\left\{\frac{m_1}{L} G_1+\frac{m_2}{L} G_2,\ (m_1,m_2)\in\{0,\cdots,L-1\}^2\right\}\label{lae}\end{equation}
with periodic boundary conditions, where
\begin{equation}
G_1=\left(\frac{2\pi}{3},\frac{2\pi}{\sqrt{3}},0\right),\mathrm\  G_2=\left(\frac{2\pi}{3},-\frac{2\pi}{\sqrt{3}},0\right).
\label{laeG}\end{equation}
It is defined in such a way that $\forall x\in\Lambda$, $\forall k\in\hat\Lambda$, 
$$e^{ikxL}=1.$$
Since the third component of vectors in $\hat\Lambda$ is always 0, we shall drop it and write vectors of $\hat\Lambda$ as elements of $\mathbb{R}^2$. In the limit $L\to\infty$, the 
set $\hat \Lambda$ tends to the torus $\hat \Lambda_\infty=\mathbb R^2/(\mathbb Z G_1+\mathbb Z G_2)$, also called the {\it Brillouin zone}. 
\par
\subseqskip

\point{Hamiltonian} Given $ x\in\Lambda$, we denote the Fermionic annihilation operators at atoms of type $a$, $\tilde b$, $\tilde a$ and $b$ 
within the elementary cell centered at $x$ respectively by $a_{ x}$, $\tilde b_{ x}$, $\tilde a_{ x-\delta_1}$ and $ b_{ x+\delta_1}$. The corresponding creation operators are their adjoint operators.\par
\bigskip

\indent We recall the Hamiltonian~(\ref{fullham})
$$\mathcal H=\mathcal H_0+\mathcal H_I$$
where $\mathcal H_0$ is the {\it free Hamiltonian} and $\mathcal H_I$ is the {\it interaction Hamiltonian}.\par
\bigskip

\subpoint{Free Hamiltonian} As was mentioned in section~\ref{introduction}, the free Hamiltonian describes the {\it hopping} of electrons from one atom to another. Here, we only consider the hoppings $\gamma_0,\gamma_1,\gamma_3$, see figure~\ref{intergraph}, so that $\mathcal H_0$ has the following expression in $x$ space:
\begin{equation}\begin{array}{r@{\ }>{\displaystyle}l}
\mathcal H_0:=&-\gamma_0\sum_{\displaystyle\mathop{\scriptstyle x\in\Lambda}_{j=1,2,3}}\left( a_{ x}^\dagger b_{ x +\delta_j}+b_{ x +\delta_j}^\dagger a_{ x}+
\tilde b_x^\dagger \tilde a_{x-\delta_j} +\tilde a_{x-\delta_j}^\dagger\tilde b_x\right)
-\gamma_1\sum_{ x\in\Lambda}\left( a_{ x}^\dagger \tilde b_{ x}+\tilde b_{ x}^\dagger a_{ x}\right)
 \\[0.2cm]
&-\gamma_3\sum_{\displaystyle\mathop{\scriptstyle x\in\Lambda}_{j=1,2,3}}\left( \tilde a_{ x-\delta_1}^\dagger b_{ x-\delta_1-\delta_j}+b_{ x-\delta_1-\delta_j}^\dagger\tilde a_{ x-\delta_1}\right)\\[0.2cm]
\end{array}\label{hamx}\end{equation}
Equation~(\ref{hamx}) can be rewritten in Fourier space as follows. We define the Fourier transform of the annihilation operators as
\begin{equation} \hat a_{k}:=\sum_{x\in\Lambda}e^{ikx}a_{x}\;,\quad 
\hat{\tilde b}_{k}:=\sum_{x\in\Lambda}e^{ikx}\hat{\tilde b}_{x+\delta_1}\;,\quad
\hat{\tilde a}_{k}:=\sum_{x\in\Lambda}e^{ikx}\hat{\tilde a}_{x-\delta_1}\;,\quad
\hat b_{k}:=\sum_{x\in\Lambda}e^{ikx}b_{x+\delta_1}\;\end{equation} 
in terms of which
\begin{equation}
\mathcal H_0=-\frac{1}{|\Lambda|}\sum_{ k\in\hat\Lambda}\hat A_{ k}^\dagger H_0( k)A_{ k}\label{hamk}
\end{equation}
where $|\Lambda|=L^2$, $\hat A_k$ is a column vector, whose transpose is $\hat A_k^T=(\hat a_{ k},\hat{\tilde b}_{ k},\hat{\tilde a}_{ k},\hat{b}_{ k})$, 
\begin{equation}H_0( k):=
\left(\begin{array}{*{4}{c}}
0&\gamma_1&0&\gamma_0\Omega^*(k)\\
\gamma_1&0&\gamma_0\Omega(k)&0\\
0&\gamma_0\Omega^*(k)&0&\gamma_3\Omega(k)e^{3ik_x}\\
\gamma_0\Omega(k)&0&\gamma_3\Omega^*(k)e^{-3ik_x}&0
\end{array}\right)\label{hmat}\end{equation}
and
$$
\Omega( k):=\sum_{j=1}^3e^{ik(\delta_j-\delta_1)}=1+2e^{-i\frac32k_x}\cos\left(\frac{\sqrt{3}}2k_y\right).
$$
We pick the energy unit in such a way that $\gamma_0=1$.
\bigskip

\subpoint{Interaction} We now define the interaction Hamiltonian. We first define the number operators $n_x^\alpha$ for 
$\alpha\in\{a,\tilde b,\tilde a,b\}$ and $x\in\Lambda$ in the following way:
\begin{equation} n_x^a=a^\dagger_x a_x\;,\quad
n^{\tilde b}_x=\tilde b_{ x}^\dagger \tilde b_x\;,\quad
n^{\tilde a}_x=\tilde a_{ x-\delta_1}^\dagger \tilde a_{ x-\delta_1}\;,\quad
n^b_x=b_{ x+\delta_1}^\dagger b_{ x+\delta_1}\;\end{equation}
and postulate the form of the interaction to be of an extended {\it Hubbard} form:
\begin{equation}
\mathcal H_I:=U\sum_{( x, y)\in\Lambda^2}\sum_{(\alpha,\alpha')\in\{a,\tilde b,\tilde a,b\}^2}v( x+d_\alpha- y-d_{\alpha'})\left( n^\alpha_{ x}-\frac{1}{2}\right)\left( n^{\alpha'}_{ y}-\frac{1}{2}\right)
\label{hamintx}\end{equation}
where the $d_\alpha$ are the vectors that give the position of each atom type with respect to the centers of the lattice $\Lambda$: $d_a:=0,\ d_{\tilde b}:=(0,0,c),\ d_{\tilde a}:=(0,0,c)-\delta_1,\ d_b:=\delta_1$ and $v$ is a bounded, rotationally invariant function, which decays exponentially fast to zero at infinity. In our spin-less case, we can assume without loss of generality that $v(0)=0$.
\subseqskip

\subsection{Schwinger function as Grassmann integrals and expectations}
\label{tracespathintssec}
\indent The aim of the present work is to compute the {\it specific free energy} and the {\it two-point Schwinger function}. These quantities are defined for finite lattices by
\begin{equation}f_\Lambda:=-\frac{1}{\beta|\Lambda|}\log\left(\mathrm{Tr}\left( e^{-\beta\mathcal H}\right)\right)\label{freeen}\end{equation}
where $\beta$ is inverse temperature and
\begin{equation}
\check s_{\alpha',\alpha}(\mathbf x_1-\mathbf x_2):=\left<\mathbf T(\alpha_{\mathbf x_1}'\alpha^\dagger_{\mathbf x_2})\right>:=\frac{\mathrm{Tr}(e^{-\beta\mathcal H}\mathbf T(\alpha'_{\mathbf x_1}\alpha_{\mathbf x_2}^\dagger))}{\mathrm{Tr}(e^{-\beta\mathcal H})}
\label{schwindef}\end{equation}
in which $(\alpha,\alpha')\in\mathcal A^2:=\{a,\tilde b,\tilde a,b\}^2$; $\mathbf x_{1,2}=(t_{1,2},x_{1,2})$ with $t_{1,2}\in[0,\beta)$; $\alpha_{\mathbf x}=e^{\mathcal Ht}\alpha_xe^{-\mathcal Ht}$; and $\mathbf T$ is the {\it Fermionic time ordering operator} defined in~(\ref{timeorderdef}). Our strategy essentially consists in deriving convergent expansions for $f_\Lambda$ and $\check s$, uniformly in $|\Lambda|$ and $\beta$, and then to take $\beta,|\Lambda|\to\infty$.\par
\bigskip

\indent However, the quantities on the right side of~(\ref{freeen}) and~(\ref{schwindef}) are somewhat difficult to manipulate. In this section, we will re-express $f_\Lambda$ and $\check s$ in terms of {\it Grassmann integrals} and {\it expectations}, and show how such quantities can be computed using a {\it determinant expansion}. This formalism will lay the groundwork for the procedure which will be used in the following to express $f_\Lambda$ and $\check s$ as series, and subsequently prove their convergence.\par
\bigskip

\point{Grassmann integral formulation} We first describe how to express~(\ref{freeen}) and~(\ref{schwindef}) as Grassmann integrals. The procedure is well known and details can be found in many references, see e.g. \cite[appendix~B]{giuOZ} and \cite{giuOZh} for a discussion adapted to the case of graphene, or \cite{genZO} for a discussion adapted to general low-dimensional Fermi systems, 
or \cite{benNFi} and \cite{salOTh} and references therein for an even more general picture. \par
\bigskip

\subpoint{Definition} We first define a Grassmann algebra and an integration procedure on it. We move to Fourier space: for every $\alpha\in\mathcal A:=\{a,\tilde b,\tilde a,b\}$, the operator $\alpha_{(t,x)}$ is associated
$$\hat\alpha_{\mathbf k=(k_0,k)}:=\frac{1}{\beta}\int_0^\beta dt\ e^{itk_0}e^{\mathcal H_0 t}\hat\alpha_ke^{-\mathcal H_0 t}$$
with $k_0\in2\pi\beta^{-1}(\mathbb Z+1/2)$ (notice that because of the $1/2$ term, $k_0\neq0$ for finite $\beta$). We notice that $\mathbf k\in\mathcal B_{\beta,L}:=(2\pi\beta^{-1}(\mathbb Z+1/2))\times\hat\Lambda$ varies in an infinite set. Since this will cause trouble when defining Grassmann integrals, we shall impose a cutoff $M\in\mathbb N$: let $\chi_0(\rho)$ be a smooth compact support function that returns $1$ if $\rho\leqslant 1/3$ and $0$ if $\rho\geqslant 2/3$, and let
$${\mathcal B}_{\beta,L}^*:=\mathcal B_{\beta,L}\cap\{(k_0,k),\ \chi_0(2^{-M}|k_0|\neq0)\}.$$
To every $(\hat\alpha_{\mathbf k},\hat\alpha_{\mathbf k}^\dagger)$ for $\alpha\in\mathcal A$ and $\mathbf k\in\mathcal B_{\beta,L}^*$, we associate a pair of {\it Grassmann variables} $(\hat\psi_{\mathbf k,\alpha}^-,\hat\psi_{\mathbf k,\alpha}^+)$, and we consider the finite Grassmann algebra (i.e. an algebra in which the $\hat\psi$ anti-commute with each other) generated by the collection $\{\hat\psi_{\mathbf k,\alpha}^\pm\}_{\mathbf k\in\mathcal B_{\beta,L}^*}^{\alpha\in\mathcal A}$. We define the Grassmann integral
$$\int
\prod^{\alpha\in{\mathcal A}}_{{\bf k}\in{\mathcal B}^*_{\beta,L}}d\hat\psi_{{\bf k},\alpha}^+ d\hat\psi_{{\bf k},\alpha}^-$$
as the linear operator on the Grassmann algebra whose action on a monomial in the variables $\hat\psi^\pm_{{\bf k},\alpha}$ is $0$ except if said monomial is $\prod^{\alpha\in{\mathcal A}}_{{\bf k}\in{\mathcal B}_{\beta,L}^*} \hat\psi^-_{{\bf k},\alpha} \hat\psi^+_{{\bf k},\alpha}$ up to a permutation of the variables, in which case the value of the integral is determined using
\begin{equation} \int
\prod^{\alpha\in{\mathcal A}}_{{\bf k}\in{\mathcal B}_{\beta,L}^*}
d\hat\psi_{{\bf k},\alpha}^+
d\hat\psi_{{\bf k},\alpha}^-
\left(\prod_{{\bf k}\in{\mathcal B}_{\beta,L}^*}^{\alpha\in{\mathcal A}}
\hat\psi^-_{{\bf k},\alpha}
\hat\psi^+_{{\bf k},\alpha}\right)=1
\label{2.1}\end{equation}
along with the anti-commutation of the $\hat\psi$.\par
\bigskip

\indent In the following, we will express the free energy and Schwinger function as {\it Grassmann integrals}, specified by a {\it propagator} and a {\it potential}.
The propagator is a $4\times4$ complex matrix $\hat g(\mathbf k)$, supported on some set $\mathcal B\subset\mathcal B_{\beta,L}^*$, and is associated with 
the {\it Gaussian Grassmann integration measure} 
\begin{equation}
P_{\hat g}(d\psi) := \left(\prod_{\mathbf k\in\mathcal B}
(\beta|\Lambda|)^{4}\det\hat g(\mathbf k)
\left(\prod_{\alpha\in\mathcal A}d\hat\psi_{\mathbf k,\alpha}^+d\hat\psi_{\mathbf k,\alpha}^-\right)\right)
\exp\left(-\frac{1}{\beta|\Lambda|}\sum_{\mathbf k\in\mathcal B}\hat\psi^{+}_{\mathbf k}\hat g^{-1}(\mathbf k)\hat\psi^{-}_{\mathbf k}\right).
\label{grassgauss}\end{equation}
Gaussian Grassmann integrals satisfy the following {\it addition principle}: given two propagators $\hat g_1$ and $\hat g_2$, and any polynomial $\mathfrak P(\psi)$ in the Grassmann variables,
\begin{equation}
\int P_{\hat g_1+\hat g_2}(d\psi)\ \mathfrak P(\psi)=\int P_{\hat g_1}(d\psi_1)\int P_{\hat g_2}(d\psi_2)\ \mathfrak P(\psi_1+\psi_2).
\label{addprop}\end{equation}
\bigskip

\subpoint{Free energy} We now express the free energy as a Grassmann integral. We define the {\it free propagator}
\begin{equation}
\hat g_{\leqslant M}({\bf k}):=\chi_0(2^{-M}|k_0|) (-ik_0\mathds 1-H_0(k))^{-1}
\label{freeprop}\end{equation}
and the Gaussian integration measure $P_{\leqslant M}(d\psi)\equiv P_{\hat g_{\leqslant M}}(d\psi)$.
One can prove (see e.g. \cite[appendix~B]{giuOZ}) that if
\begin{equation} \frac{1}{\beta|\Lambda|}\log\int P_{\leqslant M}(d\psi)\ e^{-\mathcal V(\psi)}\label{log}\end{equation}
is analytic in $U$, uniformly as $M\to\infty$, a fact we will check a posteriori, then the finite volume free energy can be written as
\begin{equation}f_\Lambda=f_{0,\Lambda}-\lim_{M\to\infty}\frac{1}{\beta|\Lambda|}\log\int P_{\leqslant M}(d\psi)\ e^{-\mathcal V(\psi)}\label{freeengrass}\end{equation}
where $f_{0,\Lambda}$ is the free energy in the $U=0$ case and, using the symbol $\int d\mathbf x$ as a shorthand for $\int_{0}^\beta dt \sum_{x\in \Lambda}$,
\begin{equation}
\mathcal V(\psi)=U\sum_{(\alpha,\alpha')\in{\mathcal A}^2}\int d\mathbf xd\mathbf y\ w_{\alpha,\alpha'}( \mathbf x-\mathbf y) \psi^+_{\mathbf x,\alpha}\psi^{-}_{\mathbf x,\alpha}
\psi^+_{\mathbf y,\alpha'}\psi^{-}_{\mathbf y,\alpha'}
\label{2.3a}\end{equation}
in which $w_{\alpha,\alpha'}(\mathbf x):=\delta(x_0)v(x+d_\alpha-d_{\alpha'})$, where $\delta(x_0)$ denotes the $\beta$-periodic Dirac delta function,
and 
\begin{equation} \psi^\pm_{\mathbf x,\alpha}:=\frac1{\beta|\Lambda|}\sum_{{\bf k}\in{\mathcal B}^*_{\beta,L}}\hat \psi^{\pm}_{{\bf k},\alpha}e^{\pm i{\bf k}\mathbf x}\;.\end{equation}
Notice that the expression of $\mathcal V(\psi)$ in~(\ref{2.3a}) is very similar to that of $\mathcal H_I$, with an added imaginary time $(x_0,y_0)$ and the $\alpha_{\mathbf x}$ replaced by $\psi_{\mathbf x,\alpha}$, except that $(\alpha_\mathbf x^\dagger\alpha_\mathbf x-1/2)$ becomes $\psi_{\mathbf x,\alpha}^+\psi_{\mathbf x,\alpha}^-$. Roughly, the reason why we ``drop the 1/2'' is because of the difference between the anti-commutation rules of $\alpha_\mathbf x$ and $\psi_{\mathbf x,\alpha}$ (i.e., $\{\alpha_\mathbf x,\alpha_\mathbf x^\dagger\}=1$, vs. $\{\psi_{\mathbf x,\alpha}^+,\psi_{\mathbf x,\alpha}^-\}=0$).
More precisely, taking $\mathbf x=(x_0,x)$ with $x_0\in(-\beta,\beta)$, 
it is easy to check that the limit as $M\to\infty$ of $g_{\leqslant M}(\mathbf x):=\int P_{\leqslant M}(d\psi)\psi^-_\mathbf x\psi^+_{\bf 0}$ is equal to $\check s(\mathbf x)$, if $\mathbf x\neq {\bf 0}$,  
and equal to $\check s({\bf 0})+1/2$, otherwise. This extra $+1/2$ accounts for the ``dropping of the 1/2" mentioned above.\par
\bigskip

\subpoint{Two-point Schwinger function} The two-point Schwinger function can be expressed as a Grassmann integral as well: under the condition that
\begin{equation}
\frac{\int P_{\leqslant M}(d\psi)\ e^{-\mathcal V(\psi)}\hat \psi_{\mathbf k,\alpha_1}^-\hat \psi_{\mathbf k,\alpha_2}^+}{\int P_{\leqslant M}(d\psi)\ e^{-\mathcal V(\psi)}}\label{sc}
\end{equation}
is analytic in $U$ uniformly in $M$, a fact we will also check a posteriori, then one can prove (see e.g. \cite[appendix~B]{giuOZ}) that the two-point Schwinger function can be written as
\begin{equation}
s_{\alpha_1,\alpha_2}(\mathbf k)=\lim_{M\to\infty}\frac{\int P_{\leqslant M}(d\psi)\ e^{-\mathcal V(\psi)}\hat \psi_{\mathbf k,\alpha_1}^-\hat \psi_{\mathbf k,\alpha_2}^+}{\int P_{\leqslant M}(d\psi)\ e^{-\mathcal V(\psi)}}.
\label{schwingrass}\end{equation}
In order to facilitate the computation of the right side of~(\ref{schwingrass}), we will first rewrite it as
\begin{equation}
s_{\alpha_1,\alpha_2}(\mathbf k)
=\lim_{M\to\infty}\int d\hat J_{\mathbf k,\alpha_1}^-d\hat J_{\mathbf k,\alpha_2}^+\log\int P_{\leqslant M}(d\psi)e^{-\mathcal V(\psi)+\hat J_{\mathbf k,\alpha_1}^+\hat\psi_{\mathbf k,\alpha_1}^-+\hat\psi_{\mathbf k,\alpha_2}^+\hat J_{\mathbf k,\alpha_2}^-}
\label{schwinform}\end{equation}
where $\hat J_{\mathbf k,\alpha}^-$ and $\hat J_{\mathbf k,\alpha'}^+$ are extra Grassmann variables introduced for the purpose of the computation (note here that the Grassmann integral over the variables $\hat J_{\mathbf k,\alpha_1}^-,\hat J_{\mathbf k,\alpha_2}^+$ acts as a functional derivative with respect to the same variables, due to the Grassmann integration/derivation rules).
We define the {\it generating functional}
\begin{equation}
\mathcal W(\psi,\hat J_{\mathbf k,\underline\alpha}):=\mathcal V(\psi)-\hat J_{\mathbf k,\alpha_1}^+\hat\psi_{\mathbf k,\alpha_1}^--\hat\psi_{\mathbf k,\alpha_2}^+\hat J_{\mathbf k,\alpha_2}^-.
\label{Weffpotdef}\end{equation}
\bigskip

\point{Expectations} We have seen that the free energy and Schwinger function can be computed as Grassmann integrals, it remains to see how one computes such integrals. 
We can write \eqref{log} as
\begin{equation}
\log\int P_{\leqslant M}(d\psi)e^{-\mathcal V(\psi)}=\sum_{N=1}^\infty\frac{(-1)^N}{N!}\mathcal E_{\leqslant M}^T(\underbrace{\mathcal V,\cdots,\mathcal V}_{N\mathrm{\ times}})=:\sum_{N=1}^\infty\frac{(-1)^N}{N!}\mathcal E_{\leqslant M}^T(\mathcal V;N).
\label{grasstruncexp}\end{equation}
where the {\it truncated expectation} is defined as
\begin{equation}\mathcal E_{\leqslant M}^T(\mathcal V_1,\cdots,\mathcal V_N):=\left.\frac{\partial^N}{\partial\lambda_1\cdots\partial\lambda_N}\log\int P_{\leqslant M}(d\psi)\ e^{\lambda_1\mathcal V_1+\cdots+\lambda_N\mathcal V_N}\right|_{\lambda_1=\cdots=\lambda_N=0}\label{truncexpdef}\;.\end{equation}
in which $(\mathcal V_1,\cdots,\mathcal V_N)$ is a collection of commuting polynomials and the index $_{\leqslant M}$ refers to the propagator of $P_{\leqslant M}(d\psi)$. A similar formula holds for~\eqref{sc}.\par
\bigskip

\indent The purpose of this rewriting is that we can compute truncated expectations in terms of a {\it determinant expansion}, also known as the Battle-Brydges-Federbush formula~\cite{brySeE, batEF}, which expresses it as the determinant of a Gram matrix. The advantage of this writing is that, provided we first re-express the propagator $\hat g_{\leqslant M}(\mathbf k)$ in $\mathbf x$-space, the afore-mentioned Gram matrix can be bounded effectively (see section~\ref{powercountingsec}). We therefore first define an $\mathbf x$-space representation for $\hat g(\mathbf k)$:
\begin{equation}
g_{\leqslant M}(\mathbf x):=\frac{1}{\beta|\Lambda|}\sum_{\mathbf k\in\mathcal B^*_{\beta,L}}e^{i\mathbf k\cdot\mathbf x}\hat g_{\leqslant M}(\mathbf k).
\label{freepropx}\end{equation}
The determinant expansion is given in the following lemma, the proof of which can be found in \cite[appendix~A.3.2]{genZO}, \cite[appendix~B]{giuOZh}.\par
\bigskip

\Theo{Lemma}\label{detexplemma}
\indent Consider a family of sets $\mathbf P=(P_1,\cdots,P_s)$ where every $P_j$ is an ordered collection of Grassmann variables, we denote the product of the elements in $P_j$ by $\Psi_{P_j}:=\prod_{\psi\in P_j}\psi$.\par
\smallskip
\indent We call a pair $(\psi_{\mathbf x,\alpha}^-,\psi_{\mathbf x',\alpha'}^+)\in\mathbf P^2$ a {\it line}, and define the set of {\it spanning trees} $\mathbf T(\mathbf P)$ as the set of collections $T$ of lines that are such that upon drawing a vertex for each $P_i$ in $\mathbf P$ and a line between the vertices corresponding to $P_i$ and to $P_j$ for each line $(\psi_{\mathbf x,\alpha}^-,\psi_{\mathbf x',\alpha'}^+)\in T$ that is such that $\psi_{\mathbf x,\alpha}^-\in P_i$ and $\psi_{\mathbf x',\alpha'}^+\in P_j$, the resulting graph is a tree that connects all of the vertices.\par
\smallskip
\indent For every spanning tree $T\in\mathbf T(\mathbf P)$, to each line $l=(\psi_{\mathbf x,\alpha}^-,\psi_{\mathbf x',\alpha'}^+)\in T$ we assign a {\it propagator} $g_l:=g_{\alpha,\alpha'}(\mathbf x-\mathbf x')$.\par
\smallskip
\indent If $\mathbf P$ contains $2(n+s-1)$ Grassmann variables, with $n\in\mathbb N$, then there exists a probability measure $dP_T(\mathbf t)$ on the set of $n\times n$ matrices of the form $\mathbf t=M^TM$ with $M$ being a matrix whose columns are unit vectors of $\mathbb R^n$, such that
\begin{equation}\mathcal E_{\leqslant M}^T(\Psi_{P_1},\cdots,\Psi_{P_s})=\sum_{T\in\mathbf T(\mathbf P)}\sigma_T\prod_{l\in T}g_l\int dP_{T}(\mathbf t)\ \det G^{(T)}(\mathbf t)\label{detexp}\end{equation}
where $\sigma_T\in\{-1,1\}$ and $G^{(T)}(\mathbf t)$ is an $n\times n$ complex matrix each of whose components is indexed by a line $l\not\in T$ and is given by
$$G^{(T)}_{l}(\mathbf t)=\mathbf t_lg_l$$
(if $s=1$, then $\mathbf T(\mathbf P)$ is empty and both the sum over $T$ and the factor $\sigma_T\prod_{l\in T}g_l$ should be dropped from the right side of~(\ref{detexp})).
\endtheo
\bigskip

\indent Lemma~\ref{detexplemma}\ gives us a formal way of computing the right side of~(\ref{grasstruncexp}). However, proving that this formal expression is correct, in the sense that it is not divergent, will require a control over the quantities involved in the right side of~(\ref{detexp}), namely the propagator $g_{\leqslant M}$.
Since, as was discussed in the introduction, $g_{\leqslant M}$ is singular, controlling the right side of~(\ref{grasstruncexp}) is a non-trivial task that will require a multiscale analysis described in section~\ref{schemesec}.\par
\subseqskip

\subsection{Symmetries of the system}
\label{symsec}
\indent In the following, we will rely heavily on the symmetries of the system, whose representation in terms of Grassmann variables is now discussed.\par
\bigskip

\indent A {\it symmetry} of the system is a map that leaves {\it both} 
\begin{equation}
h_0:=\sum_{\mathbf x,\mathbf y}\psi^+_{\mathbf x}g^{-1}(\mathbf x-\mathbf y)\psi_{\mathbf y}^-
\label{hzdef}\end{equation}
and $\mathcal V(\psi)$ invariant ($\mathcal V(\psi)$ was defined in (\ref{2.3a})). We define
\begin{equation}
\hat\xi_{\mathbf k}^+:=\left(\begin{array}{*{2}{c}}\hat\psi_{\mathbf k,a}^+&\hat\psi_{\mathbf k,\tilde b}^+\end{array}\right),\quad
\hat\xi_{\mathbf k}^-:=\left(\begin{array}{c}\hat\psi_{\mathbf k,a}^-\\\hat\psi_{\mathbf k,\tilde b}^-\end{array}\right),\quad
\hat\phi_{\mathbf k}^+:=\left(\begin{array}{*{2}{c}}\hat\psi_{\mathbf k,\tilde a}^+&\hat\psi_{\mathbf k,b}^+\end{array}\right)
\hat\phi_{\mathbf k}^-:=\left(\begin{array}{c}\hat\psi_{\mathbf k,\tilde a}^-\\\hat\psi_{\mathbf k,b}^-\end{array}\right)
\label{axiphidef}\end{equation}
as well as the Pauli matrices
$$\sigma_1:=\left(\begin{array}{*{2}{c}}0&1\\1&0\end{array}\right),\quad\sigma_2:=\left(\begin{array}{*{2}{c}}0&-i\\i&0\end{array}\right),\quad\sigma_3:=\left(\begin{array}{*{2}{c}}1&0\\0&-1\end{array}\right).$$
We now enumerate the symmetries of the system, and postpone their proofs to appendix~\ref{symapp}.\par
\bigskip

\point{Global $U(1)$} For $\theta\in\mathbb{R}/(2\pi\mathbb Z)$, the map
\begin{equation}\left\{\begin{array}l
\hat\xi_{\mathbf k}^\pm\longmapsto e^{\pm i\theta}\hat\xi_{\mathbf k}^\pm\\[0.2cm]
\hat\phi_{\mathbf k}^\pm\longmapsto e^{\pm i\theta}\hat\phi_{\mathbf k}^\pm
\end{array}\right.\label{aglobaluo}\end{equation}
is a symmetry.\par
\bigskip

\point{ $2\pi/3$ rotation} Let $T\mathbf k:=(k_0,e^{-i\frac{2\pi}3\sigma_2}k)$, $l_2:=(3/2,-\sqrt3/2)$ and $\mathcal T_{\mathbf k}:=e^{-i(l_2\cdot k)\sigma_3}$, the mapping
\begin{equation}\left\{\begin{array}l
\hat\xi_{\mathbf k}^\pm\longmapsto\hat\xi_{T\mathbf k}^\pm\\[0.2cm]
\hat\phi_{\mathbf k}^-\longmapsto\mathcal T_{T\mathbf k}\hat\phi_{T\mathbf k}^-,\ \hat\phi^+_{\mathbf k}\longmapsto\hat\phi_{T\mathbf k}^+\mathcal T_{T\mathbf k}^\dagger
\end{array}\right.\label{arotation}\end{equation}\par
is a symmetry.\par
\bigskip

\point{Complex conjugation} The map in which 
\begin{equation}\left\{\begin{array}l
\hat\xi_{\mathbf k}^\pm\longmapsto\hat\xi_{-\mathbf k}^\pm\\[0.2cm]
\hat\phi_{\mathbf k}^\pm\longmapsto\hat\phi_{-\mathbf k}^\pm.
\end{array}\right.\label{acomplexconjugation}\end{equation}
and every complex coefficient of $h_0$ and $\mathcal V$ is mapped to its complex conjugate is a symmetry. 
\bigskip

\point{ Vertical reflection} Let $R_v\mathbf k=(k_0,k_1,-k_2)$,
\begin{equation}
\left\{\begin{array}l
\hat\xi_{\mathbf k}^\pm\longmapsto \hat\xi_{R_v\mathbf k}^\pm\\[0.2cm]
\hat\phi_{\mathbf k}^\pm\longmapsto \hat\phi_{R_v\mathbf k}^\pm
\end{array}\right.\label{averticalreflection}\end{equation}
is a symmetry.
\bigskip

\point{ Horizontal reflection} Let $R_h\mathbf k=(k_0,-k_1,k_2)$,
\begin{equation}\left\{\begin{array}l
\hat\xi_{\mathbf k}^-\longmapsto \sigma_1\hat\xi_{R_h\mathbf k}^-,\ \hat\xi_{\mathbf k}^+\longmapsto\hat\xi_{R_h\mathbf k}^+\sigma_1\\[0.2cm]
\hat\phi_{\mathbf k}^-\longmapsto \sigma_1\hat\phi_{R_h\mathbf k}^-,\ \hat\phi_{\mathbf k}^+\longmapsto\hat\phi_{R_h\mathbf k}^+\sigma_1
\end{array}\right.\label{ahorizontalreflection}\end{equation}
is a symmetry.\par
\bigskip

\point{ Parity} Let $P\mathbf k=(k_0,-k_1,-k_2)$,
\begin{equation}\left\{\begin{array}l
\hat\xi_{\mathbf k}^\pm\longmapsto i(\hat\xi_{P\mathbf k}^{\mp})^T\\[0.2cm]
\hat\phi_{\mathbf k}^\pm\longmapsto i(\hat\phi_{P\mathbf k}^{\mp})^T
\end{array}\right.\label{aparticlehole}\end{equation}
is a symmetry.
\bigskip

\point{ Time inversion} Let $I\mathbf k=(-k_0,k_1,k_2)$, the mapping
\begin{equation}\left\{\begin{array}l
\hat\xi_{\mathbf k}^-\longmapsto -\sigma_3\hat\xi_{I\mathbf k}^-,\ \hat\xi_{\mathbf k}^+\longmapsto\hat\xi_{I\mathbf k}^+\sigma_3\\[0.2cm]
\hat\phi_{\mathbf k}^-\longmapsto-\sigma_3\hat\phi_{I\mathbf k}^-,\ \hat\phi_{\mathbf k}^+\longmapsto\hat\phi_{I\mathbf k}^+\sigma_3
\end{array}\right.\label{ainversiont}\end{equation}
is a symmetry.\par
\bigskip

\section{Free propagator}
\label{proppropsec}

\indent In section~\ref{tracespathintssec}, we showed how to express the free energy and the two-point Schwinger function as a formal series of truncated expectations~(\ref{grasstruncexp}). Controlling the convergence of this series is made difficult by the fact that the propagator $\hat g_{\leqslant M}$ is singular, and will require a finer analysis. In this section, we discuss which are the singularities of $\hat g_{\leqslant M}$ and how it behaves close to them, and identify three regimes in which the propagator behaves differently.\par
\subseqskip

\subsection{Fermi points}
\indent The free propagator is singular if $k_0=0$ and $k$ is such that $H_0(k)$ is not invertible. The set of such $k$'s is called the {\it Fermi surface}. 
In this subsection, we study the properties of this set. We recall the definition of $H_0$ in \eqref{hmat}, 
$$ H_0( k):=
-\left(\begin{array}{*{4}{c}}
0&\gamma_1&0&\Omega^*(k)\\
\gamma_1&0&\Omega(k)&0\\
0&\Omega^*(k)&0&\gamma_3\Omega(k)e^{3ik_x}\\
\Omega(k)&0&\gamma_3\Omega^*(k)e^{-3ik_x}&0
\end{array}\right)$$
so that, using corollary~\ref{Anokzprop} (see appendix~\ref{inversapp}),
\begin{equation}
\det H_0(k)=\left|\Omega^2(k)-\gamma_1\gamma_3\Omega^*(k)e^{-3ik_x}\right|^2.
\label{detHz}\end{equation}
It is then straightforward to compute the solutions of $\det H_0(k)=0$ (see appendix~\ref{fermiapp} for details): we find that as long as $0<\gamma_1\gamma_3<2$, there are 8 Fermi points:
\begin{equation}\left\{\begin{array}l
p_{F,0}^\omega:=\left(\frac{2\pi}{3},\omega\frac{2\pi}{3\sqrt3}\right)\\[0.5cm]
p_{F,1}^\omega:=\left(\frac{2\pi}{3},\omega\frac{2}{\sqrt3}\arccos\left(\frac{1-\gamma_1\gamma_3}{2}\right)\right)\\[0.5cm]
p_{F,2}^\omega:=\left(\frac{2\pi}{3}+\frac{2}{3}\arccos\left(\frac{\sqrt{1+\gamma_1\gamma_3}(2-\gamma_1\gamma_3)}{2}\right),\omega\frac{2}{\sqrt3}\arccos\left(\frac{1+\gamma_1\gamma_3}{2}\right)\right)\\[0.5cm]
p_{F,3}^\omega:=\left(\frac{2\pi}{3}-\frac{2}{3}\arccos\left(\frac{\sqrt{1+\gamma_1\gamma_3}(2-\gamma_1\gamma_3)}{2}\right),\omega\frac{2}{\sqrt3}\arccos\left(\frac{1+\gamma_1\gamma_3}{2}\right)\right).
\end{array}\right.\label{fermdef}\end{equation}
for $\omega\in\{-,+\}$. Note that
\begin{equation}\begin{array}c
p_{F,1}^\omega=p_{F,0}^\omega+\left(0,\omega\frac{2}{3}\gamma_1\gamma_3\right)+O(\epsilon^4),\quad
p_{F,2}^\omega=p_{F,0}^\omega+\left(\frac{1}{\sqrt3}\gamma_1\gamma_3,-\omega\frac{1}{3}\gamma_1\gamma_3\right)+O(\epsilon^4),\\[0.5cm]
p_{F,3}^\omega=p_{F,0}^\omega+\left(-\frac{1}{\sqrt3}\gamma_1\gamma_3,-\omega\frac{1}{3}\gamma_1\gamma_3\right)+O(\epsilon^4).
\end{array}\label{eq:3.2bis}\end{equation}
The points $p_{F,j}^\omega$ for $j=1,2,3$ are labeled as per figure~\ref{figferm}.\par
\subseqskip

\subsection{Behavior around the Fermi points}
\indent In this section, we compute the dominating behavior of $\hat g(\mathbf k)$ close to its singularities, that is close to $\mathbf p_{F,j}^\omega:=(0,p_{F,j}^\omega)$. We recall that $\hat A(\mathbf k):=(-ik_0\mathds1+H_0(k))$ and $\hat g(\mathbf k)=\chi_0(2^{-M}|k_0|)\hat A^{-1}(\mathbf k)$.
\bigskip

\point{First regime} We define $k':=k-p_{F,0}^\omega=(k'_x,k'_y)$, $\mathbf k':=(k_0,k')$. We have
\begin{equation}\Omega(p_{F,0}^\omega+k')=\frac{3}{2}(ik'_x+\omega k'_y)+O(|k'|^2)=:\xi+O(|k'|^2)\label{eq:omega}\end{equation}
so that, by using (\ref{detAexpr}) with $(\mathfrak a,\mathfrak b,\mathfrak c,\mathfrak x,\mathfrak z)=-(\gamma_1,\Omega(k),\gamma_3\Omega(k)e^{3ik_x}, k_0,k_0)$,
\begin{equation}
\det\hat A(\mathbf p_{F,0}^\omega+\mathbf k')=(k_0^2+|\xi|^2)^2+O(\|\mathbf k'\|_{\mathrm I}^5,\epsilon^2\|\mathbf k'\|_{\mathrm I}^2)
\label{detao}\end{equation}
where
\begin{equation}
\|\mathbf k'\|_{\mathrm I}:=\sqrt{k_0^2+|\xi|^2}
\label{normo}\end{equation}
in which the label $_{\mathrm I}$ stands for ``first regime''. If
\begin{equation}
\kappa_1\epsilon<\|\mathbf k'\|_{\mathrm I}<\bar\kappa_0
\label{odef}\end{equation}
for suitable constants $\kappa_1, \bar\kappa_0>0$, then the remainder term in \eqref{detao} is smaller than the explicit term, so that~(\ref{detao}) is adequate in this regime, which we call the ``first regime''.

We now compute the dominating part of $\hat A^{-1}$ in this regime. The computation is carried out in the following way: we neglect terms of order $\gamma_1$, $\gamma_3$ and $|k'|^2$ in $\hat A$, invert the resulting matrix using~(\ref{invAexpr}), prove that this inverse is bounded by $(\mathrm{const}.)\ \|\mathbf k'\|_{\mathrm{I}}^{-1}$, and deduce a bound on the error terms. We thus find
\begin{equation}
\hat A^{-1}(\mathbf p_{F,0}^\omega+\mathbf k')=
-\frac{1}{k_0^2+|\xi|^2}
\left(\begin{array}{*{4}{c}}
-ik_0&0&0&\xi^*\\
0&-ik_0&\xi&0\\
0&\xi^*&-ik_0&0\\
\xi&0&0&-ik_0
\end{array}\right)
\left(\mathds1+O(\|\mathbf k'\|_{\mathrm{I}},\epsilon\|\mathbf k'\|_{\mathrm{I}}^{-1})\right)
\label{freepropzo}\end{equation}
and
\begin{equation}
|\hat A^{-1}(\mathbf p_{F,0}^\omega+\mathbf k')|\leqslant(\mathrm{const}.)\ \|\mathbf k'\|_{\mathrm{I}}^{-1}.
\label{boundfreepropzo}\end{equation}
Note that, recalling that the basis in which we wrote $A^{-1}$ is $\{a,\tilde b,\tilde a,b\}$, each graphene layer is decoupled from the other in the dominating part of~(\ref{freepropzo}).\par
\bigskip

\point{Ultraviolet regime} The regime in which $\|\mathbf k'\|_{\mathrm{I}}\geqslant\bar\kappa_0$ for both $\omega=\pm$, and is called the {\it ultraviolet} regime.
For such $\mathbf k'=:\mathbf k-\mathbf p_{F,0}^\omega$, one easily checks that
\begin{equation}
|\hat A^{-1}(\mathbf k)|\leqslant(\mathrm{const}.)\ |\mathbf k|^{-1}.
\label{boundfreepropuv}\end{equation}
\bigskip

\point{Second regime} We now go beyond the first regime: we assume that $\|\mathbf k'\|_{\mathrm I}\leqslant \kappa_1\epsilon$ and, using again \eqref{eq:omega} and \eqref{detAexpr}, we write
\begin{equation}
\det\hat A(\mathbf p_{F,0}^\omega+\mathbf k')=\gamma_1^2k_0^2+|\xi|^4+O(\epsilon^{7/2}\|\mathbf k'\|_{\mathrm{II}}^{3/2},\epsilon^5\|\mathbf k'\|_{\mathrm{II}},\epsilon\|\mathbf k'\|_{\mathrm{II}}^3)\label{detat}\end{equation}
where
\begin{equation}
\|\mathbf k'\|_{\mathrm{II}}:=\sqrt{k_0^2+\frac{|\xi|^4}{\gamma_1^2}}.
\label{normt}\end{equation}
If
\begin{equation}
\kappa_2\ \epsilon^3<\|\mathbf k'\|_{\mathrm{II}}<\bar\kappa_1\ \epsilon
\label{tdef}\end{equation}
for suitable constants $\kappa_2,\bar\kappa_1>0$, then the remainder in~(\ref{detat}) is smaller than the explicit term, and we thus define the ``second regime'', for which~(\ref{detat}) is appropriate.\par
\bigskip

\indent We now compute the dominating part of $\hat A^{-1}$ in this regime. To that end, we define the dominating part $\mathfrak L_{\mathrm{II}}\hat A$ of $\hat A$ by neglecting the terms of order $\gamma_3$ and $|k'|^2$ in $\hat A$ as well as the elements $\hat A_{aa}$ and $\hat A_{\tilde b\tilde b}$ (which are both equal to $-ik_0$), block-diagonalize it using proposition~\ref{blockdiagprop} (see appendix~\ref{diagapp}) and invert it:
\begin{equation}
\left(\mathfrak L_{\mathrm{II}}\hat A(\mathbf k)\right)^{-1}=\left(\begin{array}{*{2}{c}}\mathds1&M_{\mathrm{II}}(\mathbf k)^\dagger\\0&\mathds1\end{array}\right)\left(\begin{array}{*{2}{c}}a_{\mathrm{II}}^{(M)}&0\\0&a_{\mathrm{II}}^{(m)}(\mathbf k)\end{array}\right)\left(\begin{array}{*{2}{c}}\mathds1&0\\M_{\mathrm{II}}(\mathbf k)&\mathds1\end{array}\right)
\label{LfpAt}\end{equation}
where
\begin{equation}
a_{\mathrm{II}}^{(M)}:=-\left(\begin{array}{*{2}{c}}0&\gamma_1^{-1}\\\gamma_1^{-1}&0\end{array}\right),\quad
a_{\mathrm{II}}^{(m)}(\mathbf p_{F,0}^\omega+\mathbf k'):=\frac{\gamma_1}{\gamma_1^2k_0^2+|\xi|^4}\left(\begin{array}{*{2}{c}}
i\gamma_1k_0&(\xi^*)^2\\
\xi^2&i\gamma_1k_0\end{array}\right)
\label{compsLfpAt}\end{equation}
and
\begin{equation}
M_{\mathrm{II}}(\mathbf p_{F,0}^\omega+\mathbf k'):=-\frac1{\gamma_1}\left(\begin{array}{*{2}{c}}\xi^*&0\\0&\xi\end{array}\right).
\label{diagLfpAt}\end{equation}
We then bound the right side of~(\ref{LfpAt}), and find
\begin{equation}|\left(\mathfrak L_{\mathrm{II}}\hat A(\mathbf p_{F,0}^\omega+\mathbf k')\right)^{-1}|\leqslant (\mathrm{const}.)\ \left(\begin{array}{*{2}{c}} 
\epsilon^{-1}&\epsilon^{-1/2}\|\mathbf k'\|_{\mathrm{II}}^{-1/2}\\ 
\epsilon^{-1/2}\|\mathbf k'\|_{\mathrm{II}}^{-1/2}&\|\mathbf k'\|_{\mathrm{II}}^{-1}\end{array}\right),\label{LA-1}\end{equation}
in which the bound should be understood as follows: the upper-left element in~(\ref{LA-1}) is the bound on the upper-left $2\times2$ block of $(\mathfrak L_{\mathrm{II}}\hat A)^{-1}$, and similarly for the upper-right, lower-left and lower-right. Using this bound in $$\hat A^{-1}(\mathbf k)=
\left(\mathfrak L_{\mathrm{II}}\hat A(\mathbf k)\right)^{-1}\Big[\mathds{1}+\big(\hat A(\mathbf k)-\mathfrak L_{\mathrm{II}}\hat A(\mathbf k)\big)\left(\mathfrak L_{\mathrm{II}}\hat A(\mathbf k)\right)^{-1}\Big]^{-1}$$
we deduce a bound on the error term in square brackets and find
\begin{equation}\begin{largearray}
\hat A^{-1}(\mathbf p_{F,0}^\omega+\mathbf k')=\left(\begin{array}{*{2}{c}}\mathds1&M_{\mathrm{II}}(\mathbf k)^\dagger\\0&\mathds1\end{array}\right)\left(\begin{array}{*{2}{c}}a_{\mathrm{II}}^{(M)}&0\\0&a_{\mathrm{II}}^{(m)}(\mathbf k)\end{array}\right)\left(\begin{array}{*{2}{c}}\mathds1&0\\M_{\mathrm{II}}(\mathbf k)&\mathds1\end{array}\right)\cdot\\[0.5cm]
\hfill\cdot(\mathds1+O(\epsilon^{-1/2}\|\mathbf k'\|_{\mathrm{II}}^{1/2},\epsilon^{3/2}\|\mathbf k'\|_{\mathrm{II}}^{-1/2}))
\end{largearray}\label{freepropzt}\end{equation}
which implies the analogue of \eqref{LA-1} for $\hat A^{-1}$, 
\begin{equation} 
|\hat A^{-1}(\mathbf p_{F,0}^\omega+\mathbf k')|\leqslant(\mathrm{const}.)\ \left(\begin{array}{*{2}{c}} 
\epsilon^{-1}&\epsilon^{-1/2}\|\mathbf k'\|_{\mathrm{II}}^{-1/2}\\ 
\epsilon^{-1/2}\|\mathbf k'\|_{\mathrm{II}}^{-1/2}&\|\mathbf k'\|_{\mathrm{II}}^{-1}\end{array}\right) .
\label{boundfreepropzt}\end{equation} 

\bigskip

{\bf Remark:} Using the explicit expression for $\hat A^{-1}(\mathbf p_{F,0}^\omega+\mathbf k')$ obtained by applying proposition~\ref{matinvprop} (see appendix~\ref{inversapp}), one can show that the error term on the right side 
of \eqref{freepropzt}
can be improved to $O(\epsilon^{-1}\|\mathbf k'\|_{\mathrm{II}},\epsilon^{3/2}\|\mathbf k'\|_{\mathrm{II}}^{-1/2}))$. Since we will not need this improved bound in the following, 
we do not belabor further details. 
\par
\bigskip

\point{Intermediate regime} In order to derive~(\ref{freepropzt}), we assumed that $\|\mathbf k'\|_{\mathrm{II}}<\bar\kappa_1\epsilon$ with $\bar\kappa_1$ small enough. In the intermediate regime defined by $\bar\kappa_1\epsilon<\|\mathbf k'\|_{\mathrm{II}}$ and $\|\mathbf k'\|_{\mathrm I}<\kappa_1\epsilon$, we have that $\|\mathbf k'\|_{\mathrm I}\sim \|\mathbf k'\|_{\mathrm{II}}\sim \epsilon$ (given two positive functions $a(\epsilon)$ and $b(\epsilon)$, the symbol $a\sim b$ stands for $c b\leqslant a\leqslant C b$ for some universal constants $C>c>0$). Moreover,
\begin{equation}
\det\hat A(\mathbf p_{F,0}^\omega+\mathbf k')=(k_0^2+|\xi|^2)^2+\gamma_1^2k_0^2+O(\epsilon^5)
\label{detaot}\end{equation}
therefore $|\det\hat A|>(\mathrm{const}.)\ \epsilon^4$ and 
\begin{equation}
|\hat A^{-1}(\mathbf p_{F,0}^\omega,\mathbf k')|\leqslant(\mathrm{const}.)\ \epsilon^{-1}
\label{boundfreepropzot}\end{equation}
which is identical to the bound at the end of the first regime and at the beginning of the second.\par
\bigskip

\point{Third regime} We now probe deeper, beyond the second regime, and assume that $\|\mathbf k'\|_{\mathrm{II}}\leqslant \kappa_2\epsilon^3$. 
Since we will now investigate the regime in which $|k'|<(\mathrm{const}.)\ \epsilon^2$, we will need to consider all the Fermi points $p_{F,j}^\omega$ with $j\in\{0,1,2,3\}$.\par
\bigskip

\subpoint{Around $\mathbf p_{F,0}^\omega$} We start with the neighborhood of $\mathbf p_{F,0}^\omega$: 
\begin{equation}
\det\hat A(\mathbf p_{F,0}^\omega+\mathbf k')=\gamma_1^2(k_0^2+\gamma_3^2|\xi|^2)+O(\epsilon^{-1}\|\mathbf k'\|_{\mathrm{III}}^3)
\label{detath}\end{equation}
where
\begin{equation}
\|\mathbf k'\|_{\mathrm{III}}:=\sqrt{k_0^2+\gamma_3^2|\xi|^2}.
\label{normth}\end{equation}
The third regime around $\mathbf p_{F,0}^\omega$ is defined by
\begin{equation}
\|\mathbf k'\|_{\mathrm{III}}<\bar\kappa_2\epsilon^3
\label{thdef}\end{equation}
for some $\bar\kappa_2<\kappa_2$.
The computation of the dominating part of $\hat A^{-1}$ in this regime around $\mathbf p_{F,0}^\omega$ is similar to that in the second regime, but for the fact that we only neglect the terms of order $|k'|^2$ in $\hat A$ as well as the elements $\hat A_{aa}$ and $\hat A_{\tilde b\tilde b}$. In addition, the terms that are of order $\epsilon^{-3}\|\mathbf k'\|_{\mathrm{III}}^2$ that come out of the computation of the dominating part of $\hat A$ in block-diagonal form are also put into the error term. We thus find

\begin{equation}
\hat A^{-1}(\mathbf k)=\left(\begin{array}{*{2}{c}}\mathds1&M_{\mathrm{III},0}(\mathbf k)^\dagger\\0&\mathds1\end{array}\right)\left(\begin{array}{*{2}{c}}a_{\mathrm{III},0}^{(M)}&0\\0&a_{\mathrm{III},0}^{(m)}(\mathbf k)\end{array}\right)\left(\begin{array}{*{2}{c}}\mathds1&0\\M_{\mathrm{III},0}(\mathbf k)&\mathds1\end{array}\right)
(\mathds1+O(\epsilon^{-3}\|\mathbf k'\|_{\mathrm{III}}))
\label{freepropzth}\end{equation}
where
\begin{equation}
a_{\mathrm{III},0}^{(M)}:=-\left(\begin{array}{*{2}{c}}0&\gamma_1^{-1}\\\gamma_1^{-1}&0\end{array}\right),\quad
a_{\mathrm{III},0}^{(m)}(\mathbf p_{F,0}^\omega+\mathbf k'):=-\frac{1}{k_0^2+\gamma_3^2|\xi|^2}\left(\begin{array}{*{2}{c}}
-ik_0&\gamma_3\xi\\
\gamma_3\xi^*&-ik_0\end{array}\right)
\label{compsLfpAth}\end{equation}
and
\begin{equation}
M_{\mathrm{III},0}(\mathbf p_{F,0}^\omega+\mathbf k'):=-\frac1{\gamma_1}\left(\begin{array}{*{2}{c}}\xi^*&0\\0&\xi\end{array}\right)
\label{diagLfpAth}\end{equation}
and
\begin{equation} 
|\hat A^{-1}(\mathbf p_{F,0}^\omega+\mathbf k')|\leqslant(\mathrm{const}.)\ \left(\begin{array}{*{2}{c}} 
\epsilon^{-1}&\epsilon^{-2}\\ 
\epsilon^{-2}&\|\mathbf k'\|_{\mathrm{III}}^{-1}\end{array}\right).
\label{boundfreepropzth}\end{equation} 
\bigskip

\subpoint{Around $\mathbf p_{F,1}^\omega$} We now discuss the neighborhood of $\mathbf p_{F,1}^\omega$. We define $k'_{1}:=k-p_{F,1}^\omega=(k'_{1,x},k'_{1,y})$ and $\mathbf k'_{1}:=(k_0,k'_{1})$. We have
\begin{equation}\Omega(p_{F,1}^\omega+k'_{1})=\gamma_1\gamma_3+\xi_1+O(\epsilon^2|k'_{1}|)\label{eq:O1}\end{equation}
where
$$\xi_1:=\frac{3}{2}(ik'_{1,x}+\omega k'_{1,y}).$$
Using \eqref{detAexpr} and \eqref{detnokz}, we obtain
\begin{equation} \det\hat A(\mathbf p_{F,1}^\omega+\mathbf k'_{1})=\gamma_1^2 k_0^2+|\Omega^2-\gamma_1\gamma_3\Omega^*e^{-3ik'_{1,x}}|^2+O(\epsilon^4 |k_0|^2)\end{equation}
where $\Omega$ is evaluated at $p_{F,1}^\omega+k'_{1}$. Injecting \eqref{eq:O1} into this equation, we find
\begin{equation} \det\hat A(\mathbf p_{F,1}^\omega+\mathbf k'_{1})=\gamma_1^2 (k_0^2+\gamma_3^2|x_1|^2)
+O(\epsilon^4\|\mathbf k'_{1}\|_{\mathrm{III}}^2,\epsilon^{-1}\|\mathbf k'_{1}\|_{\mathrm{III}}^3)\end{equation}
where
$$x_1:=\frac{3}{2}(3ik'_{1,x}+\omega k'_{1,y}).$$
The third regime around $p_{F,1}^\omega$ is therefore defined by
$$\|\mathbf k'_{1}\|_{\mathrm{III}}<\bar\kappa_{2}\ \epsilon^3$$
where $\bar\kappa_2$ can be assumed to be the same as in \eqref{thdef} without loss of generality.
The dominating part of $\hat A^{-1}$ in this regime around $\mathbf p_{F,1}^\omega$ is similar to that around $\mathbf p_{F,0}^\omega$, except that we neglect the terms of order $\epsilon^2k'_{1}$ in $\hat A$ as well as the elements $\hat A_{aa}$ and $\hat A_{\tilde b\tilde b}$. As around $\mathbf p_{F,0}^\omega$, the terms of order $\epsilon^{-3}\|\mathbf k'_{1}\|_{\mathrm{III}}^2$ are put into the error term. We thus find
\begin{equation}\begin{largearray}
\hat A^{-1}(\mathbf k)=\left(\begin{array}{*{2}{c}}\mathds1&M_{\mathrm{III},1}(\mathbf k)^\dagger\\0&\mathds1\end{array}\right)\left(\begin{array}{*{2}{c}}a_{\mathrm{III},1}^{(M)}&0\\0&a_{\mathrm{III},1}^{(m)}(\mathbf k)\end{array}\right)\left(\begin{array}{*{2}{c}}\mathds1&0\\M_{\mathrm{III},1}(\mathbf k)&\mathds1\end{array}\right)\cdot\\[0.5cm]
\hfill\cdot(\mathds1+O(\epsilon,\epsilon^{-3}\|\mathbf k'\|_{\mathrm{III}}))
\end{largearray}\label{freepropoth}\end{equation}
where
\begin{equation}
a_{\mathrm{III},1}^{(M)}:=-\left(\begin{array}{*{2}{c}}0&\gamma_1^{-1}\\\gamma_1^{-1}&0\end{array}\right),\quad
a_{\mathrm{III},1}^{(m)}(\mathbf p_{F,1}^\omega+\mathbf k'):=\frac{1}{k_0^2+\gamma_3^2|x_1|^2}\left(\begin{array}{*{2}{c}}
ik_0&\gamma_3 x_1^*\\
\gamma_3x_1&ik_0\end{array}\right)
\label{compsLfpAthj}\end{equation}
and
\begin{equation}
M_{\mathrm{III},1}(\mathbf p_{F,1}^\omega+\mathbf k'_{1}):=-\gamma_3\mathds1-\frac1{\gamma_1}\left(\begin{array}{*{2}{c}}\xi_1^*&0\\0&\xi_1\end{array}\right)
\label{diagLfpAthj}\end{equation}
and
\begin{equation}
|\hat A^{-1}(\mathbf p_{F,1}^\omega+\mathbf k'_{1})|\leqslant(\mathrm{const}.)\ \left(\begin{array}{*{2}{c}}
\epsilon^{2}\|\mathbf k'_{1}\|_{\mathrm{III}}^{-1}&\epsilon\|\mathbf k'_{1}\|_{\mathrm{III}}^{-1}\\
\epsilon\|\mathbf k'_{1}\|_{\mathrm{III}}^{-1}&\|\mathbf k'_{1}\|_{\mathrm{III}}^{-1}\end{array}\right).
\label{boundfreepropzthj}\end{equation}
\bigskip

\subpoint{Around $\mathbf p_{F,j}^\omega$} The behavior of $\hat g(\mathbf k)$ around $p_{F,j}^\omega$ for $j\in\{2,3\}$ can be deduced from~(\ref{freepropoth}) by using the symmetry \eqref{arotation}
under $2\pi/3$ rotations: if we define $k'_{j}:=k-p_{F,j}^\omega=(k'_{j,x},k'_{j,y})$, $\mathbf k'_{j}:=(k_0,k'_{j})$ then, for $j=2,3$ and $\omega\pm$, 
\begin{equation}
\hat A^{-1}(\mathbf k'_{j}+\mathbf p_{F,j}^\omega)=
\left(\begin{array}{*{2}{c}}\mathds1&0\\0&\mathcal T_{T\mathbf k'_{j}+\mathbf p_{F,j-\omega}^\omega}\end{array}\right)\hat A^{-1}(T\mathbf k'_{j}+\mathbf p_{F,j-\omega}^{\omega})\left(\begin{array}{*{2}{c}}\mathds1&0\\0&\mathcal T^\dagger_{T\mathbf k'_{j}+\mathbf p_{F,j-\omega}^\omega}\end{array}\right)
\label{freeproptth}\end{equation}
where $T$ and $\mathcal T_{\mathbf k}$ were defined above~(\ref{arotation}), and $\mathbf p_{F,4}^{-}\equiv \mathbf p_{F,1}^{-}$.
In addition, if $\mathbf k'_{2}$ and $\mathbf k'_{3}$ are in the third regime, then
$\mathcal T_{T\mathbf k'_{j}+\mathbf p_{F,j}^\omega}=e^{-i\omega\frac{2\pi}3\sigma_3}+O(\epsilon^2)$.
\bigskip

\point{Intermediate regime} We are left with an intermediate regime between the second and third regimes, defined by
\begin{equation}
\bar\kappa_2\epsilon^3<\|\mathbf k'\|_{\mathrm{III}}\ ,\quad\|\mathbf k'\|_{\mathrm{II}}<\kappa_2\epsilon^3\quad\mathrm{and}\quad\bar\kappa_{2}\epsilon^3<\|\mathbf k'_{j}\|_{\mathrm{III}},\
\forall j\in\{1,2,3\},
\label{inttth}\end{equation}
which implies 
$$\|\mathbf k'\|_{\mathrm{III}}\sim \|\mathbf k'\|_{\mathrm{II}}\sim \|\mathbf k'_{j}\|_{\mathrm{III}}\sim \epsilon^3$$
and 
\begin{equation}
\det\hat A(\mathbf p_{F,0}^\omega+\mathbf k')=\gamma_1^2k_0^2+\left|\xi^2-\gamma_1\gamma_3\xi^*\right|^2+O(\epsilon^{10}).
\label{detatth}\end{equation}
One can prove (see appendix~\ref{boundproptthapp}) that injecting~(\ref{inttth}) into~(\ref{detatth}) implies that $|\det\hat A|\geqslant(\mathrm{const}.)\ \epsilon^8$, which in turn implies that
\begin{equation}
|\hat A^{-1}(\mathbf p_{F,0}^\omega+\mathbf k')|\leqslant (\mathrm{const}.)\ \left(\begin{array}{*{2}{c}}
\epsilon^{-1}&\epsilon^{-2}\\
\epsilon^{-2}&\epsilon^{-3}\end{array}\right)
\label{boundfreepropztth}\end{equation}
which is identical to the bound at the end of the second regime and at the beginning of the third.\par\bigskip

\point{Summary} Let us briefly summarize this sub-section: we defined the norms
\begin{equation}
\|\mathbf k'\|_{\mathrm I}:=\sqrt{k_0^2+|\xi|^2},\quad
\|\mathbf k'\|_{\mathrm{II}}:=\sqrt{k_0^2+\frac{|\xi^4|}{\gamma_1^2}},\quad \|\mathbf k'\|_{\mathrm{III}}:=\sqrt{k_0^2+\gamma_3^2|\xi|^2},
\label{normsdef}\end{equation}
and identified an {\it ultraviolet} regime and three {\it infrared} regimes in which the free propagator $\hat g(\mathbf k)$ behaves differently:
\begin{itemize}
\item for $\| \mathbf k'\|_{\mathrm{I}}>\bar\kappa_0$, (\ref{boundfreepropuv}) holds.
\item for $\kappa_1\epsilon<\|\mathbf k'\|_{\mathrm I}<\bar\kappa_0$, (\ref{freepropzo}) holds.
\item for $\kappa_2\epsilon^3<\|\mathbf k'\|_{\mathrm{II}}<\bar\kappa_1\epsilon$, (\ref{freepropzt}) holds.
\item for $\|\mathbf k'\|_{\mathrm{III}}<\bar\kappa_2\epsilon^3$, (\ref{freepropzth}) holds, for $\|\mathbf k'_{1}\|_{\mathrm{III}}<\bar\kappa_{2}\epsilon^3$, (\ref{freepropoth}) holds,
and similarly for the $j=2,3$ cases.
\end{itemize}
\seqskip

\section{Multiscale integration scheme}
\label{schemesec}
\indent In this section, we describe the scheme that will be followed in order to compute the right side of~(\ref{grasstruncexp}). We will first define a {\it multiscale decomposition} in each regime which will play an essential role in showing that the formal series in~(\ref{grasstruncexp}) converges. In doing so, we will define {\it effective} interactions and propagators, which will be defined in $\mathbf k$-space, but since we wish to use the determinant expansion in lemma~\ref{detexplemma}\ to compute and bound the effective truncated expectations, we will have to define the effective quantities in $\mathbf x$-space as well. Once this is done, we will write bounds for the propagator in terms of scales.\par
\subseqskip

\subsection{Multiscale decomposition}
\label{multiscalesec}
\indent We will now discuss the scheme we will follow to compute the Gaussian Grassmann integrals in terms of which the free energy and two-point Schwinger function were expressed in~(\ref{freeengrass}) and~(\ref{schwinform}). The main idea is to decompose them into scales, and compute them one scale at a time. The result of the integration over one scale will then be considered as an {\it effective theory} for the remaining ones.\par
\bigskip

\indent Throughout this section, we will use a smooth cutoff function $\chi_0(\rho)$, which returns 1 for $\rho\leqslant1/3$ and 0 for $\rho\geqslant2/3$.\par
\subseqskip

\point{Ultraviolet regime} Let $\bar{\mathfrak h}_0:=\lfloor\log_2(\bar\kappa_0)\rfloor$ (in which $\bar\kappa_0$ is the constant that appeared after~(\ref{normsdef}) which defines the inferior bound of the ultraviolet regime). For $h\in\{\bar{\mathfrak h}_0,\cdots,M\}$ and $h'\in\{\bar{\mathfrak h}_0+1,\cdots,M\}$, we define
\begin{equation}\begin{array}c
f_{\leqslant h'}(\mathbf k):=\chi_0(2^{-h'}|k_0|),\quad
f_{\leqslant\bar{\mathfrak h}_0}(\mathbf k):=\sum_{\omega=\pm}\chi_0(2^{-\bar{\mathfrak h}_0}\|\mathbf k-\mathbf p_{F,0}^\omega\|_{\mathrm{I}}),\\[0.2cm]
f_{h'}(\mathbf k):=f_{\leqslant h'}(\mathbf k)-f_{\leqslant h'-1}(\mathbf k)\\[0.2cm]
\mathcal B_{\beta,L}^{(\leqslant h)}:=\mathcal B_{\beta,L}\cap\mathrm{supp}f_{\leqslant h},\quad
\mathcal B_{\beta,L}^{(h')}:=\mathcal B_{\beta,L}\cap\mathrm{supp}f_{h'},\quad
\end{array}\label{indicuvh}\end{equation}
in which $\|\cdot\|_{\mathrm{I}}$ is the norm defined in~(\ref{normsdef}). In addition, we define
\begin{equation}
\hat g_{h'}(\mathbf k):=f_{h'}(\mathbf k)\hat A^{-1}(\mathbf k),\quad \hat g_{\leqslant h}(\mathbf k):=f_{\leqslant h}(\mathbf k)\hat A^{-1}(\mathbf k)
\label{inddresspropexpuv}\end{equation}
so that, in particular,
$$\hat g_{\leqslant M}(\mathbf k)=\hat g_{\leqslant M-1}(\mathbf k)+\hat g_M(\mathbf k).$$
Furthermore, it follows from the addition property~(\ref{addprop}) that for all $h\in\{\bar{\mathfrak h}_0,\cdots,M-1\}$,
\begin{equation}\left\{\begin{array}{>{\displaystyle}l}
\int P_{\leqslant M}(d\psi)\ e^{-\mathcal V(\psi)}=e^{-\beta|\Lambda|F_{h}}\int P_{\leqslant h}(d\psi^{(\leqslant h)})\ e^{-\mathcal V^{(h)}(\psi^{(\leqslant h)})}\\[0.5cm]
\int P_{\leqslant M}(d\psi)\ e^{-\mathcal W(\psi,\hat J_{\mathbf k,\underline\alpha})}=e^{-\beta|\Lambda|F_{h}}\int P_{\leqslant h}(d\psi^{(\leqslant h)})\ e^{-\mathcal W^{(h)}(\psi^{(\leqslant h)},\hat J_{\mathbf k,\underline\alpha})}
\end{array}\right.\label{freeengrassuv}\end{equation}
where $P_{\leqslant h}(d\psi^{(\leqslant h)})\equiv P_{\hat g_{\leqslant h}}(d\psi^{(\leqslant h)})$,
\begin{equation}\begin{array}{r@{\ }>{\displaystyle}l}
-\beta|\Lambda|F_{h}-\mathcal V^{(h)}(\psi^{(\leqslant h)}):=&-\beta|\Lambda|F_{h+1}+\log\int P_{h+1}(d\psi^{(h+1)})\ e^{-\mathcal V^{(h+1)}(\psi^{(h+1)}+\psi^{(\leqslant h)})}\\[0.5cm]
=&-\beta|\Lambda|F_{h+1}+\sum_{N=1}^\infty\frac{(-1)^N}{N!}\mathcal E_{h+1}^T(\mathcal V^{(h+1)}(\psi^{(h+1)}+\psi^{(\leqslant h)});N)
\end{array}\label{effpotuv}\end{equation}
and
\begin{equation}\begin{largearray}
-\beta|\Lambda|(F_{h}-F_{h+1})-\mathcal W^{(h)}(\psi^{(\leqslant h)},\hat J_{\mathbf k,\underline\alpha})\\[0.2cm]
\hfill:=\sum_{N=1}^\infty\frac{(-1)^N}{N!}\mathcal E_{(h+1)}^T(\mathcal W^{(h+1)}(\psi^{(h+1)}+\psi^{(\leqslant h)},\hat J_{\mathbf k,\underline\alpha});N)
\end{largearray}\label{effpotWuv}\end{equation}
in which the induction is initialized by
$$\mathcal V^{(M)}:=\mathcal V,\quad\mathcal W^{(M)}:=\mathcal W,\quad F_M:=0.$$
\bigskip

\point{First regime} We now decompose the first regime into scales. The main difference with the ultraviolet regime is that we incorporate the quadratic part of the effective potential into the propagator at each step of the multiscale integration. This is necessary to get satisfactory bounds later on. The propagator will therefore be changed, or {\it dressed}, inductively at every scale, as discussed below.\par
\bigskip

\indent Let $\mathfrak h_1:=\lceil\log_2(\kappa_1\epsilon)\rceil$ (in which $\kappa_1$ is the constant that appears after~(\ref{normsdef}) which defines the inferior bound of the first regime), and $\|\cdot\|_{\mathrm I}$ be the norm defined in~(\ref{normsdef}). We define for $h\in\{\mathfrak h_1,\cdots,\bar{\mathfrak h}_0\}$ and $h'\in\{\mathfrak h_1+1,\cdots,\bar{\mathfrak h}_0\}$,
\begin{equation}
\begin{array}c
f_{\leqslant h,\omega}(\mathbf k):=\chi_0(2^{-h}\|\mathbf k-\mathbf p_{F,0}^\omega\|_{\mathrm I}),\quad f_{h',\omega}(\mathbf k):=f_{\leqslant h',\omega}(\mathbf k)-f_{\leqslant h'-1,\omega}(\mathbf k)\\[0.2cm]
\mathcal B_{\beta,L}^{(\leqslant h,\omega)}:=\mathcal B_{\beta,L}\cap\mathrm{supp}f_{\leqslant h,\omega},\quad
\mathcal B_{\beta,L}^{(h',\omega)}:=\mathcal B_{\beta,L}\cap\mathrm{supp}f_{\leqslant h',\omega}
\end{array}\label{indicoh}\end{equation}
and
\begin{equation}
\hat g_{h',\omega}(\mathbf k):=f_{h',\omega}(\mathbf k)\hat A^{-1}(\mathbf k),\quad
\hat g_{\leqslant h,\omega}(\mathbf k):=f_{\leqslant h,\omega}(\mathbf k)\hat A^{-1}(\mathbf k).
\label{indundresspropexpo}\end{equation}
For $h\in\{\mathfrak h_1,\cdots,\bar{\mathfrak h}_0-1\}$, we define
\begin{equation}\begin{largearray}
-\beta|\Lambda|(F_h-F_{h+1})-\mathcal Q^{(h)}(\psi^{(\leqslant h)})-\bar{\mathcal V}^{(h)}(\psi^{(\leqslant h)})\\[0.2cm]
\hfill:=\sum_{N=1}^\infty\frac{(-1)^N}{N!}\bar{\mathcal E}_{h+1}^T(\bar{\mathcal V}^{(h+1)}(\psi^{(h+1)}+\psi^{(\leqslant h)});N)\\[0.5cm]
\hfil\mathcal Q^{(\bar{\mathfrak h}_0)}(\psi^{(\leqslant \bar{\mathfrak h}_0)})+\bar{\mathcal V}^{(\bar{\mathfrak h}_0)}(\psi^{(\leqslant \bar{\mathfrak h}_0)}):=\mathcal V^{(\bar{\mathfrak h}_0)}(\psi^{(\leqslant\bar{\mathfrak h}_0)})
\end{largearray}\label{effpotoh}\end{equation}
and
\begin{equation}\begin{largearray}
-\beta|\Lambda|(F_h-F_{h+1})-\mathcal Q^{(h)}(\psi^{(\leqslant h)})-\bar{\mathcal W}^{(h)}(\psi^{(\leqslant h)},\hat J_{\mathbf k,\underline\alpha})\\
\hfill:=\sum_{N=1}^\infty\frac{(-1)^N}{N!}\bar{\mathcal E}_{h+1}^T(\bar{\mathcal W}^{(h+1)}(\psi^{(h+1)}+\psi^{(\leqslant h)},\hat J_{\mathbf k,\underline\alpha});N)\\[0.5cm]
\hfil\mathcal Q^{(\bar{\mathfrak h}_0)}(\psi^{(\leqslant \bar{\mathfrak h}_0)})+\bar{\mathcal W}^{(\bar{\mathfrak h}_0)}(\psi^{(\leqslant \bar{\mathfrak h}_0)},\hat J_{\mathbf k,\underline\alpha}):=\mathcal W^{(\bar{\mathfrak h}_0)}(\psi^{(\leqslant\bar{\mathfrak h}_0)},\hat J_{\mathbf k,\underline\alpha})
\end{largearray}\label{effpotWoh}\end{equation}
in which $\mathcal Q^{(h)}$ is quadratic in the $\psi$, $\bar{\mathcal V}^{(h)}$ is at least quartic and $\bar{\mathcal W}^{(h)}$ has no terms that are both quadratic in $\psi$ and constant in $\hat J_{\mathbf k,\underline\alpha}$; and $\bar{\mathcal E}^T_{h+1}$ is the truncated expectation defined from the Gaussian measure $P_{\hat{\bar g}_{h+1,+}}(d\psi_+^{(h+1)})P_{\hat{\bar g}_{h+1,-}}(d\psi_-^{(h+1)})$; in which $\hat{\bar g}_{h+1,\omega}$ is the {\it dressed propagator} and is defined as follows. Let  $\hat W_2^{(h)}(\mathbf k)$ denote the {\it kernel} of $\mathcal Q^{(h)}$ i.e.
\begin{equation}
\mathcal Q^{(h)}(\psi^{(\leqslant h)})=:\frac{1}{\beta|\Lambda|}\sum_{\omega,(\alpha,\alpha')}\sum_{\mathbf k\in\mathcal B_{\beta,L}^{(\leqslant h,\omega)}}\hat\psi^{(\leqslant h)+}_{\mathbf k,\omega,\alpha}\hat W^{(h)}_{2,(\alpha,\alpha')}(\mathbf k)\hat\psi^{(\leqslant h)-}_{\mathbf k,\omega,\alpha'}
\label{hatWtdeft}\end{equation}
(remark: the $_\omega$ index in $\hat\psi_{\mathbf k,\omega,\alpha}^\pm$ is redundant since given $\mathbf k$, it is defined as the unique $\omega$ that is such that $\mathbf k\in\mathcal B_{\beta,L}^{(\leqslant h,\omega)}$; it will however be needed when defining the $\mathbf x$-space counterpart of $\hat\psi_{\mathbf k,\omega,\alpha}^\pm$ below). We define  $\hat{\bar g}_{h,\omega}$ and $\hat{\bar g}_{\leqslant h,\omega}$ by induction: $\hat{\bar g}_{\leqslant \bar{\mathfrak h}_0,\omega}({\bf k}):=(\hat g^{-1}_{\leqslant\bar{\mathfrak h}_0,\omega}(\mathbf k)+\hat W^{(\bar{\mathfrak h}_0)}_{2}(\mathbf k))^{-1}$ and, for $h\in\{\mathfrak h_1+1,\ldots,\bar{\mathfrak h}_0\}$,
\begin{equation}\left\{\begin{array}l
\hat{\bar g}_{h,\omega}(\mathbf k):=f_{h,\omega}(\mathbf k)f_{\leqslant h,\omega}^{-1}(\mathbf k)\hat{\bar g}_{\leqslant h,\omega}(\mathbf k)\\[0.2cm]
\left(\hat{\bar g}_{\leqslant h-1,\omega}(\mathbf k)\right)^{-1}:=f^{-1}_{\leqslant h-1,\omega}(\mathbf k)\left(\hat{\bar g}_{\leqslant h,\omega}(\mathbf k)\right)^{-1}+\hat W_2^{(h-1)}(\mathbf k)
\end{array}\right.\label{inddresspropo}\end{equation}
in which $f_{\leqslant h,\omega}^{-1}(\mathbf k)$ is equal to $1/f_{\leqslant h,\omega}(\mathbf k)$ if $f_{\leqslant h,\omega}(\mathbf k)\neq0$ and to $0$ if not.\par
\bigskip

\indent The dressed propagator is thus defined so that
\begin{equation}\left\{\begin{array}{>{\displaystyle}l}
\int P_M(d\psi)\ e^{-\mathcal V(\psi)}=e^{-\beta|\Lambda|F_h}\int \bar P_{\leqslant h}(\psi^{(\leqslant h)})\ e^{-\bar{\mathcal V}^{(h)}(\psi^{(\leqslant h)})}\\[0.5cm]
\int P_M(d\psi)\ e^{-\mathcal W(\psi,\hat J_{\mathbf k,\underline\alpha})}=e^{-\beta|\Lambda|F_h}\int \bar P_{\leqslant h}(\psi^{(\leqslant h)})\ e^{-\bar{\mathcal W}^{(h)}(\psi^{(\leqslant h)},\hat J_{\mathbf k,\underline\alpha})}
\end{array}\right.\label{freeengrassoh}\end{equation}
in which $\bar P_{\leqslant h}\equiv P_{\hat{\bar g}_{\leqslant h,+}}(d\psi_+^{(\leqslant h)})P_{\hat{\bar g}_{\leqslant h,-}}(d\psi_-^{(\leqslant h)})$. Equation~(\ref{inddresspropo}) can be expanded into a more explicit form: for $h'\in\{\mathfrak h_1+1,\ldots,\bar{\mathfrak h}_0\}$ and $h\in\{\mathfrak h_1,\cdots,\bar{\mathfrak h}_0\}$,
\begin{equation}
\hat{\bar g}_{h',\omega}(\mathbf k)=f_{h',\omega}(\mathbf k)\left(\hat{\bar A}_{h',\omega}(\mathbf k)\right)^{-1},\quad
\hat{\bar g}_{\leqslant h,\omega}(\mathbf k)=f_{\leqslant h,\omega}\left(\hat{\bar A}_{h,\omega}(\mathbf k)\right)^{-1}
\label{inddresspropexpo}\end{equation}
where
\begin{equation}
\hat{\bar A}_{h,\omega}(\mathbf k):=\hat A(\mathbf k)+f_{\leqslant h,\omega}(\mathbf k)\hat W_2^{(h)}(\mathbf k)+\sum_{h'=h+1}^{\bar{\mathfrak h}_0}\hat W_2^{(h')}(\mathbf k)
\label{Adefo}\end{equation}
(in which the sum should be interpreted as zero if $h=\bar{\mathfrak h}_0$).\par
\bigskip

\point{Intermediate regime} We briefly discuss the intermediate region between regimes~1 and~2. We define
\begin{equation}
f_{\mathfrak h_1,\omega}(\mathbf k):=\chi_0(2^{-\mathfrak h_1}\|\mathbf k-\mathbf p_{F,0}^\omega\|_{\mathrm I})-\chi_0(2^{-\bar{\mathfrak h}_1}\|\mathbf k-\mathbf p_{F,0}^\omega\|_{\mathrm{II}})=:f_{\leqslant\mathfrak h_1,\omega}(\mathbf k)-f_{\leqslant\bar{\mathfrak h}_1,\omega}(\mathbf k)
\label{indicot}\end{equation}
where $\bar{\mathfrak h}_1:=\lfloor\log_2(\bar\kappa_1\epsilon)\rfloor$, from which we define $\hat{\bar g}_{\mathfrak h_1,\omega}(\mathbf k)$ and $\hat{\bar g}_{\leqslant\bar{\mathfrak h}_1,\omega}(\mathbf k)$ in the same way as in~(\ref{inddresspropexpo}) with
\begin{equation}
\hat{\bar A}_{\bar{\mathfrak h}_1,\omega}(\mathbf k):=\hat A(\mathbf k)+f_{\leqslant \bar{\mathfrak h}_1,\omega}(\mathbf k)\hat W_2^{(\bar{\mathfrak h}_1)}(\mathbf k)+\sum_{h'=\mathfrak h_1}^{\bar{\mathfrak h}_0}\hat W_2^{(h')}(\mathbf k).
\label{Adefot}\end{equation}
The analogue of \eqref{freeengrassoh} holds here as well.\par
\bigskip

\point{Second regime} We now define a multiscale decomposition for the integration in the second regime. Proceeding as we did in the first regime, we define $\mathfrak h_2:=\lceil\log_2(\kappa_2\epsilon^3)\rceil$, for $h\in\{\mathfrak h_2,\cdots,\bar{\mathfrak h}_1\}$ and $h'\in\{\mathfrak h_2+1,\cdots,\bar{\mathfrak h}_1\}$, we define
\begin{equation}\begin{array}c
f_{\leqslant h,\omega}(\mathbf k):=\chi_0(2^{-h}\|\mathbf k-\mathbf p_{F,0}^\omega\|_{\mathrm{II}}),\quad f_{h',\omega}(\mathbf k):=f_{\leqslant h',\omega}(\mathbf k)-f_{\leqslant h'-1,\omega}(\mathbf k)\\[0.2cm]
\mathcal B_{\beta,L}^{(\leqslant h,\omega)}:=\mathcal B_{\beta,L}\cap\mathrm{supp}f_{\leqslant h,\omega},\quad
\mathcal B_{\beta,L}^{(h',\omega)}:=\mathcal B_{\beta,L}\cap\mathrm{supp}f_{\leqslant h',\omega}.
\end{array}\label{indicth}\end{equation}
The analogues of \eqref{freeengrassoh}, and~(\ref{inddresspropexpo}) hold with
\begin{equation}
\hat{\bar A}_{h-1,\omega}(\mathbf k):=\hat A(\mathbf k)+ f_{\leqslant h-1,\omega}(\mathbf k)\hat W_2^{(h-1)}(\mathbf k)+\sum_{h'=h}^{\bar{\mathfrak h}_1}\hat W_2^{(h')}(\mathbf k)+\sum_{h'=\mathfrak h_1}^{\bar{\mathfrak h}_0}\hat W_2^{(h')}(\mathbf k).
\label{Adeft}\end{equation}
\bigskip

\point{Intermediate regime} The intermediate region between regimes~2 and~3 is defined in analogy with that between regimes~1 and~2: we let 
\begin{equation}\begin{array}{>{\displaystyle}c}
f_{\mathfrak h_2,\omega}(\mathbf k):=\chi_0(2^{-\mathfrak h_2}\|\mathbf k-\mathbf p_{F,0}^\omega\|_{\mathrm{II}})-\sum_{j\in\{0,1,2,3\}}\chi_0(2^{-\bar{\mathfrak h}_2}\|\mathbf k-\mathbf p_{F,j}^\omega\|_{\mathrm{III}})\\[0.5cm]
f_{\leqslant\bar{\mathfrak h}_2,\omega,j}(\mathbf k):=\chi_0(2^{-\bar{\mathfrak h}_2}\|\mathbf k'_{\omega,j}\|_{\mathrm{III}})
\end{array}\label{indictth}\end{equation}
where $\bar{\mathfrak h}_2:=\lfloor\log_2(\bar\kappa_2\epsilon^3)\rfloor$ from which we define $\hat{\bar g}_{\mathfrak h_2,\omega}(\mathbf k)$ and $\hat{\bar g}_{\leqslant\bar{\mathfrak h}_2,\omega}(\mathbf k)$ in the same way as in~(\ref{inddresspropexpo}) with
\begin{equation}
\hat{\bar A}_{\bar{\mathfrak h}_2,\omega}(\mathbf k):=\hat A(\mathbf k)+f_{\leqslant \bar{\mathfrak h}_2,\omega}(\mathbf k)\hat W_2^{(\bar{\mathfrak h}_2)}(\mathbf k)+\sum_{h'=\mathfrak h_2}^{\bar{\mathfrak h}_1}\hat W_2^{(h')}(\mathbf k)+\sum_{h'=\mathfrak h_1}^{\bar{\mathfrak h}_0}\hat W_2^{(h')}(\mathbf k).
\label{Adeftth}\end{equation}
The analogue of \eqref{freeengrassoh} holds here as well.\par
\bigskip

\point{Third regime} There is an extra subtlety in the third regime: we will see in section~\ref{thsec} that the singularities of the dressed propagator are slightly different from those of the bare (i.e. non-interacting) propagator: at scale $h$ the effective Fermi points $\mathbf p_{F,j}^{\omega}$ with $j=1,2,3$ are moved to $\tilde{\mathbf p}_{F,j}^{(\omega,h)}$, with 
\begin{equation} \label{eq4.20bis}\|\tilde{\mathbf p}_{F,j}^{(\omega,h)}-\mathbf p_{F,j}^{\omega}\|_{\mathrm{III}}\leqslant (\mathrm{const}.)\ |U|\epsilon^3.\end{equation}
The central Fermi points, $j=0$, are left invariant by the interaction. For notational uniformity we set $\tilde{\mathbf p}_{F,0}^{(\omega,h)}\equiv \mathbf p_{F,0}^{\omega}$.
Keeping this in mind, we then proceed in a way reminiscent of the first and second regimes: let $\mathfrak h_\beta:=\lfloor\log_2(\pi/\beta)\rfloor$, for $h\in\{\mathfrak h_\beta,\cdots,\bar{\mathfrak h}_2\}$ and $h'\in\{\mathfrak h_\beta+1,\cdots,\bar{\mathfrak h}_2\}$, we define
\begin{equation}
\begin{array}c
f_{\leqslant h,\omega,j}(\mathbf k):=\chi_0(2^{-h}\|\mathbf k-\tilde{\mathbf p}_{F,j}^{(\omega,h+1)}\|_{\mathrm{III}}),\quad f_{h',\omega,j}(\mathbf k):=f_{\leqslant h',\omega,j}(\mathbf k)-f_{\leqslant h'-1,\omega,j}(\mathbf k)\\[0.2cm]
\mathcal B_{\beta,L}^{(\leqslant h,\omega,j)}:=\mathcal B_{\beta,L}\cap\mathrm{supp}f_{\leqslant h,\omega,j},\quad
\mathcal B_{\beta,L}^{(h',\omega,j)}:=\mathcal B_{\beta,L}\cap\mathrm{supp}f_{\leqslant h',\omega,j}
\end{array}\label{indicthh}\end{equation}
and the analogues of \eqref{freeengrassoh}, and~(\ref{inddresspropexpo}) hold with
\begin{equation}\begin{largearray}
\hat{\bar A}_{h-1,\omega,j}(\mathbf k):=\hat A(\mathbf k)+f_{\leqslant h-1,\omega,j}(\mathbf k)\hat W_2^{(h-1)}(\mathbf k)\\[0.2cm]
\hfill+\sum_{h'=h}^{\bar{\mathfrak h}_2}\hat W_2^{(h')}(\mathbf k)+\sum_{h'=\mathfrak h_2}^{\bar{\mathfrak h}_1}\hat W_2^{(h')}(\mathbf k)+\sum_{h'=\mathfrak h_1}^{\bar{\mathfrak h}_0}\hat W_2^{(h')}(\mathbf k).
\end{largearray}\label{Adefth}\end{equation}
\par
\bigskip

\point{Last scale} Recalling that $|k_0|\geqslant\pi/\beta$, the smallest possible scale is $\mathfrak h_\beta:=\lfloor\log_2(\pi/\beta)\rfloor$. The last integration is therefore that on scale $h=\mathfrak h_\beta+1$, after which, we are left with 
\begin{equation}\left\{\begin{array}{>{\displaystyle}l}
\int P_{\leqslant M}(d\psi)\ e^{-\mathcal V(\psi)}=e^{-\beta|\Lambda|F_{\mathfrak h_\beta}}\\[0.5cm]
\int P_{\leqslant M}(d\psi)\ e^{-\mathcal W(\psi,\hat J_{\mathbf k,\underline\alpha})}=e^{-\beta|\Lambda|F_{\mathfrak h_\beta}}e^{-\mathcal W^{(\mathfrak h_\beta)}(\hat J_{\mathbf k,\underline\alpha})}.
\end{array}\right.\label{freeengrassend}\end{equation}
The subsequent sections are dedicated to the proof of the fact that both $F_{\mathfrak h_\beta}$ and $\mathcal W^{(\mathfrak h_\beta)}$ are analytic in $U$, uniformly in $L$, $\beta$ and $\epsilon$. We will do this by studying each regime, one at a time, performing a {\it tree expansion} in each of them in order to bound the terms of the series (see section~\ref{treeexpsec} and following).\par
\subseqskip

\subsection{$\mathbf x$-space representation of the effective potentials}
\label{xspacesec}
\indent We will compute the truncated expectations arising in~(\ref{effpotuv}), (\ref{effpotWuv}), (\ref{effpotoh}) and~(\ref{effpotWoh}) using a determinant expansion (see lemma~\ref{detexplemma}) which, as was mentioned above, is only useful if the propagator and effective potential are expressed in $\mathbf x$-space. We will discuss their definition in this section. We restrict our attention to the effective potentials $\mathcal V^{(h)}$ since, in order to compute the two-point Schwinger function in the regimes we are interested in, we will not need to express the kernels of $\mathcal W^{(h)}$ in $\mathbf x$-space. 
\par
\bigskip

\point{Ultraviolet regime} We first discuss the ultraviolet regime, which differs from the others in that the propagator does not depend on the index $\omega$. We write $\mathcal V^{(h)}$ in terms of its {\it kernels} (anti-symmetric in the exchange of their indices), defined as
\begin{equation}\begin{largearray}
\mathcal V^{(h)}(\psi^{(\leqslant h)})=:\sum_{l=1}^\infty\frac{1}{(\beta|\Lambda|)^{2l-1}}\sum_{\underline\alpha=(\alpha_1,\cdots,\alpha_{2l})}\sum_{\displaystyle\mathop{\scriptstyle (\mathbf k_1,\cdots,\mathbf k_{2l})\in\mathcal B_{\beta,L}^{(\leqslant h)2l}}_{\mathbf k_1-\mathbf k_2+\cdots+\mathbf k_{2l-1}-\mathbf k_{2l}=0}}\hskip-15pt\hat W_{2l,\underline\alpha}^{(h)}(\mathbf k_1,\cdots,\mathbf k_{2l-1})\cdot\\[1.5cm]
\hfill\cdot\hat\psi^{(\leqslant h)+}_{\mathbf k_1,\alpha_1}\hat\psi^{(\leqslant h)-}_{\mathbf k_2,\alpha_2}\cdots\hat\psi^{(\leqslant h)+}_{\mathbf k_{2l-1},\alpha_{2l-1}}\hat\psi^{(\leqslant h)-}_{\mathbf k_{2l},\alpha_{2l}}.
\end{largearray}\label{hatWdefo}\end{equation}
The $\mathbf x$-space expression for $\hat\psi_{\mathbf k,\alpha}^{(\leqslant h)\pm}$ is defined as
\begin{equation}
\psi_{\mathbf x,\alpha}^{(\leqslant h)\pm}:=\frac{1}{\beta|\Lambda|}\sum_{\mathbf k\in\mathcal B_{\beta,L}^{(\leqslant h)}}e^{\pm i\mathbf k\cdot\mathbf x}\hat\psi_{\mathbf k,\alpha}^{(\leqslant h)\pm}
\label{psixdefo}\end{equation}
so that the propagator's formulation in $\mathbf x$-space is
\begin{equation}
g_{h}(\mathbf x-\mathbf y):=\frac{1}{\beta|\Lambda|}\sum_{\mathbf k\in\mathcal B_{\beta,L}^{(\leqslant h)}}e^{i\mathbf k\cdot(\mathbf x-\mathbf y)}\hat g_h(\mathbf k)
\label{propxuv}\end{equation}
and similarly for $g_{\leqslant h}$, and the effective potential~(\ref{hatWdefo}) becomes
\begin{equation}\begin{largearray}
\mathcal V^{(h)}(\psi^{(\leqslant h)})=\sum_{l=1}^\infty\sum_{\underline\alpha}\frac{1}{\beta|\Lambda|}\int d\mathbf x_1\cdots d\mathbf x_{2l}\ W^{(h)}_{2l,\underline\alpha}(\mathbf x_1-\mathbf x_{2l},\cdots,\mathbf x_{2l-1}-\mathbf x_{2l})\cdot\\[0.5cm]
\hfill\cdot\psi^{(\leqslant h)+}_{\mathbf x_1,\alpha_1}\psi^{(\leqslant h)-}_{\mathbf x_2,\alpha_2}\cdots\psi^{(\leqslant h)+}_{\mathbf x_{2l-1},\alpha_{2l-1}}\psi^{(\leqslant h)-}_{\mathbf x_{2l},\alpha_{2l}}
\end{largearray}\label{Wdefo}\end{equation}
with
\begin{equation}
W^{(h)}_{2l,\underline\alpha}(\mathbf u_1,\cdots,\mathbf u_{2l-1})\\[0.2cm]
\hfill:=\frac{1}{(\beta|\Lambda|)^{2l-1}}\sum_{(\mathbf k_1,\cdots,\mathbf k_{2l-1})\in\mathcal B_{\beta,L}^{2l-1}}\hskip-25pt e^{i(\sum_{i=1}^{2l-1}(-1)^i\mathbf k_i\cdot\mathbf u_i)}\hat W^{(h)}_{2l,\underline\alpha}(\mathbf k_1,\cdots,\mathbf k_{2l-1}).
\label{xspaceWuv}\end{equation}
\bigskip

{\bf Remark}: From \eqref{hatWdefo}, $\hat W_{2l,\underline\alpha}^{(h)}(\underline{\mathbf k})$ is not defined for $\mathbf k_i\not\in\mathcal B_{\beta,L}^{(\leqslant h)}$, however, one can easily check that~(\ref{xspaceWuv}) holds for any extension of $\hat W_{2l,\underline\alpha}^{(h)}$ to $\mathcal B_{\beta,L}^{2l-1}$, thanks to the compact support properties of 
$\psi^{(\leqslant h)}$ in momentum space. In order to get satisfactory bounds on $W_{2l,\underline\alpha}^{(h)}(\underline{\mathbf x})$, that is in order to avoid Gibbs phenomena, we define the extension of $\hat W_{2l,\underline\alpha}^{(h)}(\underline{\mathbf k})$ similarly to~(\ref{hatWdefo}) by relaxing the condition that $\psi^{(\leqslant h)}$ is supported on $\mathcal B_{\beta,L}^{(\leqslant h)}$ and iterating~(\ref{effpotuv}). In other words, we let $\hat W_{2l,\underline\alpha}^{(h)}(\underline{{\bf k}})$ for $\underline{{\bf k}}\in \mathcal B_{\beta,L}^{2l-1}$ be the kernels of $\mathcal V^{*(h)}$ defined inductively by
\begin{equation}
-\beta|\Lambda|\mathfrak e_h-\mathcal V^{*(h)}(\Psi):=\sum_{N=1}^\infty\frac{(-1)^N}{N!}\mathcal E_{h+1}^T(\mathcal V^{*(h+1)}(\psi^{(h+1)}+\Psi);N)
\label{extendedVuv}\end{equation}
in which $\{\hat\Psi_{\mathbf k,\alpha}\}_{\mathbf k\in\mathcal B_{\beta,L},\alpha\in\mathcal A}$ is a collection of {\it external fields} (in reference to the fact that, contrary to $\psi^{(\leqslant h)}$, they have a non-compact support in momentum space). The use of this specific extension can be justified {\it ab-initio} by re-defining the cutoff function $\chi$ in such a way that its support is $\mathbb{R}$, e.g. using exponential tails that depend on a parameter $\epsilon_\chi$ in such a way that the support tends to be compact as $\epsilon_\chi$ goes to 0. Following this logic, we could first define $\hat W$ using the non-compactly supported cutoff function and then take the $\epsilon_\chi\to0$ limit, thus recovering \eqref{extendedVuv}.
Such an approach is dicussed in~\cite{bmZT}. From now on, with some abuse of notation, we shall 
identify $\mathcal V^{*(h)}$ with $\mathcal V^{(h)}$ and denote them by the same symbol $\mathcal V^{(h)}$, which is justified by the fact that 
their kernels are (or can be chosen, from what said above, to be) the same.\par
\bigskip

\point{First and second regimes} We now discuss the first and second regimes (the third regime is very slightly different in that the index $\omega$ is complemented by an extra index $j$ and the Fermi points are shifted). Similarly to~(\ref{hatWdefo}), we define the {\it kernels} of $\bar{\mathcal V}$:
\begin{equation}\begin{largearray}
\bar{\mathcal V}^{(h)}(\psi^{(\leqslant h)})=:\sum_{l=2}^\infty\frac{1}{(\beta|\Lambda|)^{2l-1}}\sum_{\underline\omega,\underline\alpha}\sum_{\displaystyle\mathop{\scriptstyle (\mathbf k_1,\cdots,\mathbf k_{2l})\in\mathcal B_{\beta,L}^{(\leqslant h,\underline\omega)}}_{\mathbf k_1-\mathbf k_2+\cdots+\mathbf k_{2l-1}-\mathbf k_{2l}=0}}\hskip-15pt\hat W_{2l,\underline\alpha}^{(h)}(\mathbf k_1,\cdots,\mathbf k_{2l-1})\cdot\\[1.5cm]
\hfill\cdot\hat\psi^{(\leqslant h)+}_{\mathbf k_1,\alpha_1,\omega_1}\hat\psi^{(\leqslant h)-}_{\mathbf k_2,\alpha_2,\omega_2}\cdots\hat\psi^{(\leqslant h)+}_{\mathbf k_{2l-1},\alpha_{2l-1},\omega_{2l-1}}\hat\psi^{(\leqslant h)-}_{\mathbf k_{2l},\alpha_{2l},\omega_{2l}}.
\end{largearray}\label{hatWdeft}\end{equation}
where $\mathcal B_{\beta,L}^{(\leqslant h,\underline\omega)}=\mathcal B_{\beta,L}^{(\leqslant h,\omega_1)}\times\cdots\times\mathcal B_{\beta,L}^{(\leqslant h,\omega_{2l})}$. Note that the kernel $\hat W_{2l,\underline\alpha}^{(h)}$ is independent of $\underline\omega$, which can be easily proved using the symmetry $\omega_i\mapsto-\omega_i$. The $\mathbf x$-space expression for $\hat\psi_{\mathbf k,\alpha,\omega}^{(\leqslant h)\pm}$ is
\begin{equation}
\psi_{\mathbf x,\alpha,\omega}^{(\leqslant h)\pm}:=\frac{1}{\beta|\Lambda|}\sum_{\mathbf k\in\mathcal B_{\beta,L}^{(\leqslant h,\omega)}}e^{\pm i(\mathbf k-\mathbf p_{F,0}^\omega)\cdot\mathbf x}\hat\psi_{\mathbf k,\alpha,\omega}^{(\leqslant h)\pm}.
\label{psixdeft}\end{equation}
\bigskip

{\bf Remark}: Unlike $\hat\psi_{\mathbf k,\alpha,\omega}$, the $_\omega$ index in $\psi_{\mathbf x,\alpha,\omega}^{(\leqslant h)\pm}$ is {\it not} redundant. Keeping track of this dependence is required to prove properties of $\hat W_{2l}(\mathbf k)$ and $\hat{\bar g}_h(\mathbf k)$ close to $\mathbf p_{F,0}^\omega$ while working in $\mathbf x$-space. Such considerations were first discussed in~\cite{benNZ} in which $\psi_{\mathbf x,\alpha,\omega}$ were called {\it quasi-particle} fields.\par
\bigskip

\indent We then define the propagator in $\mathbf x$-space:
\begin{equation}
\hat g_{h,\omega}(\mathbf x-\mathbf y):=\frac{1}{\beta|\Lambda|}\sum_{\mathbf k\in\mathcal B_{\beta,L}^{(\leqslant h,\omega)}}e^{i(\mathbf k-\mathbf p_{F,0}^\omega)\cdot(\mathbf x-\mathbf y)}\hat{\bar g}_{h,\omega}(\mathbf k)
\label{propxoh}\end{equation}
and similarly for $\bar g_{\leqslant h,\omega}$, and the effective potential~(\ref{hatWdeft}) becomes
\begin{equation}\begin{largearray}
\bar{\mathcal V}^{(h)}(\psi^{(\leqslant h)})=\sum_{l=2}^\infty\sum_{\underline\omega,\underline\alpha}\frac{1}{\beta|\Lambda|}\int d\mathbf x_1\cdots d\mathbf x_{2l}\ W^{(h)}_{2l,\underline\alpha,\underline\omega}(\mathbf x_1-\mathbf x_{2l},\cdots,\mathbf x_{2l-1}-\mathbf x_{2l})\cdot\\[0.5cm]
\hfill\cdot\psi^{(\leqslant h)+}_{\mathbf x_1,\alpha_1,\omega_1}\psi^{(\leqslant h)-}_{\mathbf x_2,\alpha_2,\omega_2}\cdots\psi^{(\leqslant h)+}_{\mathbf x_{2l-1},\alpha_{2l-1},\omega_{2l-1}}\psi^{(\leqslant h)-}_{\mathbf x_{2l},\alpha_{2l},\omega_{2l}}
\end{largearray}\label{Wdeft}\end{equation}
and
\begin{equation}
\mathcal Q^{(h)}(\psi^{(\leqslant h)})=\sum_{\omega,(\alpha,\alpha')}\int d\mathbf xd\mathbf y\ \psi^{(\leqslant h)+}_{\mathbf x,\omega,\alpha}W^{(h)}_{2,\omega,(\alpha,\alpha')}(\mathbf x-\mathbf y)\psi^{(\leqslant h)-}_{\mathbf y,\omega,\alpha'}
\label{Wtdeft}\end{equation}
in which
\begin{equation}\begin{largearray}
W^{(h)}_{2l,\underline\alpha,\underline\omega}(\mathbf u_1,\cdots,\mathbf u_{2l-1})\\[0.5cm]
\hfill:=\frac{\delta_{0,\sum_{j=1}^{2l}(-1)^{j}\mathbf p_{F,0}^{\omega_j}}}{(\beta|\Lambda|)^{2l-1}}\sum_{(\mathbf k_1,\cdots,\mathbf k_{2l-1})\in\mathcal B_{\beta,L}^{2l-1}} e^{i(\sum_{j=1}^{2l-1}(-1)^j(\mathbf k_j-\mathbf p_{F,0}^{\omega_j})\cdot\mathbf u_j)}\hat W^{(h)}_{2l,\underline\alpha}(\mathbf k_1,\cdots,\mathbf k_{2l-1}).
\end{largearray}\label{xspaceWo}\end{equation}
As in the ultraviolet, the definition of $\hat W_{2l,\underline\alpha}^{(h)}(\underline{\mathbf k})$ is extended to $\mathcal B_{\beta,L}^{2l-1}$ by defining it as the kernel of $\mathcal V^{*(h)}$:
\begin{equation}
-\beta|\Lambda|\mathfrak e_h-\mathcal V^{*(h)}(\Psi):=\sum_{N=1}^\infty\frac{(-1)^N}{N!}\bar{\mathcal E}_{h+1}^T(\mathcal V^{*(h+1)}(\psi^{(h+1)}+\Psi);N)
\label{extendedVo}\end{equation}
in which $\{\hat\Psi_{\mathbf k,\alpha}\}_{\mathbf k\in\mathcal B_{\beta,L},\alpha\in\mathcal A}$ is a collection of {\it external fields}. The definition~(\ref{xspaceWo}) suggests a definition for $\bar A_{h,\omega}$ (see~(\ref{Adefo}) and~(\ref{Adeft})):
\begin{equation}
\bar A_{h,\omega}(\mathbf x):=\frac{1}{\beta|\Lambda|}\sum_{\mathbf k\in\mathcal B_{\beta,L}}e^{i(\mathbf k-\mathbf p_{F,0}^\omega)\cdot\mathbf x}\hat{\bar A}_{h,\omega}(\mathbf k).
\label{Ahoox}\end{equation}
\bigskip

\point{Third regime} We now turn our attention to the third regime. As discussed in section~\ref{multiscalesec}, in addition to there being an extra index $j$, the Fermi points are also shifted in the third regime. The kernels of $\bar{\mathcal V}$ and $\mathcal Q$ are defined as in~(\ref{hatWdeft}), but with $\omega$ replaced by $(\omega,j)$. The $\mathbf x$-space representation of $\hat\psi_{\mathbf k,\alpha,\omega,j}^{(\leqslant h)\pm}$ is defined as
\begin{equation}
\psi_{\mathbf x,\alpha,\omega,j}^{(\leqslant h)\pm}:=\frac{1}{\beta|\Lambda|}\sum_{\mathbf k\in\mathcal B_{\beta,L}^{(\leqslant h,\omega,j)}}e^{\pm i(\mathbf k-\tilde{\mathbf p}_{F,j}^{(\omega,h)})\cdot\mathbf x}\hat\psi_{\mathbf k,\alpha,\omega,j}^{(\leqslant h)\pm}
\label{psixdefth}\end{equation}
and the $\mathbf x$-space expression of the propagator and the kernels of $\bar{\mathcal V}$ and $\mathcal Q$ are defined by analogy with the first regime:
\begin{equation}
\hat g_{h,\omega,j}(\mathbf x-\mathbf y):=\frac{1}{\beta|\Lambda|}\sum_{\mathbf k\in\mathcal B_{\beta,L}^{(\leqslant h,\omega,j)}}e^{i(\mathbf k-\tilde{\mathbf p}_{F,j}^{(\omega,h)})\cdot(\mathbf x-\mathbf y)}\hat{\bar g}_{h,\omega,j}(\mathbf k)
\label{propxothh}\end{equation}
and
\begin{equation}\begin{largearray}
W^{(h)}_{2l,\underline\alpha,\underline\omega,\underline j}(\mathbf u_1,\cdots,\mathbf u_{2l-1})
:=\frac{\delta_{0,\sum_{n=1}^{2l}(-1)^{n}\tilde{\mathbf p}_{F,j}^{(h,\omega_n)}}}{(\beta|\Lambda|)^{2l-1}}\cdot\\[0.5cm]
\hfill\cdot\sum_{(\mathbf k_1,\cdots,\mathbf k_{2l-1})\in\mathcal B_{\beta,L}^{2l-1}} e^{i(\sum_{n=1}^{2l-1}(-1)^n(\mathbf k_n-\tilde{\mathbf p}_{F,j}^{(\omega_n,h)})\cdot\mathbf u_j)}\hat W^{(h)}_{2l,\underline\alpha}(\mathbf k_1,\cdots,\mathbf k_{2l-1}).
\end{largearray}\label{xspaceWth}\end{equation}
In addition
\begin{equation}
\bar A_{h,\omega,j}(\mathbf x):=\frac{1}{\beta|\Lambda|}\sum_{\mathbf k\in\mathcal B_{\beta,L}}e^{i(\mathbf k-\tilde{\mathbf p}_{F,j}^{(\omega,h)})\cdot\mathbf x}\hat{\bar A}_{h,\omega,j}(\mathbf k).
\label{Ahojx}\end{equation}
\subseqskip

\subsection{Estimates of the free propagator}
\label{estpropsubsec}
\indent Before moving along with the tree expansion, we first compute a bound on $\hat g_{h}$ in the different regimes, which will be used in the following.\par
\bigskip

\point{Ultraviolet regime} We first study the ultraviolet regime, i.e. $h\in\{1,\cdots,M\}$.\par
\medskip
\subpoint{Fourier space bounds} We have
$$\hat A(\mathbf k)^{-1}:=-(ik_0\mathds1+H_0(k))^{-1}=-\frac{1}{ik_0}\left(\mathds1+\frac{H_0(k)}{ik_0}\right)^{-1}$$
and 
$$|\hat g_{h}(\mathbf k)|=|f_h(\mathbf k)\hat A^{-1}(\mathbf k)|\leqslant (\mathrm{const.})\ 2^{-h},$$
where $|\cdot|$ is  the operator norm.
Therefore
\begin{equation}
\frac{1}{\beta|\Lambda|}\sum_{\mathbf k\in\mathcal B_{\beta,L}^*}|\hat g_{h}(\mathbf k)|\leqslant (\mathrm{const}.).
\label{grambounduv}\end{equation}
Furthermore, for all $m_0+m_k\leqslant7$ (we choose the constant $7$ in order to get adequate bounds on the real-space decay of the free propagator, 
good enough for performing the localization and renormalization procedure described below; any other larger constant would yield identical results),
\begin{equation}
|2^{hm_0}\partial_{k_0}^{m_0}\partial_k^{m_k}\hat g_{h}(\mathbf k)|\leqslant (\mathrm{const.})\ 2^{-h}
\label{estgkuv}\end{equation}
in which $\partial_{k_0}$ denotes the discrete derivative with respect to $k_0$ and, with a slightly abusive notation, $\partial_k$ the discrete derivative with respect to either $k_1$ or $k_2$. Indeed the derivatives over $k$ land on $ik_0\hat A^{-1}$, which does not change the previous estimate, and the derivatives over $k_0$ either land on $f_h$, $1/(ik_0)$, or $ik_0\hat A^{-1}$, which yields an extra $2^{-h}$ in the estimate.\par
\medskip

{\bf Remark}: The previous argument implicitly uses the Leibnitz rule, which must be used carefully since the derivatives are discrete. However, since the estimate is purely dimensional,
we can replace the discrete discrete derivative with a continuous one without changing the order of magnitude of the resulting bound.\par
\bigskip

\subpoint{Configuration space bounds} We now prove that the inverse Fourier transform of $\hat g_{h}$
\begin{equation} g_{h}(\mathbf x):=\frac{1}{\beta|\Lambda|}\sum_{\mathbf k\in\mathcal B^*_{\beta,L}}\ e^{-i\mathbf k\cdot\mathbf x}\hat g_{h}(\mathbf k)\label{5.41bis}\end{equation}
satisfies the following estimate: for all $m_0+m_k\leqslant3$,
\begin{equation}
\int d\mathbf x\ x_0^{m_0}x^{m_k}|g_{h}(\mathbf x)|\leqslant (\mathrm{const.})\ 2^{-h-m_0h},
\label{estguv}\end{equation}
where we recall that $\int d\mathbf x$ is a shorthand for $\int_{0}^\beta dt \sum_{x\in \Lambda}$.
Indeed, note that the right side of \eqref{5.41bis} can be thought of as the Riemann sum approximation of 
\begin{equation} \int_{\mathbb R}\frac{dk_0}{2\pi}\int_{\hat\Lambda_\infty} \frac{d{ k}}{|\hat\Lambda_\infty|} e^{-i\mathbf k\cdot\mathbf x}\hat g_{h}(\mathbf k)\label{5.42bis}\end{equation}
where $\hat\Lambda_\infty=\{t_1 G_1+t_2 G_2: t_i\in[0,1)\}$ is the limit as $L\to\infty$ of $\hat \Lambda$, see \eqref{lae} and following lines. The dimensional estimates one finds using this continuum approximation are the same as those using \eqref{5.41bis} therefore, integrating \eqref{5.42bis} 7 times by parts and using (\ref{estgkuv}) we find
$$|g_{h}(\mathbf x)|\leqslant\frac{(\mathrm{const.})}{1+(2^{h}|x_0|+|x|)^7}$$
so that by changing variables in the integral over $x_0$ to $2^hx_0$, and using
$$\int d\mathbf x\ \frac{x_0^{m_0}x^{m_k}}{1+(|x_0|+|x|)^7}<(\mathrm{const.})$$
we find~(\ref{estguv}).\par
\bigskip

\point{First regime} We now consider the first regime, i.e. $h\in\{\mathfrak h_1+1,\cdots,\bar{\mathfrak h}_0\}$.\par
\medskip

\subpoint{Fourier space bounds} From~(\ref{freepropzo}) we find
$$|\hat g_{h,\omega}(\mathbf k)|\leqslant (\mathrm{const.})\ 2^{-h}$$
therefore
\begin{equation}
\frac{1}{\beta|\Lambda|}\sum_{\mathbf k\in\mathcal B_{\beta,L}^*}|\hat g_{h,\omega}(\mathbf k)|\leqslant (\mathrm{const}.)\ 2^{2h}
\label{gramboundo}\end{equation}
and for $m\leqslant7$,
\begin{equation}
|2^{mh}\partial_{\mathbf k}^m\hat g_{h,\omega}(\mathbf k)|\leqslant (\mathrm{const.})\ 2^{-h}
\label{estgko}\end{equation}
in which we again used the slightly abusive notation of writing $\partial_{\mathbf k}$ to mean any derivative with respect to $k_0$, $k_1$ or $k_2$. Equation~(\ref{estgko}) then follows from similar considerations as those in the ultraviolet regime.\par
\bigskip

\subpoint{Configuration space bounds} We estimate the real-space counterpart of $\hat g_{h,\omega}$, 
$$g_{h,\omega}(\mathbf x):=\frac{1}{\beta|\Lambda|}\sum_{\mathbf k\in\mathcal B_{\beta,L}^{(h,\omega)}}e^{-i(\mathbf k-\mathbf p_{F,0}^\omega)\cdot\mathbf x}\hat g_{h,\omega}(\mathbf k),$$
and find that for $m\leqslant3$,
\begin{equation}
\int d\mathbf x\ |\mathbf x^mg_{h,\omega}(\mathbf x)|\leqslant (\mathrm{const.})\ 2^{-(1+m)h}
\label{estgo}\end{equation}
which follows from very similar considerations as the ultraviolet estimate.\par
\bigskip

\point{Second regime} We treat the second regime, i.e. $h\in\{\mathfrak h_2+1,\cdots,\bar{\mathfrak h}_1\}$ in a very similar way (we skip the intermediate regime which can be treated in the same way as either the first or second regimes):
\begin{equation}
\frac{1}{\beta|\Lambda|}\sum_{\mathbf k\in\mathcal B_{\beta,L}^*}|\hat g_{h,\omega}(\mathbf k)|\leqslant (\mathrm{const}.)\ 2^{h+h_\epsilon}
\label{gramboundt}\end{equation}
and for all $m_0+m_k\leqslant7$,
\begin{equation}
\big|2^{m_0h}\partial_{k_0}^{m_0}2^{m_k\frac{h+h_\epsilon}2}\partial_k^{m_k}\hat g_{h,\omega}(\mathbf k)\big|\leqslant (\mathrm{const.})\ 2^{-h}
\label{estgkt}\end{equation}
where $h_\epsilon:=\log_2(\epsilon)$. Therefore for all $m_0+m_k\leqslant 3$,
\begin{equation}
\int d\mathbf x\ |x_0^{m_0}x^{m_k}g_{h,\omega}(\mathbf x)|\leqslant (\mathrm{const.})\ 2^{-h-m_0h-m_k\frac{h+h_\epsilon}2}.
\label{estgt}\end{equation}
\bigskip

\point{Third regime} Finally, the third regime, i.e. $h\in\{\mathfrak h_3+1,\cdots,\bar{\mathfrak h}_2\}$:
\begin{equation}
\frac{1}{\beta|\Lambda|}\sum_{\mathbf k\in\mathcal B_{\beta,L}^*}|\hat g_{h,\omega}(\mathbf k)|\leqslant (\mathrm{const}.)\ 2^{2h-2h_\epsilon}
\label{gramboundth}\end{equation}
and for all $m_0+m_k\leqslant7$,
\begin{equation}
|2^{m_0h}\partial_{k_0}^{m_0}2^{m_k(h-h_\epsilon)}\hat g_{h,\omega,j}(\mathbf k)|\leqslant (\mathrm{const.})\ 2^{-h}.
\label{estgkth}\end{equation}
Therefore for all $m_0+m_k\leqslant 3$,
\begin{equation}
\int d\mathbf x\ |x_0^{m_0}x^{m_k}g_{h,\omega,j}(\mathbf x)|\leqslant (\mathrm{const.})\ 2^{-h-m_0h-m_k(h-h_\epsilon)}
\label{estgth}\end{equation}
where
$$g_{h,\omega,j}(\mathbf x):=\frac{1}{\beta|\Lambda|}\sum_{\mathbf k\in\mathcal B_{\beta,L}^{(h,\omega,j)}}e^{-i(\mathbf k-\tilde{\mathbf p}_{F,j}^{(\omega,h+1)})\cdot\mathbf x}\hat g_{h,\omega}(\mathbf k).$$
\seqskip

\section{Tree expansion and constructive bounds}
\label{treeexpsec}
\indent In this section, we shall define the Gallavotti-Nicol\`o tree expansion~\cite{galEFi}, and show how it can be used to compute bounds for the $\mathfrak e_h$, $\mathcal V^{(h)}$,
$\mathcal Q^{(h)}$ and $\bar{\mathcal V}^{(h)}$ defined above in~(\ref{effpotuv}) and (\ref{effpotoh}), using the estimates~(\ref{estguv}), (\ref{estgo}), (\ref{estgt}) and~(\ref{estgth}). We follow~\cite{benNZ, genZO, giuOZ}.
We conclude the section by showing how to compute the terms in $\bar{\mathcal W}^{(h)}$ that are quadratic in $\hat J_{\mathbf k,\underline\alpha}$ from $\mathcal V^{(h)}$ and $\hat{\bar g}_h$.\par
\bigskip

\indent The discussion in this section is meant to be somewhat general, in order to be applied to the ultraviolet, first, second and third regimes (except for lemma~\ref{powercountinglemma} which does not apply to the ultraviolet regime).\par
\subseqskip

\subsection{Gallavotti-Nicol\`o Tree expansion}
\indent In this section, we will define a tree expansion to re-express equations {\it of the type}
\begin{equation}
-v^{(h)}(\psi^{(\leqslant h)})-\mathcal V^{(h)}(\psi^{(\leqslant h)})=\sum_{N=1}^\infty\frac{(-1)^N}{N!}\mathcal E^T_{h+1}\left(\mathcal V^{(h+1)}(\psi^{(\leqslant h)}+\psi^{(h+1)});N\right)
\label{treeindeq}\end{equation}
for $h\in\{h_2^*,\cdots,h_1^*-1\}$ (in the ultraviolet regime $h_2^*=\bar{\mathfrak h}_0$, $h_1^*=M$; in the first $h_2^*=\mathfrak h_1$, $h_1^*=\bar{\mathfrak h}_0$; in the second $h_2^*=\mathfrak h_2$, $h_1^*=\bar{\mathfrak h}_1$; 
and in the third, $h_2^*=\mathfrak h_\beta$, $h_1^*=\bar{\mathfrak h}_2$), with
\begin{equation} \left\{\begin{array}{>{\displaystyle}l}
\mathcal V^{(h)}(\psi^{(\leqslant h)})=\sum_{l=q}^\infty\sum_{\underline\varpi}\int d\underline{\mathbf x}\ W_{2l,\underline\varpi}^{(h)}(\underline{\mathbf x})\psi^{(\leqslant h)+}_{\mathbf x_1,\varpi_1}\psi^{(\leqslant h)-}_{\mathbf x_2,\varpi_2}\cdots\psi^{(\leqslant h)+}_{\mathbf x_{2l-1},\varpi_{2l-1}}\psi^{(\leqslant h)-}_{\mathbf x_{2l},\varpi_{2l}}\\[0.5cm]
v^{(h)}(\psi^{(\leqslant h)})=\sum_{l=0}^{q-1}\sum_{\underline\varpi}\int d\underline{\mathbf x}\ W_{2l,\underline\varpi}^{(h)}(\underline{\mathbf x})\psi^{(\leqslant h)+}_{\mathbf x_1,\varpi_1}\psi^{(\leqslant h)-}_{\mathbf x_2,\varpi_2}\cdots\psi^{(\leqslant h)+}_{\mathbf x_{2l-1},\varpi_{2l-1}}\psi^{(\leqslant h)-}_{\mathbf x_{2l},\varpi_{2l}}
\end{array}\right.\label{defW}\end{equation}
($q=1$ in the ultraviolet regime and $q=2$ in the first, second and third) in which $\underline\varpi$ and $\underline{\mathbf x}$ are shorthands for $(\varpi_1,\cdots,\varpi_{2l})$ and $(\mathbf x_1,\cdots,\mathbf x_{2l})$; $\varpi$ denotes a collection of indices: $(\alpha,\omega)$ in the first and second regimes, $(\alpha,\omega,j)$ in the third, and $(\alpha)$ in the ultraviolet; and $W_{2l,\underline\varpi}^{(h)}(\underline{\mathbf x})$ is a function that only depends on the differences $\mathbf x_i-\mathbf x_j$.
The propagator associated with $\mathcal E^T_{h+1}$ will be denoted $g_{(h+1),(\varpi,\varpi')}(\mathbf x-\mathbf x')$ and is to be interpreted as the dressed propagator $\bar g_{(h+1,\omega),(\alpha,\alpha')}$ in the first and second regimes, 
and as $\bar g_{(h+1,\omega,j),(\alpha,\alpha')}$ in the third. Note in particular that in the first and second regimes the propagator is diagonal in the $\omega$ indices, and is diagonal in $(\omega,j)$ in the third. In all cases, we write
\begin{equation} g_{(h+1),(\varpi,\varpi')}(\mathbf x-\mathbf x')=\frac1{\beta|\Lambda|}\sum_{{\bf k}\in\mathcal B_{\beta,L}}e^{-i({\bf k}-\mathbf p^{(h+1)}_{\varpi})(\mathbf x-\mathbf x')}\hat g_{(h+1),(\varpi,\varpi')}({\bf k})\;,\label{6.3}\end{equation}
where $\mathbf p^{(h+1)}_{\varpi}$ should be interpreted as ${\bf 0}$ in the ultraviolet regime, as $\mathbf p_{F,0}^\omega$ in the first and second, and as $\tilde{\mathbf p}_{F,j}^{(\omega,h+1)}$ in the third, see \eqref{eq4.20bis}. 
\par
\bigskip

{\bf Remark}: The usual way of computing expressions of the form~(\ref{treeindeq}) is to write the right side as a sum over Feynman diagrams. The tree expansion detailed below provides a way of identifying the sub-diagrams that scale in the same way (see the remark at the end of this section). In the proofs below, there will be no mention of Feynman diagrams, since a diagramatic expansion would yield insufficient bounds.\par
\bigskip

\indent We will now be a bit rough for a few sentences, in order to carry the main idea of the tree expansion across: equation~(\ref{treeindeq}) is an inductive equation for the $\mathcal V^{(h)}$, which we will pictorially think of as the {\it merging} of a selection of $N$ potentials $\mathcal V^{(h+1)}$ via a truncated expectation. If we iterate~(\ref{treeindeq}) all the way to scale $h^*_2$, then we get a set of {\it merges} that {\it fit} into each other, creating a tree structure. The sum over the choice of $N$'s at every step will be expressed as a sum over Gallavotti-Nicol\`o trees, which we will now define precisely.

\begin{figure}
\hfil\includegraphics[width=11cm]{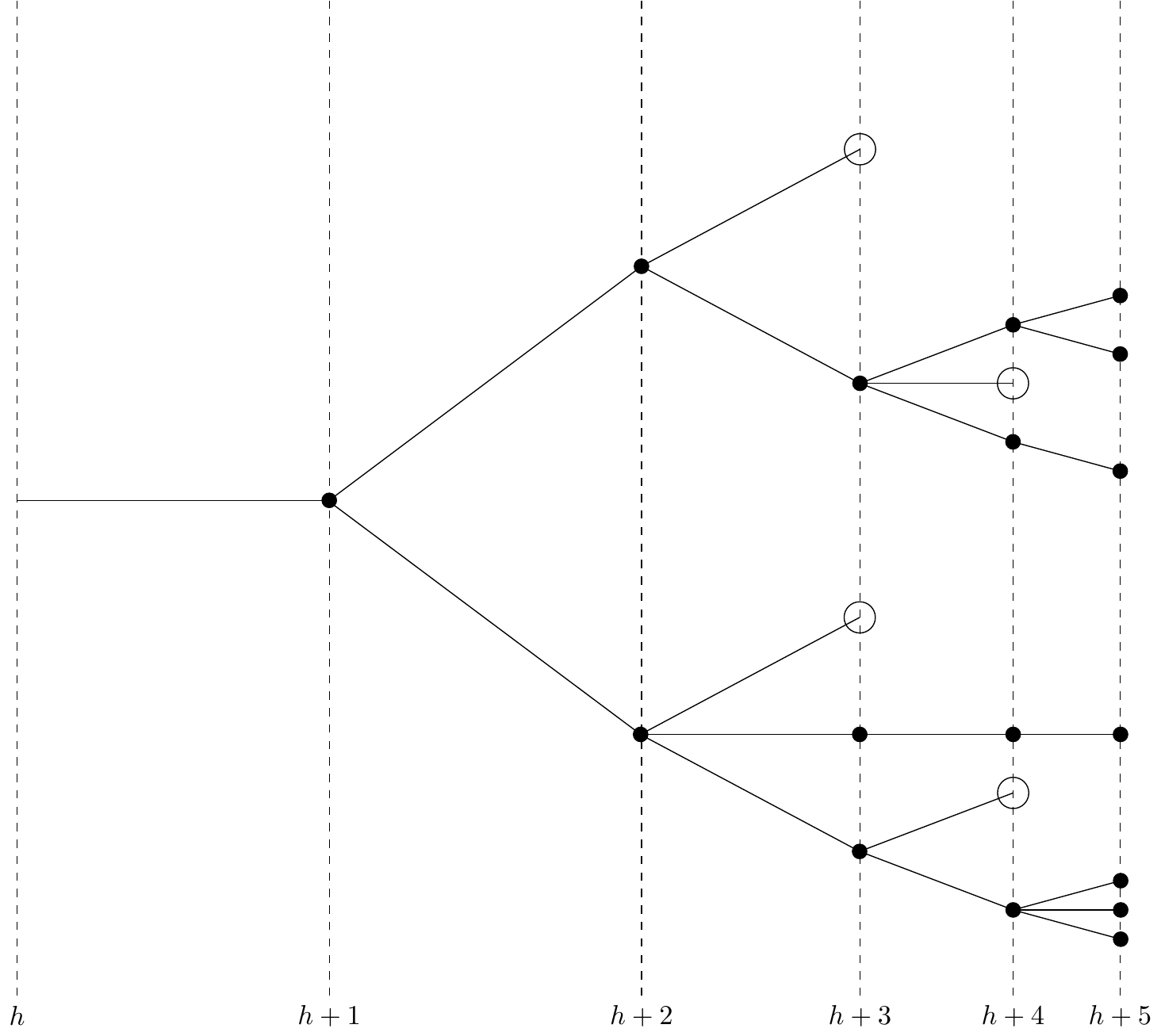}\par
\caption{example of a tree on scale $h$ up to scale $h^*_1+1=h+5$ with 11 leaves, 5 of which are local and 6 irrelevant. Local leaves are represented as empty circles, whereas irrelevant leaves are represented as full circles.}
\label{treefig}\end{figure}

Given a scale $h\in\{h^*_2,\cdots,h^*_1-1\}$
and an integer $N\geqslant 1$, we define the set $\mathcal T_{N}^{(h)}$ of Gallavotti-Nicol\`o (GN) trees as a set of {\it labeled} rooted trees with $N$ leaves in the following way.
\begin{itemize}
\item We define the set of unlabeled trees inductively: we start with a {\it root}, that is connected to a node $v_0$ that we will call the {\it first node}
of the tree; every node is assigned an ordered set of child nodes. $v_0$ must have at least one child, while the other nodes may be childless. 
We denote the parent-child partial ordering by $v'\prec v$ ($v'$ is the parent of $v$). The nodes that have no children are called {\it leaves} or {\it endpoints}. By convention, the root is not considered to be a node, but we will still call it the parent of $v_0$.
\item Each node is assigned a {\it scale label} $h'\in\{h+1,\cdots,h^*_1+1\}$ and the root is assigned the {\it scale label} $h$, in such a way that the children of the root or of a node on scale $h'$ are on scale $h'+1$ (keep in mind that it is possible for a node to have a single child).
\item The leaves whose scale is $\leqslant h_1^*$ are called {\it local}. The leaves on scale $h_1^*+1$ can either be {\it local} or {\it irrelevant} (see figure~\ref{treefig}).
\item Every local leaf must be preceeded by a {\it branching node}, i.e. a node with at least two children. In other words, every local leaf must have at least one sibling.

\item We denote the set of nodes of a tree $\tau$ by $\bar{\mathfrak V}(\tau)$, the set of nodes that are not leaves by $\mathfrak V(\tau)$ and the set of leaves by $\mathfrak E(\tau)$.
\end{itemize}
\bigskip

{\bf Remark}: Local leaves are called ``local'' because those nodes are usually applied a {\it localization} operation (see e.g. \cite{benNFi}). In the present case, such a step is not needed, due to the super-renormalizable nature of the first and third regimes.\par
\bigskip

\indent Every node of a Gallavotti-Nicol\`o tree $\tau$ corresponds to a truncated expectation of effective potentials of the form~(\ref{treeindeq}). If one expands the product of factors of the form $(\psi_{\mathbf x,\varpi}^{(\leqslant h)\pm}+\psi_{\mathbf x,\varpi}^{(h+1)\pm})$ in every term in the right side of~(\ref{treeindeq}), then one finds a sum over {\it choices} between $\psi^{(\leqslant h)}$ and $\psi^{(h)}$ for every $(\mathbf x,\varpi,\pm)$. We will express this sum as a sum over a set of {\it external field labels} (corresponding to the labels of $\psi^{(\leqslant h)}$ which are called external because they can be factored out of the truncated expectation) defined in the following way. Given an integer $\ell_0\geqslant q$, whose purpose will become clear in lemma~\ref{powercountinglemma}\ (we will choose $\ell_0$ to be $=1$ in the ultraviolet regime, and $=2,3,2$ in the first, second, third infrared regimes, respectively), a tree $\tau\in\mathcal T_N^{(h)}$ whose endpoints are denoted by $(v_1,\cdots,v_N)$, as well as a collection of integers $\underline l_\tau:=(l_{v_1},\cdots,l_{v_N})\in\mathbb N^N$ such that $l_{v_i}\geqslant q$ and, if $v_i$ is a local leaf, $l_{v_i}<\ell_{0}$ (in particular, if $\ell_0=q$ there are no local leaves), we introduce an ordered collection of {\it fields}, i.e. triplets
\begin{equation}
F=\left((\mathbf x_1,\varpi_1,+),(\mathbf x_2,\varpi_2,-),\cdots,(\mathbf x_{2L-1},\varpi_{2L-1},+),(\mathbf x_{2L},\varpi_{2L},-)\right).
\label{Ftriplets}\end{equation}
where $L:=l_{v_1}+\cdots+l_{v_N}$. We then define the set of {\it external field labels} of each endpoint $v_i$ as the following ordered collections of integers
$$I_{v_1}:=\left(1,\cdots,2l_{v_1}\right),\cdots,
I_{v_N}:=\left(2l_{v_{N-1}}+1,\cdots,2l_{v_N}\right).$$
We define the set $\mathcal P_{\tau,\underline l_\tau,\ell_0}$ of {\it external field labels} compatible with a tree $\tau\in\mathcal T_N^{(h)}$ as the set of all the collections $\mathbf P=\{P_v\}_{v\in\mathfrak V(\tau)}$ where $P_v$ are themselves collections of integers that satisfy the following constraints:
\begin{itemize}
\item For every $v\in\mathfrak V(\tau)$ whose children are $(v_1,\cdots,v_s)$, $P_v\subset P_{v_1}\cup\cdots\cup P_{v_s}$ in which, by convention, if $v_i$ is an endpoint then $P_{v_i}=I_{v_i}$; and the order of the elements of $P_v$ is that of $P_{v_1}$ through $P_{v_s}$ (in particular the integers coming from $P_{v_1}$ precede those from $P_{v_2}$ and so forth).
\item For all $v\in\mathfrak V(\tau)$, $P_v$ must contain as many even integers as odd ones (even integers correspond to fields with a $-$, and odd ones to a $+$).
\item If $v$ has more than one child, then $P_v\neq P_{v'}$ for all $v'\succ v$
\item For all $v\in\bar{\mathfrak V}(\tau)\setminus\{v_0\}$ which is not a local leaf, the cardinality of $P_v$ must satisfy $|P_v|\geqslant2\ell_0$. 
\end{itemize}
Furthemore, given a node $v$ whose children are $(v_1,\cdots,v_s)$, we define $R_v:=\bigcup_{i=1}^sP_{v_i}\setminus P_v$.\par
\bigskip

\indent We associate a {\it value} to each node $v$ of such a tree in the following way. If $v$ is a leaf, then its value is
\begin{equation}
\rho_v:=W_{|P_v|,\underline\varpi_v}^{(h_v-1)}(\underline{\mathbf x}_v)
\label{valend}\end{equation}
where $|P_v|$ denotes the cardinality of $P_v$, and $\underline\varpi_v$ and $\underline{\mathbf x}_v$ are the {\it field labels} (i.e. elements of $F$) specified by the indices in $P_v$. If $v$ is not a leaf and $R_v\neq \emptyset$, then its value is
\begin{equation}
\rho_v:=\sum_{T_v\in\mathbf T(\mathbf R_v)}\sigma_{T_v}\prod_{l\in T_v}g_{(h_v),l}\int dP_{T_v}(\mathbf t)\ \det G^{(T_v,h_v)}(\mathbf t)=:\sum_{T_v\in\mathbf T(\mathbf R_v)}\rho_v^{(T_v)}
\label{valmid}\end{equation}
where $\mathbf T(\mathbf R_v)$, $g_{(h_v),l}$, $dP_{T_v}(\mathbf t)$ and $G^{(T_v,h_v)}$ are defined as in lemma~\ref{detexplemma}\ with $g$ replaced by $g_{h_v}$, and if the children of $v$ are denoted by $(v_1,\cdots,v_s)$, then $\mathbf R_v:=(P_{v_1}\setminus P_v,\ldots, P_{v_s}\setminus P_v)$. If $v$ is not a leaf and $R_v=\emptyset$, then it has exactly one child and we let its value be $\rho_v=1$.
\par
\subseqskip

\Theo{Lemma}\label{treexplemma}
Equation~(\ref{treeindeq}) can be re-written as
\begin{equation}
-v^{(h)}(\psi^{(\leqslant h)})-\mathcal V^{(h)}(\psi^{(\leqslant h)})=\sum_{N=1}^\infty\sum_{\tau\in\mathcal T_{N}^{(h)}}\sum_{\underline l_\tau}\sum_{\underline\varpi_\tau}\int d\underline{\mathbf x}_\tau\ \sum_{\mathbf P\in\mathcal P_{\tau,\underline l_\tau,\ell_0}}\Psi^{(\leqslant h)}_{P_{v_0}}\prod_{v\in\bar{\mathfrak V}(\tau)}\frac{(-1)^{s_v}}{s_v!}\rho_v
\label{treeexp}\end{equation}
where $\underline l_\tau:=(l_{v_1},\cdots,l_{v_N})$ (see above),
$\underline\varpi_\tau$ and $\underline{\mathbf x}_\tau$ are the field labels in $F$,
$s_v$ is the number of children of $v$, $\rho_v$ was defined above in~(\ref{valend}) and~(\ref{valmid}), $v_0$ is the first node of $\tau$ and
$$\Psi^{(\leqslant h)}_{P_{v_0}}:=\prod_{i\in P_{v_0}}\psi^{(\leqslant h)\epsilon_i}_{\mathbf x_i,\varpi_i}$$
where $\epsilon_i$ is the third component of the $i$-th triplet in $F$.
\endtheo
{\bf Remark}: The sum over $\mathbf P\in\mathcal P_{\tau,\underline l_\tau,\ell_0}$ is a sum over the assignement of $P_v$ for nodes that are not endpoints. The sets $I_v$ are not summed over, instead they are fixed by $\underline l_\tau$. Furthermore, if $\mathcal P_{\tau,\underline l_\tau,\ell_0}=\emptyset$ (e.g. if $\ell_0=q$ and $\tau$ contains local leaves), then the sum should be interpreted as 0.\par
\bigskip

By injecting~(\ref{valmid}) into~(\ref{treeexp}), we can re-write
\begin{equation}
\begin{largearray}
-v^{(h)}(\psi^{(\leqslant h)})-\mathcal V^{(h)}(\psi^{(\leqslant h)})\\[0.2cm]
\hfill=\sum_{N=1}^\infty\sum_{\tau\in\mathcal T_{N}^{(h)}}\sum_{T\in\mathbf T(\tau)}\sum_{\underline l_\tau}\sum_{\underline\varpi_\tau}\int d\underline{\mathbf x}_\tau\ \sum_{\mathbf P\in\mathcal P_{\tau,\underline l_\tau,\ell_0}}\Psi^{(\leqslant h)}_{P_{v_0}}\prod_{v\in\bar{\mathfrak V}(\tau)}\frac{(-1)^{s_v}}{s_v!}\rho_v^{(T_v)}
\end{largearray}
\label{treeexpT}\end{equation}
where $\mathbf T(\tau)$ is the set of collections of $(T_v\in\mathbf T(\mathbf R_v))_{v\in\mathfrak V(\tau)}$. Moreover, while $\rho^{(T_v)}_v$ was defined in~(\ref{valmid}) if $v\in\mathfrak V(\tau)$, it stands for $\rho_v$ if $v\in\mathfrak E(\tau)$ (note that in this case $T_v=\emptyset$).\par
\bigskip

\indent\underline{Idea of the proof}: The proof of this lemma can easily be reconstructed from the schematic description below. We do not present it in full detail here because its
proof has already been discussed in several references, among which~\cite{benNFi,genZO,giuOZh}.\par
\bigskip

\indent The lemma follows from an induction on $h$, in which we write the truncated expectation in the right side of~(\ref{treeindeq}) as
$$\begin{largearray}
\sum_{l_1,\cdots,l_N}\sum_{\underline\varpi_1,\cdots,\underline\varpi_N}\int d\underline{\mathbf x}_1\cdots d\underline{\mathbf x}_N\ 
W_{2l_1,\underline\varpi_1}^{(h+1)}(\underline{\mathbf x}_1)\cdots W_{2l_N,\underline\varpi_N}^{(h+1)}(\underline{\mathbf x}_N)\cdot\\[0.5cm]
\hskip1cm\cdot\mathcal E_{h+1}^T\left(\prod_{j=1}^{l_1}(\psi_{x_{1,2j-1},\varpi_{1,2j-1}}^{(\leqslant h)+}+\psi_{x_{1,2j-1},\varpi_{1,2j-1}}^{(h+1)+})(\psi_{x_{1,2j},\varpi_{1,2j}}^{(\leqslant h)-}+\psi_{x_{1,2j},\varpi_{1,2j}}^{(h+1)-}),\cdots\right.\\[0.5cm]
\hfill\left.\cdots,\prod_{j=1}^{l_N}(\psi_{x_{N,2j-1},\varpi_{N,2j-1}}^{(\leqslant h)+}+\psi_{x_{N,2j-1},\varpi_{N,2j-1}}^{(h+1)+})(\psi_{x_{N,2j},\varpi_{N,2j}}^{(\leqslant h)-}+\psi_{x_{N,2j},\varpi_{N,2j}}^{(h+1)-})\right)
\end{largearray}$$
which yields a sum over the choices between $\psi^{(\leqslant h)}$ and $\psi^{(h+1)}$, with each choice corresponding to an instance of $P_v$: each $\psi^{(\leqslant h)\epsilon}_{\mathbf x,\varpi}$ ``creates'' the element $(\mathbf x,\varpi,\epsilon)$ in $P_v$. The remaining truncated expectation is then computed by applying lemma~\ref{detexplemma}. Finally, the $W_{2l_j,\underline\varpi_j}^{(h+1)}$ with $l_j<\ell_0$ are left as such, and yield a {\it local leaf} in the tree expansion, the others are expanded using the inductive hypothesis.\par
\bigskip

\begin{figure}
\hfil\includegraphics[height=6cm]{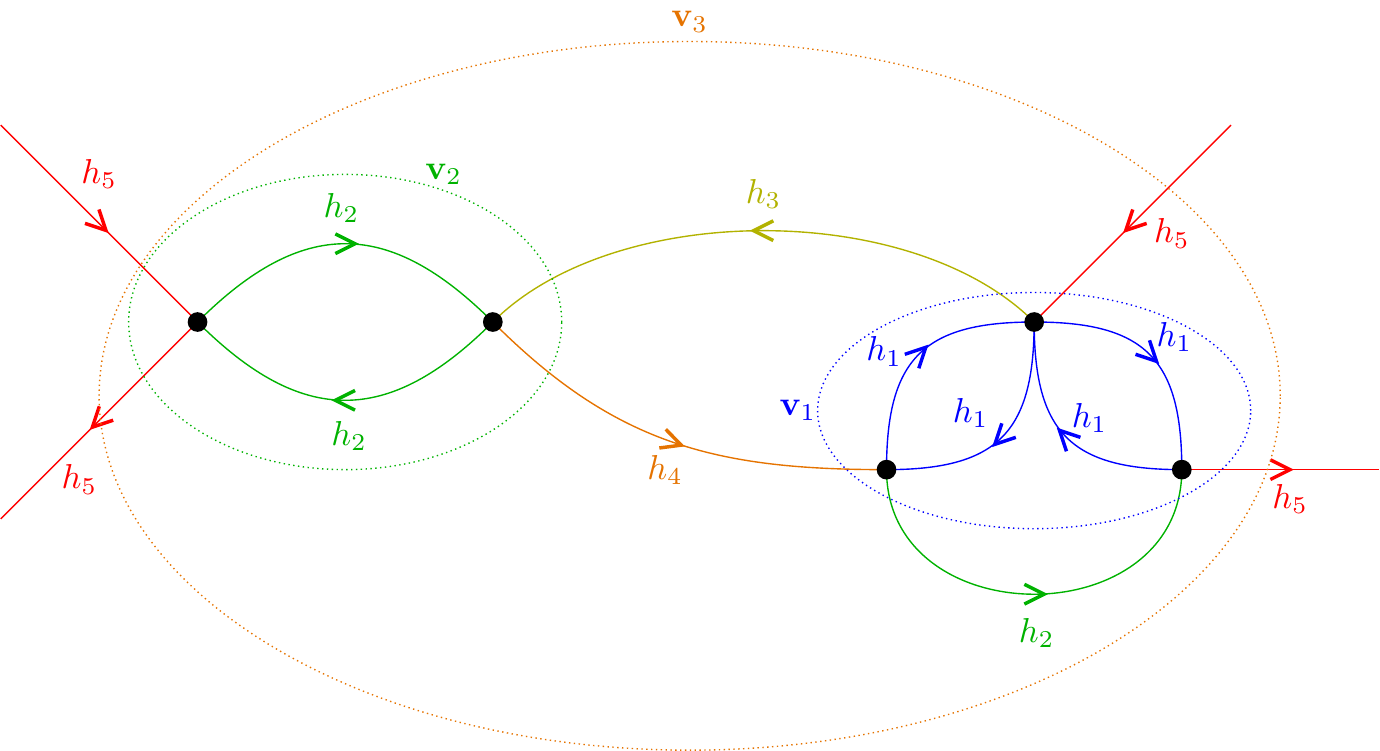}\par
\bigskip
\hfil\includegraphics[height=6cm]{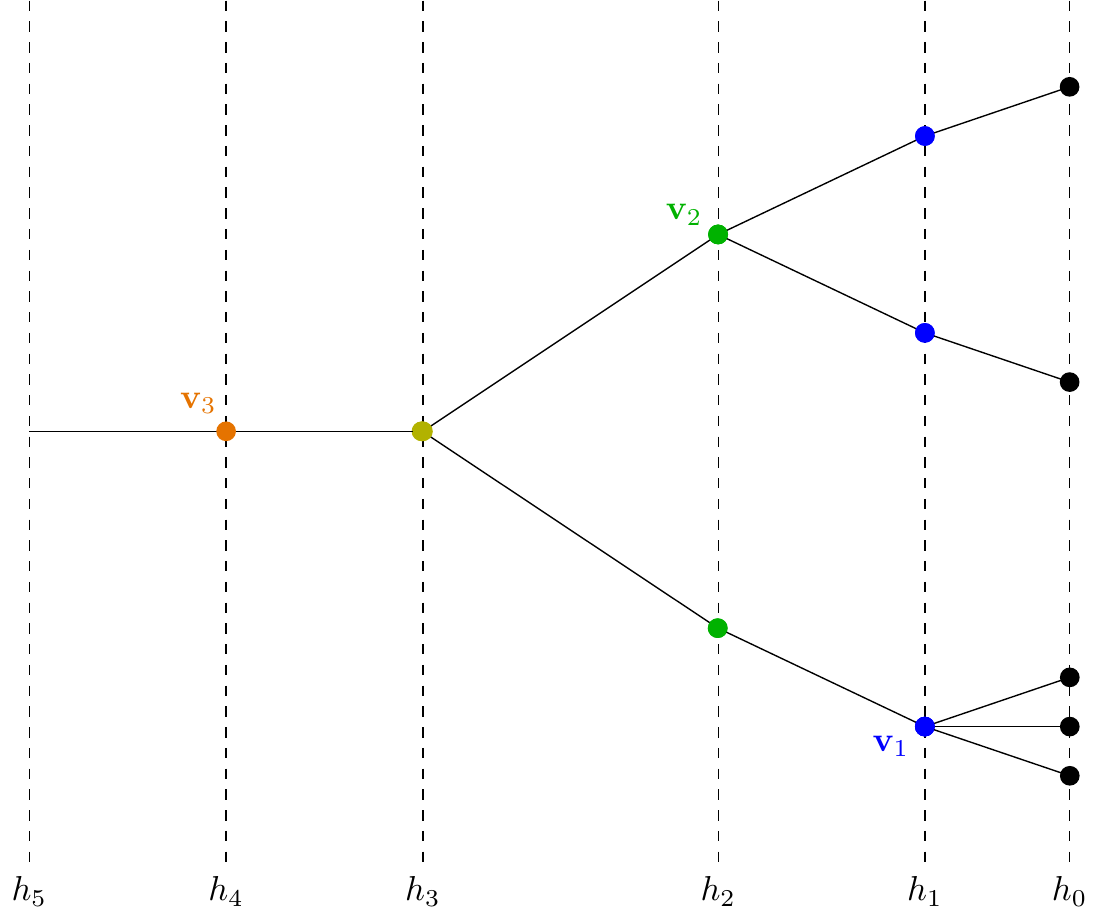}
\caption{Example of a labeled Feynman diagram and its corresponding tree. Three clusters, denoted by $\mathbf v_1$, $\mathbf v_2$ and $\mathbf v_3$, on scale $h_1$, $h_2$ and $h_4$ respectively, are explicitely drawn as dotted ellipses. There are 4 more clusters (2 on scale $h_1$, 1 on scale $h_2$ and 1 on scale $h_3$) which are not represented. The scales are drawn in different colors (color online): {\color{\colorfi}red} for $h_5$, {\color{\colorf}orange} for $h_4$, {\color{\colorth}yellow} for $h_3$, {\color{\colort}green} for $h_2$ and {\color{\coloro}blue} for $h_1$.}
\label{feyntreefig}\end{figure}

{\bf Remark}: For readers who are familiar with Feynman diagram expansions, it may be worth pointing out that a Gallavotti-Nicol\`o tree paired up with a set of external field labels $\mathbf P$
represents a class of labeled 
Feynman diagrams (the labels being the scales attached to the lines, or equivalently to the propagators)
with similar scaling properties. In fact, given a labeled Feynman diagram, one defines a tree and a set of external field labels by the following procedure. For every $h$, we define the {\it clusters on scale $h$} as the connected components of the diagram one obtains by removing the lines with a scale label that is $< h$. We assign a node with scale label $h$ to every cluster on scale $h$. The set $P_v$ contains the indices of the legs of the Feynman diagram that exit the corresponding cluster. If a cluster on scale $h$ contains a cluster on scale $h+1$, then we draw a branch between the two corresponding nodes. See figure~\ref{feyntreefig} for an example.\par
\indent Local leaves correspond to clusters that have {\it few} external legs. They are considered as ``black boxes'': the clusters on larger scales contained inside them are discarded.\par
\indent A more detailed discussion of this correspondence can be found in~\cite[section~5.2]{genZO} among other references.
\subseqskip

\subsection{Power counting lemma}
\label{powercountingsec}
\indent We will now state and prove the {\it power counting lemma}, which is an important step in bounding the elements in the tree expansion~(\ref{treeexpT}) in the first, second and third regimes.\par
\bigskip

\indent In the following, we will use a slightly abusive notation: given $\underline{\mathbf x}=(\mathbf x_1,\ldots,\mathbf x_{n})$, we will write $\underline{\mathbf x}^m$ to mean ``any of the products of the following form''
$$x_{j_1,i_1}\cdots x_{j_m,i_m}$$
where $i_\nu\in\{0,1,2\}$ indexes the components of $\mathbf x$ and $j_\nu\in\{1,\cdots,n\}$ indexes the components of $\underline{\mathbf x}$. We will also denote the translate of $\underline{\mathbf x}$ by $\mathbf y$ by $\underline{\mathbf x}-\mathbf y\equiv(\mathbf x_1-\mathbf y,\ldots,\mathbf x_{n}-\mathbf y)$. Furthermore, given $\underline{\mathbf x}^m$, we define the vector $\underline m$ whose $i$-th component is the number of occurrences of $x_{\cdot,i}$ in the product $\underline{\mathbf x}^m$ (note that $m_0+m_1+m_2=m$).
\par
\bigskip

\indent The power counting lemma will be stated as an inequality on the so-called {\it beta function} of the renormalization group flow, defined as
\begin{equation}
B_{2l,\underline\varpi}^{(h)}(\underline{\mathbf x}):=\left\{\begin{array}l
W_{2l,\underline\varpi}^{(h)}(\underline{\mathbf x})-W_{2l,\underline\varpi}^{(h+1)}(\underline{\mathbf x})\quad\mathrm{if\ }l\geqslant q\\[0.5cm]
W_{2l,\underline\varpi}^{(h)}(\underline{\mathbf x})\quad\mathrm{if\ }l<q.\end{array}\right.
\label{betafundef}\end{equation}
In terms of the tree expansion~(\ref{treeexp}), $B_{2l}^{(h)}$ is the sum of the contributions to $W_{2l}^{(h)}$ whose field label assignment $\mathbf P$ is such that every node $v\in\mathfrak V(\tau)\setminus\{v_0\}$ that is connected to the root by a chain of nodes with only one child satisfies $|P_v|>2l$. We denote the set of such field label assignments by $\tilde{\mathcal P}_{\tau,\underline l_\tau,\ell_0}$ for any given $\tau$, $\underline l_\tau$ and $\ell_0$. In other words, $B_{2l}^{(h)}$ contains all the contributions that have at least one propagator on scale $h+1$. If $l<q$, then all the contributions have a propagator on scale $h+1$, so $B_{2l}=W_{2l}$.\par
\bigskip

\Theo{Lemma}\label{powercountinglemma}
Assume that the propagator $g_{(h),(\varpi,\varpi')}(\mathbf x-\mathbf x')$ can be written as in \eqref{6.3}.
Given $h\in \{h_2^*,\cdots,h^*_1-1\}$, 
if $\forall m\in\{0,1,2,3\}$, and
\begin{equation}\left\{\begin{array}{>{\displaystyle}l}
\int d\mathbf x\ |\mathbf x^mg_{h'}(\mathbf x)|\leqslant C_g2^{-c_gh'}\mathfrak F_{h'}(\underline m)\\[0.5cm]
\frac{1}{\beta|\Lambda|}\sum_{{\bf k}\in\mathcal B_{\beta,L}}|\hat g_{h'}({\bf k})|\leqslant C_G2^{(c_k-c_g)h'}
\end{array}\right.,\qquad \forall h'\in\{h+1,\cdots,h^*_1\},\label{assumprop}\end{equation}
where $c_g$, $c_k$, $C_g$ and $C_G$ are constants, independent of $h$, and $\mathfrak F_{h'}(\underline m)$ is a shorthand for
$$A_0^{m_0}A_1^{m_1}A_2^{m_2}2^{-h'(d_0m_0+d_1m_1+d_2m_2)}$$
in which $A_0,A_1,A_2>0$, $d_0,d_1,d_2\geqslant0$, and $m_i$ is the number of times any of the $x_{j,i}$ appears in $\mathbf x^m$;
if
\begin{equation} \ell_0>\frac{c_k}{c_k-c_g}\label{6.ell0}\end{equation}
and
\begin{equation}
\begin{largearray}
\frac{1}{\beta|\Lambda|}\int d\underline{\mathbf x}\ \left|(\underline{\mathbf x}-\mathbf x_{2l})^mW_{2l,\underline\varpi}^{(h')}(\underline{\mathbf x})\right|\leqslant\mathfrak C_{2l}|U|^{\max(1,l-1)}2^{h'(c_k-(c_k-c_g)l)}\mathfrak F_{h'}(\underline m),\\[0.2cm]
\hfill \forall h'\in\{h+1,\cdots,h^*_1\}
\end{largearray}\label{assumW}\end{equation}
where $q\leqslant l<\ell_0$ for $h'<h^*_1$, $l\geqslant q$ for $h'=h_1^*$
(in particular, if $q\geqslant\ell_0$, then $h'=h_1^*$),
and $\mathfrak C_{2l}$ are constants,
then
\begin{equation}\begin{largearray}
\frac{1}{\beta|\Lambda|}\int d\underline{\mathbf x}\ \left|(\underline{\mathbf x}-\mathbf x_{2l})^mB_{2l,\underline\varpi}^{(h)}(\underline{\mathbf x})\right|\leqslant 2^{h(c_k-(c_k-c_g)l)}\mathfrak F_h(\underline m)(C_3C_G^{-1})^l\sum_{N=1}^\infty\sum_{\tau\in\mathcal T_N^{(h)}}\sum_{\underline l_\tau}\\[0.2cm]
\hfill\sum_{\displaystyle\mathop{\scriptstyle \mathbf P\in\tilde{\mathcal P}_{\tau,\underline l_\tau,\ell_0}}_{|P_{v_0}|=2l}}\kern-10pt C_1^N(C_gC_G^{-1})^{N-1}\left(\prod_{v\in\mathfrak V(\tau)}2^{(c_k-(c_k-c_g)\frac{|P_v|}2)}\right)\left(\prod_{v\in\mathfrak E(\tau)}(C_2C_G)^{l_v}\mathfrak C_{2l_v}|U|^{\max(1,l_v-1)}\right)
\end{largearray}\label{powercountingl}\end{equation}
where $C_1$, $C_2$ and $C_3$ are constants, independent of $c_g$, $c_k$, $C_g$, $C_G$ and $h$.
\endtheo
\bigskip

{\bf Remarks}: Here are a few comments about this lemma.
\begin{itemize}
\item Combining this lemma with~(\ref{betafundef}) yields a bound on $W_{2l,\varpi}^{(h)}(\underline{\mathbf x})$. In particular, if $l\geqslant\ell_0$ and $h<h^*_1$, then
\end{itemize}
\begin{equation}\begin{largearray}
\frac{1}{\beta|\Lambda|}\int d\underline{\mathbf x}\ \left|(\underline{\mathbf x}-\mathbf x_{2l})^mW_{2l,\underline\varpi}^{(h)}(\underline{\mathbf x})\right|\leqslant 2^{h(c_k-(c_k-c_g)l)}\mathfrak F_h(\underline m)(C_3C_G^{-1})^l\sum_{N=1}^\infty\sum_{\tau\in\mathcal T_N^{(h)}}\sum_{\underline l_\tau}\\[0.2cm]
\hfill\sum_{\displaystyle\mathop{\scriptstyle \mathbf P\in\mathcal P_{\tau,\underline l_\tau,\ell_0}}_{|P_{v_0}|=2l}}\kern-10pt C_1^N(C_gC_G^{-1})^{N-1}\left(\prod_{v\in\mathfrak V(\tau)}2^{(c_k-(c_k-c_g)\frac{|P_v|}2)}\right)\left(\prod_{v\in\mathfrak E(\tau)}(C_2C_G)^{l_v}\mathfrak C_{2l_v}|U|^{\max(1,l_v-1)}\right).
\end{largearray}\label{powercountinglW}\end{equation}
\begin{itemize}
\item The lemma cannot be used in this form in the ultraviolet regime, since in that case the right side of 
\eqref{6.ell0} is infinite, because $c_k=c_g=1$. 
In the ultraviolet we will need to re-organize the tree expansion, in order to derive convergent bounds on the series, as discussed in section~\ref{uvsec} below. 

\item The lemma gives a bound on the $m$-th derivative of $\hat W^{(h)}_{2l,\underline\varpi}(\underline{\mathbf k})$, which we will need in order to write the dominating behavior of the two-point Schwinger function as stated in Theorems~\ref{theoo}, \ref{theot}, \ref{theoth}; however, we will never need to take $m$ larger than $3$, which is important because the bound~(\ref{powercountingl}),
if generalized to larger values of $m$, would diverge faster than $m!$ as $m\to\infty$.

\item Recall that the propagator $g_{h}$ appearing in the statement should be interpreted as
the dressed propagator $\bar g_{h}$ in the first, second and third regimes. Since $\bar g_{h}$ depends on $W_{2l,\underline\varpi}^{(h')}$ for $h'\geqslant h$, we will have to apply the lemma inductively, proving at each step that the dressed propagator satisfies the bounds~(\ref{assumprop}). 

\item Similarly, the bounds~(\ref{assumW}) will have to be proved inductively.

\item In this lemma, the purpose of $\ell_0$, which up until now may have seemed like an arbitrary definition, is made clear. In fact, the condition that $\ell_0>c_k/(c_k-c_g)$ implies that $c_k-(c_k-c_g)|P_v|/2<0$, $\forall v\in \mathfrak V(\tau)\setminus\{v_0\}$. If this were not the case, then the weight of each tree $\tau$ could increase with the size of the tree, making the right side of~(\ref{powercountingl}) divergent. 

\item The combination $c_k-(c_k-c_g)|P_v|/2$ is called the {\it scaling dimension} of the cluster $v$. Under the assumptions of the lemma,
the scaling dimension is negative, $\forall v\in \mathfrak V(\tau)\setminus\{v_0\}$.
The clusters with non-negative scaling dimensions are necessarily leaves, and condition \eqref{assumW}
corresponds to the requirement that we can control the size of these dangerous clusters. 
Essentially, what this lemma shows is that the only terms that are potentially problematic are those with non-negative scaling dimension.
This prompts the following definitions: a node with negative scaling dimension will be called {\it irrelevant}, one with vanishing scaling dimension {\it marginal} and one with positive 
scaling dimension {\it relevant}.

\item We will show that in the first and third regimes $c_k=3$ and $c_g=1$, so that the scaling dimension is $3-|P_v|$. Therefore, the nodes with $|P_v|=2$ are relevant whereas all the others are irrelevant. In the second regime,
$c_k=2$ and $c_g=1$, so that the scaling dimension is $2-|P_v|/2$. Therefore, the nodes with $|P_v|=2$ are relevant, those with $|P_v|=4$ are marginal, and all other nodes are irrelevant.

\item The purpose of the factor $\mathfrak F_h(\underline m)$ is to take into account the dependence of the order of magnitude of the different components $k_0$, $k_1$ and $k_2$ in the different regimes. In other words, as was shown in~(\ref{estguv}), (\ref{estgo}), (\ref{estgt}) and~(\ref{estgth}), the effect of multiplying $g$ by $x_{j,i}$ depends on $i$, which is a fact the lemma must take into account.

\item The reason why we have stated this bound in $\mathbf x$-space is because of the estimate of $\det(G^{(h_v,T_v)})$ detailed below, which is very inefficient in $\mathbf k$-space.
\end{itemize}

\bigskip

\indent\underline{Proof}: The proof proceeds in five steps: first  we estimate the determinant appearing in~(\ref{valmid}) using the Gram-Hadamard inequality; then we perform a change of variables in the integral over $\underline{\mathbf x}_\tau$ in the right side of~(\ref{treeexpT}) in order to re-express it as an integration on differences $\mathbf x_i-\mathbf x_j$; we then decompose $(\underline{\mathbf x}-\mathbf x_{2l})^m$; and then compute a bound, which we re-arrange; and finally we use a bound on the number of spanning trees $\mathbf T(\tau)$ to conclude the proof.\par
\bigskip

\point{Gram bound} We first estimate $|\det G^{(T_v,h_v)}|$.\par
\bigskip

\subpoint{Gram-Hadamard inequality} We shall make use of the Gram-Hadamard inequality, which states that the determinant of a matrix $M$ whose components are given by $M_{i,j}=\mathbf A_i\odot \mathbf B_j$ where $(\mathbf A_i)$ and $(\mathbf B_i)$ are vectors in some Hilbert space with scalar product $\odot$ (writing $M$ as a scalar product is called writing it in {\it Gram form}) can be bounded by
\begin{equation}
|\det(M)|\leqslant\prod_i\sqrt{\mathbf A_i\odot \mathbf A_i}\sqrt{\mathbf B_i\odot \mathbf B_i}.
\label{gramhadamard}\end{equation}
The proof of this inequality is based on applying a Gram-Schmidt process to turn $(\mathbf A_i)$ and $(\mathbf B_i)$ into orthonormal families, at which point the inequality follows trivially.
We recall that 
$G^{(T_v,h_v)}$ is an $(n_v-(s_v-1))\times(n_v-(s_v-1))$ matrix in which $s_v$ denotes the number of children of $v$ and 
if we denote the children of $v$ by $(v_1,\cdots,v_{s_v})$, then $n_v=|R_v|/2=(\sum_{i=1}^{s_v}|P_{v_i}|-|P_v|)/2$. Its components are of the form $\mathbf t_\ell g_{(h_v),\ell}$ (see lemma~\ref{detexplemma}), with $\mathbf t_{(i,j)}=u_i\cdot u_j$ in which the $u_i$ are unit vectors.\par
\bigskip

\subpoint{Gram form} We now put $(g_{(h),(\alpha,\alpha')}(\mathbf x-\mathbf x'))_{(\mathbf x,\alpha),(\mathbf x',\alpha')}$ in Gram form by using the $\mathbf k$-space representation of $g_{h}$ in~(\ref{6.3}). Let $\mathcal H=\ell_2(\mathcal B_{\beta,L}\times\{a,\tilde b,\tilde a,b\})$ denote the Hilbert space of square summable sequences indexed by $(\mathbf k,\alpha)\in\mathcal B_{\beta,L}\times\{a,\tilde b,\tilde a,b\}$. For every $h\in\{h^*_2,\cdots,h^*_1-1\}$ and $(\mathbf x,\alpha)\in([0,\beta)\times\Lambda)\times\{a,\tilde b,\tilde a,b\}$, we define a pair of vectors $(\mathbf A_\alpha^{(h)}(\mathbf x),\mathbf B_\alpha^{(h)}(\mathbf x))\in\mathcal H^2$ by
\begin{equation}\left\{\begin{array}{>{\displaystyle}l}
(\mathbf A_\alpha^{(h)}(\mathbf x))_{\mathbf k,\alpha'}:=\frac{1}{\sqrt{\beta|\Lambda|}}e^{-i\mathbf k\cdot\mathbf x}\hat V^{(h)}_{\alpha',\alpha}(\mathbf k)\sqrt{\hat\lambda_{\alpha'}^{(h)}(\mathbf k)}\\[0.5cm]
(\mathbf B_\alpha^{(h)}(\mathbf x))_{\mathbf k,\alpha'}:=\frac{1}{\sqrt{\beta|\Lambda|}}e^{-i\mathbf k\cdot\mathbf x}\hat U_{\alpha,\alpha'}^{(h)}(\mathbf k)\ \sqrt{\hat\lambda_{\alpha'}^{(h)}(\mathbf k)}
\end{array}\right.
\label{AB}\end{equation}
where $\hat\lambda_{\alpha'}^{(h)}(\mathbf k)$ denotes the $\alpha$-th eigenvalue of $\sqrt{\hat g_h^\dagger(\mathbf k)\hat g_{h}(\mathbf k)}$ (i.e. the {\it singular values} of $\hat g_{h}(\mathbf k)$) and $\hat V^{(h)}(\mathbf k)$ and $\hat U^{(h)}(\mathbf k)$ are unitary matrices that are such that
$$\hat g_{h}(\mathbf k)=\hat V^{(h)\dagger}(\mathbf k)\hat D^{(h)}({\bf k})\hat U^{(h)}(\mathbf k),$$
where $\hat D^{(h)}({\bf k})$ is the diagonal matrix with entries $\hat \lambda^{(h)}_\alpha({\bf k})$.
We can now write $g_{h}$ as
\begin{equation}
g_{(h),(\alpha,\alpha')}(\mathbf x-\mathbf x')=\mathbf A_\alpha^{(h)}(\mathbf x)\odot \mathbf B_{\alpha'}^{(h)}(\mathbf x')
\label{gramg}\end{equation}
where $\odot$ denotes the scalar product on $\mathcal H$. Furthermore, recalling that $|\hat g_{h}(\mathbf k)|$ is the operator norm of $\hat g_{h}(\mathbf k)$, so that $|\hat g_{h}(\mathbf k)|=\mathrm{max}\ \mathrm{spec}\sqrt{\hat g_h^\dagger(\mathbf k)\hat g_{h}(\mathbf k)}$, we have
\begin{equation}
\mathbf A_\alpha^{(h)}(\mathbf x)\odot \mathbf A_\alpha^{(h)}(\mathbf x)=\mathbf B_\alpha^{(h)}(\mathbf x)\odot \mathbf B_\alpha^{(h)}(\mathbf x)\leqslant \frac1{\beta|\Lambda|}\sum_{\mathbf k\in\mathcal B_{\beta,L}}|\hat g_{h}(\mathbf k)|\leqslant C_G2^{(c_k-c_g)h}
\label{grambound}\end{equation}

The Gram form for $G^{(T_v,h_v)}$ is then
\begin{equation} t_{(i,j)} g_{(h),(\varpi_i,\varpi_j)}(\mathbf x_i-\mathbf x_j)
=(u_i\cdot u_j)(\mathbf A_{\varpi_i}(\mathbf x_i)\odot \mathbf B_{\varpi_j}(\mathbf x_j))
\end{equation}
so that, using \eqref{gramhadamard} and \eqref{grambound},
\begin{equation}|\det G^{(T_v,h_v)}|\leqslant(C_G2^{(c_k-c_g)h_v})^{n_v-(s_v-1)}.\label{boundGt}\end{equation}
\par
\bigskip

\begin{figure}
\hfil\includegraphics{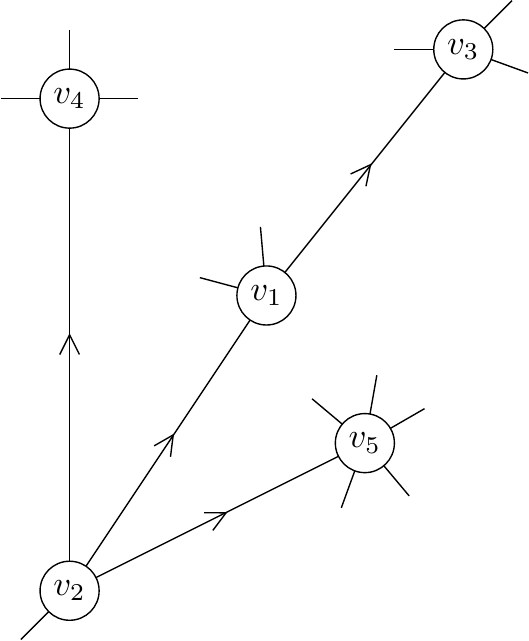}\par
\caption{example of a spanning tree with $s_v=5$ and $|P_{v_1}|=|P_{v_2}|=|P_{v_3}|=|P_{v_4}|=4$, $|P_{v_5}|=6$; whose root is $v_2$.}
\label{spanningfig}\end{figure}

\point{Change of variables} We change variables in the integration over $\underline{\mathbf x}_\tau$. For every $v\in\bar{\mathfrak V}(\tau)$, let $P_{v}=:(j_1^{(v)},\cdots,j_{2l_{v}}^{(v)})$. We recall that a spanning tree $T\in\mathbf T(\tau)$ is a diagram connecting the fields specified by the $I_v$'s for $v\in\mathfrak E(\tau)$: more precisely, if we draw a vertex for each $v\in\mathfrak E(\tau)$ with $|I_v|$ half-lines attached to it that are labeled by the elements of $I_v$, then $T\in\mathbf T(\tau)$ is a pairing of some of the half-lines that results in a tree called a {\it spanning tree} (not to be confused with a Gallavotti-Nicol\`o tree) (for an example, see figure~\ref{spanningfig}). The vertex $v_r$ of a spanning tree that contains the {\it last external field}, i.e. that is such that $j_{2l_{v_0}}^{(v_0)}\in I_{v_r}$, is defined as its root, which allows us to unambiguously define a parent-child partial order, so that we can dress each branch with an arrow that is directed away from the root. For every $v\in\mathfrak E(\tau)$ that is not the root of $T$, we define $J^{(v)}\in I_v$ as the index of the field in which $T$ {\it enters}, i.e. the index of the half-line of $T$ with an arrow pointing towards $v$. We also define $J^{(v_r)}:=j_{2l_{v_0}}^{(v_0)}$.
Now, for every $v\in\mathfrak E(\tau)$, we define
$$\mathbf z_{j^{(v)}}:=\mathbf x_{j^{(v)}}-\mathbf x_{J^{(v)}}$$
for all $j^{(v)}\in I_v\setminus\{J^{(v)}\}$, and given a line of $T$ connecting $j^{(v)}$ to $J^{(v')}$, we define
$$\mathbf z_{J^{(v')}}:=\mathbf x_{J^{(v')}}-\mathbf x_{j^{(v)}}.$$
We have thus defined $(\sum_{v\in\mathfrak E(\tau)}|I_v|)-1$ variables $\mathbf z$, so that we are left with $\mathbf x_{J^{(v_r)}}$, which we call $\mathbf x_0$. It follows directly from the definitions that 
the change of variables from $\underline{\mathbf x}_\tau$ to $\{\mathbf x_0,\{\mathbf z_{j}\}_{j\in{\mathcal I}_{\tau}\setminus \{J^{(v_r)}\}}\}$, where ${\mathcal I}_{\tau}=\bigcup_{v\in \mathfrak E(\tau)}I_v$, has Jacobian equal to $1$. \par
\bigskip

\point{Decomposing $(\underline{\mathbf x}-\mathbf x_{2l})^m$} We now decompose the $(\underline{\mathbf x}-\mathbf x_{2l})^m$ factor in \eqref{powercountingl} in the following way (note that  in terms of the indices in $P_{v_0}$, $\mathbf x_{2l}\equiv\mathbf x_{J^{(v_r)}}$):
$(\underline{\mathbf x}-\mathbf x_{2l})^m$ is a product of terms of the form $(x_{j,i}-x_{J^{(v_r)},i})$ which we rewrite as a sum of $z_{j',i}$'s for $v\in \mathfrak E(\tau)$ {\it on the path from $J^{(v_r)}$ to $j$}, a concept we will now make more precise. $j$ and $J^{(v_r)}$ are in $I_{v(j)}$ and $I_{v_r}$ respectively, where $v(j)$ is the unique node in $\mathfrak E(\tau)$ such that $j\in I_{v(j)}$. There exists a unique sequence of lines of $T$ that links $v_r$ to $v(j)$, which we denote by $((j_1,j_1'),\cdots,(j_\rho,j_\rho'))$, the convention being that the line 
$(j,j')$ is oriented from $j$ to $j'$. The {\it path from $J^{(v_r)}$ to $j$} is the sequence $\mathbf z_{j_1}, \mathbf z_{j_1'},\mathbf z_{j_2},\cdots$ and so forth, until $j$ is reached. We can therefore write
$$x_{j,i}-x_{J^{(v_r)},i}=\sum_{p=1}^{\rho}(z_{j_p,i}+z_{j_p',i}).$$
\bigskip

\point{Bound in terms of number of spanning trees} Let us now turn to the object of interest, namely the left side of \eqref{powercountingl}. It follows from 
\eqref{defW} and \eqref{treeexpT} that 
\begin{equation} B_{2l,\underline\varpi}^{(h)}(\underline{\mathbf x})=\sum_{N=1}^\infty\sum_{\tau\in\mathcal T_{N}^{(h)}}\sum_{T\in\mathbf T(\tau)}\sum_{\underline l_\tau}\sum_{\underline\varpi_\tau}\int d\underline{\mathbf x}_\tau\ \sum_{\mathop{\scriptstyle \mathbf P\in\tilde{\mathcal P}_{\tau,\underline l_\tau,\ell_0}}_{|P_{v_0}|=2l}}\prod_{v\in\bar{\mathfrak V}(\tau)}\frac{(-1)^{s_v}}{s_v!}\rho_v^{(T_v)}.\end{equation}
Therefore, using the bound~(\ref{boundGt}), the change of variables defined above and the decomposition of $(\underline{\mathbf x}-\mathbf x_{2l})^m$ described above, we find
\begin{equation}\begin{largearray}
\frac{1}{\beta|\Lambda|}\sum_{\underline\varpi}\int d\underline{\mathbf x}\ \left|(\underline{\mathbf x}-\mathbf x_{2l})^mB_{2l,\underline\varpi}^{(h)}(\underline{\mathbf x})\right|
\leqslant\frac{1}{\beta|\Lambda|}\sum_{N=1}^\infty\sum_{\tau\in\mathcal T_N^{(h)}}\sum_{T\in\mathbf T(\tau)}\sum_{\underline l_\tau}\sum_{\underline\varpi_\tau}\int d\mathbf x_0\sum_{\displaystyle\mathop{\scriptstyle \mathbf P\in\tilde{\mathcal P}_{\tau,\underline l_\tau,\ell_0}}_{|P_{v_0}|=2l}}\\[0.5cm]
\hskip0.5cm\sum_{\displaystyle\mathop{\scriptstyle (m_\ell)_{\ell\in T},(m_v)_{v\in\mathfrak E(\tau)}}_{\sum (m_\ell+m_v)=m}}
\prod_{v\in\mathfrak V(\tau)}\left(\frac{1}{s_v!}
\left( C_G2^{(c_k-c_g)h_v}\right)^{n_v-(s_v-1)}
\prod_{\ell\in T_v}\left(\int d\mathbf z_\ell\ \left|\mathbf z_\ell^{m_\ell} g_{(h_v),\ell}(\mathbf z_\ell)\right|\right)
\right)\cdot\\[1cm]
\hfill\cdot\prod_{v\in\mathfrak E(\tau)}\int d\underline{\mathbf z}^{(v)}\ \left|(\underline{\mathbf z}^{(v)})^{m_v}W_{2l_v,\underline\varpi_v}^{(h_v-1)}(\underline{\mathbf z}^{(v)})\right|
\end{largearray}\label{ineqpcexpl}\end{equation}
(we recall that by definition, if $v\in\mathfrak E(\tau)$, $I_v=P_v$ and $|I_v|=2l_v$) in which we inject~(\ref{assumprop}) and~(\ref{assumW}) to find
\begin{equation}
\begin{array}{>{\displaystyle}r@{\ }>{\displaystyle}l}
\frac{1}{\beta|\Lambda|}\int d\underline{\mathbf x}\ \left|(\underline{\mathbf x}-\mathbf x_{2l})^mB_{2l,\underline\varpi}^{(h)}(\underline{\mathbf x})\right|
\leqslant&\sum_{N=1}^\infty\sum_{\tau\in\mathcal T_N^{(h)}}\sum_{T\in\mathbf T(\tau)}\sum_{\underline l_\tau}\sum_{\displaystyle\mathop{\scriptstyle \mathbf P\in\tilde{\mathcal P}_{\tau,\underline l_\tau,\ell_0}}_{|P_{v_0}|=2l}}c_1^N\mathfrak F_h(\underline m)\cdot\\[0.5cm]
&\cdot\prod_{v\in\mathfrak V(\tau)}\frac{1}{s_v!}C_G^{n_v-s_v+1}C_g^{s_v-1}2^{h_v((c_k-c_g)n_v-c_k(s_v-1))}\cdot\\[0.5cm]
&\cdot\prod_{v\in\mathfrak E(\tau)}c_2^{2l_v}\mathfrak C_{2l_v}|U|^{\max(1,l_v-1)}2^{(h_v-1)(c_k-(c_k-c_g)l_v)}
\end{array}
\label{boundpre}\end{equation}
in which $C_1^N$ is an upper bound on the number of terms in the sum over $(m_l)$ and $(m_v)$ in the previous equation,
and $c_2$ denotes the number of elements in the sum over $\varpi_v$. Recalling that 
$n_v=|R_v|/2=(\sum_{i=1}^{s_v}|P_{v_i}|-|P_v|)/2$, we re-arrange~(\ref{boundpre}) by using
$$\left\{\begin{array}{>{\displaystyle}l}
\sum_{v\in\mathfrak V(\tau)}h_v|R_v|=-h|P_{v_0}|-\sum_{v\in\mathfrak V(\tau)}|P_v|+\sum_{v\in\mathfrak E(\tau)}(h_v-1)|I_v|\\[0.5cm]
\sum_{v\in\mathfrak V(\tau)}h_v(s_v-1)=-h-\sum_{v\in\mathfrak V(\tau)}1+\sum_{v\in\mathfrak E(\tau)}(h_v-1)
\end{array}\right.$$
and
$$\left\{\begin{array}{>{\displaystyle}l}
\sum_{v\in\mathfrak V(\tau)}|R_v|=|I_{v_0}|-|P_{v_0}|\\[0.5cm]
\sum_{v\in\mathfrak V(\tau)}(s_v-1)=N-1
\end{array}\right.$$
to find
\begin{equation}
\begin{largearray}
\frac{1}{\beta|\Lambda|}\int d\underline{\mathbf x}\ \left|(\underline{\mathbf x}-\mathbf x_{2l})^mB_{2l,\underline\varpi}^{(h)}(\underline{\mathbf x})\right|
\leqslant C_G^{-l}\sum_{N=1}^\infty\sum_{\tau\in\mathcal T_N^{(h)}}\sum_{T\in\mathbf T(\tau)}\sum_{\underline l_\tau}\sum_{\displaystyle\mathop{\scriptstyle \mathbf P\in\tilde{\mathcal P}_{\tau,\underline l_\tau,\ell_0}}_{|P_{v_0}|=2l}}\kern-10ptC_1^N(C_gC_G^{-1})^{N-1}\cdot\\[0.5cm]
\hfill\cdot2^{h(c_k-(c_k-c_g)l)}\mathfrak F_h(\underline m)\prod_{v\in\mathfrak V(\tau)}\frac{1}{s_v!}2^{c_k-(c_k-c_g)\frac{|P_v|}2}\prod_{v\in\mathfrak E(\tau)}(c_2^2C_G)^{l_v}\mathfrak C_{2l_v}|U|^{\max(1,l_v-1)}.
\end{largearray}
\label{boundt}\end{equation}
\bigskip

\point{Bound on the number of spanning trees} Finally, the number of choices for $T$ can be bounded (see \cite[lemma~A.5]{genZO})
\begin{equation}\sum_{T\in\mathbf T(\tau)}1\leqslant \prod_{v\in\mathfrak V(\tau)}c_3^{\frac{|R_v|}2}s_v!\label{boundnrspan}\end{equation}
so that by injecting~(\ref{boundnrspan}) into~(\ref{boundt}), we find~(\ref{powercountingl}), with $C_2=c_2^2c_3$ and $C_3=c_3^{-1}$.\penalty10000\hfill\penalty10000$\square$\par
\subseqskip

\subsection{Schwinger function from the effective potential}
\label{schwinsec}
\indent In this section we show how to compute $\bar{\mathcal W}^{(h)}$ in a similarly general setting as above: consider
\begin{equation}\begin{largearray}
-\beta|\Lambda|\mathfrak e_h-\mathcal Q^{(h)}(\psi^{(\leqslant h)})-\bar{\mathcal W}^{(h)}(\psi^{(\leqslant h)},\hat J_{\mathbf k,\underline\alpha})\\
\hfill=\sum_{N=1}^\infty\frac{(-1)^N}{N!}\mathcal E^T_{h+1}\left(\bar{\mathcal W}^{(h+1)}(\psi^{(\leqslant h)}+\psi^{(h+1)},\hat J_{\mathbf k,\underline\alpha});N\right)
\end{largearray}\label{Windeq}\end{equation}
for $h\in\{h_2^*,\cdots,h_1^*-1\}$. This discussion will not be used in the ultraviolet regime, so we can safely discard the cases in which the propagator is not renormalized. Unlike~(\ref{treeindeq}), it is necessary to separate the $\alpha$ indices from the $(\omega,j)$ indices, so we write the propagator of $\mathcal E_{h+1}^T$ as $g_{(h+1,\varpi),(\alpha,\alpha')}$ where $\varpi$ stands for $\omega$ in the first and second regimes, and $(\omega,j)$ in the third. \par
\bigskip

\indent We now rewrite the terms in the right side of~(\ref{Windeq}) in terms of the effective potential $\mathcal V^{(h)}$. Let
\begin{equation}
\mathcal X^{(h)}(\psi,\hat J_{\mathbf k,\underline\alpha}):=\mathcal V^{(h)}(\psi)-\bar{\mathcal W}^{(h)}(\psi,\hat J_{\mathbf k,\underline\alpha}).
\label{Qdef}\end{equation}
Note that the terms in $\mathcal X^{(h)}$ are either linear or quadratic in $\hat J_{\mathbf k,\underline\alpha}$, simply because the two $J$ variables we have at our disposal, $\hat J_{{\bf k},\alpha_1}^+,\hat J^-_{{\bf k},\alpha_2}$, are Grassmann variables and square to zero. We define the functional derivative of $\mathcal V^{(h)}$ with respect to $\hat\psi^{\pm}_{\mathbf k,\alpha}$:
$$
\partial_{\mathbf k,\alpha}^\pm\mathcal V^{(h)}(\psi):=\int d\hat\psi_{\mathbf k,\alpha}^{\pm}\ \mathcal V^{(h)}(\psi).
$$
\bigskip

\Theo{Lemma}\label{schwinlemma} Assume that, for $h=h^*_1$,
\begin{equation}\begin{array}{r@{\ }>{\displaystyle}l}
{\mathcal X}^{(h)}(\psi,\hat J_{\mathbf k,\underline\alpha})=&\hat J^+_{\mathbf k,\alpha_1}s_{\alpha_1,\alpha_2}^{(h)}(\mathbf k)\hat J^-_{\mathbf k,\alpha_2}
+\sum_{\alpha'}(\hat J^+_{\mathbf k,\alpha_1}q_{\alpha_1,\alpha'}^{+(h)}(\mathbf k)\hat\psi^{-}_{\mathbf k,\alpha'}+\hat\psi^{+}_{\mathbf k,\alpha'}q_{\alpha',\alpha_2}^{-(h)}(\mathbf k)\hat J^-_{\mathbf k,\alpha_2})\\[0.5cm]
&+\sum_{\alpha'}\left(\partial^-_{\mathbf k,\alpha'}\mathcal V^{(h)}(\psi)\bar G_{\alpha',\alpha_2}^{-(h)}(\mathbf k)\hat J^-_{\mathbf k,\alpha_2}-\hat J^+_{\mathbf k,\alpha_1}\bar G_{\alpha_1,\alpha'}^{+(h)}(\mathbf k)\partial^+_{\mathbf k,\alpha'}\mathcal V^{(h)}(\psi)\right)\\[0.5cm]
&+\sum_{\alpha',\alpha''}\left(\hat J^+_{\mathbf k,\alpha_1}\bar G_{\alpha_1,\alpha'}^{+(h)}(\mathbf k)\partial^+_{\mathbf k,\alpha'}\partial^-_{\mathbf k,\alpha''}\mathcal V^{(h)}(\psi)\bar G_{\alpha'',\alpha_2}^{-(h)}(\mathbf k)\hat J^-_{\mathbf k,\alpha_2}\right)
\end{array}\label{exprdomorderren}\end{equation}
for some $s_{\alpha_1,\alpha_2}^{(h^*_1)}(\mathbf k)$, $q_{\alpha,\alpha'}^{\pm(h^*_1)}(\mathbf k)$, $\bar G_{\alpha,\alpha'}^{(h^*_1)}({\bf k})$. Then \eqref{exprdomorderren}
holds for $h\in\{h^*_2,\ldots,h^*_1-1\}$ as well, with 
\begin{equation}\left\{\begin{array}{>{\displaystyle}l}
\bar G^{+(h)}_{\alpha,\alpha'}(\mathbf k):=\bar G^{+(h+1)}_{\alpha,\alpha'}(\mathbf k)+\sum_{\alpha'',\varpi}q_{\alpha,\alpha''}^{+(h+1)}(\mathbf k)\hat g_{(h+1,\varpi),(\alpha'',\alpha')}(\mathbf k)\\[0.5cm]
\bar G^{-(h)}_{\alpha,\alpha'}(\mathbf k):=\bar G^{-(h+1)}_{\alpha,\alpha'}(\mathbf k)+\sum_{\alpha'',\varpi}\hat g_{(h+1,\varpi),(\alpha,\alpha'')}(\mathbf k)q_{\alpha'',\alpha'}^{-(h+1)}(\mathbf k)
\end{array}\right.\label{indGdefren}\end{equation}
\begin{equation}\left\{\begin{array}{>{\displaystyle}l}
q_{\alpha,\alpha'}^{+(h)}(\mathbf k):=q_{\alpha,\alpha'}^{+(h+1)}(\mathbf k)-\sum_{\alpha''}\bar G_{\alpha,\alpha''}^{+(h)}(\mathbf k)\hat W_{2,(\alpha'',\alpha')}^{(h)}(\mathbf k)\\[0.5cm]
q_{\alpha,\alpha'}^{-(h)}(\mathbf k):=q_{\alpha,\alpha'}^{-(h+1)}(\mathbf k)-\sum_{\alpha''}\hat W_{2,(\alpha,\alpha'')}^{(h)}(\mathbf k)\bar G_{\alpha'',\alpha'}^{-(h)}(\mathbf k)
\end{array}\right.\label{indqdefren}\end{equation}
and
\begin{equation}\begin{largearray}
s_{\alpha_1,\alpha_2}^{(h)}(\mathbf k):=s_{\alpha_1,\alpha_2}^{(h+1)}(\mathbf k)+\sum_{\alpha',\alpha'',\omega}q_{\alpha_1,\alpha'}^{+(h+1)}(\mathbf k)\hat g_{(h+1,\varpi),(\alpha',\alpha'')}(\mathbf k)q_{\alpha'',\alpha_2}^{-(h+1)}(\mathbf k)\\[0.5cm]
\hfill-\sum_{\alpha',\alpha''}\bar G^{+(h)}_{\alpha_1,\alpha'}(\mathbf k)\hat W_{2,(\alpha',\alpha'')}^{(h)}(\mathbf k)\bar G^{-(h)}_{\alpha'',\alpha_2}(\mathbf k)
\end{largearray}\label{indsdefren}\end{equation}
in which the sums over $\alpha$ are sums over the indices of $g$.
\endtheo
\bigskip

\indent The (inductive) proof of lemma~\ref{schwinlemma}\ is straightforward, although it requires some bookkeeping, and is left to the reader.\par
\bigskip

{\bf Remark}: It follows from~(\ref{freeengrassend}) and~(\ref{schwinform}) that the two-point Schwinger function $s_2(\mathbf k)$ is given by $s_2(\mathbf k)=s^{(\mathfrak h_\beta)}(\mathbf k)$ (indeed, once all of the fields have been integrated, $\mathcal X^{(\mathfrak h_\beta)}=\hat J_{\mathbf k}^+s^{(h)}(\mathbf k)\hat J_{\mathbf k}^-$). Therefore~(\ref{indsdefren}) is an inductive formula for the two-point Schwinger function.\par
\seqskip

\section{Ultraviolet integration}
\label{uvsec}
\indent We now detail the integration over the ultraviolet regime. We start from the tree expansion in the general form discussed in section~\ref{treeexpsec}, with $q=\ell_0=1$
and $h^*_1=M$; note that by construction these trees have no local leaves. 
As mentioned in the first remark after lemma~\ref{powercountinglemma}, we cannot apply that lemma to prove convergence of the tree expansion: however, 
as we shall see in a moment, a simple re-organization of it will allow to derive uniformly convergent bounds. 
We recall the estimates~(\ref{estguv}) and~(\ref{grambounduv}) of $\hat g_{h}$ in the ultraviolet regime: for $m_0+m_k\leqslant3$
\begin{equation}\left\{\begin{array}{>{\displaystyle}l}
\int d\mathbf x\ x_0^{m_0}x^{m_k}|g_{h}(\mathbf x)|\leqslant (\mathrm{const.})\ 2^{-h-m_0h}\\[0.5cm]
\frac{1}{\beta|\Lambda|}\sum_{\mathbf k\in\mathcal B_{\beta,L}}|\hat g_{h}(\mathbf k)|\leqslant(\mathrm{const.}).
\end{array}\right.\label{8.sys}\end{equation}
Equation~(\ref{8.sys}) has the same form as~(\ref{assumprop}), with 
$$c_g=c_k=1,\quad \mathfrak F_h(\underline m)=2^{-m_0h}.$$
\bigskip

\indent We now move on to the power counting estimate. The first remark to be made is that the values of the leaves have a much better dimensional estimate than the one assumed in lemma~\ref{powercountinglemma}. In fact, 
the value of any leaf, called $W^{(M)}_{4,\underline \alpha}
(\underline{\mathbf x})$, 
is the antisymmetric part of
\begin{equation}
\delta_{\alpha_1,\alpha_2}\delta_{\alpha_3,\alpha_4}\delta(\mathbf x_1-\mathbf x_2)\delta(\mathbf x_3-\mathbf x_4)Uw_{\alpha_1,\alpha_3}(\mathbf x_1-\mathbf x_3)
\label{Wfline}\end{equation}
so that 
\begin{equation}
\frac{1}{\beta|\Lambda|}\int d\underline{\mathbf x}\ |(\underline{\mathbf x}-\mathbf x_4)^mW^{(M)}_{4,\underline\alpha}(\underline{\mathbf x})|\leqslant\mathfrak C_4'|U|.
\label{boundWuv}\end{equation}
\bigskip

\point{Resumming trivial branches} Next, we re-sum the branches of Gallavotti-Nicol\`o trees that are only followed by a single endpoint: the naive dimensional bound 
on the value of these branches tends to diverge logarithmically as $M\to\infty$, but one can easily exhibit a cancellation that improves their estimate, as explained below. Consider a tree $\tau$ made of a single branch, with a root on scale $h$ and a single leaf on scale $M+1$ with value $W^{(M)}_4$. The 4-field kernel 
associated with such a tree is $K_{4,\underline\alpha}^{(h)}(\underline{\mathbf x}):=W^{(M)}_{4,\underline\alpha}(\underline{\mathbf x})$.
The 2-field kernel associated with $\tau$, once summed over the choices of $P_v$ and over the field labels it indexes for $h+1<h_v\leqslant M$, keeping $P_{v_0}$ and its field labels fixed, 
 can be computed explicitly:
\begin{equation}
K_{2,(\alpha,\alpha')}^{(h)}(\mathbf x)=2U\sum_{h'=h+1}^M \left( w_{\alpha,\alpha'}(\mathbf x)g_{\alpha,\alpha'}^{(h')}(\mathbf x)-\delta_{\alpha,\alpha'}\delta(\mathbf x)\sum_{\alpha_2}\int d\mathbf y\ w_{\alpha,\alpha_2}(\mathbf y)g_{\alpha_2,\alpha_2}^{(h')}(\mathbf 0)
\right).
\label{Wtline}\end{equation}
If one were to bound the right side of~(\ref{Wtline}) term by term in the sum over $h'$ using the dimensional estimates on the propagator (see ~(\ref{estguv}) and following),
one would find a {\it logarithmic} divergence for $\int d\mathbf x|K_{2,(\alpha,\alpha')}^{(h)}(\mathbf x)|$, i.e. a bound proportional to $M-h$. However, the right side of~(\ref{Wtline}) depends on propagators evaluated at $x_0=0$ (because $w(\mathbf x)$ is proportional to $\delta(x_0)$), so we can use an improved bound on the propagator $g_{h'}$:
 the dominant terms in $\hat g_{h}(\mathbf k)$ are odd in $k_0$, so they cancel when considering
$$\sum_{k_0\in\frac{2\pi}\beta(\mathbb Z+\frac12)}\hat g_{h}(\mathbf k).$$
From this idea, we compute an improved bound for $|g_{h}(\mathbf x)|$ with $x_0=0$:
$$|g_{h}(0,x_1,x_2)|\leqslant \sum_{k_1,k_2}\left|\sum_{k_0}\hat g_{h}(\mathbf k)\right|\leqslant (\mathrm{const.})\ 2^{-h}.$$
All in all, we find
\begin{equation}
\int d\mathbf x\ |\mathbf x^mK^{(h)}_{2,(\alpha,\alpha')}(\mathbf x)|\leqslant\mathfrak C_4|U|,\quad
\frac{1}{\beta|\Lambda|}\int d\underline{\mathbf x}\ |(\underline{\mathbf x}-\mathbf x_4)^mK^{(h)}_{4,\underline{\alpha}}(\underline{\mathbf x})|\leqslant\mathfrak C_4|U|
\label{boundWline}\end{equation}
for some constant $\mathfrak C_4$. We then re-organize the right side of~(\ref{treeexp}) by: \begin{enumerate}\item summing over the set of {\it contracted} trees $\tilde{\mathcal T}_N^{(h)}$, which is defined like $\mathcal T_N^{(h)}$ but for the fact that every node $v\succ v_0$ that is not an endpoint must have at least two endpoints following it, and the endpoints can be on any scale in $[h+2, M+1]$; \item re-defining the value of the endpoints to be $\tilde\rho_v=K^{(h_v-1)}_{2l_v}$, with $l_v=1,2$.\end{enumerate}
\bigskip

\point{Contracted tree expansion} We can now estimate the ``contracted tree'' expansion, by 
repeating the steps of the proof of lemma~\ref{powercountinglemma}, thus finding
\begin{equation}
\begin{largearray}
\frac{1}{\beta|\Lambda|}\int d\underline{\mathbf x}\ \left|(\underline{\mathbf x}-\mathbf x_{2l})^mW_{2l,\underline\alpha}^{(h)}(\underline{\mathbf x})\right|
\leqslant\sum_{N=1}^\infty\sum_{\tau\in\tilde{\mathcal T}_N^{(h)}}\sum_{T\in\mathbf T(\tau)}\sum_{\underline l_\tau}\sum_{\displaystyle\mathop{\scriptstyle \mathbf P\in\mathcal P_{\tau,\underline l_\tau,1}^{(h)}}_{|P_{v_0}|=2l}}c_1^N\cdot\\[0.5cm]
\hfill\cdot\prod_{v\in\mathfrak V(\tau)}\frac{1}{s_v!}2^{-h_v(s_v-1)}
\prod_{v\in\mathfrak E(\tau)}c_2^4\mathfrak C_{4}|U|
\end{largearray}\label{boundpreuv}\end{equation}
for two constants $c_1$ and $c_2$ in which the sum over $\underline l_\tau$ is a sum over the $l_v\in\{1,2\}$. It then follows from the following equation
$$
\sum_{v\in\mathfrak V(\tau)}h_v(s_v-1)=h(N-1)+\sum_{v\in\mathfrak V(\tau)}(N_v-1)
$$
in which $N_v$ denotes the number of endpoints following $v\in\tau$, which can be proved by induction, that
\begin{equation}\begin{largearray}
\frac{1}{\beta|\Lambda|}\int d\underline{\mathbf x}\ \left|(\underline{\mathbf x}-\mathbf x_{2l})^mW_{2l,\underline\alpha}^{(h)}(\underline{\mathbf x})\right|\\[0.5cm]
\hfill\leqslant\sum_{N=1}^\infty(|U|c_3)^N
2^{-h(N-1)}\sum_{\tau\in\tilde{\mathcal T}_N^{(h)}}\sum_{\underline l_\tau}\sum_{\displaystyle\mathop{\scriptstyle \mathbf P\in\mathcal P_{\tau,\underline l_\tau,1}^{(h)}}_{|P_{v_0}|=2l}}\prod_{v\in\mathfrak V(\tau)}2^{-(N_v-1)}.
\end{largearray}\label{boundpreuvt}\end{equation}
Furthermore, we notice that by the definition of $\mathcal P_{\tau,\underline l_\tau,1}^{(h)}$, $|P_v|\leqslant 2N_v+2$.
In particular, for $v=v_0$, $2l\leqslant 2N+2$, so the sum over $N$ actually starts at $\max\{1,l-1\}$:
\begin{equation}
\begin{largearray}
\frac{1}{\beta|\Lambda|}\int d\underline{\mathbf x}\ \left|(\underline{\mathbf x}-\mathbf x_{2l})^mW_{2l,\underline\alpha}^{(h)}(\underline{\mathbf x})\right|\\[0.5cm]
\hfill\leqslant\sum_{N=\max\{1,l-1\}}^\infty(|U|c_3)^N
2^{-h(N-1)}\sum_{\tau\in\tilde{\mathcal T}_N^{(h)}}\sum_{\underline l_\tau}\sum_{\displaystyle\mathop{\scriptstyle \mathbf P\in\mathcal P_{\tau,\underline l_\tau,1}^{(h)}}_{|P_{v_0}|=2l}}\prod_{v\in\mathfrak V(\tau)}2^{-(N_v-1)}.
\end{largearray}\label{boundprerefuv}\end{equation}
\bigskip

\point{Bound on the contribution at fixed $N$} We temporarily restrict to the case $N>1$. We bound
$$\mathfrak T_{N}:=\sum_{\tau\in\tilde{\mathcal T}_{N}^{(h)}}\sum_{\underline l_\tau}\sum_{\mathbf P\in\mathcal P_{\tau,\underline l_\tau,1}}\prod_{v\in\mathfrak V(\tau)}2^{-(N_v-1)}.$$
Since $N_v\geqslant2$ and $|P_v|\leqslant 2N_v+2$, $\forall\mu\in(0,1)$,
$$-(N_v-1)\leqslant \min\big\{2-\frac{|P_v|}{2},-1\big\}\leqslant (1-\mu)\min\big\{2-\frac{|P_v|}{2},-1\big\}-\mu\leqslant-(1-\mu)\frac{|P_v|}{6}-\mu$$
so that
$$\mathfrak T_N\leqslant\sum_{\tau\in\tilde{\mathcal T}_{N}^{(h)}}\sum_{\underline l_\tau}\sum_{\mathbf P\in\mathcal P_{\tau,\underline l_\tau,1}}\prod_{v\in\mathfrak V(\tau)}2^{-(1-\mu)\frac{|P_v|}6}2^{-\mu}.$$
\bigskip

\subpoint{Bound on the field label assignments} We bound
$$\sum_{\mathbf P\in\mathcal P_{\tau,\underline l_\tau,1}}\prod_{v\in\mathfrak V(\tau)}2^{-(1-\mu)\frac{|P_v|}6}.$$
We proceed by induction: if $v_0$ denotes the first node of $\tau$ (i.e. the node immediately following the root), $(v_1,\cdots,v_s)$ its children, and $(\tau_1,\cdots,\tau_s)$ the sub-trees with first node $(v_1,\cdots,v_s)$, then
$$\begin{largearray*}2{>{\displaystyle}r@{\ }>{\displaystyle}l}
\sum_{\mathbf P\in\mathcal P_{\tau,\underline l_\tau,1}}\prod_{v\in\mathfrak V(\tau)}2^{-(1-\mu)\frac{|P_v|}6}\leqslant&
\sum_{\mathbf P_1\in\mathcal P(\tau_1)}\cdots\sum_{\mathbf P_s\in\mathcal P(\tau_s)}\sum_{p_{v_0}=0}^{|P_{v_1}|+\cdots+|P_{v_s}|}{|P_{v_1}|+\cdots+|P_{v_s}|\choose p_{v_0}}\cdot\\[0.5cm]
&\hfill\cdot2^{-\frac{1-\mu}6p_{v_0}}\prod_{i=1}^s\prod_{v\in\mathfrak V(\tau_i)}2^{-(1-\mu)\frac{|P_v|}6}\\[0.5cm]
=&\prod_{i=1}^s\left(\sum_{\mathbf P_i\in\mathcal P(\tau_i)}(1+2^{-\frac{1-\mu}6})^{|P_{v_i}|}\prod_{v\in\mathfrak V(\tau_i)}2^{-(1-\mu)\frac{|P_v|}6}\right)
\end{largearray*}$$
so that by iterating this step down to the leaves, we find
\begin{equation}
\sum_{\mathbf P\in\mathcal P_{\tau,\underline l_\tau,1}}\prod_{v\in\mathfrak V(\tau)}2^{-(1-\mu)\frac{|P_v|}6}\leqslant\left(\sum_{p=0}^{M-h}2^{-\frac{1-\mu}6p}\right)^{4N}\leqslant C_P^N
\label{boundP}\end{equation}
for some constant $C_P$.\par
\bigskip

\subpoint{Bound on trees} Finally, we bound
$$\sum_{\tau\in\mathcal T_N^{(h)}}\prod_{v\in\mathfrak V(\tau)}2^{-\mu}.$$
We can re-express the sum over $\tau$ as a sum over trees with no scale labels that are such that each node that is not a leaf has at least two children, and a sum over scale labels:
$$\sum_{\tau\in\mathcal T_{N}^{(h)}}=\sum_{\tau^*\in\mathcal T_N^*}\sum_{\mathbf h\in\mathbf H_h(\tau^*)}$$
in which $\mathcal T_N^*$ denotes the set of unlabeled rooted trees with $N$ endpoints and $\mathbf H_h(\tau^*)$ denotes the set of scale labels {\it compatible} with $\tau^*$. Therefore
$$\sum_{\tau\in\mathcal T_{N}^{(h)}}\prod_{v\in\mathfrak V(\tau)}2^{-\mu}=\sum_{\tau^*\in\mathcal T_N^*}\sum_{\mathbf h\in\mathbf H_h(\tau^*)}\prod_{v\in\mathfrak V(\tau^*)}2^{-\mu(h_v-h_{p(v)})}$$
in which $p(v)$ denotes the parent of $v$, so that
$$\sum_{\tau\in\mathcal T_{N}^{(h)}}\prod_{v\in\mathfrak V(\tau)}2^{-\mu}\leqslant\sum_{\tau^*\in\mathcal T_N^*}\prod_{v\in\mathfrak V(\tau^*)}\sum_{q=1}^\infty2^{-\mu q}\leqslant\sum_{\tau^*\in\mathcal T_N^*}C_{T,1}^N$$
for some constant $C_{T,1}$, in which we used the fact that $|\mathfrak V(\tau^*)|\leqslant N$. Furthermore, it is a well known fact that $\sum_{\tau^*}1\leqslant 4^N$ (see e.g. \cite[lemma~A.1]{genZO}, the proof is based on constructing an injective map to the set of random walks with $2N$ steps: given a tree, consider a walker that starts at the root, and then travels over branches towards the right until it reaches a leaf, and then travels left until it can go right again on a different branch). Therefore
\begin{equation}
\sum_{\tau\in\mathcal T_{N}^{(h)}}\prod_{v\in\mathfrak V(\tau)}2^{-\mu}\leqslant C_T^N
\label{boundT}\end{equation}
for some constant $C_T$.\par
\bigskip

\subpoint{Conclusion of the proof} Therefore, by combining~(\ref{boundP}) and~(\ref{boundT}) with the trivial estimate $\sum_{\underline l_\tau}1\leqslant 2^N$, we find
\begin{equation}
\mathfrak T_N\leqslant (\mathrm{const.})^N.
\label{boundtreeuv}\end{equation}
Equation~(\ref{boundtreeuv}) trivially holds for $N=1$ as well. If we inject~(\ref{boundtreeuv}) into~(\ref{boundprerefuv}) we get:
\begin{equation}
\frac{1}{\beta|\Lambda|}\int d\underline{\mathbf x}\ \left|(\underline{\mathbf x}-\mathbf x_{2l})^mW_{2l,\underline\alpha}^{(h)}(\underline{\mathbf x})\right|
\leqslant\sum_{N=\max\{1,l-1\}}^\infty(|U|C')^N
2^{-h(N-1)}
\label{bounduvh}\end{equation}
for some constant $C'$ and $h\geqslant 0$. In conclusion, if $|U|$ is small enough (uniformly in $h$ and $l$),
\begin{equation}
\frac{1}{\beta|\Lambda|}\int d\underline{\mathbf x}\ \left|(\underline{\mathbf x}-\mathbf x_{2l})^mW_{2l,\underline\alpha}^{(h)}(\underline{\mathbf x})\right|
\leqslant(|U|C_0)^{\max\{1,l-1\}}
2^{-h({\max\{1,l-1\}}-1)}
\label{pcuv}\end{equation}
for some constant $C_0>0$.\par
\seqskip

\section{First regime}
\label{osec}
\indent We now study the first regime. We consider the tree expansion in the general form discussed in section~\ref{treeexpsec}, with $h^*_1=\bar{\mathfrak h}_0$ and $q=\ell_0=2$, so that there are no local leaves, i.e., all leaves are irrelevant, on scale $\bar{\mathfrak h}_0+1$. Recall that the truncated expectation $\mathcal E^T_{h+1}$ in the right side of \eqref{treeindeq} is with respect to the dressed propagator $\bar g_{h+1}$ in \eqref{inddresspropexpo}, so that \eqref{treeindeq} is to be interpreted as \eqref{effpotoh}. A non trivial aspect of the analysis is that we do not have a priori bounds on the dressed propagator, but just on the ``bare'' one 
$g_{h,\omega}$, see \eqref{estgo}, \eqref{gramboundo}. The goal is to show inductively on $h$ that the same qualitative bounds are valid for $\bar g_{h,\omega}$, namely
\begin{equation}\left\{\begin{array}{>{\displaystyle}l}
\int d\mathbf x\ |\mathbf x^m\bar g_{h,\omega}(\mathbf x)|\leqslant C_g2^{-h}2^{-mh}\\[0.5cm]
\frac{1}{\beta|\Lambda|}\sum_{{\bf k}\in\mathcal B_{\beta,L}}|\hat {\bar g}_{h,\omega}({\bf k})|\leqslant C_G 2^{2h}
\end{array}\right.\label{assumdpropo}\end{equation}
which in terms of the hypotheses of lemma~\ref{powercountinglemma}\ means
$$c_k=3,\ c_g=1,\quad\mathfrak F_h(\underline m)=2^{-mh}.$$
Note that $\ell_0=\lceil c_k/(c_k-c_g)\rceil>c_k/(c_k-c_g)$, as desired. \par
\subseqskip

\subsection{Power counting in the first regime}
\label{powercountingosec}
\indent It follows from lemma~\ref{powercountinglemma} and~(\ref{pcuv}) that
\begin{equation}\begin{largearray}
\frac{1}{\beta|\Lambda|}\int d\underline{\mathbf x}\ \left|(\underline{\mathbf x}-\mathbf x_{2l})^mB_{2l,\underline\omega,\underline\alpha}^{(h)}(\underline{\mathbf x})\right|\\[0.2cm]
\hfill\leqslant 2^{h(3-2l)}2^{-mh}\sum_{N=1}^\infty\sum_{\tau\in\mathcal T_N^{(h)}}\sum_{\underline l_\tau}\sum_{\displaystyle\mathop{\scriptstyle \mathbf P\in\tilde{\mathcal P}_{\tau,\underline l_\tau,2}^{(h)}}_{|P_{v_0}|=2l}}C_1'{}^N\prod_{v\in\mathfrak V(\tau)}2^{(3-|P_v|)}\prod_{v\in\mathfrak E(\tau)}C_1''{}^{l_v}|U|^{\max(1,l_v-1)}
\end{largearray}\label{powercountingol}\end{equation}
for two constants $C_1'$ and $C_1''$.\par
\bigskip

\point{Bounding the sum on trees} First, we notice that the sum over $\underline l_\tau$ can be written as a sum over $l_1,\cdots,l_N$, so that it can be moved before $\sum_\tau$. 
We focus on the sum 
\begin{equation}
\sum_{\tau\in\mathcal T_N^{(h)}}\sum_{\displaystyle\mathop{\scriptstyle \mathbf P\in\tilde{\mathcal P}_{\tau,\underline l_\tau,2}^{(h)}}_{|P_{v_0}|=2l}}\prod_{v\in\mathfrak V(\tau)}2^{(3-|P_v|)}.
\label{9.sum}\end{equation}
We first consider the case $l\geqslant 2$. For all $\theta\in(0,1)$,
$$\sum_{\tau\in\mathcal T_N^{(h)}}\sum_{\displaystyle\mathop{\scriptstyle \mathbf P\in\tilde{\mathcal P}_{\tau,\underline l_\tau,2}^{(h)}}_{|P_{v_0}|=2l}}\prod_{v\in\mathfrak V(\tau)}2^{(3-|P_v|)}=\sum_{\tau\in\mathcal T_N^{(h)}}\sum_{\displaystyle\mathop{\scriptstyle \mathbf P\in\tilde{\mathcal P}_{\tau,\underline l_\tau,2}^{(h)}}_{|P_{v_0}|=2l}}\prod_{v\in\mathfrak V(\tau)}2^{(\theta+(1-\theta))(3-|P_v|)}$$
and since $\ell_0=2$, $|P_v|\geqslant4$ for every node $v$ that is not the first node or a leaf, so that $3-|P_v|\leqslant-|P_v|/4$. Now, if $N\geqslant2$, then given $\tau$, let $v^*_\tau$ be the node with at least two children that is closest to the root, and $h_\tau^*$ its scale. Using the fact that $|P_v|\geqslant 2l+2$ for all $v\prec v_\tau^*$ and the fact that $\tau$ has at least two branches on scales $\geqslant h_\tau^*$, we have
$$\prod_{v\in\mathfrak V(\tau)}2^{\theta(3-|P_v|)}\leqslant 2^{\theta(2l-1)(h-h_\tau^*)}2^{2\theta h_\tau^*}.$$
If $N=1$, we let $h_\tau^*:=0$, and note that the same estimate holds. 
Therefore
$$\begin{largearray}
\sum_{\tau\in\mathcal T_N^{(h)}}\sum_{\displaystyle\mathop{\scriptstyle \mathbf P\in\tilde{\mathcal P}_{\tau,\underline l_\tau,2}^{(h)}}_{|P_{v_0}|=2l}}\prod_{v\in\mathfrak V(\tau)\setminus\{v_0\}}2^{(\theta+(1-\theta))(3-|P_v|)}\\[0.5cm]
\hfill\leqslant
\sum_{h_\tau^*=h+1}^{0}2^{\theta(2l-1)(h-h_\tau^*)+2\theta h_\tau^*}
\sum_{\tau\in\mathcal T_{N}^{(h)}}\sum_{\mathbf P\in\mathcal P_{\tau,\underline l_\tau,2}}\prod_{v\in\mathfrak V(\tau)\setminus\{v_0\}}2^{-(1-\theta)\frac{|P_v|}4}
\end{largearray}$$
which we bound in the same way as in the proof of~(\ref{boundtreeuv}), i.e. splitting
$$(1-\theta)\frac{|P_v|}{2}=(1-\theta)(1-\mu)\frac{|P_v|}{4}+(1-\theta)\mu\frac{|P_v|}{4}\geqslant(1-\theta)(1-\mu)\frac{|P_v|}{4}+(1-\theta)\mu$$
for all $\mu\in(0,1)$ and bounding
$$\sum_{\mathbf P\in\mathcal P_{\tau,\underline l_\tau,2}}\prod_{v\in\mathfrak V(\tau)\setminus\{v_0\}}2^{-(1-\theta)(1-\mu)\frac{|P_v|}4}\leqslant C_P^{\sum_{i=1}^Nl_i}$$
and
$$\sum_{\tau\in\mathcal T_{N}^{(h)}}\prod_{v\in\mathfrak V(\tau)\setminus\{v_0\}}2^{-(1-\theta)\mu}\leqslant C_T^N.$$
Therefore if $l\geqslant2$, then
\begin{equation}
\sum_{\tau\in\mathcal T_N^{(h)}}\sum_{\displaystyle\mathop{\scriptstyle \mathbf P\in\tilde{\mathcal P}_{\tau,\underline l_\tau,2}^{(h)}}_{|P_{v_0}|=2l}}\prod_{v\in\mathfrak V(\tau)\setminus\{v_0\}}2^{(\theta+(1-\theta))(3-|P_v|)}
\leqslant
2^{2\theta h}C_T^N\prod_{i=1}^NC_P^{l_i}.
\label{ineqtreesumf}\end{equation}
Consider now the case with $l=1$. If $N=1$ then the sum over $\tau$ is trivial, i.e., $\mathcal T_1^{(h)}$ consists of a single element, and the sum over ${\mathbf P}$ can be bounded as
\begin{equation} \sum_{\displaystyle\mathop{\scriptstyle \mathbf P\in\tilde{\mathcal P}_{\tau,l_1,2}^{(h)}}_{|P_{v_0}|=2}}\prod_{v\in\mathfrak V(\tau)}2^{(3-|P_v|)}\leqslant 2^h\sum_{\displaystyle\mathop{\scriptstyle \mathbf P\in\tilde{\mathcal P}_{\tau,l_1,2}^{(h)}}_{|P_{v_0}|=2}}\prod_{\substack{v\in\mathfrak V(\tau):\\ v\succeq v'}}2^{4-|P_v|},\label{9.16}\end{equation}
where $v'$ is, if it exists, the leftmost node such that $|P_v|>4$, in which case $4-|P_v|\leqslant -|P_v|/3$; otherwise, we interpret the product over $v$ as 1. Proceeding as in the case $l\geqslant 2$, we bound the right side of \eqref{9.16} by 
\begin{equation} 2^h C^{l_1}\sum_{h_{v'}=h+2}^{0} 2^{2\theta h_{v'}}\leqslant 2^h C' C^{l_1}.\label{9.r8}\end{equation}
If $N\geqslant 2$, then we denote by $\tau^*$ the subtree with $v^*_\tau:v^*$ as first node, and $\tau'_{}$ the linear tree with root on scale $h$ and the endpoint on scale $h^*$, so that $\tau\tau'\cup \tau^*$.
We split \eqref{9.sum} as
\begin{equation} \sum_{l^*=2}^{\sum_{i=1}^Nl_i-N+1}\sum_{h^*=h}^{-2}\ \sum_{\substack{{\bf P}\in\tilde{\mathcal P}_{\tau'_{},l^*,2}:\\ |P_{v_0}|=2}}\Big(\prod_{v\in\mathfrak V(\tau')}2^{3-|P_v|}\Big)\Big(\sum_{\tau^*\in\mathcal T^{(h^*)}_N}\sum_{\substack{{\mathbf P}\in\tilde{\mathcal P}_{\tau^*,\underline l_{\tau^*},2}:\\ |P_{v^*}|=2l^*}}\prod_{v\in\mathfrak V(\tau^*)}2^{3-|P_v|}\Big).\end{equation}
The sum in the last parentheses can be bounded as in the case $l\geqslant 2$, yielding $C^{\sum_il_i}2^{2\theta h^*}$. The remaining sum can be bounded as in 
\eqref{9.16}-\eqref{9.r8} so that, in conclusion, 
\begin{equation}\begin{array}{>{\displaystyle}r@{\ }>{\displaystyle}l}
\sum_{\tau\in\mathcal T_N^{(h)}}\sum_{\displaystyle\mathop{\scriptstyle \mathbf P\in\tilde{\mathcal P}_{\tau,\underline l_\tau,2}^{(h)}}_{|P_{v_0}|=2}}\prod_{v\in\mathfrak V(\tau)}2^{(3-|P_v|)}\leqslant& (C')^{\sum_{i=1}^Nl_i}
\sum_{h^*=h}^{-2} 2^{h-h^*}\sum_{h'=h+2}^{h^*} 2^{2\theta(h'-h^*)}2^{2\theta h^*}\\ \leqslant& (C'')^{\sum_{i=1}^Nl_i}2^h.
\end{array}\label{ineqtreesumo}\end{equation}
\bigskip

\pspoint{$l=1$} Therefore, if $l=1$, (\ref{powercountingol}) becomes (we recall that $q=2>1$ so that $B_2=W_2$, see~(\ref{betafundef}))
\begin{equation}
\int d\mathbf x\ \left|\mathbf x^mW_{2,\omega,\underline\alpha}^{(h)}(\mathbf x)\right|\\[0.5cm]
\hfill\leqslant 2^{2h}2^{-mh}\sum_{N=1}^\infty\,\sum_{l_1,\cdots,l_N\geqslant2}^\infty (C_1'''|U|)^{\sum_{i=1}^N\max(1,l_{i}-1)}
\label{powercountingnoto}\end{equation}
Assuming $|U|$ is small enough and using the subadditivity of the $\max$ function, we rewrite~(\ref{powercountingnoto}) as
\begin{equation}
\int d\mathbf x\ \left|\mathbf x^mW_{2,\omega,\underline\alpha}^{(h)}(\mathbf x)\right|
\leqslant 2^{2h}2^{-mh}C_1|U|
\label{pco}\end{equation}
which we recall holds for $m\leqslant3$.\par
\bigskip

\subpoint{$l\geqslant 2$} Similarly, if $l\geqslant 2$,
\begin{equation}\begin{largearray}
\frac{1}{\beta|\Lambda|}\int d\underline{\mathbf x}\ \left|(\underline{\mathbf x}-\mathbf x_{2l})^mB_{2l,\underline\omega,\underline\alpha}^{(h)}(\underline{\mathbf x})\right|\\[0.5cm]
\hfill\leqslant 2^{h(3-2l+2\theta)}2^{-mh}\sum_{N=1}^\infty\sum_{\displaystyle\mathop{\scriptstyle l_1,\cdots,l_N\geqslant2}_{(l_1-1)+\cdots+(l_N-1)\geqslant l-1+\delta_{N,1}}}^\infty (C_1'''|U|)^{\sum_{i=1}^N\max(1,l_{i}-1)}
\end{largearray}\label{powercountingnotol}\end{equation}
in which the constraint on $l_1,\cdots,l_N$ arises from the fact that, if $N>1$, 
$$|P_{v_0}|\leqslant |I_{v_0}|-2(N-1),$$
while, if $N=1$, $|P_{v_0}|<|I_{v_0}|$. Therefore, assuming that $|U|$ is small enough and summing~(\ref{powercountingnotol}) over $h$, we find
\begin{equation}
\left\{\begin{array}{>{\displaystyle}l}
\frac{1}{\beta|\Lambda|}\int d\underline{\mathbf x}\ \left|(\underline{\mathbf x}-\mathbf x_{4})^mW_{4,\underline\omega,\underline\alpha}^{(h)}(\underline{\mathbf x})\right|
\leqslant 2^{-mh}C_1|U|\\[0.5cm]
\frac{1}{\beta|\Lambda|}\int d\underline{\mathbf x}\ \left|(\underline{\mathbf x}-\mathbf x_{2l})^mW_{2l,\underline\omega,\underline\alpha}^{(h)}(\underline{\mathbf x})\right|
\leqslant 2^{h(3-2l+2\theta)}2^{-mh}(C_1|U|)^{l-1}
\end{array}\right.\label{pcol}\end{equation}
for $l\geqslant3$ and $m\leqslant3$.\par
\bigskip
{\bf Remark:} The estimates \eqref{powercountingol} and \eqref{ineqtreesumo} imply the convergence of 
the tree expansion \eqref{treeexpT}, thus providing a convergent expansion of $W_{2l,\underline\omega,\underline\alpha}^{(h)}$ in $U$. 
\subseqskip

\subsection{The dressed propagator}
\label{dressproposec}
\indent We now prove the estimate~(\ref{assumdpropo}) on the dressed propagator by induction. We recall~(\ref{inddresspropexpo})
\begin{equation}
\left(\hat{\bar g}_{h,\omega}(\mathbf k)\right)^{-1}=f_{h,\omega}^{-1}(\mathbf k)\hat{\bar A}^{(h,\omega)}(\mathbf k)
\label{gbaroA}\end{equation}
with
$$\hat{\bar A}^{(h,\omega)}(\mathbf k):=\hat A(\mathbf k)+
f_{\leqslant h,\omega}(\mathbf k)\hat W_2^{(h)}(\mathbf k)+\sum_{h'=h+1}^{\bar{\mathfrak h}_0}\hat W_2^{(h')}(\mathbf k)$$
whose inverse Fourier transform is denoted by $\bar A^{(h,\omega)}$. Note that (\ref{pco}) on its own does not suffice to prove~(\ref{assumdpropo}) because the bound on
\begin{equation}
f_{\leqslant h,\omega}(\mathbf k)\hat W_2^{(h)}(\mathbf k)+\sum_{h'=h+1}^{\bar{\mathfrak h}_0}\hat W_2^{(h')}(\mathbf k)
\label{Wing}\end{equation}
that it would yield is $(\mathrm{const}.)\ |U|$ whereas on the support of $f_{h,\omega}$, $\hat g^{-1}\sim2^h$, which we cannot 
compare with $|U|$ unless we impose an $\epsilon$-dependent smallness condition on $U$, which we do not want. 
In addition, even if~(\ref{Wing}) were bounded by $(\mathrm{const}.)\ |U|2^h$, we would have to face an extra difficulty to bound $\bar g$ in $\mathbf x$-space: indeed, the naive approach we have used so far (see e.g.~(\ref{estguv})) to bound
$$\int d\mathbf x\ |\mathbf x^m\bar g_{h,\omega}(\mathbf x)|$$
would require a bound on $\partial^n_{\bf k}\hat{\bar g}_{h,\omega}({\bf k})$ with $n>m+3$ (we recall that the integral over $\mathbf x$ is $3$-dimensional), which would in turn require an estimate on
$$\int d\mathbf x\ |\mathbf x^n\bar g_{h',\omega}(\mathbf x)|$$
for $h'>h$, which we do not have (and if we tried to prove it by induction, we would immediately find that the estimate would be required to be uniform in $n$, which we cannot expect to be true).\par
\bigskip

\indent In order to overcome both of the previously mentioned difficulties, we will expand $\hat W_2^{(h')}$ at first order around $\mathbf p_{F,0}^\omega$. The contributions up to first order in ${\bf k}-\mathbf p_{F,0}^\omega$ will be called the {\it local part} of $\hat W_2^{(h')}$. Through symmetry considerations, we will write the local part in terms of constants which we can control, and then use~(\ref{pco}) to bound the remainder. In particular, we will prove that $\hat W_2^{(h)}(\mathbf p_{F,0}^\omega)=0$ from which we will deduce an improved bound for~(\ref{Wing}). Furthermore, since the $\mathbf k$-dependance of the local part is explicit, we will be able to bound all of its derivatives and bound $\bar g$ in $\mathbf x$-space.\par
\bigskip

\point{Local and irrelevant contributions} We define a {\it localization} operator:
\begin{equation}
\mathcal L:\bar A_{h,\omega}(\mathbf x)\longmapsto\delta(\mathbf x)\int d\mathbf y\ \bar A_{h,\omega}(\mathbf y)-\partial_\mathbf x\delta(\mathbf x)\cdot\int d\mathbf y\ \mathbf y \bar A_{h,\omega}(\mathbf y)
\label{Wreldefo}\end{equation}
where $\delta(\mathbf x):=\delta(x_0)\delta_{x_1,0}\delta_{x_2,0}$ and in the second term, as usual, the derivative with respect to $x_1$ and $x_2$ is discrete; as well as
the corresponding {\it irrelevator}:
\begin{equation}
\mathcal R:=\mathds1-\mathcal L.
\label{Rodef}\end{equation}
The action of $\mathcal L$ on functions on $\mathbf k$-space is
(up to finite size corrections coming from the fact that $L<\infty$ that do not change the dimensional estimates computed in this section and that we neglect for  the sake simplicity)
\begin{equation}
\mathcal L\hat{ \bar A}_{h,\omega}(\mathbf k)=\hat{ \bar A}_{h,\omega}(\mathbf p_{F,0}^\omega)+(\mathbf k-\mathbf p_{F,0}^\omega)\cdot\partial_{\mathbf k}\hat{ \bar A}_{h,\omega}(\mathbf p_{F,0}^\omega).
\label{Wrelk}\end{equation}

{\bf Remark}: The reason why $\mathcal L$ is defined as the first order Taylor expansion, is that its role is to separate the {\it relevant} and {\it marginal} parts of $\hat W_2^{(h')}$ from the {\it irrelevant} ones. Indeed, we recall the definition of the {\it scaling dimension} associated to a kernel $\hat W_2^{(h')}$ (see one of the remarks after lemma~\ref{powercountinglemma})
$$c_k-(c_k-c_g)=1$$
which, roughly, means that $\hat W_2^{(h')}$ is bounded by $2^{(c_k-(c_k-c_g))h'}=2^{h'}$. As was remarked above, this bound is insufficient since it does not constrain $\sum_{h'\geqslant h}\hat W_2^{(h')}$ to be smaller than $2^h\sim\hat g^{-1}$. Note that, while $\hat W_2^{(h')}({\bf k})$ is bounded by $2^{h'}$, irrespective of ${\bf k}$, 
$(\mathbf k-\mathbf p_{F,0}^\omega)\cdot\partial_{\mathbf k}\hat W_2^{(h')}({\bf k})$ has an improved dimensional bound, proportional to $2^{h-h'}2^{h'}$, where $2^h\sim |{\bf k}-\mathbf p_{F,0}^\omega|$;
in this sense, we can think of the operator $(\mathbf k-\mathbf p_{F,0}^\omega)\cdot\partial_{\mathbf k}$ as scaling like $2^{h-h'}$. Therefore, the remainder of the first order Taylor expansion is bounded by $2^{2(h-h')}2^{h'}=2^{2h-h'}$ and thereby has a scaling dimension of $-1$ (with respect to $h'$). Thus, by defining $\mathcal L$ as the first order Taylor expansion, we take the focus away from the remainder, which can be bounded easily because it is irrelevant (i.e., it has negative scaling dimension), and concentrate our attention on the relevant and marginal contributions of $\hat W_2^{(h')}$. See~\cite[chapter~8]{benNFi} for details.\par
\bigskip

\indent We then rewrite~(\ref{gbaroA}) as
\begin{equation}
\hat {\bar g}_{h,\omega}(\mathbf k)=f_{h,\omega}(\mathbf k)\left(\mathcal L\hat{\bar A}_{h,\omega}(\mathbf k)\right)^{-1}
\left(\mathds1+\left(\mathcal R\hat{\bar A}_{h,\omega}(\mathbf k)\right)\left(\mathbb L\hat{\bar{\mathfrak g}}_{[h],\omega}(\mathbf k)\right)\right)^{-1}
\label{gseplino}\end{equation}
where $\mathbb L\hat{\bar{\mathfrak g}}_{[h],\omega}$ is a shorthand for
$$\mathbb L\hat{\bar{\mathfrak g}}_{[h],\omega}(\mathbf k):=(f_{\leqslant h+1,\omega}({\bf k})-f_{\leqslant h-2,\omega}({\bf k}))\left(\mathcal L\hat{\bar A}_{h,\omega}(\mathbf k)\right)^{-1}$$
(we can put in the $(f_{\leqslant h+1,\omega}({\bf k})-f_{\leqslant h-2,\omega}({\bf k}))$ factor for free because of the initial $f_{h,\omega}(\mathbf k)$).\par
\bigskip

\point{Local part} We first compute $\mathcal L\hat{\bar A}_{h,\omega}(\mathbf k)$.\par
\bigskip

\subpoint{Non-interacting components} As a first step, we write the local part of the free inverse propagator as
\begin{equation}
\mathcal L\hat A(\mathbf k)=-\left(\begin{array}{*{4}{c}}i k_0&\gamma_1&0&\xi^*\\
\gamma_1&i k_0&\xi&0\\
0&\xi^*&ik_0&\gamma_3\xi\\
\xi&0&\gamma_3\xi^*&ik_0
\end{array}\right)
\label{Lgo}\end{equation}
where
\begin{equation}
\xi:=\frac{3}{2}(ik'_x+\omega k'_y).
\label{xixi2}\end{equation}
\bigskip

\subpoint{Interacting components} We now turn to the terms coming from the interaction. We first note that $\mathcal V^{(h')}$ satisfies the same symmetries as the {\it initial} potential $\mathcal V$ (\ref{2.3a}), listed in section~\ref{symsec}. Indeed, $\mathcal V^{(h')}$ is a function of $\mathcal V$ and a quantity similar to~(\ref{hzdef}) but with an extra cutoff function, which satisfies the symmetries~(\ref{aglobaluo}) through~(\ref{ainversiont}). Therefore
\begin{equation}\begin{largearray}
\hat W_2^{(h')}(\mathbf k)= \hat W_2^{(h')}(-\mathbf k)^*= \hat W_2^{(h')}(R_v\mathbf k)=\sigma_1 \hat W_2^{(h')}(R_h\mathbf k)\sigma_1=-\sigma_3 \hat W_2^{(h')}(I\mathbf k)\sigma_3\\[0.2cm]
\hfill=\hat W_2^{(h')}(P\mathbf k)^T=\left(\begin{array}{*{2}{c}}\mathds1&0\\0&\mathcal T_{\mathbf k}^\dagger\end{array}\right)\hat W_2^{(h')}(T^{-1}\mathbf k)\left(\begin{array}{*{2}{c}}\mathds1&0\\0&\mathcal T_{\mathbf k}\end{array}\right).
\end{largearray}\label{invsymsWo}\end{equation}
This imposes a number of restrictions on $\mathcal L\hat W_2^{(h')}$: indeed, it follows from propositions~\ref{symprop} and~\ref{symfourprop} (see appendix~\ref{constWtapp}) that, since
\begin{equation}
\mathbf p_{F,0}^\omega=-\mathbf p_{F,0}^{-\omega}=R_v\mathbf p_{F,0}^{-\omega}=R_h\mathbf p_{F,0}^\omega=I\mathbf p_{F,0}^{\omega}=P\mathbf p_{F,0}^{-\omega}=T\mathbf p_{F,0}^\omega
\label{pfzsyms}\end{equation}
in which $R_v$, $R_h$, $I$, $P$ and $T$ were defined in section~\ref{symsec}, we have
\begin{equation}
\mathcal L\hat W_2^{(h')}(\mathbf k'+\mathbf p_{F,0}^\omega)=-\left(\begin{array}{*{4}{c}}i\tilde\zeta_{h'}k_0&\gamma_1\tilde\mu_{h'}&0&\nu_{h'}\xi^*\\
\gamma_1\tilde\mu_{h'}&i\tilde\zeta_{h'}k_0&\nu_{h'}\xi&0\\
0&\nu_{h'}\xi^*&i\zeta_{h'}k_0&\gamma_3\tilde\nu_{h'}\xi\\
\nu_{h'}\xi&0&\gamma_3\tilde \nu_{h'}\xi^*&i\zeta_{h'}k_0
\end{array}\right),
\label{Wrelo}\end{equation}
with $(\tilde\zeta_{h'}, \tilde\mu_{h'}, \tilde\nu_{h'},\zeta_{h'},\nu_{h'})\in\mathbb{R}^5$. Furthermore, it follows from (\ref{pco}) that if $h'\leqslant\bar{\mathfrak h}_0$, then
\begin{equation}\begin{array}c
|\tilde\zeta_{h'}|\leqslant (\mathrm{const}.)\  |U|2^{h'},\quad
|\zeta_{h'}|\leqslant (\mathrm{const}.)\  |U|2^{h'},\quad |\tilde\mu_{h'}|\leqslant (\mathrm{const}.)\  |U|2^{2h'-h_\epsilon},
\\
|\nu_{h'}|\leqslant (\mathrm{const}.)\  |U|2^{h'}, \quad 
|\tilde\nu_{h'}|\leqslant (\mathrm{const}.)\  |U|2^{h'-h_\epsilon}.
\end{array}\label{boundzvo}\end{equation}
Injecting~(\ref{Lgo}) and~(\ref{Wrelo}) into~(\ref{Adefo}), we find that
\begin{equation}
\mathcal L\hat{\bar A}_{h,\omega}(\mathbf k'+\mathbf p_{F,0}^\omega)=-\left(\begin{array}{*{4}{c}}i\tilde z_{h}k_0&\gamma_1\tilde m_{h}&0&v_{h}\xi^*\\
\gamma_1\tilde m_{h}&i\tilde z_{h}k_0&v_{h}\xi&0\\
0&v_{h}\xi^*&i z_{h}k_0&\gamma_3 \tilde v_{h}\xi\\
v_{h}\xi&0&\gamma_3 \tilde v_{h}\xi^*&i z_{h}k_0
\end{array}\right)
\label{LAo}\end{equation}
where
\begin{equation}\begin{array}{>{\displaystyle}c}
\tilde z_h:=1+\sum_{h'=h}^{\bar{\mathfrak h}_0}\tilde\zeta_{h'},\quad
\tilde m_h:=1+\sum_{h'=h}^{\bar{\mathfrak h}_0}\tilde\mu_{h'},\quad
\tilde v_h:=1+\sum_{h'=h}^{\bar{\mathfrak h}_0}\tilde\nu_{h'},\\[0.2cm]
z_h:=1+\sum_{h'=h}^{\bar{\mathfrak h}_0}\zeta_{h'},\quad
v_h:=1+\sum_{h'=h}^{\bar{\mathfrak h}_0}\nu_{h'}.
\end{array}\label{summedzvmo}\end{equation}
By injecting~(\ref{boundzvo}) into~(\ref{summedzvmo}), we find
\begin{equation}\begin{array}c
|\tilde m_h-1|\leqslant(\mathrm{const}.)\ |U|,\quad
|\tilde z_h-1|\leqslant(\mathrm{const}.)\ |U|,\quad
|z_h-1|\leqslant(\mathrm{const}.)\ |U|,\\[0.2cm]
|\tilde v_h-1|\leqslant(\mathrm{const}.)\ |U|,\quad
|v_h-1|\leqslant(\mathrm{const}.)\ |U|.
\end{array}\label{boundzvtop}\end{equation}
\bigskip

\subpoint{Dominant part of $\mathcal L\hat{\bar A}_{h,\omega}$} Furthermore, we notice that the terms proportional to $\tilde m_h$ or $\tilde v_h$ are sub-dominant:
\begin{equation}
\mathcal L\hat{\bar A}_{h,\omega}(\mathbf k'+\mathbf p_{F,0}^\omega)=
\mathfrak L\hat{\bar A}_{h,\omega}(\mathbf k'+\mathbf p_{F,0}^\omega)(\mathds1+\sigma_1(\mathbf k'))
\label{sepLAodom}\end{equation}
where
\begin{equation}
\mathfrak L\hat{\bar A}_{h,\omega}(\mathbf k'+\mathbf p_{F,0}^\omega)=-\left(\begin{array}{*{4}{c}}i\tilde z_h k_0&0&0&v_{h}\xi^*\\
0&i\tilde z_h k_0&v_{h}\xi&0\\
0&v_{h}\xi^*&i z_{h}k_0&0\\
v_{h}\xi&0&0&i z_{h}k_0
\end{array}\right)
\label{LAodom}\end{equation}
Before bounding $\sigma_1$, we compute the inverse of~(\ref{LAodom}): using proposition~\ref{matinvprop} (see appendix~\ref{inversapp}), we find that if we define
\begin{equation}
\bar k_0:=z_h k_0,\quad
\tilde k_0:=\tilde z_h k_0,\quad
\bar \xi:=v_h\xi
\label{bardefso}\end{equation}
then
\begin{equation}
\det\mathfrak L\hat{\bar A}_{h,\omega}^{-1}(\mathbf k)(\mathbf k'+\mathbf p_{F,0}^\omega)=\left(\tilde k_0\bar k_0+|\bar\xi|^2\right)^2
\label{detLAodom}\end{equation}
and
\begin{equation}
\mathfrak L\hat{\bar A}_{h,\omega}^{-1}(\mathbf k)(\mathbf k'+\mathbf p_{F,0}^\omega)
=-\frac{(\tilde k_0\bar k_0+|\bar\xi|^2)}{\det\mathfrak L\hat{\bar A}_{h,\omega}}
\left(\begin{array}{*{4}{c}}
-i\bar k_0&0&0&\bar\xi^*\\
0&-i\bar k_0&\bar\xi&0\\
0&\bar\xi^*&-i\tilde k_0&0\\
\bar\xi&0&0&-i\tilde k_0
\end{array}\right).
\label{invLAodom}\end{equation}
In particular, this implies that
\begin{equation}
|\mathfrak L\hat{\bar A}_{h,\omega}^{-1}(\mathbf k'+\mathbf p_{F,0}^{\omega})|\leqslant(\mathrm{const}.)\ 2^{-h}
\label{boundinvLAodom}\end{equation}
which in turn implies
\begin{equation}
|\sigma_1(\mathbf k')|\leqslant(\mathrm{const}.)\ 2^{h_\epsilon-h}.
\label{boundsigmaoo}\end{equation}
\bigskip

\point{Irrelevant part} We now focus on the remainder term $\mathcal R\hat{\bar A}_{h,\omega}(\mathbf k)\,\mathbb L\hat{\bar{\mathfrak g}}_{[h],\omega}(\mathbf k)$ in \eqref{gseplino}, which we now show to be small. The estimates are carried out in $\mathbf x$ space.
We have
$$\begin{largearray}
\int d\mathbf x\left|\mathcal RW_{2,\omega}^{(h')}\ast \mathbb L\bar{\mathfrak g}_{[h],\omega}(\mathbf x)\right|\\[0.5cm]
\hfill=\int d\mathbf x\left|\int d\mathbf y\ W_{2,\omega}^{(h')}(\mathbf y)\left( \mathbb L\bar{\mathfrak g}_{[h],\omega}(\mathbf x-\mathbf y)-\mathbb L\bar{\mathfrak g}_{[h],\omega}(\mathbf x)+\mathbf y\partial_{\mathbf x}\mathbb L\bar{\mathfrak g}_{[h],\omega}(\mathbf x)\right)\right|
\end{largearray}$$
which, by Taylor's theorem, yields
$$\begin{largearray}\int d\mathbf x\left|\mathcal RW_{2,\omega}^{(h')}\ast \mathbb L\bar{\mathfrak g}_{[h],\omega}(\mathbf x)\right|\leqslant \frac92\max_{i,j} \int d\mathbf y\ \left|y_iy_j W_{2,\omega}^{(h')}(\mathbf y)\right|
\cdot\\[0.5cm]
\hfill\cdot\max_{i,j}
\int d\mathbf x\ \left|\partial_{x_i}\partial_{x_j}\mathbb L\bar{\mathfrak g}_{[h],\omega}(\mathbf x)\right|\end{largearray}$$
in which we inject~(\ref{pco}) and~(\ref{estgko}) to find, 
\begin{equation}
\int d\mathbf x\left|\mathcal RW_{2,\omega}^{(h')}\ast \mathbb L\bar{\mathfrak g}_{[h],\omega}(\mathbf x)\right|\leqslant 2^{h}\ (\mathrm{const.})\ |U|.
\label{ineqRtgpre}\end{equation}
Similarly, we find that for all $m\leqslant3$,
\begin{equation}
\int d\mathbf x\left|\mathbf x^m\mathcal RW_{2,\omega}^{(h')}\ast \mathbb L\bar{\mathfrak g}_{[h],\omega}(\mathbf x)\right|\leqslant2^{h}2^{-mh}(\mathrm{const}.)\ |U|.
\label{ineqRtg}\end{equation}
This follows in a straightforward way from
$$
\int d\mathbf y\ \mathbf y\mathcal RW_{2,\omega}^{(h')}(\mathbf y)\mathbb L\bar{\mathfrak g}_{[h],\omega}(\mathbf x-\mathbf y)
=\int d\mathbf y\ \mathbf y W_{2,\omega}^{(h')}(\mathbf y)\left( \mathbb L\bar{\mathfrak g}_{[h],\omega}(\mathbf x-\mathbf y)-\mathbb L\bar{\mathfrak g}_{[h],\omega}(\mathbf x)\right)
$$
and, for $2\leqslant m\leqslant3$,
$$
\int d\mathbf y\ \mathbf y^m\mathcal RW_{2,\omega}^{(h')}(\mathbf y)\mathbb L\bar{\mathfrak g}_{[h],\omega}(\mathbf x-\mathbf y)
=\int d\mathbf y\ \mathbf y^m W_{2,\omega}^{(h')}(\mathbf y) \mathbb L\bar{\mathfrak g}_{[h],\omega}(\mathbf x-\mathbf y).
$$
{\bf Remark}: The estimate \eqref{ineqRtg}, as compared to the dimensional estimate without $\mathcal R$, is 
better by a factor $2^{2(h-h')}$. This is a fairly general argument, and could be repeated with $\mathbb L\bar{\mathfrak g}_{[h],\omega}$ replaced by the inverse Fourier transform of $f_{h,\omega}$:
\begin{equation}
\int d\mathbf x\left|\mathbf x^m\mathcal RW_{2,\omega}^{(h',1)}\ast\check f_{h,\omega}(\mathbf x)\right|\leqslant 2^{2h-mh}\ (\mathrm{const.})\ |U|.
\label{ineqRtf}\end{equation}
\bigskip

Finally, using (\ref{ineqRtg}) and the explicit expression of $\hat g$, we obtain
\begin{equation}
\int d\mathbf x\left|\mathbf x^m\mathcal R\hat{\bar A}^{(h,\omega)}\ast \mathbb L\bar{\mathfrak g}_{[h],\omega}(\mathbf x)\right|\leqslant2^{h}2^{-mh}(\mathrm{const}.)\ (1+|U||h|).
\label{ineqRAo}\end{equation}
\bigskip

\point{Conclusion of the proof} The proof of the first of~(\ref{assumdpropo}) is then completed by injecting~(\ref{LAodom}), (\ref{boundsigmaoo}), (\ref{sepLAodom}), (\ref{boundzvtop}) and (\ref{ineqRAo}) into~(\ref{gseplino}) and its corresponding $\mathbf x$-space representation. The second of~(\ref{assumdpropo}) follows from the first.\par
\subseqskip

\subsection{Two-point Schwinger function}
\label{schwinosec}
\indent We now compute the dominant part of the two-point Schwinger function for $\mathbf k$ {\it well inside} the first regime, i.e.
$${\bf k}\in \mathcal B_{\rm I}^{(\omega)}:=\bigcup_{h=\mathfrak h_1+1}^{\bar{\mathfrak h}_0-1}\mathrm{supp}f_{h,\omega}.$$
Let
$$h_{\mathbf k}:=\max\{h: f_{h,\omega}(\mathbf k)\neq0\}$$
so that if $h\not\in\{h_{\bf k},h_{{\bf k}}-1\}$, then $f_{h,\omega}({\bf k})=0$.\par
\bigskip

\point{Schwinger function in terms of dressed propagators} Since $h_{\mathbf k}\leqslant\bar{\mathfrak h}_0$, the source term $\hat J_{\mathbf k,\alpha_1}^+\hat\psi_{\mathbf k,\alpha_1}^-+\hat\psi_{\mathbf k,\alpha_2}^+\hat J_{\mathbf k,\alpha_2}^-$ is constant with respect to the ultraviolet fields, so that the effective source term $\mathcal X^{(h)}$ defined in \eqref{Qdef} is given, for $h=\bar{\mathfrak h}_0$, by 
\begin{equation}
\mathcal X^{(\bar{\mathfrak h}_0)}(\psi,\hat J_{\mathbf k,\underline\alpha})=\hat J_{\mathbf k,\alpha_1}^+\hat\psi_{\mathbf k,\alpha_1}^-+\hat\psi_{\mathbf k,\alpha_2}^+\hat J_{\mathbf k,\alpha_2}^-
\label{initXo}\end{equation}
which implies that $\mathcal X^{(\bar{\mathfrak h}_0)}$ is in the form~(\ref{exprdomorderren}) with
$$q^{\pm(\bar{\mathfrak h}_0)}=\mathds1,\quad s^{(\bar{\mathfrak h}_0)}({\bf k})=0,\quad  \bar G^{\pm(\bar{\mathfrak h}_0)}=0.$$
Therefore, we can compute $\mathcal X^{(h)}$ for $h\in\{\mathfrak h_1,\cdots,\bar{\mathfrak h}_0-1\}$ inductively using lemma~\ref{schwinlemma}. By using the fact that the support of $\hat{\bar g}_{h,\omega}$ is compact, we find that $\bar G^{(h)}({\bf k})$ no longer depends on $h$ as soon as $h\leqslant h_{\mathbf k}-2$, i.e., 
$\bar G^{(h)}({\bf k})=\bar G^{(h_{{\bf k}}-2)}$, $\forall h\leqslant h_{\bf k}-2$.
Moreover, if $h\leqslant h_{\bf k}-2$, the iterative equation for $s^{(h)}({\bf k})$ \eqref{indsdefren} simplifies into 
\begin{equation}
s_{\alpha_1,\alpha_2}^{(h)}(\mathbf k):=s_{\alpha_1,\alpha_2}^{(h+1)}(\mathbf k)-\sum_{\alpha',\alpha''}\bar G^{+(h_{\bf k}-2)}_{\alpha_1,\alpha'}(\mathbf k)\hat W_{2,(\alpha',\alpha'')}^{(h)}(\mathbf k)\bar G^{-(h_{\bf k}-2)}_{\alpha'',\alpha_2}(\mathbf k).
\end{equation}
We can therefore write out~(\ref{indsdefren}) quite explicitly: for $\mathfrak h_1\leqslant h\leqslant h_{\mathbf k}-2$
\begin{equation}\begin{largearray*}2{r@{\ }>{\displaystyle}l}
 s^{(h)}({\bf k})=&\hat{\bar g}_{h_{\mathbf k},\omega}-\hat{\bar g}_{h_{\mathbf k},\omega}\hat W_{2}^{(h_{\mathbf k}-1)}\hat{\bar g}_{h_{\mathbf k},\omega}\\[0.5cm]
&+\left(\mathds1-\hat{\bar g}_{h_{\mathbf k},\omega}\hat W_{2}^{(h_{\mathbf k}-1,\omega)}\right)\hat{\bar g}_{h_{\mathbf k}-1,\omega}\left(\mathds1-\hat W_{2}^{(h_{\mathbf k}-1)}\hat{\bar g}_{h_{\mathbf k},\omega}\right)\\[0.5cm]
&-\left( \hat{\bar g}_{h_{\mathbf k},\omega}+\hat{\bar g}_{h_{\mathbf k}-1,\omega}-\hat{\bar g}_{h_{\mathbf k},\omega}\hat W_2^{(h_{\mathbf k}-1)}\hat{\bar g}_{h_{\mathbf k}-1,\omega}\right)\left(\sum_{h'=h}^{h_{\mathbf k}-2}\hat W_2^{(h')}\right)\cdot\\[0.5cm]
&\hfill\cdot\left( \hat{\bar g}_{h_{\mathbf k},\omega}+\hat{\bar g}_{h_{\mathbf k}-1,\omega}-\hat{\bar g}_{h_{\mathbf k}-1,\omega}\hat W_2^{(h_{\mathbf k}-1)}\hat{\bar g}_{h_{\mathbf k},\omega}\right)
\end{largearray*}\label{schwinexpr}\end{equation}
where all the functions in the right side are evaluated at ${\bf k}$. Note that in order to get the two-point function defined in section~\ref{introduction}, we must integrate down to $h=\mathfrak h_\beta$: $s_2({\bf k})=s^{(\mathfrak h_\beta)}({\bf k})$. This requires an analysis of the second and third regimes (see sections~\ref{schwintsec} and~\ref{schwinthsec} below). We thus find
\begin{equation}
s_2(\mathbf k)=\left(\hat{\bar g}_{h_{\mathbf k},\omega}(\mathbf k)+\hat{\bar g}_{h_{\mathbf k}-1,\omega}(\mathbf k)\right)\left(\mathds1-\sigma(\mathbf k)-\sigma_{<h_{\mathbf k}}(\mathbf k)\right)
\label{schwinexpro}\end{equation}
where
\begin{equation}
\sigma(\mathbf k):=
\hat W_2^{(h_{\mathbf k}-1)}\hat{\bar g}_{h_{\mathbf k},\omega}
+(\hat{\bar g}_{h_{\mathbf k},\omega}+\hat{\bar g}_{h_{\mathbf k}-1,\omega})^{-1}\hat{\bar g}_{h_{\mathbf k},\omega}\hat W_2^{(h_{\mathbf k}-1)}\hat{\bar g}_{h_{\mathbf k}-1,\omega}(\mathds1-\hat W_2^{(h_{\mathbf k}-1)}\hat{\bar g}_{h_{\mathbf k},\omega})
\label{sigmaodef}\end{equation}
and
\begin{equation}\begin{largearray}
\sigma_{<h_{\mathbf k}}(\mathbf k):=\left(\mathds1-\left( \hat{\bar g}_{h_{\mathbf k},\omega}+\hat{\bar g}_{h_{\mathbf k}-1,\omega}\right)^{-1}\hat{\bar g}_{h_{\mathbf k},\omega}\hat W_2^{(h_{\mathbf k}-1)}\hat{\bar g}_{h_{\mathbf k}-1,\omega}\right)
\left(\sum_{h'=\mathfrak h_\beta}^{h_{\mathbf k}-2}\hat W_2^{(h')}\right)\cdot\\[0.5cm]
\hfill\cdot\left( \hat{\bar g}_{h_{\mathbf k},\omega}+\hat{\bar g}_{h_{\mathbf k}-1,\omega}-\hat{\bar g}_{h_{\mathbf k}-1,\omega}\hat{W}_{2,\omega}^{(h_{\mathbf k}-1)}\hat{\bar g}_{h_{\mathbf k},\omega}\right)
\end{largearray}\label{sigmalodef}\end{equation}
in which $\hat W_2^{(h')}$ with $h'\in\{\bar{\mathfrak h}_2+1,\cdots,\mathfrak h_2-1\}\cup\{\bar{\mathfrak h}_1+1,\cdots,\mathfrak h_1-1\}$ should be interpreted as 0.\par
\bigskip

\point{Bounding the error terms} We then use~(\ref{assumdpropo}), (\ref{pco}) as well as the bound
\begin{equation}
|(\hat{\bar g}_{h_{\mathbf k},\omega}+\hat{\bar g}_{h_{\mathbf k}-1,\omega})^{-1}|\leqslant (\mathrm{const}.)\ 2^{h_{\mathbf k}}
\label{boundginvo}\end{equation}
which follows from~(\ref{LAodom}) and~(\ref{boundzvtop}), in order to bound $\sigma(\mathbf k)$:
\begin{equation}
|\sigma(\mathbf k)|\leqslant (\mathrm{const}.)\  2^{h_{\mathbf k}}|U|.
\label{boundsigmao}\end{equation}
Furthermore, if we assume that
\begin{equation}
\left|\sum_{h'=\mathfrak h_\beta}^{\bar{\mathfrak h}_1}\hat W_2^{(h')}(\mathbf k)\right|\leqslant (\mathrm{const}.)\  2^{2h_\epsilon}|U|
\label{sumWo}\end{equation}
which will be proved when studying the second and third regimes~(\ref{boundtsumW}) and~(\ref{boundthsumW}), then
\begin{equation}
|\sigma_{<h_{\mathbf k}}(\mathbf k)|\leqslant (\mathrm{const}.)\  2^{h_{\mathbf k}}|U|.
\label{boundsigmalk}\end{equation}
\bigskip

\point{Dominant part of the dressed propagators} Furthermore, it follows from~(\ref{invLAodom}) that
\begin{equation}
\hat{\bar g}_{h_{\mathbf k},\omega}(\mathbf k)+\hat{\bar g}_{h_{\mathbf k}-1,\omega}(\mathbf k)=-\frac{1}{\tilde k_0\bar k_0+|\bar\xi|^2}\left(\begin{array}{*{4}{c}}
-i\bar k_0&0&0&\bar\xi^*\\
0&-i\bar k_0&\bar\xi&0\\
0&\bar\xi^*&-i\tilde k_0&0\\
\bar\xi&0&0&-i\tilde k_0\end{array}\right)
(\mathds1+\sigma')
\label{schwinxprgo}\end{equation}
where we recall~(\ref{bardefso})
\begin{equation}
\bar k_0:=z_{h_{\mathbf k}} k_0,\quad
\tilde k_0:=\tilde z_{h_{\mathbf k}} k_0,\quad
\bar \xi:=v_{h_{\mathbf k}}\xi
\label{renko}\end{equation}
in which $\tilde z_{h_{\mathbf k}}$, $z_{h_{\mathbf k}}$ and $v_{h_{\mathbf k}}$ were defined in~(\ref{summedzvmo}) and satisfy (see~(\ref{boundzvtop}))
$$|1-\tilde z_{h_{\mathbf k}}|\leqslant \tilde C_1^{(z)}|U|,\quad
|1-z_{h_{\mathbf k}}|\leqslant C_1^{(z)}|U|,\quad
|1-v_{h_{\mathbf k}}|\leqslant C_1^{(v)}|U|$$
where $\tilde C_1^{(z)}$, $C_1^{(z)}$ and $C_1^{(v)}$ are constants (independent of $h_{\mathbf k}$, $U$ and $\epsilon$). Finally the error term $\sigma'$ is bounded using~(\ref{ineqRAo}) and~(\ref{boundsigmaoo})
\begin{equation}
|\sigma'({\bf k})|\leqslant (\mathrm{const}.)\ ((1+|U||h_{\mathbf k}|)2^{h_{\bf k}}+2^{h_\epsilon-h_{\mathbf k}}).
\label{boundsigmapo}\end{equation}
\bigskip

\point{Proof of Theorem~\ref{theoo}} We now conclude the proof of Theorem~\ref{theoo}, {\it under the assumption} \eqref{sumWo}: we define
$$
z_1:=z_{\mathfrak h_1},\quad
\tilde z_1:=\tilde z_{\mathfrak h_1},\quad
v_1:=v_{\mathfrak h_1}
$$
and use~(\ref{boundzvo}) to bound
$$
|z_{h_{\mathbf k}}-z_1|\leqslant(\mathrm{const}.)\ |U|2^{h_{\mathbf k}},\quad
|\tilde z_{h_{\mathbf k}}-\tilde z_1|\leqslant(\mathrm{const}.)\ |U|2^{h_{\mathbf k}},\quad
|v_{h_{\mathbf k}}-v_1|\leqslant(\mathrm{const}.)\ |U|2^{h_{\mathbf k}}
$$
which we inject into~(\ref{schwinxprgo}), which, in turn, combined with~(\ref{schwinexpro}), (\ref{boundsigmao}), (\ref{boundsigmalk}) and~(\ref{boundsigmapo}) yields~(\ref{schwino}).\par
\subseqskip

\subsection{Intermediate regime: first to second}
\label{otsec}
\point{Integration over the intermediate regime} The integration over the intermediate regime between scales $\mathfrak h_1$ and $\bar{\mathfrak h}_1$ can be performed in a way that is entirely analogous to that in the bulk of the first regime, with the difference that it is performed in a single step. The outcome is that, in particular, the effective potential on scale $\bar{\mathfrak h}_1$ satisfies
an estimate analogous to \eqref{pco} (details are left to the reader):
\begin{equation}
\left\{\begin{array}{>{\displaystyle}l}
\int d\mathbf x\ \left|\mathbf x^mW_{2,\omega,\underline\alpha}^{(\bar{\mathfrak h}_1)}(\mathbf x)\right|\leqslant \bar C_{1}2^{2\bar{\mathfrak h}_1}2^{-m \bar{\mathfrak h}_1}|U|\\[0.5cm]
\frac{1}{\beta|\Lambda|}\int d\underline{\mathbf x}\ \left|(\underline{\mathbf x}-\mathbf x_{4})^mW_{4,\underline\omega,\underline\alpha}^{(\bar{\mathfrak h}_1)}(\underline{\mathbf x})\right|\leqslant \bar C_{1}2^{-\bar{\mathfrak h}_1m}|U|\\[0.5cm]
\frac{1}{\beta|\Lambda|}\int d\underline{\mathbf x}\ \left|(\underline{\mathbf x}-\mathbf x_{2l})^mW_{2l,\underline\omega,\underline\alpha}^{(\bar{\mathfrak h}_1)}(\underline{\mathbf x})\right|
\leqslant 2^{\bar{\mathfrak h}_1(3-2l+2\theta-m)}(\bar C_{1}|U|)^{l-1}
\end{array}\right. \label{boundot}\end{equation}
for $l\geqslant3$ and $m\leqslant3$.\par
\bigskip

\par
\point{Improved estimate on inter-layer terms} In order to treat the second regime, we will need an improved estimate on
\begin{equation}
\int d\mathbf x\ \mathbf x^mW_{2,\omega,(\alpha,\alpha')}^{(h')}(\mathbf x)
\label{9.45}\end{equation}
where $(i,j)$ are in {\it different layers}, i.e. $(\alpha,\alpha')\in\{a,b\}\times\{\tilde a,\tilde b\}$ or $(\alpha,\alpha')\in\{\tilde a,\tilde b\}\times\{a,b\}$, $h'\geqslant\bar{\mathfrak h}_1$. Note that since $W_{4,(\alpha_1,\alpha_1',\alpha_2,\alpha_2')}^{(M)}$ is proportional to $\delta_{\alpha_1,\alpha_1'}\delta_{\alpha_2,\alpha_2'}$, any contribution to $W_{2,\omega,(\alpha,\alpha')}^{(h',2)}$ must contain at least one propagator between different layers, i.e. $\bar g_{(h'',\omega),(\bar\alpha,\bar\alpha')}$ with $h'<h''\leqslant\bar{\mathfrak h}_0$ or $g_{(h''),(\bar\alpha,\bar\alpha')}$ with $h''\geqslant0$, and $(\bar\alpha,\bar\alpha')\in\{a,b\}\times\{\tilde a,\tilde b\}\cup\{\tilde a,\tilde b\}\times\{a,b\}$. This can be easily proved using the fact that if the inter-layer hoppings were neglected (i.e. $\gamma_1=\gamma_3=0$), then the system would be symmetric under
$$
\psi_{\mathbf k,a}\mapsto\psi_{\mathbf k,a},\quad
\psi_{\mathbf k,\tilde b}\mapsto-\psi_{\mathbf k,\tilde b},\quad
\psi_{\mathbf k,\tilde a}\mapsto-\psi_{\mathbf k,\tilde a},\quad
\psi_{\mathbf k,b}\mapsto\psi_{\mathbf k,b}
$$
which would imply that $W_{2,\omega,(\alpha,\alpha')}^{(h')}=0$. The presence of at least one propagator between different layers allows us to obtain a dimensional gain, induced by an improved estimate on each such propagator.  To prove an improved estimate on the inter-layer propagator, let us start by considering the bare one, $g_{(h'',\omega),(\bar\alpha,\bar\alpha')}$ with $(\bar\alpha,\bar\alpha')\in\{a,b\}\times\{\tilde a,\tilde b\}\cup\{\tilde a,\tilde b\}\times\{a,b\}$ and $h'<h''\leqslant\bar{\mathfrak h}_0$ (similar considerations are valid for the ultraviolet counterpart): using the explicit expression (\ref{freeprop}) it is straightforward to check that it is bounded as in \eqref{estgo}, \eqref{estgko}, times an extra factor $\epsilon2^{-h''}$. We now proceed as in section~\ref{powercountingosec} and prove by induction that the same dimensional gain is associated with the {\it dressed} propagator $\bar g_{(h'',\omega),(\bar\alpha,\bar\alpha')}$, with $(\bar\alpha,\bar\alpha')\in\{a,b\}\times\{\tilde a,\tilde b\}\cup\{\tilde a,\tilde b\}\times\{a,b\}$, and, therefore, with \eqref{9.45} itself. \par
\bigskip

\subpoint{Trees with a single endpoint} We first consider the contributions $\mathfrak A_{2,\omega,(\alpha,\alpha')}^{(h')}$ to $W_{2,\omega,(\alpha,\alpha')}^{(h')}$ from trees $\tau\in\mathcal T_1^{(h)}$ with a single endpoint. The $\mathfrak F_h(\underline m)$ factor in the estimate~(\ref{boundpre}) can be removed for these contributions using the fact that they have an empty spanning tree (i.e. $\mathbf T(\tau)=\emptyset$), which implies that the $\mathbf z^m$'s in the right side of~(\ref{ineqpcexpl}) are all $\underline{\mathbf z}^{(v)}$'s and not $\mathbf z_\ell$'s, and can be estimated dimensionally by a constant instead of $\mathfrak F_h(\underline m)$. Therefore, combining this fact with the gain associated to the propagator, we find that for all $m\leqslant3$,
\begin{equation}
\int d\mathbf x\ \left|\mathbf x^m\mathfrak A_{2,\omega,(\alpha,\alpha')}^{(h')}(\mathbf x)\right|\leqslant(\mathrm{const}.)\ \epsilon2^{h'}|U|.
\label{improvestoneep}\end{equation}
\bigskip

\subpoint{Trees with at least two endpoints} We now consider the contributions $\mathfrak B_{2,\omega,(\alpha,\alpha')}^{(h')}$ to $W_{2,\omega,(\alpha,\alpha')}^{(h')}$ from trees with $\geqslant2$ endpoints. Let $v^*_\tau$ be the node that has at least two children that is closest to the root and let $h^*_\tau$ be its scale. Repeating the reasoning leading to~(\ref{powercountingnoto}), and using the fact that the $\mathbf x^m$ falls on a node on scale $\geqslant h^*_\tau$, we find
$$\int d\mathbf x\ \left|\mathbf x^m\mathfrak B_{2,\omega,(\alpha,\alpha')}^{(h')}(\mathbf x)\right|\leqslant(\mathrm{const.})\ \epsilon\sum_{h_\tau^*=h'+1}^{0}2^{-mh_\tau^*}2^{(h'-h_\tau^*)}2^{2\theta h_{\tau}^*}|U|^2$$
for any $\theta\in(0,1)$, so that
\begin{equation}
\int d\mathbf x\ \left|\mathbf x^m\mathfrak B_{2,\omega,(\alpha,\alpha')}^{(h')}(\mathbf x)\right|\leqslant(\mathrm{const}.)\ \epsilon2^{\theta'h'+\min(0,1-m)h'}|U|^2
\label{improvestbetaep}\end{equation}
where $\theta':=2\theta-1>1$.\par
\bigskip

\indent Combining~(\ref{improvestoneep}) and~(\ref{improvestbetaep}), and repeating the argument in section~\ref{dressproposec}, we conclude the proof of the desired improvement on the estimate of $\bar g$, and that
\begin{equation}
\int d\mathbf x\ \left|\mathbf x^mW_{2,\omega,(\alpha,\alpha')}^{(h')}(\mathbf x)\right|\leqslant(\mathrm{const}.)\ \epsilon2^{\theta'h}|U|(1+2^{\min(0,1-m)h}|U|)
\label{ineqvh}\end{equation}
for $m\leqslant3$.\par

\seqskip

\section{Second regime}
\label{tsec}
\indent We now perform the multiscale integration in the second regime. As in the first regime, we shall inductively prove that $\bar g_{h,\omega}$ satisfies the same estimate as $g_{h,\omega}$ (see~(\ref{estgt}) and~(\ref{gramboundt})): for all $m\leqslant3$,
\begin{equation}\left\{\begin{array}{>{\displaystyle}l}
\int d\mathbf x\ |x_0^{m_0}x^{m_k}\bar g_{h,\omega}(\mathbf x)|\leqslant (\mathrm{const.})\ 2^{-h-m_0h-m_k\frac{h+h_\epsilon}2}\\[0.5cm]
\frac{1}{\beta|\Lambda|}\sum_{\mathbf k\in\mathcal B_{\beta,L}^{(h,\omega)}}|\hat{\bar g}_{h,\omega}(\mathbf k)|\leqslant(\mathrm{const.})\ 2^{h+h_\epsilon}
\end{array}\right.\label{assumdpropt}\end{equation}
which in terms of the hypotheses of lemma~\ref{powercountinglemma}\ means
$$c_k=2,\ c_g=1,\quad\mathfrak F_h(m_0,m_1,m_2)=2^{-m_0h-(m_1+m_2)\frac{h+h_\epsilon}2},$$
$C_g=(\mathrm{const}.)\ $ and $C_G=(\mathrm{const}.)\  2^{h_\epsilon}$.\par
\bigskip

{\bf Remark}: As can be seen from~(\ref{boundfreepropzt}), different components of $g_{h,\omega}$ scale in different ways. In order to highlight this fact, we call the $\{a,\tilde b\}$ components {\it massive} and the $\{\tilde a,b\}$ components {\it massless}. It follows from~(\ref{boundfreepropzt}) that the $L_1$ norm of the massive-massive sub-block of $g_{h,\omega}(\mathbf x)$ is bounded by $(\mathrm{const}.)\ 2^{-h_\epsilon}$ (instead of $2^{-h}$, compare 
with \eqref{assumdpropt}) and that the massive-massless sub-blocks are bounded by $(\mathrm{const}.)\ 2^{-(h+h_\epsilon)/2}$. In the following, in order to simplify the discussion, we will ignore these improvements, even though the bounds we will thus derive for the non-local corrections may not be optimal.\par
\bigskip

\indent In addition, in order to apply lemma~\ref{powercountinglemma}, we have to ensure that hypothesis~(\ref{assumW}) is satisfied, so we will also prove a bound on the 4-field kernels by induction ($\ell_0=3$ in this regime, so~(\ref{assumW}) must be satisfied by the 4-field kernels): for all $m\leqslant3$,
\begin{equation}
\frac{1}{\beta|\Lambda|}\int d\underline{\mathbf x}\ |(\underline{\mathbf x}-\mathbf x_4)^mW_{4,\omega,\underline\alpha}^{(h)}(\underline{\mathbf x})|\leqslant C_\mu'|U|\mathfrak F_h(\underline m)\label{assumWft}\end{equation}
where $C_\mu'$ is a constant that will be defined below. Note that in this regime,
$$\ell_0=3>\frac{c_k}{c_k-c_g}=2$$
as desired.
\subseqskip

\subsection{Power counting in the second regime}
\label{pcosec}
\point{Power counting estimate} It follows from lemma~\ref{powercountinglemma}\ and~(\ref{boundot}) that for all $m\leqslant3$ and some $c_1,c_2>0$, 
$$\begin{largearray}
\frac{1}{\beta|\Lambda|}\int d\underline{\mathbf x}\ \left|(\underline{\mathbf x}-\mathbf x_{2l})^mB_{2l,\underline\omega,\underline\alpha}^{(h)}(\underline{\mathbf x})\right|\leqslant 2^{h(2-l)}\mathfrak F_h(\underline m)2^{-lh_\epsilon}\sum_{N=1}^\infty\sum_{\tau\in\mathcal T_N^{(h)}}\sum_{\underline l_\tau}\sum_{\displaystyle\mathop{\scriptstyle \mathbf P\in\tilde{\mathcal P}_{\tau,\underline l_\tau,3}^{(h)}}_{|P_{v_0}|=2l}}\\[0.5cm]
\hfill(c_12^{-h_\epsilon})^{N-1}\prod_{v\in\mathfrak V(\tau)}2^{(2-\frac{|P_v|}2)}\prod_{v\in\mathfrak E(\tau)}(c_22^{h_\epsilon})^{l_v}|U|^{l_v-1}2^{\mathbf1_{l_v>2}(2-l_v+\theta')h_\epsilon}
\end{largearray}$$
where $\mathbf 1_{l_v>2}$ is equal to 1 if $l_v>2$ and 0 otherwise, and $\theta':=2\theta-1>0$, so that
\begin{equation}\begin{largearray}
\frac{1}{\beta|\Lambda|}\int d\underline{\mathbf x}\ \left|(\underline{\mathbf x}-\mathbf x_{2l})^mB_{2l,\underline\omega,\underline\alpha}^{(h)}(\underline{\mathbf x})\right|\leqslant 2^{h(2-l)}\mathfrak F_h(\underline m)2^{-(l-1)h_\epsilon}\cdot\\[0.5cm]
\hfill\cdot\sum_{N=1}^\infty\sum_{\tau\in\mathcal T_N^{(h)}}\sum_{\underline l_\tau}\sum_{\displaystyle\mathop{\scriptstyle \mathbf P\in\tilde{\mathcal P}_{\tau,\underline l_\tau,3}^{(h)}}_{|P_{v_0}|=2l}}c_1^{N-1}2^{Nh_\epsilon}\prod_{v\in\mathfrak V(\tau)}2^{(2-\frac{|P_v|}2)}\prod_{v\in\mathfrak E(\tau)}c_2^{l_v}|U|^{l_v-1}2^{\mathbf1_{l_v>2}\theta'h_\epsilon}.
\end{largearray}\label{powercountingt}\end{equation}
\bigskip

\point{Bounding the sum on trees} By repeating the computation that leads to~(\ref{boundtreeuv}), noticing that if $\ell_0=3$, then for $v\in\mathfrak V(\tau)$ we have 
$2-|P_v|/2\leqslant-|P_v|/6$, we bound
\begin{equation}
\sum_{\tau\in\mathcal T_{N}^{(h)}}\sum_{\displaystyle\mathop{\scriptstyle \mathbf P\in\mathcal P_{\tau,\underline l_\tau,3}^{(h)}}_{|P_{v_0}|=2l}}\prod_{v\in\mathfrak V(\tau)}2^{2-\frac{|P_v|}2}\leqslant c_3^N
\label{ineqtreesumt}\end{equation}
for some constant $c_3>0$. Thus~(\ref{powercountingt}) becomes 
\begin{equation}\begin{largearray}
\frac{1}{\beta|\Lambda|}\int d\underline{\mathbf x}\ \left|(\underline{\mathbf x}-\mathbf x_{2l})^mB_{2l,\underline\omega,\underline\alpha}^{(h)}(\underline{\mathbf x})\right|\leqslant 2^{h(2-l)}\mathfrak F_h(\underline m)2^{-(l-1)h_\epsilon}\cdot\\[0.5cm]
\hfill\cdot\sum_{N\geqslant 1}\sum_{\substack{l_1,\ldots,l_N\geqslant 2\\ \sum_{i=1}^N(l_i-1)\geqslant l-1+\delta_{N,1}}}2^{Nh_\epsilon}(c_4|U|)^{\sum_{i=1}^N(l_i-1)}
\label{10.12}\end{largearray}\end{equation}
for some $c_4>0$. Note that, if $l=2$, the contribution with $N=1$ to the left side admits an improved bound of the form $c_4\mathfrak F_h(\underline m)2^{\theta' h}|U|^2$, which is better 
than the corresponding term in the right side of \eqref{10.12}. This implies 
\begin{equation}
\int d{\mathbf x}\left| {\mathbf x}^mW_{2,\underline\omega,\underline\alpha}^{(h)}({\mathbf x})\right| \leqslant c_52^{h+h_\epsilon}\mathfrak F_h(\underline m)|U|
\label{pct}\end{equation}
and
\begin{equation}
\left\{\begin{array}{>{\displaystyle}l}
\frac{1}{\beta|\Lambda|}\int d\underline{\mathbf x}\ \left|(\underline{\mathbf x}-\mathbf x_{4})^mB_{4,\underline\omega,\underline\alpha}^{(h)}(\underline{\mathbf x})\right|\leqslant c_5\mathfrak F_h(\underline m)(2^{h_\epsilon}+2^{\theta'h})|U|^2\\[0.5cm]
\frac{1}{\beta|\Lambda|}\int d\underline{\mathbf x}\ \left|(\underline{\mathbf x}-\mathbf x_{2l})^mB_{2l,\underline\omega,\underline\alpha}^{(h)}(\underline{\mathbf x})\right|\leqslant 2^{(h+h_\epsilon)(2-l)}\mathfrak F_h(\underline m)(c_5|U|)^{l-1}
\end{array}\right.\label{powercountingttB}\end{equation}
for some $c_5>0$, with $l\geqslant3$. By summing the previous two inequalities, we find
\begin{equation}
\left\{\begin{array}{>{\displaystyle}l}
\frac{1}{\beta|\Lambda|}\int d\underline{\mathbf x}\ \left|(\underline{\mathbf x}-\mathbf x_{4})^mW_{4,\underline\omega,\underline\alpha}^{(h)}(\underline{\mathbf x})\right|\leqslant C_\mu\mathfrak F_h(\underline m)|U|(1+c_6|U|(\epsilon (h_\epsilon-h)+\epsilon^{\theta'}))\\[0.5cm]
\frac{1}{\beta|\Lambda|}\int d\underline{\mathbf x}\ \left|(\underline{\mathbf x}-\mathbf x_{2l})^mW_{2l,\underline\omega,\underline\alpha}^{(h)}(\underline{\mathbf x})\right|\leqslant 2^{(h+h_\epsilon)(2-l)}\mathfrak F_h(\underline m)(c_6|U|)^{l-1}
\end{array}\right.\label{powercountingtt}\end{equation}
for some $c_6>0$, which, in particular, recalling that in the second regime $h_\epsilon-h\leqslant -2h_\epsilon+C$, for some constant $C$ independent of $\epsilon$, 
implies~(\ref{assumWft}) with
$$C_\mu':=C_\mu(1+c_7\mathop{\mathrm{sup}}_{|U|<U_0,\epsilon<\epsilon_0}|U|(\epsilon|\log\epsilon|+\epsilon^{\theta'}))$$
for some $c_7>0$.
\bigskip

{\bf Remark}: The estimates \eqref{powercountingt} and \eqref{ineqtreesumt} imply the convergence of 
the tree expansion \eqref{treeexpT}, thus providing a convergent expansion of $W_{2l,\underline\omega,\underline\alpha}^{(h)}$ in $U$. 

\bigskip

{\bf Remark}: The first of~(\ref{powercountingtt}) exhibits a tendency to grow {\it logarithmically}
in $2^{-h}$. This is not an artifact of the bounding procedure: indeed the second-order flow, computed in~\cite{vafOZ}, exhibits the same logarithmic growth. 
However, the presence of the $\epsilon$ factor in front of $(h_\epsilon-h)\leqslant2|\log\epsilon|$ ensures this growth is benign: 
it is cut off before it has a chance to be realized. 
\par
\subseqskip

\subsection{The dressed propagator}
\label{dressproptsec}
\indent We now turn to the inductive proof of \eqref{assumdpropt}. We recall that (see~(\ref{Adeft}))
\begin{equation}
\hat{\bar g}_{h,\omega}(\mathbf k)
=f_{h,\omega}(\mathbf k)\hat{\bar A}_{h,\omega}^{-1}(\mathbf k)
\label{dresspropt}\end{equation}
where
$$
\hat{\bar A}_{h,\omega}(\mathbf k):=\hat A(\mathbf k)+ f_{\leqslant h,\omega}(\mathbf k)\hat W_2^{(h)}(\mathbf k)+\sum_{h'=h+1}^{\bar{\mathfrak h}_1}\hat W_2^{(h')}(\mathbf k)+\sum_{h'=\mathfrak h_1}^{\bar{\mathfrak h}_0}\hat W_2^{(h')}(\mathbf k).
$$
We will separate the {\it local} part of $\bar A$ from the remainder by using the localization operator defined in~(\ref{Wreldefo}) (see the remark at the end of this section for an explanation of why we can choose the same localization operator as in the first regime even though the scaling dimension is different) and rewrite~(\ref{dresspropt}) as
\begin{equation}
\hat{\bar g}_{h,\omega}(\mathbf k)=f_{h,\omega}(\mathbf k)\left(\mathcal L\hat{\bar A}_{h,\omega}(\mathbf k)\right)^{-1}
\left(\mathds1+\mathcal R\hat{\bar A}_{h,\omega}(\mathbf k)\left(\mathbb L\hat{\bar{\mathfrak g}}_{[h],\omega}(\mathbf k)\right)\right)^{-1}
\label{gseplint}\end{equation}
where $\mathbb L\hat{\bar{\mathfrak g}}_{[h],\omega}$ is a shorthand for
$$\mathbb L\hat{\bar{\mathfrak g}}_{[h],\omega}(\mathbf k):=(f_{\leqslant h+1,\omega}({\bf k})-f_{\leqslant h-2,\omega}({\bf k}))\left(\mathcal L\hat{\bar A}_{h,\omega}(\mathbf k)\right)^{-1}.$$
Similarly to the first regime, we now compute $\mathcal L\hat{\bar A}_{h,\omega}(\mathbf k)$ and bound $\mathcal R\hat{\bar A}_{h,\omega}(\mathbf k)\mathbb L\hat{\bar{\mathfrak g}}_{[h],\omega}(\mathbf k)$. We first write the local part of the non-interacting contribution:
\begin{equation}
\mathcal L\hat A(\mathbf k)=-\left(\begin{array}{*{4}{c}}i k_0&\gamma_1&0&\xi^*\\
\gamma_1&i k_0&\xi&0\\
0&\xi^*&ik_0&\gamma_3\xi\\
\xi&0&\gamma_3\xi^*&ik_0
\end{array}\right)
\label{Lgt}\end{equation}
where
\begin{equation}
\xi:=\frac{3}{2}(ik'_x+\omega k'_y).
\label{xixi3}\end{equation}
\bigskip

\point{Local part} The symmetries discussed in the first regime (see~(\ref{invsymsWo}) and~(\ref{pfzsyms})) still hold in this regime, so that~(\ref{Wrelo}) still holds:
\begin{equation}
\mathcal L\hat W_2^{(h')}(\mathbf k'+\mathbf p_{F,0}^\omega)=-\left(\begin{array}{*{4}{c}}i\tilde\zeta_{h'}k_0&\gamma_1\tilde\mu_{h'}&0&\nu_{h'}\xi^*\\
\gamma_1\tilde\mu_{h'}&i\tilde\zeta_{h'}k_0&\nu_{h'}\xi&0\\
0&\nu_{h'}\xi^*&i\zeta_{h'}k_0&\gamma_3\tilde\nu_{h'}\xi\\
\nu_{h'}\xi&0&\gamma_3\tilde \nu_{h'}\xi^*&i\zeta_{h'}k_0
\end{array}\right),
\label{Wrelt}\end{equation}
with $(\tilde\zeta_{h'}, \tilde\mu_{h'}, \tilde\nu_{h'},\zeta_{h'},\nu_{h'})\in\mathbb{R}^5$. Furthermore, it follows from~(\ref{pct}) that if $h'\leqslant\bar{\mathfrak h}_1$, then
\begin{equation}\begin{array}c
|\tilde\zeta_{h'}|\leqslant (\mathrm{const}.)\  |U|2^{h_\epsilon},\quad
|\zeta_{h'}|\leqslant (\mathrm{const}.)\  |U|2^{h_\epsilon},\quad |\tilde\mu_{h'}|\leqslant (\mathrm{const}.)\  |U|2^{h'},
\\
|\nu_{h'}|\leqslant (\mathrm{const}.)\  |U|2^{\frac {h'}2+\frac{h_\epsilon}2}, \quad 
|\tilde\nu_{h'}|\leqslant (\mathrm{const}.)\  |U|2^{\frac {h'}2-\frac{h_\epsilon}2}.
\end{array}\label{boundzvtt}\end{equation}
If $\mathfrak h_1\leqslant h'\leqslant\bar{\mathfrak h}_0$, then it follows from~(\ref{pco}) that
\begin{equation}
|\tilde\zeta_{h'}|\leqslant (\mathrm{const}.)\  |U|2^{h'},\quad
|\zeta_{h'}|\leqslant (\mathrm{const}.)\  |U|2^{h'},\quad
|\nu_{h'}|\leqslant (\mathrm{const}.)\  |U|2^{h'},
\label{boundzvto}\end{equation}
and from~(\ref{ineqvh}) that
\begin{equation}
|\tilde\mu_{h'}|\leqslant(\mathrm{const}.)\ 2^{\theta h'}|U|,\quad
|\tilde\nu_{h'}|\leqslant(\mathrm{const}.)\ 2^{\theta' h'}|U|
\label{boundvot}\end{equation}
for some $\theta'\in(0,1)$. Injecting~(\ref{Lgt}) and~(\ref{Wrelt}) into~(\ref{Adeft}), we find that
\begin{equation}
\mathcal L\hat{\bar A}_{h,\omega}(\mathbf k'+\mathbf p_{F,0}^\omega)=-\left(\begin{array}{*{4}{c}}i\tilde z_{h}k_0&\gamma_1\tilde m_{h}&0&v_{h}\xi^*\\
\gamma_1\tilde m_{h}&i\tilde z_{h}k_0&v_{h}\xi&0\\
0&v_{h}\xi^*&i z_{h}k_0&\gamma_3 \tilde v_{h}\xi\\
v_{h}\xi&0&\gamma_3 \tilde v_{h}\xi^*&i z_{h}k_0
\end{array}\right)
\label{LAt}\end{equation}
where
\begin{equation}\begin{array}{>{\displaystyle}c}
\tilde z_h:=1+\sum_{h'=h}^{\bar{\mathfrak h}_0}\tilde\zeta_{h'},\quad
\tilde m_h:=1+\sum_{h'=h}^{\bar{\mathfrak h}_0}\tilde\mu_{h'},\quad
\tilde v_h:=1+\sum_{h'=h}^{\bar{\mathfrak h}_0}\tilde\nu_{h'},\\[0.2cm]
z_h:=1+\sum_{h'=h}^{\bar{\mathfrak h}_0}\zeta_{h'},\quad
v_h:=1+\sum_{h'=h}^{\bar{\mathfrak h}_0}\nu_{h'}
\end{array}\label{summedzvmt}\end{equation}
in which $\tilde\zeta_{h'}$, $\tilde\mu_{h'}$, $\tilde\nu_{h'}$, $\zeta_{h'}$ and $\nu_{h'}$ with $h'\in\{\bar{\mathfrak h}_1+1,\cdots,\mathfrak h_1-1\}$ are to be interpreted as 0.
By injecting~(\ref{boundzvtt}) through~(\ref{boundvot}) into~(\ref{summedzvmt}), we find
\begin{equation}\begin{array}c
|\tilde m_h-1|\leqslant(\mathrm{const}.)\ |U|,\quad
|\tilde z_h-1|\leqslant(\mathrm{const}.)\ |U|,\quad
|z_h-1|\leqslant(\mathrm{const}.)\ |U|,\\[0.2cm]
|\tilde v_h-1|\leqslant(\mathrm{const}.)\ |U|,\quad
|v_h-1|\leqslant(\mathrm{const}.)\ |U|.
\end{array}\label{boundzvttp}\end{equation}
\bigskip

\point{Dominant part of $\mathcal L\hat{\bar A}_{h,\omega}$} Furthermore, we notice that the terms proportional to $\tilde z_h$ or $\tilde v_h$ are sub-dominant:
\begin{equation}
\mathcal L\hat{\bar A}_{h,\omega}(\mathbf k'+\mathbf p_{F,0}^\omega)=
\mathfrak L\hat{\bar A}_{h,\omega}(\mathbf k'+\mathbf p_{F,0}^\omega)(\mathds1+\sigma_3(\mathbf k'))
\label{sepLAtdom}\end{equation}
where
\begin{equation}
\mathfrak L\hat{\bar A}_{h,\omega}(\mathbf k'+\mathbf p_{F,0}^\omega)=-\left(\begin{array}{*{4}{c}}0&\gamma_1\tilde m_{h}&0&v_{h}\xi^*\\
\gamma_1\tilde m_{h}&0&v_{h}\xi&0\\
0&v_{h}\xi^*&i z_{h}k_0&0\\
v_{h}\xi&0&0&i z_{h}k_0
\end{array}\right)
\label{LAtdom}\end{equation}
Before bounding $\sigma_3$, we compute the inverse of~(\ref{LAtdom}), which is elementary once it is put in block-diagonal form: using proposition~\ref{blockdiagprop} (see appendix~\ref{diagapp}), we find that if we define
\begin{equation}
\bar\gamma_1:=\tilde m_h \gamma_1,\quad
\bar k_0:=z_h k_0,\quad
\bar \xi:=v_h\xi
\label{bardefst}\end{equation}
then
\begin{equation}
\left(\mathfrak L\hat{\bar A}_{h,\omega}(\mathbf k)\right)^{-1}=
\left(\begin{array}{*{2}{c}}\mathds1&\bar M_{h}(\mathbf k)^\dagger\\0&\mathds1\end{array}\right)\left(\begin{array}{*{2}{c}}\bar a_{h}^{(M)}&0\\0&\bar a_{h}^{(m)}(\mathbf k)\end{array}\right)\left(\begin{array}{*{2}{c}}\mathds1&0\\\bar M_{h}(\mathbf k)&\mathds1\end{array}\right)
\label{invLAtdom}\end{equation}
where
\begin{equation}
\bar a_{h}^{(M)}:=-\left(\begin{array}{*{2}{c}}0&\bar\gamma_1^{-1}\\\bar\gamma_1^{-1}&0\end{array}\right),\quad
\bar a_{h}^{(m)}(\mathbf p_{F,0}^\omega+\mathbf k'):=\frac{\bar\gamma_1}{\bar\gamma_1^2\bar k_0^2+|\bar\xi|^4}\left(\begin{array}{*{2}{c}}
i\bar\gamma_1\bar k_0&(\bar\xi^*)^2\\
\bar\xi^2&i\bar\gamma_1\bar k_0\end{array}\right)
\label{bat}\end{equation}
and
\begin{equation}
\bar M_{h}(\mathbf p_{F,0}^\omega+\mathbf k'):=-\frac1{\bar\gamma_1}\left(\begin{array}{*{2}{c}}\bar\xi^*&0\\0&\bar\xi\end{array}\right).
\label{Mt}\end{equation}
In particular, this implies that
\begin{equation}
|\mathfrak L\hat{\bar A}_{h,\omega}^{-1}(\mathbf k'+\mathbf p_{F,0}^{\omega})|\leqslant(\mathrm{const}.)\ \left(\begin{array}{*{2}{c}}2^{-h_\epsilon}&2^{-\frac{h+h_\epsilon}2}\\2^{-\frac{h+h_\epsilon}2}&2^{-h}\end{array}\right)
\label{boundinvLAtdom}\end{equation}
in which the bound should be understood as follows: the upper-left element in~(\ref{boundinvLAtdom}) is the bound on the upper-left $2\times2$ block of $\mathfrak L\hat{\bar A}_{h,\omega,0}^{-1}$, and similarly for the upper-right, lower-left and lower-right. In turn, (\ref{boundinvLAtdom}) implies
\begin{equation}
|\sigma_3(\mathbf k')|\leqslant(\mathrm{const}.)\ \left(2^{\frac{h-h_\epsilon}2}+2^{\frac{3h_\epsilon-h}2}\right).
\label{boundsigmatht}\end{equation}
\bigskip

\point{Irrelevant part} The irrelevant part is bounded in the same way as in the first regime (see~(\ref{ineqRtg})): using~(\ref{LAt}) and the bounds~(\ref{boundzvtt}) through~(\ref{boundvot}), we find that for $m\leqslant3$ and ${\mathfrak h}_2\leqslant h\leqslant h'\leqslant \bar{\mathfrak h}_1$, 
\begin{equation}
\int d\mathbf x\left|\mathbf x^m\mathcal RW_{2,\omega}^{(h')}\ast\mathbb L\bar{\mathfrak g}_{[h],\omega}(\mathbf x)\right|\leqslant 2^{h_\epsilon}\mathfrak F_h(\underline m)(\mathrm{const.})\ |U|
\label{ineqRtgt}\end{equation}
and for ${\mathfrak h}_2\leqslant h\leqslant\mathfrak h_1\leqslant h'\leqslant\bar{\mathfrak h}_0$,
\begin{equation}
\int d\mathbf x\left|\mathbf x^m\mathcal RW_{2,\omega}^{(h')}\ast\mathbb L\bar{\mathfrak g}_{[h],\omega}(\mathbf x)\right|\leqslant 2^{h_\epsilon}\mathfrak F_h(\underline m)(\mathrm{const.})\ |U|.
\label{ineqRogt}\end{equation}
Therefore, using the fact that
$$\int d\mathbf x\ \left|\mathbf x^m\mathcal R(g^{-1})\ast \mathbb L\bar{\mathfrak g}_{[h],\omega}(\mathbf x)\right|\leqslant2^{h_\epsilon}\mathfrak F_h(\underline m)(\mathrm{const}.)\ $$
we find
\begin{equation}
\int d\mathbf x\left|\mathbf x^m\mathcal R\bar A_{h,\omega}\ast\mathbb L\bar{\mathfrak g}_{[h],\omega}(\mathbf x)\right|\leqslant 2^{h_\epsilon}\mathfrak F_h(\underline m)(\mathrm{const.})\ (1+|h| |U|).
\label{ineqRAt}\end{equation}
\bigskip

\point{Conclusion of the proof} The proof of~(\ref{assumdpropt}) is then concluded by injecting~(\ref{LAtdom}), (\ref{boundsigmatht}), (\ref{sepLAtdom}) and~(\ref{ineqRAt}) into~(\ref{gseplint}).\par
\bigskip

{\bf Remark}: By following the rationale explained in the remark following~(\ref{Wrelk}), one may notice that the ``correct'' localization operator in the second regime is different from that in the first. Indeed, in the second regime, $(k-p_{F,0}^\omega)\partial_k$ scales like $2^{\frac12(h-h')}$ instead of $2^{h-h'}$ in the first. This implies that the remainder of the first order Taylor expansion of $\hat W_2^{(h')}$ is bounded by $2^{h}$ instead of $2^{2h-h'}$ in the first regime, and is therefore {\it marginal}. However, this is not a problem in this case since the effect of the ``marginality'' of the remainder is to produce the $|h|$ factor in~(\ref{ineqRAt}), which, since the second regime is cut off at scale $3h_\epsilon$ and the integration over the super-renormalizable first regime produced an extra $2^{h_\epsilon}$ (see~(\ref{ineqRAt})), is of little consequence. If one were to do things ``right'', one would define the localization operator for the {\it massless} fields as the Taylor expansion to {\it second} order in $k$ and first order in $k_0$, and find that the $|h|$ factor in (\ref{ineqRAt}) can be dropped. We have not taken this approach here, since it complicates the definition of $\mathcal L$ (which would differ between massive and massless blocks) as well as the symmetry discussion that we used in~(\ref{LAt}).\par
\subseqskip

\subsection{Two-point Schwinger function}
\label{schwintsec}
\indent We now compute the dominant part of the two-point Schwinger function for $\mathbf k$ {\it well inside} the second regime, i.e.
$${\bf k}\in \mathcal B_{\rm II}^{(\omega)}:=\bigcup_{h=\mathfrak h_2+1}^{\bar{\mathfrak h}_1-1}\mathrm{supp}f_{h,\omega}.$$
Let
$$h_{\mathbf k}:=\max\{h: f_{h,\omega}(\mathbf k)\neq0\}$$
so that if $h\not\in\{h_{\bf k},h_{{\bf k}}-1\}$, then $f_{h,\omega}({\bf k})=0$.\par
\bigskip

\point{Schwinger function in terms of dressed propagators} Recall that the two-point Schwinger function can be computed in terms of the effective source term $\mathcal X^{(h)}$ defined in \eqref{Qdef}, see the comment after lemma~\ref{schwinlemma}. 
Since $h_{\mathbf k}\leqslant\bar{\mathfrak h}_1$, $\mathcal X^{(h)}$ is left invariant by the integration over the ultraviolet and the first regime, in the sense that $\mathcal X^{(\bar{\mathfrak h}_1)}=\mathcal X^{(\bar{\mathfrak h}_0)}$, with $\mathcal X^{(\bar{\mathfrak h}_0)}$ given by \eqref{initXo}. We can therefore compute $\mathcal X^{(h)}$ for $h\in\{\mathfrak h_2,\cdots,\bar{\mathfrak h}_1-1\}$ inductively using lemma~\ref{schwinlemma}, and find, similarly to~(\ref{schwinexpro}), that
\begin{equation}
s_2(\mathbf k)=\left(\hat{\bar g}_{h_{\mathbf k},\omega}(\mathbf k)+\hat{\bar g}_{h_{\mathbf k}-1,\omega}(\mathbf k)\right)\left(\mathds1-\sigma(\mathbf k)-\sigma_{<h_{\mathbf k}}(\mathbf k)\right)
\label{schwinexprt}\end{equation}
where
\begin{equation}
\sigma(\mathbf k):=
\hat W_2^{(h_{\mathbf k}-1)}\hat{\bar g}_{h_{\mathbf k},\omega}
+(\hat{\bar g}_{h_{\mathbf k},\omega}+\hat{\bar g}_{h_{\mathbf k}-1,\omega})^{-1}\hat{\bar g}_{h_{\mathbf k},\omega}\hat W_2^{(h_{\mathbf k}-1)}\hat{\bar g}_{h_{\mathbf k}-1,\omega}(\mathds1-\hat W_2^{(h_{\mathbf k}-1)}\hat{\bar g}_{h_{\mathbf k},\omega})
\label{sigmatdef}\end{equation}
and
\begin{equation}\begin{largearray}
\sigma_{<h_{\mathbf k}}(\mathbf k):=\left(\mathds1-\left( \hat{\bar g}_{h_{\mathbf k},\omega}+\hat{\bar g}_{h_{\mathbf k}-1,\omega}\right)^{-1}\hat{\bar g}_{h_{\mathbf k},\omega}\hat W_2^{(h_{\mathbf k}-1)}\hat{\bar g}_{h_{\mathbf k}-1,\omega}\right)
\left(\sum_{h'=\mathfrak h_\beta}^{h_{\mathbf k}-2}\hat W_2^{(h')}\right)\cdot\\[0.5cm]
\hfill\cdot\left( \hat{\bar g}_{h_{\mathbf k},\omega}+\hat{\bar g}_{h_{\mathbf k}-1,\omega}-\hat{\bar g}_{h_{\mathbf k}-1,\omega}\hat W_{2,\omega}^{(h_{\mathbf k}-1)}\hat{\bar g}_{h_{\mathbf k},\omega}\right).
\end{largearray}\label{sigmaltdef}\end{equation}
\bigskip

\point{Bounding the error terms} We now bound $\sigma(\mathbf k)$ and $\sigma_{<h_{\mathbf k}}(\mathbf k)$. We first note that 
\begin{equation}
\left|(\hat{\bar g}_{h_{\mathbf k},\omega}+\hat{\bar g}_{h_{\mathbf k}-1,\omega})^{-1}\hat{\bar g}_{h_{\mathbf k},\omega}\right|\leqslant(\mathrm{const}.)\ 
\label{ineqginvgt}\end{equation}
which can be proved as follows: using \eqref{dresspropt}, we write $\hat{\bar g}_{h_{\mathbf k},\omega}=f_{h_{\bf k}}\hat{\bar A}_{h_{\bf k},\omega}^{-1}$ and 
$$\hat{\bar g}_{h_{\mathbf k}-1,\omega}=f_{h_{\bf k}-1}\hat{\bar A}_{h_{\bf k},\omega}^{-1}(\mathds{1}+f_{\leqslant h_{\bf k}-1}\hat W_{2}^{(h_{\bf k}-1)}\hat{\bar A}_{h_{{\bf k}},\omega}^{-1})^{-1}$$
Therefore, noting that  $f_{h_{\bf k}}({\bf k})+f_{h_{{\bf k}-1}}({\bf k})=1$, we obtain
\begin{equation} (\hat{\bar g}_{h_{\mathbf k},\omega}+\hat{\bar g}_{h_{\mathbf k}-1,\omega})^{-1}\hat{\bar g}_{h_{\mathbf k},\omega}=f_{h_{\bf k}}\Big[\mathds{1}+f_{h_{\bf k}-1}
\Big((\mathds{1}+f_{\leqslant h_{\bf k}-1}\hat W_{2}^{(h_{\bf k}-1)}\hat{\bar A}_{h_{{\bf k}},\omega}^{-1})^{-1}-\mathds{1}\Big)\Big]^{-1}.
\end{equation}
Now, by~(\ref{pct}), we see that $|\hat W_{2}^{(h_{\bf k}-1)}({\bf k})\hat{\bar A}_{h_{{\bf k}},\omega}^{-1}({\bf k})|\leqslant (\mathrm{const}.)\  2^{h_\epsilon}$, which implies (\ref{ineqginvgt}). By inserting 
(\ref{ineqginvgt}), (\ref{pct}) and~(\ref{assumdpropt}) into~(\ref{sigmatdef}), we obtain
\begin{equation}
|\sigma(\mathbf k)|\leqslant (\mathrm{const}.)\  2^{h_\epsilon}|U|.
\label{boundsigmat}\end{equation}
Moreover, if we assume that
\begin{equation}
\left|\sum_{h'=\mathfrak h_\beta}^{\bar{\mathfrak h}_2}\hat W_2^{(h')}(\mathbf k)\right|\leqslant (\mathrm{const.})\ 2^{4h_\epsilon}|U|
\label{sumWt}\end{equation}
which will be proved after studying the third regime~(\ref{boundthsumW}), then, since $3h_\epsilon\leqslant\mathfrak h_2 \leqslant h_{\bf k}$,
\begin{equation}
|\sigma_{<h_{\mathbf k}}(\mathbf k)|\leqslant (\mathrm{const}.)\  2^{h_\epsilon}|U|.
\label{boundsigmalkt}\end{equation}
\bigskip

\point{Dominant part of the dressed propagators} We now compute $\hat{\bar g}_{h_{\mathbf k},\omega}+\hat{\bar g}_{h_{\mathbf k}-1,\omega}$: it follows from~(\ref{gseplint}), (\ref{sepLAtdom}) and~(\ref{invLAtdom}) that
\begin{equation}\begin{largearray}
\hat{\bar g}_{{h_{\mathbf k}},\omega}(\mathbf k)+\hat{\bar g}_{{h_{\mathbf k}}-1,\omega}(\mathbf k)\\
\hfill=\left(\begin{array}{*{2}{c}}\mathds1&\bar M_{{h_{\mathbf k}}}^\dagger(\mathbf k)\\0&\mathds1\end{array}\right)\left(\begin{array}{*{2}{c}}\bar a_{{h_{\mathbf k}}}^{(M)}&0\\0&\bar a_{{h_{\mathbf k}}}^{(m)}(\mathbf k)\end{array}\right)\left(\begin{array}{*{2}{c}}\mathds1&0\\\bar M_{{h_{\mathbf k}}}(\mathbf k)&\mathds1\end{array}\right)
(\mathds1+\sigma'(\mathbf k))
\end{largearray}\label{schwinxprgtp}\end{equation}
where $\bar M_{{h_{\mathbf k}}}$, $\bar a_{{h_{\mathbf k}}}^{(M)}$ and $\bar a_{{h_{\mathbf k}}}^{(m)}$ were defined in~(\ref{Mt}) and~(\ref{bat}), and the error term $\sigma'$ can be bounded using~(\ref{ineqRAt}) and~(\ref{boundsigmatht}):
\begin{equation}
|\sigma'(\mathbf k)|\leqslant (\mathrm{const}.)\  \left(2^{\frac{{h_{\mathbf k}}-h_\epsilon}2}+2^{\frac{3h_\epsilon-{h_{\mathbf k}}}2}+|U||h_\epsilon|2^{h_\epsilon}\right).
\label{boundsigmapt}\end{equation}
\bigskip

\point{Proof of Theorem~\ref{theot}} We now conclude the proof of Theorem~\ref{theot}, {\it under the assumption} \eqref{sumWt}. We define
$$
B_{h_{\mathbf k}}(\mathbf k):=(\mathds1+\sigma'(\mathbf k))\left(\hat{\bar g}_{h_{\mathbf k},\omega}(\mathbf k)+\hat{\bar g}_{h_{\mathbf k}-1,\omega}(\mathbf k)\right)^{-1}
$$
(i.e. the inverse of the matrix on the right side of~(\ref{schwinxprgtp}), whose explicit expression is similar to the right side of~(\ref{LAtdom})), and
$$
\tilde m_2:=\tilde m_{\mathfrak h_2},\quad
z_2:=z_{\mathfrak h_2},\quad
v_2:=v_{\mathfrak h_2}
$$
and use~(\ref{boundzvtt}) to bound
$$\begin{array}c
|\tilde m_{h_{\mathbf k}}-\tilde m_2|\leqslant(\mathrm{const}.)\ |U|2^{h_{\mathbf k}},\quad
|z_{h_{\mathbf k}}-z_2|\leqslant(\mathrm{const}.)\ |U||h_\epsilon|2^{h_\epsilon},\\[0.2cm]
|v_{h_{\mathbf k}}-v_2|\leqslant(\mathrm{const}.)\ |U|2^{\frac12(h_{\mathbf k}+h_\epsilon)}
\end{array}$$
so that
$$
\left|(B_{\mathfrak h_2}(\mathbf k)-B_{h_{\mathbf k}}(\mathbf k))B_{\mathfrak h_2}^{-1}(\mathbf k)\right|\leqslant(\mathrm{const}.)\ |U||h_\epsilon|2^{h_\epsilon}
$$
which implies
\begin{equation}
B_{h_{\mathbf k}}^{-1}(\mathbf k)=B_{\mathfrak h_2}^{-1}(\mathbf k)(\mathds1+O(|U||h_\epsilon|2^{h_\epsilon})).
\label{approxBht}\end{equation}
We inject~(\ref{approxBht}) into~(\ref{schwinxprgtp}), which we then combine with~(\ref{schwinexprt}), (\ref{boundsigmat}), (\ref{boundsigmalkt}) and~(\ref{boundsigmapt}), and find an expression for $s_2$ which is similar to the right side of~(\ref{schwinxprgtp}) but with $h_{\mathbf k}$ replaced by $\mathfrak h_2$. This concludes the proof of~(\ref{schwint}). Furthermore, the estimate~(\ref{ineqrcct}) follows from~(\ref{boundzvttp}), which concludes the proof of Theorem~\ref{theot}.\par
\bigskip

\point{Partial proof of (\ref{sumWo})} Before moving on to the third regime, we bound part of the sum on the left side of~(\ref{sumWo}), which we recall was assumed to be true to prove~(\ref{schwino}) (see section~\ref{schwinosec}). It follows from~(\ref{pct}) that
\begin{equation}
\left|\sum_{h'=\mathfrak h_2}^{\bar{\mathfrak h}_1}\hat W_2^{(h')}(\mathbf k)\right|\leqslant(\mathrm{const}.)\ 2^{2h_\epsilon}|U|.
\label{boundtsumW}\end{equation}
\subseqskip

\subsection{Intermediate regime: second to third}
\label{tthsec}
\indent In the intermediate regime, we integrate over the first scales for which the effect of the extra Fermi points $\mathbf p_{F,j}^\omega$ cannot be neglected. As a consequence, the local part of $\hat{\bar A}_{\mathfrak h_2,\omega}(\mathbf k)$ is not dominant, so that the proof of the inductive assumption~(\ref{assumdpropt}) for $h=\mathfrak h_2$ must be discussed anew. In addition, we will see that dressing the propagator throughout the integrations over the first and second regimes will have shifted the Fermi points away from $\mathbf p_{F,j}^\omega$ by a small amount. Such an effect has not been seen so far because the position of $\mathbf p_{F,0}^\omega$ is fixed by symmetry.\par
\bigskip

\point{Power counting estimate} We first prove that
\begin{equation}
\int d\mathbf x\ |\mathbf x^m\bar g_{\mathfrak h_2,\omega}(\mathbf x)|\leqslant(\mathrm{const}.)\  2^{-\mathfrak h_2}\mathfrak F_{\mathfrak h_2}(\underline m).
\label{assumdproptth}\end{equation}
The proof is slightly different from the proof in section~\ref{dressproptsec}: instead of splitting $\hat{\bar g}_{\mathfrak h_2,\omega}$ according to~(\ref{gseplint}), we rewrite it as
\begin{equation}
\hat{\bar g}_{\mathfrak h_2,\omega}(\mathbf k)=f_{\mathfrak h_2,\omega}(\mathbf k)\left(\hat A(\mathbf k)+\mathcal L\hat{\bar{\mathfrak W}}_{\mathfrak h_2,\omega}(\mathbf k)\right)^{-1}
\left(\mathds1+\left(\mathcal R\hat{\bar{\mathfrak W}}_{\mathfrak h_2,\omega}(\mathbf k)\right)\left(\mathbb L\hat{\mathfrak g}_{[\mathfrak h_2],\omega}(\mathbf k)\right)\right)^{-1}
\label{gseplintth}\end{equation}
(this decomposition suggests that the dominant part of $\hat{\bar A}_{\mathfrak h_2,\omega}$ is $\hat A+\mathcal L\hat{\bar{\mathfrak W}}_{\mathfrak h_2,\omega}$ instead of $\mathcal L\hat{\bar A}_{\mathfrak h_2,\omega}$) in which we recall that $\hat A\equiv\hat{\bar A}_{\mathfrak h_2,\omega}\big|_{U=0}$,
$$\hat{\bar{\mathfrak W}}_{\mathfrak h_2,\omega}(\mathbf k):=\hat{\bar A}_{\mathfrak h_2,\omega}(\mathbf k)-\hat A(\mathbf k)$$
and
$$\mathbb L\hat{\mathfrak g}_{[\mathfrak h_2],\omega}(\mathbf k):=\left(f_{\leqslant\mathfrak h_2+1,\omega}({\bf k})-\sum_{j\in\{0,1,2,3\}}f_{\leqslant\mathfrak h_2-2,\omega,j}({\bf k})\right)\left(\hat A(\mathbf k)+\mathcal L\hat{\bar{\mathfrak W}}_{\mathfrak h_2,\omega}(\mathbf k)\right)^{-1}(\mathbf k).$$
We want to estimate the behavior of \eqref{gseplintth} in $\mathcal B_{\beta,L}^{(\mathfrak h_2,\omega)}$, which we recall is a ball with four holes around each $\mathbf p_{F,j}^{\omega}$, $j=0,1,2,3$. The splitting in \eqref{gseplintth} is convenient in that it is easy to see that $\hat A(\mathbf k)+\mathcal L\hat{\bar{\mathfrak W}}_{\mathfrak h_2,\omega}(\mathbf k)$ satisfies the 
same estimates as $\hat A({\bf k})$; in particular, via proposition~\ref{matinvprop} (see appendix~\ref{inversapp}), we see that  $\det\big(\hat A(\mathbf k)+\mathcal L\hat{\bar{\mathfrak W}}_{\mathfrak h_2,\omega}(\mathbf k)\big)\geqslant \det \hat A({\bf k}) \cdot (1+O(U))$  on $\mathcal B_{\beta,L}^{(\mathfrak h_2,\omega)}$, so that for all $n\leqslant7$ and $\mathbf k\in\mathcal B_{\beta,L}^{(\mathfrak h_2,\omega)}$,
\begin{equation}
\left|\partial_{\mathbf k}^n\left(\hat A(\mathbf k)+\mathcal L\hat{\bar{\mathfrak W}}_{\mathfrak h_2,\omega}(\mathbf k)\right)^{-1}\right|\leqslant (\mathrm{const}.)\  2^{-\mathfrak h_2}\mathfrak F_{\mathfrak h_2}(\underline n)
\label{ineqLAtth}\end{equation}
and, moreover, for $m\leqslant 3$, 
\begin{equation}
\int d\mathbf x\ \left|\mathbf x^m\mathcal R\bar{\mathfrak W}_{\mathfrak h_2,\omega}\ast\mathbb L\mathfrak g_{[\mathfrak h_2],\omega}(\mathbf x)\right|\leqslant (\mathrm{const}.)\ |U| |h_\epsilon|2^{h_\epsilon}\mathfrak F_{\mathfrak h_2}(\underline m).
\label{ineqRAtth}\end{equation}
The proof of~(\ref{assumdproptth}) is then concluded by injecting~(\ref{ineqLAtth}) and~(\ref{ineqRAtth}) into~(\ref{gseplintth}). We can then use the discussion in section~\ref{pcosec} to bound
\begin{equation}\left\{\begin{array}{>{\displaystyle}l}
\int d\mathbf x\ \left|\mathbf x^mW_{2,\omega,\underline\alpha}^{(\bar{\mathfrak h}_2)}(\mathbf x)\right|\leqslant \bar C_22^{\bar{\mathfrak h}_2+h_\epsilon}\mathfrak F_{\bar{\mathfrak h}_2}(\underline m)|U|\\[0.5cm]
\frac{1}{\beta|\Lambda|}\int d\underline{\mathbf x}\ \left|(\underline{\mathbf x}-\mathbf x_{4})^mW_{4,\underline\omega,\underline\alpha}^{(\bar{\mathfrak h}_2)}(\underline{\mathbf x})\right|\leqslant\bar C_{2}\mathfrak F_{\bar{\mathfrak h}_2}(\underline m)|U|\\[0.5cm]
\frac{1}{\beta|\Lambda|}\int d\underline{\mathbf x}\ \left|(\underline{\mathbf x}-\mathbf x_{2l})^mW_{2l,\underline\omega,\underline\alpha}^{(\bar{\mathfrak h}_2)}(\underline{\mathbf x})\right|\leqslant 2^{(\bar{\mathfrak h}_2+h_\epsilon)(2-l)}\mathfrak F_{\bar{\mathfrak h}_2}(\underline m)(\bar C_{2}|U|)^{l-1}
\end{array}\right.
\label{boundtth}\end{equation}
for some constant $\bar C_2>1$.\par
\bigskip

\point{Shift in the Fermi points} We now discuss the shift of the Fermi points, and show that $\hat{\bar g}_{\leqslant\mathfrak h_2,\omega}$ has {\it at least} 8 singularities: $\mathbf p_{F,0}^\omega$ and $\tilde{\mathbf p}_{F,j}^{(\omega,\mathfrak h_2)}$ for $j\in\{1,2,3\}$ where
\begin{equation}
\tilde{\mathbf p}_{F,1}^{(\omega,\mathfrak h_2)}=\mathbf p_{F,1}^\omega+(0,0,\omega\Delta_{\mathfrak h_2})
\label{shiftttho}\end{equation}
and
\begin{equation}
\tilde{\mathbf p}_{F,2}^{(\omega,\mathfrak h_2)}=T^{-\omega}\tilde{\mathbf p}_{F,1}^{(\omega,\mathfrak h_2)},\quad
\tilde{\mathbf p}_{F,3}^{(\omega,\mathfrak h_2)}=T^\omega\tilde{\mathbf p}_{F,1}^{(\omega,\mathfrak h_2)}
\label{shiftttht}\end{equation}
in which $T^\pm$ denotes the spatial rotation by $\pm2\pi/3$; and that
\begin{equation}
|\Delta_{\mathfrak h_2}|\leqslant(\mathrm{const}.)\ \epsilon^2|U|
\label{ineqshifttth}\end{equation}
(note that~(\ref{shiftttht}) follows immediately from the rotation symmetry~(\ref{arotation}), so we can restrict our discussion to $j=1$).
\bigskip

{\bf Remark}: Actually, we could prove in this section that $\hat{\bar g}_{\leqslant\mathfrak h_2,\omega}$ has {\it exactly} 8 singularities, but this fact follows automatically from the discussion in section~\ref{thsec}, for the same reason that the proof that the splittings~(\ref{gseplino}) and~(\ref{gseplint}) are well defined in the first and second regimes implies that no additional singularity can appear in those regimes. Since the third regime extends to $h\to-\infty$, proving that the splitting~(\ref{gseplint}) is well defined in the third regime will imply that there are 8 Fermi points.\par
\bigskip

\indent We will be looking for $\tilde{\mathbf p}_{F,1}^{(\omega,\mathfrak h_2)}$ in the form~(\ref{shiftttho}). In particular, its $k_0$ component vanishes, so that, by corollary~\ref{Anokzprop} (see appendix~\ref{inversapp}), $\Delta_{\mathfrak h_2}$ solves
\begin{equation}
\hat{\bar D}_{\mathfrak h_2,\omega}(\Delta_{\mathfrak h_2}):=
\hat{\bar A}_{\mathfrak h_2,\omega,(b,a)}^2(\tilde{\mathbf p}_{F,1}^{(\omega,\mathfrak h_2)})-\hat{\bar A}_{\mathfrak h_2,\omega,(\tilde b,a)}(\tilde{\mathbf p}_{F,1}^{(\omega,\mathfrak h_2)})\hat{\bar A}_{\mathfrak h_2,\omega,(b,\tilde a)}(\tilde{\mathbf p}_{F,1}^{(\omega,\mathfrak h_2)})=0.
\label{DdeftthN}\end{equation}
In order to solve~(\ref{DdeftthN}), we can use a Newton iteration, so we expand $\hat{\bar D}_{\mathfrak h_2,\omega}$ around $0$: it follows from the symmetries~(\ref{averticalreflection}) and~(\ref{ahorizontalreflection}) that
\begin{equation}
\hat{\bar D}_{\mathfrak h_2,\omega}(\Delta_{\mathfrak h_2})=M_{\mathfrak h_2}+\omega Y_{\mathfrak h_2}\Delta_{\mathfrak h_2}+\Delta_{\mathfrak h_2}^2R_{\mathfrak h_2,\omega}^{(2)}(\Delta_{\mathfrak h_2})
\label{detAordero}\end{equation}
with $(M_{\mathfrak h_2},Y_{\mathfrak h_2})\in\mathbb{R}^2$, independent of $\omega$. Furthermore by injecting~(\ref{pco}) and~(\ref{pct}) into~(\ref{DdeftthN}), we find that
\begin{equation}
Y_{\mathfrak h_2}=\frac{3}{2}\gamma_1\gamma_3+O(\epsilon^2|U|)+O(\epsilon^4),\quad
M_{\mathfrak h_2}=O(\epsilon^4|U|)
\label{estYMtth}\end{equation}
and
\begin{equation}
\left|R_{\mathfrak h_2,\omega}^{(2)}(\Delta_{\mathfrak h_2})\right|\leqslant(\mathrm{const}.)\ .
\label{boundRttth}\end{equation}
Therefore, by using a Newton scheme, one finds a root $\Delta_{\mathfrak h_2}$ of (\ref{DdeftthN}) and, by~(\ref{estYMtth}) and~(\ref{boundRttth}),
\begin{equation}
|\Delta_{\mathfrak h_2}|\leqslant(\mathrm{const}.)\ \epsilon^2|U|.
\label{boundimpfunfinal}\end{equation}
\bigskip
This concludes the proof of~(\ref{shiftttho}) and~(\ref{ineqshifttth}).\par
\seqskip

\section{Third regime}
\label{thsec}
\indent Finally, we perform the multiscale integration in the third regime. Similarly to the first and second regimes, we prove by induction that $\bar g_{h,\omega,j}$ satisfies the same estimate as $g_{h,\omega,j}$ (see~(\ref{estgth}) and~(\ref{gramboundth})): for all $m\leqslant3$,
\begin{equation}
\left\{\begin{array}{>{\displaystyle}l}
\int d\mathbf x\ |x_0^{m_0}x^{m_k}\bar g_{h,\omega,j}(\mathbf x)|\leqslant (\mathrm{const.})\ 2^{-h-m_0h-m_k(h-h_\epsilon)}\\[0.5cm]
\frac{1}{\beta|\Lambda|}\sum_{\mathbf k\in\mathcal B_{\beta,L}^{(h,\omega,j)}}|\hat{\bar g}_{h,\omega,j}(\mathbf k)|\leqslant (\mathrm{const.})\ 2^{2h-2h_\epsilon}.
\end{array}\right.\label{assumdpropth}\end{equation}
which in terms of the hypotheses of lemma~\ref{powercountinglemma}\ means
$$c_k=3,\ c_g=1,\quad\mathfrak F_h(m_0,m_1,m_2)=2^{-m_0h-(m_1+m_2)(h-h_\epsilon)},$$
$C_g=(\mathrm{const}.)\ $ and $C_G=(\mathrm{const}.)\  2^{-2h_\epsilon}$.\par\bigskip

{\bf Remark}: As in the second regime, the estimates~(\ref{assumdpropth}) are not optimal because the massive components scale differently from the massless ones.\par
\bigskip

\indent Like in the first regime,
$$\ell_0=2>\frac{c_k}{c_k-c_g}=\frac{3}{2}.$$
\subseqskip

\subsection{Power counting in the third regime}

\point{Power counting estimate} By lemma~\ref{powercountinglemma} and \eqref{boundtth}, we find that for all $m\leqslant3$
$$\begin{largearray}
\frac{1}{\beta|\Lambda|}\int d\underline{\mathbf x}\ \left|(\underline{\mathbf x}-\mathbf x_{2l})^mB_{2l,\underline\omega,\underline j,\underline\alpha}^{(h)}(\underline{\mathbf x})\right|\leqslant 2^{h(3-2l)}\mathfrak F_h(\underline m)2^{2lh_\epsilon}
\sum_{N=1}^\infty\sum_{\tau\in\mathcal T_N^{(h)}}\sum_{\underline l_\tau}\sum_{\displaystyle\mathop{\scriptstyle \mathbf P\in\tilde{\mathcal P}_{\tau,\underline l_\tau,2}^{(h)}}_{|P_{v_0}|=2l}}\\[0.5cm]
\hfill(c_12^{2h_\epsilon})^{N-1}\prod_{v\in\mathfrak V(\tau)}2^{(3-|P_v|)}\prod_{v\in\mathfrak E(\tau)}(c_22^{-2h_\epsilon})^{l_v}|U|^{\max(1,l_v-1)}2^{(2l_v-1)h_\epsilon}
\end{largearray}$$
so that
\begin{equation}\begin{largearray}
\frac{1}{\beta|\Lambda|}\int d\underline{\mathbf x}\ \left|(\underline{\mathbf x}-\mathbf x_{2l})^mB_{2l,\underline\omega,\underline j,\underline\alpha}^{(h)}(\underline{\mathbf x})\right|\leqslant 2^{h(3-2l)}\mathfrak F_h(\underline m)2^{2(l-1)h_\epsilon}
\cdot\\[0.5cm]
\hfill\cdot\sum_{N=1}^\infty\sum_{\tau\in\mathcal T_N^{(h)}}\sum_{\underline l_\tau}\sum_{\displaystyle\mathop{\scriptstyle \mathbf P\in\tilde{\mathcal P}_{\tau,\underline l_\tau,2}^{(h)}}_{|P_{v_0}|=2l}}c_1^{N-1}2^{Nh_\epsilon}\prod_{v\in\mathfrak V(\tau)}2^{(3-|P_v|)}\prod_{v\in\mathfrak E(\tau)}c_2^{l_v}|U|^{\max(1,l_v-1)}.
\end{largearray}\label{powercountingth}\end{equation}
\bigskip

\point{Bounding the sum of trees} We then bound the sum over trees as in the first regime (see~(\ref{ineqtreesumf}) and~(\ref{ineqtreesumo})): if $l\geqslant2$ then for $\theta\in(0,1)$ and recalling that $\bar{\mathfrak h}_2=3h_\epsilon+{\rm const}$, 
\begin{equation}
\sum_{\tau\in\mathcal T_N^{(h)}}\sum_{\displaystyle\mathop{\scriptstyle \mathbf P\in\bar{\mathcal P}_{\tau,\underline l_\tau,2}^{(h)}}_{|P_{v_0}|=2l}}\prod_{v\in\mathfrak V(\tau)\setminus\{v_0\}}2^{(3-|P_v|)}
\leqslant
2^{2\theta(h-3h_\epsilon)}C_T^N\prod_{i=1}^NC_P^{2l_i}.
\label{ineqtreesumfth}\end{equation}
and if $l=1$ then
\begin{equation}
\sum_{\tau\in\mathcal T_N^{(h)}}\sum_{\displaystyle\mathop{\scriptstyle \mathbf P\in\bar{\mathcal P}_{\tau,\underline l_\tau,2}^{(h)}}_{|P_{v_0}|=2}}\prod_{v\in\mathfrak V(\tau)\setminus\{v_0\}}2^{(3-|P_v|)}
\leqslant
2^{h-3h_\epsilon}C_T^N\prod_{i=1}^NC_P^{2l_i}.
\label{ineqtreesumoth}\end{equation}
Therefore, proceeding as in the proof of~(\ref{pco}) and~(\ref{pcol}) we find that
\begin{equation}
\int d\mathbf x\ \left|\mathbf x^mW_{2,\omega,j,\underline\alpha}^{(h)}(\mathbf x)\right|
\leqslant 2^{2(h-h_\epsilon)}\mathfrak F_h(\underline m)C_1|U|
\label{pcth}\end{equation}
and
\begin{equation}
\left\{\begin{array}l
\frac{1}{\beta|\Lambda|}\int d\underline{\mathbf x}\ \left|(\underline{\mathbf x}-\mathbf x_{4})^mW_{4,\underline\omega,\underline j,\underline\alpha}^{(h)}(\underline{\mathbf x})\right|
\leqslant \mathfrak F_h(\underline m)C_1|U|\\[0.5cm]
\frac{1}{\beta|\Lambda|}\int d\underline{\mathbf x}\ \left|(\underline{\mathbf x}-\mathbf x_{2l})^mW_{2l,\underline\omega,\underline j,\underline\alpha}^{(h)}(\underline{\mathbf x})\right|
\leqslant 2^{(3-2l)h+2\theta(h-3h_\epsilon)+(2l-1)h_\epsilon}\mathfrak F_h(\underline m)(C_1|U|)^{l-1}
\end{array}\right.\label{pcthl}\end{equation}
for $l\geqslant3$ and $m\leqslant3$.\par
\bigskip

{\bf Remark}: The estimates \eqref{powercountingth}, \eqref{ineqtreesumfth} and \eqref{ineqtreesumoth} imply the convergence of 
the tree expansion \eqref{treeexpT}, thus providing a convergent expansion of $W_{2l,\underline\omega,\underline\alpha}^{(h)}$ in $U$. 
\subseqskip

\subsection{The dressed propagator}
\indent We now prove \eqref{assumdpropth}. We recall that (see~(\ref{Adefth}))
\begin{equation}
\hat{\bar g}_{h,\omega,j}(\mathbf k)
=f_{h,\omega,j}(\mathbf k)\hat{\bar A}_{h,\omega,j}^{-1}(\mathbf k)
\label{dresspropth}\end{equation}
where
$$\begin{largearray}
\hat{\bar A}_{h,\omega,j}(\mathbf k)
=\hat A(\mathbf k)+f_{\leqslant h,\omega,j}(\mathbf k)\hat W_2^{(h)}(\mathbf k)+\sum_{h'=h+1}^{\bar{\mathfrak h}_2}\hat W_2^{(h')}(\mathbf k)\\
\hfill+\sum_{h'=\mathfrak h_2}^{\bar{\mathfrak h}_1}\hat W_2^{(h')}(\mathbf k)+\sum_{h'=\mathfrak h_1}^{\bar{\mathfrak h}_0}\hat W_2^{(h')}(\mathbf k).
\end{largearray}$$

\point{$j=0$ case} We first study the $j=0$ case, which is similar to the discussion in the second regime. We use the localization operator defined in~(\ref{Wreldefo}) and split $\hat{\bar g}_{h,\omega,0}$ in the same way as in~(\ref{gseplint}). We then compute $\mathcal L\hat W_2^{(h')}$ and bound $\mathcal R\hat{\bar A}_{h,\omega,0}\mathbb L\hat{\bar{\mathfrak g}}_{[h],\omega,0}$.
\bigskip

\subpoint{Local part} The symmetry considerations of the first and second regime still hold (see~(\ref{invsymsWo}) and~(\ref{pfzsyms})) so that~(\ref{Wrelo}) still holds:
\begin{equation}
\mathcal L\hat W_2^{(h')}(\mathbf k'+\mathbf p_{F,0}^\omega)=-\left(\begin{array}{*{4}{c}}i\tilde\zeta_{h'}k_0&\gamma_1\tilde\mu_{h'}&0&\nu_{h'}\xi^*\\
\gamma_1\tilde\mu_{h'}&i\tilde\zeta_{h'}k_0&\nu_{h'}\xi&0\\
0&\nu_{h'}\xi^*&i\zeta_{h'}k_0&\gamma_3\tilde\nu_{h'}\xi\\
\nu_{h'}\xi&0&\gamma_3\tilde \nu_{h'}\xi^*&i\zeta_{h'}k_0
\end{array}\right),
\label{Wrelth}\end{equation}
with $(\tilde\zeta_{h'}, \tilde\mu_{h'}, \tilde\nu_{h'},\zeta_{h'},\nu_{h'})\in\mathbb{R}^5$. The estimates~(\ref{boundzvtt}) through~(\ref{boundvot}) hold, and it follows from~(\ref{pcth}) that if $h'\leqslant\bar{\mathfrak h}_2$, then
\begin{equation}\begin{array}c
|\tilde\zeta_{h'}|\leqslant (\mathrm{const}.)\  |U|2^{h'-2h_\epsilon},\quad
|\zeta_{h'}|\leqslant (\mathrm{const}.)\  |U|2^{h'-2h_\epsilon},\quad |\tilde\mu_{h'}|\leqslant (\mathrm{const}.)\  |U|2^{2h'-3h_\epsilon},
\\
|\nu_{h'}|\leqslant (\mathrm{const}.)\  |U|2^{h'-h_\epsilon}, \quad 
|\tilde\nu_{h'}|\leqslant (\mathrm{const}.)\  |U|2^{h'-2h_\epsilon}.
\end{array}\label{boundzvtth}\end{equation}
Therefore
\begin{equation}
\mathcal L\hat{\bar A}_{h,\omega,0}(\mathbf k'+\mathbf p_{F,0}^\omega)=-\left(\begin{array}{*{4}{c}}
i\tilde z_{h}k_0&\gamma_1\tilde m_{h}&0&v_{h}\xi^*\\
\gamma_1\tilde m_{h}&i\tilde z_{h}k_0&v_{h}\xi&0\\
0&v_{h}\xi^*&i z_{h}k_0&\gamma_3 \tilde v_{h}\xi\\
v_{h}\xi&0&\gamma_3 \tilde v_{h}\xi^*&i z_{h}k_0
\end{array}\right)
\label{LAth}\end{equation}
where $z_h$, $\tilde z_h$, $m_h$, $v_h$ and $\tilde v_h$ are defined as in~(\ref{summedzvmt}). and are bounded as in~(\ref{boundzvttp}):
\begin{equation}\begin{array}c
|\tilde m_h-1|\leqslant(\mathrm{const}.)\ |U|,\quad
|\tilde z_h-1|\leqslant(\mathrm{const}.)\ |U|,\quad
|z_h-1|\leqslant(\mathrm{const}.)\ |U|,\\[0.2cm]
|\tilde v_h-1|\leqslant(\mathrm{const}.)\ |U|,\quad
|v_h-1|\leqslant(\mathrm{const}.)\ |U|.
\end{array}\label{boundzvtthp}\end{equation}
\bigskip

\subpoint{Dominant part of $\mathcal L\hat{\bar A}_{h,\omega,0}$} Furthermore, we notice that the terms proportional to $\tilde z_h$ are sub-dominant:
\begin{equation}
\mathcal L\hat{\bar A}_{h,\omega,0}(\mathbf k'+\mathbf p_{F,0}^\omega)=\mathfrak L\hat{\bar A}_{h,\omega,0}(\mathbf k'+\mathbf p_{F,0}^\omega)(\mathds1+\sigma_4(\mathbf k'))
\label{sepLAthdom}\end{equation}
where
\begin{equation}
\mathfrak L\hat{\bar A}_{h,\omega,0}(\mathbf k'+\mathbf p_{F,0}^\omega):=-\left(\begin{array}{*{4}{c}}
0&\gamma_1\tilde m_{h}&0&v_{h}\xi^*\\
\gamma_1\tilde m_{h}&0&v_{h}\xi&0\\
0&v_{h}\xi^*&i z_{h}k_0&\gamma_3 \tilde v_{h}\xi\\
v_{h}\xi&0&\gamma_3 \tilde v_{h}\xi^*&i z_{h}k_0
\end{array}\right).
\label{LAthdom}\end{equation}
Before bounding $\sigma_4$, we compute the inverse of~(\ref{LAthdom}) by block-diagonalizing it using proposition~\ref{blockdiagprop} (see appendix~\ref{diagapp}): if we define
\begin{equation}
\bar k_0:=z_hk_0,\quad
\bar\gamma_1:=\tilde m_h\gamma_1,\quad
\tilde\xi:=\tilde v_h\xi,\quad
\bar\xi:=v_h\xi
\label{bardefsth}\end{equation}
then for $\mathbf k\in\mathcal B_{\beta,L}^{(h,\omega,0)}$,
\begin{equation}
\left(\mathfrak L\hat{\bar A}_{h,\omega,0}(\mathbf k)\right)^{-1}=\kern-5pt
\left(\begin{array}{*{2}{c}}\mathds1&\kern-5pt\bar M_{h,0}^\dagger(\mathbf k)\\0&\mathds1\end{array}\right)\left(\begin{array}{*{2}{c}}\bar a_{h,0}^{(M)}&0\\0&\kern-10pt\bar a_{h,0}^{(m)}(\mathbf k)\end{array}\right)\left(\begin{array}{*{2}{c}}\mathds1&\kern-5pt 0\\\bar M_{h,0}(\mathbf k)&\kern-5pt \mathds1\end{array}\right)(\mathds1+O(2^{h-3h_\epsilon}))
\label{invLAthdom}\end{equation}
where
\begin{equation}
\bar a_{h,0}^{(M)}:=-\left(\begin{array}{*{2}{c}}0&\bar\gamma_1^{-1}\\\bar\gamma_1^{-1}&0\end{array}\right),\quad
\bar a_{h,0}^{(m)}(\mathbf p_{F,0}^\omega+\mathbf k'):=-\frac{1}{\bar k_0^2+\gamma_3^2|\tilde\xi|^2}\left(\begin{array}{*{2}{c}}
-i\bar k_0&\gamma_3\tilde\xi\\
\gamma_3\tilde\xi^*&-i\bar k_0\end{array}\right)
\label{bath}\end{equation}
(the $O(2^{h-3h_\epsilon})$ term comes from the terms we neglected from $\bar a^{(m)}$ that are of order $2^{-3h_\epsilon}$) and
\begin{equation}
\bar M_{h,0}(\mathbf p_{F,0}^\omega+\mathbf k'):=-\frac1{\bar\gamma_1}\left(\begin{array}{*{2}{c}}\bar\xi^*&0\\0&\bar\xi\end{array}\right).
\label{Mth}\end{equation}
In particular, this implies that, if $(\mathbf k'+\mathbf p_{F,0}^{\omega})\in\mathcal B_{\beta,L}^{(h,\omega,0)}$, then
\begin{equation}
|\mathfrak L\hat{\bar A}_{h,\omega,0}^{-1}(\mathbf k'+\mathbf p_{F,0}^{\omega})|\leqslant(\mathrm{const}.)\ \left(\begin{array}{*{2}{c}}2^{-h_\epsilon}&2^{-2h_\epsilon}\\2^{-2h_\epsilon}&2^{-h}\end{array}\right)
\label{boundinvLAthdom}\end{equation}
in which the bound should be understood as follows: the upper-left element in~(\ref{boundinvLAthdom}) is the bound on the upper-left $2\times2$ block of $\mathfrak L\hat{\bar A}_{h,\omega,0}^{-1}$, and similarly for the upper-right, lower-left and lower-right. In turn, (\ref{boundinvLAthdom}) implies
\begin{equation}
|\sigma_{4}(\mathbf k')|\leqslant(\mathrm{const}.)\ 2^{h-2h_\epsilon}.
\label{boundsigmafth}\end{equation}
\bigskip

\subpoint{Irrelevant part} We now bound $\mathcal RW_{2,\omega,0}^{(h')}\ast\mathbb L\bar{\mathfrak g}_{[h,\omega,0]}$ in the same way as in the second regime, and find that for $m\leqslant3$, if $h\leqslant h'\leqslant\bar{\mathfrak h}_0$, then
\begin{equation}
\int d\mathbf x\left|\mathbf x^m\mathcal RW_{2,\omega,0}^{(h')}\ast\mathbb L\bar{\mathfrak g}_{[h],\omega,0}(\mathbf x)\right|\leqslant 2^{h-2h_\epsilon}\mathfrak F_h(\underline m)(\mathrm{const.})\ |U|
\label{ineqRthgth}\end{equation}
so that
\begin{equation}
\int d\mathbf x\left|\mathbf x^m\mathcal R\bar A_{h,\omega,0}\ast\mathbb L\bar{\mathfrak g}_{[h],\omega,0}(\mathbf x)\right|\leqslant 2^{h-2h_\epsilon}\mathfrak F_h(\underline m)(\mathrm{const.})\ (1+|h| |U|).
\label{ineqRAth}\end{equation}
This concludes the proof of~(\ref{assumdpropth}) for $j=0$.\par
\bigskip

\point{$j=1$ case} We now turn to the case $j=1$ ($j=2,3$ will then follow by using the $2\pi/3$-rotation symmetry). Again, we split $\hat{\bar g}_{h,\omega,1}$ in the same way as in~(\ref{gseplint}), then we compute $\mathcal L\hat W_2^{(h')}$ and bound $\mathcal R\hat{\bar A}_{h,\omega,1}\mathbb L\hat{\bar{\mathfrak g}}_{[h],\omega,1}$. 
Before computing $\mathcal L\hat{\bar A}_{h,\omega,1}$ and bounding $\mathcal R\hat{\bar A}_{h,\omega,1}\mathbb L\hat{\bar{\mathfrak g}}_{[h],\omega,1}$, we first discuss the shift in the Fermi points $\tilde{\mathbf p}_{F,1}^{(\omega,h)}$ (i.e., the singularities of 
$\hat{\bar A}_{h,\omega,1}^{-1}({\bf k})$ in the vicinity of ${\mathbf p}_{F,1}^{(\omega,h)}$), 
due to the renormalization group flow.\par
\bigskip

\subpoint{Shift in the Fermi points} We compute the position of the shifted Fermi points in the form
\begin{equation}
\tilde{\mathbf p}_{F,1}^{(\omega,h)}=\mathbf p_{F,1}^\omega+(0,0,\omega\Delta_{h})
\label{shifttho}\end{equation}
and show that
\begin{equation}
|\Delta_{h}|\leqslant(\mathrm{const}.)\ \epsilon^2|U|.
\label{ineqshiftth}\end{equation}
The proof goes along the same lines as that in section~\ref{tthsec}.\par
\bigskip

\indent Similarly to~(\ref{DdeftthN}), $\Delta_{h}$ is a solution of
\begin{equation}
\hat{\bar D}_{h,\omega,1}(\Delta_{h}):=
\hat{\bar A}_{h,\omega,1,(b,a)}^2(\tilde{\mathbf p}_{F,1}^{(\omega,h)})-\hat{\bar A}_{h,\omega,1,(\tilde b,a)}(\tilde{\mathbf p}_{F,1}^{(\omega,h)})\hat{\bar A}_{h,\omega,1,(b,\tilde a)}(\tilde{\mathbf p}_{F,1}^{(\omega,h)})=0.
\label{DdefthN}\end{equation}
We expand $\hat{\bar D}_{h,\omega,1}$ around $\Delta_{h+1}$: it follows from the symmetries~(\ref{averticalreflection}) and~(\ref{ahorizontalreflection}) that
\begin{equation}
\hat{\bar D}_{h,\omega,1}(\Delta_{h})=M_{h}+\omega Y_{h}(\Delta_{h}-\Delta_{h+1})+(\Delta_{h}-\Delta_{h+1})^2R_{h,\omega,1}^{(2)}(\Delta_{h})
\label{detAordert}\end{equation}
with $(M_{h},Y_{h})\in\mathbb{R}^2$, independent of $\omega$. Furthermore,
$$
M_h=\hat{\bar D}_{h,\omega,1}(\Delta_{h+1})=\hat{\bar D}_{h,\omega,1}(\Delta_{h+1})-\hat{\bar D}_{h+1,\omega,1}(\Delta_{h+1})
$$
so that, by injecting~(\ref{pcth}), (\ref{pco}) and~(\ref{pct}) into~(\ref{DdefthN}) and using the symmetry structure of $\hat{\bar A}_{h,\omega,1}({\bf k})$ (which imply, in particular, that 
$|\hat{\bar A}_{h,\omega,1}({\bf k})|\leqslant (\mathrm{const}.)\ \epsilon$ in $\mathcal B_{\beta,L}^{(\leqslant h,\omega,1)}$), we find
\begin{equation}
|M_h|\leqslant(\mathrm{const}.)\ 2^{2h-3h_\epsilon}\epsilon^2|U|
\label{estMth}\end{equation}
and
\begin{equation}
Y_{h}=\frac{3}{2}\gamma_1\gamma_3+O(\epsilon^2|U|)+O(\epsilon^4).
\label{estYth}\end{equation}
as well as
\begin{equation}
\left|R_{h,\omega,1}^{(2)}(\Delta_{h})\right|\leqslant(\mathrm{const}.)\ (1+\epsilon|U||h|).
\label{boundRtth}\end{equation}
Therefore, by using a Newton scheme, we compute $\Delta_{h}$ satisfying~(\ref{DdefthN}) and, by~(\ref{estMth}), (\ref{estYth}) and~(\ref{boundRtth}),
\begin{equation}
|\Delta_{h}-\Delta_{h+1}|\leqslant(\mathrm{const}.)\ 2^{2h-3h_\epsilon}|U|.
\label{boundimpfunfinalt}\end{equation}
\bigskip
This concludes the proof of~(\ref{shifttho}) and~(\ref{ineqshiftth}).\par
\bigskip

\subpoint{Local part} We now compute $\mathcal L\hat{\bar A}_{h,\omega,1}$. The computation is similar to the $j=0$ case, though it is complicated slightly by the presence of constant terms in $\hat{\bar A}_{h,\omega,1}$. Recall the $\mathbf x$-space representation of $\bar A_{h,\omega,1}$~(\ref{Ahojx}). The localization operator has the same definition as~(\ref{Wreldefo}), but because of the shift by $\tilde{\mathbf p}_{F,1}^{(\omega,h)}$ in the Fourier transform, its action in $\mathbf k$-space becomes
$$
\mathcal L\hat{ \bar A}_{h,\omega,1}(\mathbf k)=\hat{\bar A}_{h,\omega,1}(\tilde{\mathbf p}_{F,1}^{(\omega,h)})+(\mathbf k-\tilde{\mathbf p}_{F,1}^{(\omega,h)})\cdot\partial_{\mathbf k}\hat{\bar A}_{h,\omega,1}(\tilde{\mathbf p}_{F,1}^{(\omega,h)}).
$$
In order to avoid confusion, we will denote the localization operator in $\mathbf k$ space around $\tilde{\mathbf p}_{F,1}^{(\omega,h)}$ by $\hat{\mathcal L}_h$.\par
\bigskip

\subsubpoint{Non-interacting local part} As a preliminary step, we discuss the action of $\hat{\mathcal L}$ on the {\it undressed} inverse propagator $\hat A({\bf k})$. Let us first 
split $\hat A({\bf k})$ into $2\times2$ blocks:
$$\hat A(\mathbf k)=:\left(\begin{array}{*{2}{c}}
\hat A^{\xi\xi}(\mathbf k)&\hat A^{\xi\phi}(\mathbf k)\\
\hat A^{\phi\xi}(\mathbf k)&\hat A^{\phi\phi}(\mathbf k)\\
\end{array}\right)$$
in terms of which
\begin{equation}\begin{largearray}
\hfil\hat{\mathcal L}_h\hat A^{\xi\xi}(\mathbf k'_{1}+\tilde{\mathbf p}_{F,1}^{(\omega,h)})=-\left(\begin{array}{*{2}{c}}
ik_0&\gamma_1\\
\gamma_1&ik_0\end{array}\right)\hfil\\[0.5cm]
\hat{\mathcal L}_h\hat A^{\xi\phi}(\mathbf k'_{1}+\tilde{\mathbf p}_{F,1}^{(\omega,h)})=\hat{\mathcal L}_h\hat A^{\phi\xi}(\mathbf k'_{1}+\tilde{\mathbf p}_{F,1}^{(\omega,h)})\\[0.5cm]
\hfill=-\left(\begin{array}{*{2}{c}}
0&m_h^{(0)}+(-iv_h^{(0)}k'_{1,x}+\omega w_h^{(0)}k'_{1,y})\\
m_h^{(0)}+(iv_h^{(0)}k'_{1,x}+\omega w_h^{(0)}k'_{1,y})&0\end{array}\right)\\[1.5cm]
\hat{\mathcal L}_h\hat A^{\phi\phi}(\mathbf k'_{1}+\tilde{\mathbf p}_{F,1}^{(\omega,h)})\\[0.5cm]
\hfill=-\left(\begin{array}{*{2}{c}}
ik'_{\omega,1,0}&\gamma_3(m_h^{(0)}+(i\tilde v_h^{(0)}k'_{1,x}+\omega w_h^{(0)}k'_{1,y}))\\
\gamma_3(m_h^{(0)}+(-i\tilde v_h^{(0)}k'_{1,x}+\omega w_h^{(0)}k'_{1,y}))&ik'_{\omega,1,0}\end{array}\right)
\end{largearray}\label{Lgthj}\end{equation}
where
\begin{equation}\begin{array}{>{\displaystyle}c}
m_h^{(0)}=\gamma_1\gamma_3+O(\Delta_h),\quad
v_h^{(0)}=\frac{3}{2}+O(\epsilon^2,\Delta_h),\\[0.5cm]
\tilde v_h^{(0)}=\frac{3}{2}+O(\epsilon^2,\Delta_h),\quad
w_h^{(0)}=\frac{3}{2}+O(\epsilon^2,\Delta_h).
\end{array}\label{LfreeAo}\end{equation}
\bigskip

\subsubpoint{Local part of $\hat W_2$} We now turn our attention to $\hat{\mathcal L}_h\hat W_2^{(h')}$. In order to reduce the size of the coming equations, we split $\hat W_2^{(h')}$ into $2\times2$ blocks:
$$\hat W_2^{(h')}=:\left(\begin{array}{*{2}{c}}
\hat W_2^{(h')\xi\xi}&\hat W_2^{(h')\xi\phi}\\
\hat W_2^{(h')\phi\xi}&\hat W_2^{(h')\phi\phi}\\
\end{array}\right).$$
The symmetry structure around $\tilde{\mathbf p}_{F,1}^{(\omega,h)}$ is slightly different from that around $\mathbf p_{F,0}^\omega$. Indeed~(\ref{invsymsWo}) still holds, but $\tilde{\mathbf p}_{F,1}^{(\omega,h)}$ is not invariant under rotations, so that~(\ref{pfzsyms}) becomes
\begin{equation}
\tilde{\mathbf p}_{F,1}^{(\omega,h)}=-\tilde{\mathbf p}_{F,1}^{(-\omega,h)}=R_v\tilde{\mathbf p}_{F,1}^{(-\omega,h)}=R_h\tilde{\mathbf p}_{F,1}^{(\omega,h)}=I\tilde{\mathbf p}_{F,1}^{(\omega,h)}=P\tilde{\mathbf p}_{F,1}^{(-\omega,h)}.
\label{pfosyms}\end{equation}
It then follows from proposition~\ref{symprop} (see appendix~\ref{constWtapp}) that for all $(f,f')\in\{\xi,\phi\}^2$,
\begin{equation}\begin{largearray}
\hat{\mathcal L}_{h}\hat W_{2}^{(h')ff'}(\mathbf k'_{1}+\tilde{\mathbf p}_{F,1}^{(\omega,h)})\\[0.5cm]
\hfill=-\left(\begin{array}{*{2}{c}}i\zeta_{h',1}^{ff'}k_0&\mu_{h',1}^{ff'}+(i\nu_{h',1}^{ff'}k'_{1,x}+\omega\varpi_{h',1}^{ff'} k'_{1,y})\\
\mu_{h',1}^{ff'}+(-i\nu_{h',1}^{ff'}k'_{1,x}+\omega\varpi_{h',1}^{ff'} k'_{1,y})&i\zeta_{h',1}^{ff'}k_0
\end{array}\right)
\end{largearray}\label{Wrelthzwconst}\end{equation}
with $(\mu_{h',1}^{ff'},\zeta_{h',1}^{ff'},\nu_{h',1}^{ff'},\varpi_{h',1}^{ff'})\in\mathbb{R}^4$.
In addition, by using the parity symmetry, it follows from~(\ref{eqrotsym}) (see appendix~\ref{constWtapp}) that the $\xi\phi$ block is equal to the $\phi\xi$ block. 
Furthermore, it follows from~(\ref{pcth}) that for $h'\leqslant\bar{\mathfrak h}_2$,
\begin{equation}\begin{array}c
|\mu_{h',1}^{ff'}|\leqslant (\mathrm{const}.)\ |U| 2^{2(h'-h_\epsilon)},\quad |\zeta_{h',1}^{ff'}|\leqslant (\mathrm{const}.)\ |U| 2^{h'-2h_\epsilon},\\[0.2cm]
|\nu_{h',1}^{ff'}|\leqslant (\mathrm{const}.)\ |U|2^{h'-h_\epsilon},\quad |\varpi_{h',1}^{ff'}|\leqslant (\mathrm{const}.)\ |U|2^{h'-h_\epsilon}.
\end{array}\label{boundzvthowconst}\end{equation}
If $\mathfrak h_2\leqslant h'\leqslant\bar{\mathfrak h}_1$, then it follows from~(\ref{pct}) that
\begin{equation}\begin{array}c
|\zeta_{h',1}^{ff'}|\leqslant (\mathrm{const}.)\ |U| 2^{h_\epsilon},\\[0.2cm]
|\nu_{h',1}^{ff'}|\leqslant (\mathrm{const}.)\ |U|2^{\frac12(h'+h_\epsilon)},\quad |\varpi_{h',1}^{ff'}|\leqslant (\mathrm{const}.)\ |U|2^{\frac12(h'+h_\epsilon)}
\end{array}\label{boundzvthowconsto}\end{equation}
and because $\hat W_{2}^{(h')}(\mathbf p_{F,0}^\omega)=0$ and $|\tilde{\mathbf p}_{F,1}^{(\omega,h)}-\mathbf p_{F,0}^\omega|\leqslant(\mathrm{const}.)\ 2^{2h_\epsilon}$, by expanding $\hat W_2^{(h')}$ to first order around $\mathbf p_{F,0}^\omega$, we find that it follows from~(\ref{pct}) that
\begin{equation}
|\mu_{h',1}^{ff'}|\leqslant(\mathrm{const}.)\ 2^{\frac12(h'+h_\epsilon)}2^{2h_\epsilon}|U|.
\label{boundmthowconsto}\end{equation}
Finally, if $\mathfrak h_1\leqslant h'\leqslant\bar{\mathfrak h}_0$, then it follows from~(\ref{pco}) that
\begin{equation}\begin{array}c
|\zeta_{h',1}^{ff'}|\leqslant(\mathrm{const}.)\  |U|2^{h'},\\[0.2cm]
|\nu_{h',1}^{ff'}|\leqslant(\mathrm{const}.)\  |U|2^{h'},\quad
|\varpi_{h',1}^{ff'}|\leqslant(\mathrm{const}.)\  |U|2^{h'}
\end{array}\label{boundzvthowconstt}\end{equation}
and by expanding $\hat W_2^{(h')}$ to first order around $\mathbf p_{F,0}^\omega$, we find that
\begin{equation}
|\mu_{h',1}^{ff'}|\leqslant(\mathrm{const}.)\  |U|2^{h'+2h_\epsilon}.
\label{boundmzvthowconsto}\end{equation}
By using the improved estimate~(\ref{ineqvh}), we can refine these estimates for the inter-layer components, thus finding:
\begin{equation}\begin{array}c
|\mu_{h',1}^{ff}|\leqslant(\mathrm{const}.)\  |U|2^{\theta h'+3h_\epsilon},\\[0.2cm]
|\nu_{h',1}^{ff}|\leqslant(\mathrm{const}.)\  |U|2^{\theta h'+h_\epsilon},\quad
|\varpi_{h',1}^{ff}|\leqslant(\mathrm{const}.)\  |U|2^{\theta h'+h_\epsilon},\\[0.2cm]
|\zeta_{h',1}^{\phi\xi}|=|\zeta_{h',1}^{\xi\phi}|\leqslant(\mathrm{const}.)\  |U|2^{\theta h'+h_\epsilon}
\end{array}\label{boundmzvthowconstoimp}\end{equation}
for all $f\in\{\phi,\xi\}$.\par
\bigskip

\subsubpoint{Interacting local part} Therefore, putting~(\ref{Wrelthzwconst}) together with~(\ref{Lgthj}), we find
\begin{equation}\begin{largearray}
\hat{\mathcal L}_h\hat{\bar A}_{h,\omega,1}(\mathbf k'_{1}+\tilde{\mathbf p}_{F,1}^{(\omega,h)})\\[0.5cm]
\hfill=-\left(\begin{array}{*{4}{c}}
iz_{h,1}^{\xi\xi}k_0&\gamma_1(m_{h,1}^{\xi\xi}+K_{h,1}^{*\xi\xi})&iz_{h,1}^{\xi\phi}k_0&m_{h,1}^{\xi\phi}+K_{h,1}^{*\xi\phi}\\
\gamma_1(m_{h,1}^{\xi\xi}+K_{h,1}^{\xi\xi})&iz_{h,1}^{\xi\xi}k_0&m_{h,1}^{\xi\phi}+K_{h,1}^{\xi\phi}&iz_{h,1}^{\xi\phi}k_0\\
iz_{h,1}^{\xi\phi}k_0&m_{h,1}^{\xi\phi}+K_{h,1}^{*\xi\phi}&iz_{h,1}^{\phi\phi}k_0&\gamma_3(m_{h,1}^{\phi\phi}+K_{h,1}^{\phi\phi})\\
m_{h,1}^{\xi\phi}+K_{h,1}^{\xi\phi}&iz_{h,1}^{\xi\phi}k_0&\gamma_3(m_{h,1}^{\phi\phi}+K_{h,1}^{*\phi\phi})&iz_{h,1}^{\phi\phi}k_0
\end{array}\right)
\end{largearray}\label{LAthj}\end{equation}
with
$$
K_{h,1}^{ff'}:=iv_{h,1}^{ff'}k'_{1,x}+\omega w_{h,1}^{ff'}k'_{1,y}
$$
for $(f,f')\in\{\phi,\xi\}^2$, and
\begin{equation}\begin{array}{>{\displaystyle}c}
m_{h,1}^{\phi\phi}:=m_h^{(0)}+\frac{1}{\gamma_3}\sum_{h'=h}^{\bar{\mathfrak h}_0}\mu_{h',1}^{\phi\phi},\quad
m_{h,1}^{\xi\xi}:=1+\frac{1}{\gamma_1}\sum_{h'=h}^{\bar{\mathfrak h}_0}\mu_{h',1}^{\xi\xi},\quad
m_{h,1}^{\xi\phi}:=m_h^{(0)}+\sum_{h'=h}^{\bar{\mathfrak h}_0}\mu_{h',1}^{\xi\phi},\\[0.5cm]
z_{h,1}^{ff}:=1+\sum_{h'=h}^{\bar{\mathfrak h}_0}\zeta_{h',1}^{ff},\quad
z_{h,1}^{\xi\phi}:=\sum_{h'=h}^{\bar{\mathfrak h}_0}\zeta_{h',1}^{\xi\phi},\\[0.5cm]
v_{h,1}^{\phi\phi}:=\tilde v_h^{(0)}+\frac{1}{\gamma_3}\sum_{h'=h}^{\bar{\mathfrak h}_0}\nu_{h',1}^{\phi\phi},\quad
v_{h,1}^{\xi\xi}:=-\frac{1}{\gamma_1}\sum_{h'=h}^{\bar{\mathfrak h}_0}\nu_{h',1}^{\xi\xi},\quad
v_{h,1}^{\xi\phi}:=v_h^{(0)}-\sum_{h'=h}^{\bar{\mathfrak h}_0}\nu_{h',1}^{\xi\phi},\\[0.5cm]
w_{h,1}^{\phi\phi}:=w_h^{(0)}+\frac{1}{\gamma_3}\sum_{h'=h}^{\bar{\mathfrak h}_0}\varpi_{h',1}^{\phi\phi},\quad
w_{h,1}^{\xi\xi}:=\frac{1}{\gamma_1}\sum_{h'=h}^{\bar{\mathfrak h}_0}\varpi_{h',1}^{\xi\xi},\quad
w_{h,1}^{\xi\phi}:=w_h^{(0)}+\sum_{h'=h}^{\bar{\mathfrak h}_0}\varpi_{h',1}^{\xi\phi}.
\end{array}\label{summedmzvwthj}\end{equation}
Furthermore, using the bounds \eqref{boundzvthowconst} through \eqref{boundmzvthowconstoimp}, 
\begin{equation}\begin{array}c
|m_{h,1}^{\phi\phi}-m_h^{(0)}|+
|m_{h,1}^{\xi\xi}-1|+
|m_{h,1}^{\xi\phi}-m_h^{(0)}|\leqslant(\mathrm{const}.)\ \epsilon^2|U|,\\[0.5cm]
|z_{h,1}^{ff}-1|\leqslant(\mathrm{const}.)\ |U|,\quad
|z_{h,1}^{\xi\phi}|\leqslant(\mathrm{const}.)\ |\log\epsilon|\epsilon|U|,\\[0.5cm]
|v_{h,1}^{\phi\phi}-\tilde v_h^{(0)}|+
|v_{h,1}^{\xi\xi}|+
|v_{h,1}^{\xi\phi}-v_h^{(0)}|\leqslant(\mathrm{const}.)\ |U|,\\[0.5cm]
|w_{h,1}^{\phi\phi}-w_h^{(0)}|+
|w_{h,1}^{\xi\xi}|+
|w_{h,1}^{\xi\phi}-w_h^{(0)}|\leqslant(\mathrm{const}.)\ |U|.
\end{array}\label{sboundmzvwthj}\end{equation}
\bigskip

\subsubpoint{Dominant part of $\hat{\mathcal L}_h\hat{\bar A}_{h,\omega,1}$} Finally, we notice that the terms in~(\ref{LAthj}) that are proportional to $z_{h,1}^{\xi\xi}$, $z_{h,1}^{\xi\phi}$ or $K_{h,1}^{\xi\xi}$ are subdominant:
\begin{equation}
\hat{\mathcal L}_h\hat{\bar A}_{h,\omega,1}(\mathbf k'_{1}+\tilde{\mathbf p}_{F,1}^{(\omega,h)})=\hat{\mathfrak L}_h\hat{\bar A}_{h,\omega,1}(\mathbf k'_{1}+\tilde{\mathbf p}_{F,1}^{(\omega,h)})(\mathds1+\sigma_{4,1}(\mathbf k'_{1}))
\label{sepLAthjdom}\end{equation}
where
\begin{equation}\begin{largearray}
\hat{\mathfrak L}_h\hat{\bar A}_{h,\omega,1}(\mathbf k'_{1}+\tilde{\mathbf p}_{F,1}^{(\omega,h)})\\
\hfill:=-\left(\begin{array}{*{4}{c}}
0&\gamma_1m_{h,1}^{\xi\xi}&0&m_{h,1}^{\xi\phi}+K_{h,1}^{*\xi\phi}\\
\gamma_1m_{h,1}^{\xi\xi}&0&m_{h,1}^{\xi\phi}+K_{h,1}^{\xi\phi}&0\\
0&m_{h,1}^{\xi\phi}+K_{h,1}^{*\xi\phi}&iz_{h,1}^{\phi\phi}k_0&\gamma_3(m_{h,1}^{\phi\phi}+K_{h,1}^{\phi\phi})\\
m_{h,1}^{\xi\phi}+K_{h,1}^{\xi\phi}&0&\gamma_3(m_{h,1}^{\phi\phi}+K_{h,1}^{*\phi\phi})&iz_{h,1}^{\phi\phi}k_0
\end{array}\right).
\end{largearray}\label{LAthjdom}\end{equation}
Before bounding $\sigma_{4,1}$, we compute the inverse of~(\ref{LAthjdom}) by block-diagonalizing it using proposition~\ref{blockdiagprop} (see appendix~\ref{diagapp}): if we define
\begin{equation}
\bar k_0:=z_{h,1}^{\phi\phi}k_0,\quad
\bar\gamma_1:=m_{h,1}^{\xi\xi}\gamma_1,\quad
\bar\Xi_1:=m_{h,1}^{\xi\phi}+K^{\xi\phi}_{h,1},\quad
\bar x_1:=\frac{2m_{h,1}^{\xi\phi}}{\bar\gamma_1\gamma_3}K_{h,1}^{\xi\phi}-K_{h,1}^{\phi\phi}
\end{equation}
then for $\mathbf k\in\mathcal B_{\beta,L}^{(h,\omega,1)}$,
\begin{equation}
\left(\mathfrak L\hat{\bar A}_{h,\omega,1}(\mathbf k)\right)^{-1}=\kern-5pt
\left(\begin{array}{*{2}{c}}\mathds1&\kern-5pt\bar M_{h,1}^\dagger(\mathbf k)\\0&\mathds1\end{array}\right)\left(\begin{array}{*{2}{c}}\bar a_{h,1}^{(M)}&0\\0&\kern-10pt\bar a_{h,1}^{(m)}(\mathbf k)\end{array}\right)\left(\begin{array}{*{2}{c}}\mathds1&\kern-5pt 0\\\bar M_{h,1}(\mathbf k)&\kern-5pt \mathds1\end{array}\right)(\mathds1+O(2^{h-3h_\epsilon}))
\label{invLAthjdom}\end{equation}
where
\begin{equation}
\bar a_{h,1}^{(M)}:=-\left(\begin{array}{*{2}{c}}0&\bar\gamma_1^{-1}\\\bar\gamma_1^{-1}&0\end{array}\right),\quad
\bar a_{h,1}^{(m)}(\mathbf p_{F,1}^\omega+\mathbf k'_{1}):=\frac{1}{\bar k_0^2+\gamma_3^2|\bar x_1|^2}\left(\begin{array}{*{2}{c}}
i\bar k_0&\gamma_3\bar x_1^*\\
\gamma_3\bar x_1&i\bar k_0\end{array}\right)
\label{bathj}\end{equation}
(the $O(2^{h-3h_\epsilon})$ term comes from the terms in $\bar a^{(m)}$ of order $2^{-3h_\epsilon}$) and
\begin{equation}
\bar M_{h,1}(\mathbf p_{F,1}^\omega+\mathbf k'_{1}):=-\frac1{\bar\gamma_1}\left(\begin{array}{*{2}{c}}\bar\Xi_1^*&0\\0&\bar\Xi_1\end{array}\right).
\label{Mthj}\end{equation}
In particular, this implies that if $(\mathbf k'_{1}+\tilde{\mathbf p}_{F,1}^{(\omega,h)})\in\mathcal B_{\beta,L}^{(h,\omega,1)}$, then
\begin{equation}
\big|\big[\hat{\mathfrak L}_h\hat{\bar A}_{h,\omega,1}(\mathbf k'_{1}+\tilde{\mathbf p}_{F,1}^{(\omega,h)})\big]^{-1}\big|\leqslant(\mathrm{const}.)\ \left(\begin{array}{*{2}{c}}2^{2h_\epsilon-h}&2^{h_\epsilon-h}\\2^{h_\epsilon-h}&2^{-h}\end{array}\right)
\label{boundinvLAthjdom}\end{equation}
in which the bound should be understood as follows: the upper-left element in~(\ref{boundinvLAthjdom}) is the bound on the upper-left $2\times2$ block of $\hat{\mathfrak L}_h\hat{\bar A}_{h,\omega,1}^{-1}$, and similarly for the upper-right, lower-left and lower-right. In turn, using (\ref{boundinvLAthjdom}) we obtain 
\begin{equation}
|\sigma_{4,1}(\mathbf k'_{1})|\leqslant(\mathrm{const}.)\ \epsilon(1+|\log\epsilon||U|).
\label{boundsigmaftho}\end{equation}
\bigskip

\subpoint{Irrelevant part} Finally, we are left with bounding $\mathcal R\bar A_{h,\omega,1}\mathbb L\bar{\mathfrak g}_{[h],\omega,1}$, which we show is small.  The bound
 is identical to~(\ref{ineqRAth}): indeed, it follows from~(\ref{invLAthjdom}) and \eqref{boundinvLAthjdom} that for all $m\leqslant3$,
$$
\int d\mathbf x\ \left|\mathbf x^m\mathbb L\bar{\mathfrak g}_{[h],\omega,1}(\mathbf x)\right|\leqslant(\mathrm{const}.)\ 2^{-h}\mathfrak F_{h}(\underline m)
$$
so that
\begin{equation}
\int d\mathbf x\left|\mathbf x^m\mathcal R\bar A_{h,\omega,1}\ast\mathbb L\bar{\mathfrak g}_{[h],\omega,1}(\mathbf x)\right|\leqslant 2^{h-2h_\epsilon}\mathfrak F_h(\underline m)(\mathrm{const.})\ (1+|h| |U|).
\label{ineqRAthj}\end{equation}
\bigskip

\point{$j=2,3$ cases} The cases with $j=2,3$ follow from the $2\pi/3$-rotation symmetry~(\ref{arotation}):
\begin{equation}
\hat{\bar g}_{h,\omega,j}(\mathbf k'_{j}+\tilde{\mathbf p}_{F,j}^{(\omega,h)})=
\left(\begin{array}{*{2}{c}}\mathds1&0\\0&\mathcal T_{T\mathbf k'_{j}+\tilde{\mathbf p}_{F,j-\omega}^{(\omega,h)}}\end{array}\right)\hat{\bar g}_{h,\omega,j}(T\mathbf k'_{j}+\tilde{\mathbf p}_{F,j-\omega}^{(\omega,h)})\left(\begin{array}{*{2}{c}}\mathds1&0\\0&\mathcal T^\dagger_{T\mathbf k'_{j}+\tilde{\mathbf p}_{F,j-\omega}^{(\omega,h)}}\end{array}\right)
\label{Wreltht}\end{equation}
where $T$ and $\mathcal T_{\mathbf k}$ were defined above~(\ref{arotation}), and $\tilde{\mathbf p}_{F,4}^{(-,h)}\equiv \tilde{\mathbf p}_{F,1}^{(-,h)}$.
\subseqskip

\subsection{Two-point Schwinger function}
\label{schwinthsec}
\indent We now compute the dominant part of the two-point Schwinger function for $\mathbf k$ {\it well inside} the third regime, i.e.
$${\bf k}\in \mathcal B_{\mathrm{III}}^{(\omega,j)}:=\bigcup_{h=\mathfrak h_\beta+1}^{\bar{\mathfrak h}_2-1}\mathrm{supp}f_{h,\omega,j}.$$
Let
$$h_{\mathbf k}:=\max\{h: f_{h,\omega,j}(\mathbf k)\neq0\}$$
so that if $h\not\in\{h_{\bf k},h_{{\bf k}}-1\}$, then $f_{h,\omega,j}({\bf k})=0$.\par
\bigskip

\point{Schwinger function in terms of dressed propagators} Recall that the two-point Schwinger function can be computed in terms of the effective source term $\mathcal X^{(h)}$, see \eqref{Qdef} and comment after Lemma~\ref{schwinlemma}. 
Since $h_{\mathbf k}\leqslant\bar{\mathfrak h}_2$, $\mathcal X^{(h)}$ is left invariant by the integration over the ultraviolet, the first and the second regimes, in the sense that $\mathcal X^{(\bar{\mathfrak h}_2)}=\mathcal X^{(\bar{\mathfrak h}_0)}$, with $\mathcal X^{(\bar{\mathfrak h}_0)}$ given by \eqref{initXo}. Therefore, we can compute $\mathcal X^{(h)}$ for $h\in\{\mathfrak h_\beta,\cdots,\bar{\mathfrak h}_2-1\}$ inductively using lemma~\ref{schwinlemma}, and find, similarly to~(\ref{schwinexpro}) and~(\ref{schwinexprt}), that
\begin{equation}
s_2(\mathbf k)=\left(\hat{\bar g}_{h_{\mathbf k},\omega,j}(\mathbf k)+\hat{\bar g}_{h_{\mathbf k}-1,\omega,j}(\mathbf k)\right)\left(\mathds1-\sigma(\mathbf k)-\sigma_{<h_{\mathbf k}}(\mathbf k)\right)
\label{schwinexprth}\end{equation}
where
\begin{equation}
\sigma(\mathbf k):=
\hat W_2^{(h_{\mathbf k}-1)}\hat{\bar g}_{h_{\mathbf k},\omega,j}
+(\hat{\bar g}_{h_{\mathbf k},\omega,j}+\hat{\bar g}_{h_{\mathbf k}-1,\omega,j})^{-1}\hat{\bar g}_{h_{\mathbf k},\omega,j}\hat W_2^{(h_{\mathbf k}-1)}\hat{\bar g}_{h_{\mathbf k}-1,\omega,j}(\mathds1-\hat W_2^{(h_{\mathbf k}-1)}\hat{\bar g}_{h_{\mathbf k},\omega,j})
\label{sigmathdef}\end{equation}
and
\begin{equation}\begin{largearray}
\sigma_{<h_{\mathbf k}}(\mathbf k):=\left(\mathds1-\left( \hat{\bar g}_{h_{\mathbf k},\omega,j}+\hat{\bar g}_{h_{\mathbf k}-1,\omega,j}\right)^{-1}\hat{\bar g}_{h_{\mathbf k},\omega,j}\hat W_2^{(h_{\mathbf k}-1)}\hat{\bar g}_{h_{\mathbf k}-1,\omega,j}\right)
\left(\sum_{h'=\mathfrak h_\beta}^{h_{\mathbf k}-2}\hat W_2^{(h')}\right)\cdot\\[0.5cm]
\hfill\cdot\left( \hat{\bar g}_{h_{\mathbf k},\omega,j}+\hat{\bar g}_{h_{\mathbf k}-1,\omega,j}-\hat{\bar g}_{h_{\mathbf k}-1,\omega,j}\hat W_{2,\omega}^{(h_{\mathbf k}-1)}\hat{\bar g}_{h_{\mathbf k},\omega,j}\right).
\end{largearray}\label{sigmalthdef}\end{equation}
Similarly to~(\ref{ineqginvgt}), we have
\begin{equation}
\left|(\hat{\bar g}_{h_{\mathbf k},\omega,j}+\hat{\bar g}_{h_{\mathbf k}-1,\omega,j})^{-1}\hat{\bar g}_{h_{\mathbf k},\omega,j}\right|\leqslant(\mathrm{const}.)\ 
\label{ineqginvgth}\end{equation}
and, by (\ref{pcth}) and~(\ref{assumdpropth}), we have
\begin{equation}\left\{\begin{array}l
|\sigma(\mathbf k)|\leqslant (\mathrm{const}.)\  2^{h_{\bf k}-2h_\epsilon}|U|\\[0.2cm]
|\sigma_{<h_{\mathbf k}}(\mathbf k)|\leqslant (\mathrm{const}.)\  2^{h_{\bf k}-2h_\epsilon}|U|.
\end{array}\right.\label{boundsigmalkth}\end{equation}
\bigskip

\point{Dominant part of the dressed propagators} We now compute $\hat{\bar g}_{h_{\mathbf k},\omega,j}+\hat{\bar g}_{h_{\mathbf k}-1,\omega,j}$.\par
\bigskip

\subpoint{$j=0$ case} We first treat the case $j=0$. It follows from (the analogue of) (\ref{gseplint}), (\ref{sepLAthdom}) and (\ref{invLAthdom}), that
\begin{equation}\begin{largearray}
\hat{\bar g}_{{h_{\mathbf k}},\omega,0}(\mathbf k)+\hat{\bar g}_{{h_{\mathbf k}}-1,\omega,0}(\mathbf k)\\
\hfill=\left(\begin{array}{*{2}{c}}\mathds1&\bar M_{{h_{\mathbf k}},0}^\dagger(\mathbf k)\\0&\mathds1\end{array}\right)\left(\begin{array}{*{2}{c}}\bar a_{{h_{\mathbf k}},0}^{(M)}&0\\0&\bar a_{{h_{\mathbf k}},0}^{(m)}(\mathbf k)\end{array}\right)\left(\begin{array}{*{2}{c}}\mathds1& 0\\\bar M_{{h_{\mathbf k}},0}(\mathbf k)&\mathds1\end{array}\right)(\mathds1+\sigma'_0(\mathbf k))
\end{largearray}\label{schwinxprgthp}\end{equation}
where $\bar M_{{h_{\mathbf k}},0}$, $\bar a_{{h_{\mathbf k}},0}^{(M)}$ and $\bar a_{{h_{\mathbf k}},0}^{(m)}$ were defined in~(\ref{Mth}) and~(\ref{bath}), and the error term $\sigma'_0$ can be bounded using~(\ref{ineqRAth}) and~(\ref{boundsigmafth}):
\begin{equation}
|\sigma'_0(\mathbf k)|\leqslant (\mathrm{const}.)\  2^{{h_{\mathbf k}}-2h_\epsilon}(2^{-h_\epsilon}+|h_{\bf k}||U|).
\label{boundsigmapth}\end{equation}
\bigskip

\subpoint{$j=1$ case} We now consider $j=1$. It follows from~(\ref{sepLAthjdom}) and~(\ref{invLAthjdom}) that
\begin{equation}\begin{largearray}
\hat{\bar g}_{{h_{\mathbf k}},\omega,0}(\mathbf k)+\hat{\bar g}_{{h_{\mathbf k}}-1,\omega,0}(\mathbf k)=\\
\hfill\left(\begin{array}{*{2}{c}}\mathds1&\bar M_{{h_{\mathbf k}},1}^\dagger(\mathbf k)\\0&\mathds1\end{array}\right)\left(\begin{array}{*{2}{c}}\bar a_{{h_{\mathbf k}},1}^{(M)}&0\\0&\bar a_{{h_{\mathbf k}},1}^{(m)}(\mathbf k)\end{array}\right)\left(\begin{array}{*{2}{c}}\mathds1&0\\\bar M_{{h_{\mathbf k}},1}(\mathbf k)&\mathds1\end{array}\right)(\mathds1+\sigma'_1(\mathbf k))
\end{largearray}\label{schwinxprgthjp}\end{equation}
where $\bar M_{{h_{\mathbf k}},1}$, $\bar a_{{h_{\mathbf k}},1}^{(M)}$ and $\bar a_{{h_{\mathbf k}},1}^{(m)}$ were defined in~(\ref{Mthj}) and~(\ref{bathj}), and the error term $\sigma'_1$ can be bounded using~(\ref{ineqRAthj}) and~(\ref{boundsigmaftho}):
\begin{equation}
|\sigma'_1(\mathbf k)|\leqslant (\mathrm{const}.)\  \left(2^{h_\epsilon}(1+|h_\epsilon||U|)+2^{{h_{\mathbf k}}-2h_\epsilon}(2^{-h_\epsilon}+|h_{\mathbf k}||U|)\right).
\label{boundsigmapthj}\end{equation}

\subpoint{$j=2,3$ cases} The cases with $j=2,3$ follow from the $2\pi/3$-rotation symmetry~(\ref{arotation}) (see~(\ref{Wreltht})).\par
\bigskip

\point{Proof of Theorem~\ref{theoth}} We now conclude the proof of Theorem~\ref{theoth}. We focus our attention on $j=0,1$ since the cases with $j=2,3$ follow by symmetry. Similarly to section~\ref{schwintsec}, we define
$$
B_{h_{\mathbf k},j}(\mathbf k):=(\mathds1+\sigma'_j(\mathbf k))\left(\hat{\bar g}_{h_{\mathbf k},\omega,j}(\mathbf k)+\hat{\bar g}_{h_{\mathbf k}-1,\omega,j}(\mathbf k)\right)^{-1}
$$
(i.e. the inverse of the matrix on the right side of~(\ref{schwinxprgthp}) for $j=0$, (\ref{schwinxprgthjp}) for $j=1$, whose explicit expression is similar to the right side of~(\ref{LAthdom}) and~(\ref{LAthjdom})), and
$$\begin{array}c
\tilde m_{3,0}:=\tilde m_{\mathfrak h_\beta},\quad
z_{3,0}:=z_{\mathfrak h_2},\quad
v_{3,0}:=v_{\mathfrak h_2},\quad
\tilde v_{3,0}:=\tilde v_{\mathfrak h_2},\\[0.2cm]
\tilde m_{3,1}:=m^{\xi\xi}_{\mathfrak h_\beta,1},\quad
\bar m_{3,1}:=m^{\phi\phi}_{\mathfrak h_\beta,1},\quad
m_{3,1}:=m^{\xi\phi}_{\mathfrak h_\beta,1},\quad
z_{3,1}:=z^{\phi\phi}_{\mathfrak h_\beta,1},\\[0.2cm]
\bar v_{3,1}:=v^{\xi\phi}_{\mathfrak h_\beta,1},\quad
\bar w_{3,1}:=w^{\xi\phi}_{\mathfrak h_\beta,1},\quad
\tilde v_{3,1}:=v^{\phi\phi}_{\mathfrak h_\beta,1},\quad
\tilde w_{3,1}:=w^{\phi\phi}_{\mathfrak h_\beta,1}
\end{array}$$
and use~(\ref{boundzvtth}) and~(\ref{boundzvthowconst}) to bound
$$\begin{array}c
|\tilde m_{h_{\mathbf k}}-\tilde m_{3,0}|+|m^{\xi\xi}_{h_{\mathbf k},1}-\tilde m_{3,1}|+|m^{\phi\phi}_{h_{\mathbf k},1}-\bar m_{3,1}|\leqslant(\mathrm{const}.)\ |U|2^{2h_{\mathbf k}-3h_\epsilon},\\[0.2cm]
|m^{\xi\phi}_{h_{\mathbf k},1}-m_{3,1}|\leqslant(\mathrm{const}.)\ |U|2^{2h_{\mathbf k}-2h_\epsilon},\\[0.2cm]
|z_{h_{\mathbf k}}-z_{3,1}|+|z^{\phi\phi}_{h_{\mathbf k},1}-z_{3,0}|\leqslant(\mathrm{const}.)\ |U|2^{h_{\mathbf k}-2h_\epsilon},\\[0.2cm]
|v_{h_{\mathbf k}}-v_{3,0}|+|v^{\xi\phi}_{h_{\mathbf k},1}-v_{3,1}|+|w^{\xi\phi}_{h_{\mathbf k},1}-w_{3,1}|\leqslant(\mathrm{const}.)\ |U|2^{h_{\mathbf k}-h_\epsilon},\\[0.2cm]
|\tilde v_{h_{\mathbf k}}-\tilde v_{3,0}|+|v^{\phi\phi}_{h_{\mathbf k},1}-\tilde v_{3,1}|+|w^{\phi\phi}_{h_{\mathbf k},1}-\tilde w_{3,1}|\leqslant(\mathrm{const}.)\ |U|2^{h_{\mathbf k}-2h_\epsilon}
\end{array}$$
so that
$$
\left|(B_{\mathfrak h_2,j}(\mathbf k)-B_{h_{\mathbf k},j}(\mathbf k))B_{\mathfrak h_2,j}^{-1}(\mathbf k)\right|\leqslant(\mathrm{const}.)\ |U|2^{h_{\mathbf k}-2h_\epsilon}
$$
which implies
\begin{equation}
B_{h_{\mathbf k}}^{-1}(\mathbf k)=B_{\mathfrak h_2}^{-1}(\mathbf k)(\mathds1+O(|U|2^{h_{\mathbf k}-2h_\epsilon})).
\label{approxBhth}\end{equation}
We inject~(\ref{approxBhth}) into~(\ref{schwinxprgthp}) and~(\ref{schwinxprgthjp}), which we then combine with~(\ref{schwinexprth}), (\ref{boundsigmalkth}), (\ref{boundsigmapth}) and~(\ref{boundsigmapthj}), and find an expression for $s_2$ which is similar to the right side of~(\ref{schwinxprgthp}) and~(\ref{schwinxprgthjp}) but with $h_{\mathbf k}$ replaced by $\mathfrak h_2$. This concludes the proof of~(\ref{schwinth}). Furthermore, the estimate~(\ref{ineqrccthj}) follows from~(\ref{boundzvtthp}) and~(\ref{sboundmzvwthj}) as well as (\ref{LfreeAo}) and~(\ref{ineqshiftth}), which concludes the proof of Theorem~\ref{theoth}.\par

\bigskip

\point{Proof of (\ref{sumWo}) and (\ref{sumWt})} In order to conclude the proofs of Theorems~\ref{theoo} and~\ref{theot} as well as the Main Theorem, we still have to bound the sums on the left side of~(\ref{sumWo}) and of~(\ref{sumWt}), which we recall were assumed to be true to prove~(\ref{schwino}) and~(\ref{schwint}) (see sections~\ref{schwinosec} and~\ref{schwintsec}). It follows from~(\ref{pcth}) that
\begin{equation}
\left|\sum_{h'=\mathfrak h_\beta}^{\bar{\mathfrak h}_2}\hat W_2^{(h')}(\mathbf k)\right|\leqslant(\mathrm{const}.)\ 2^{4h_\epsilon}|U|.
\label{boundthsumW}\end{equation}\par
This, along with~(\ref{boundtsumW}) concludes the proofs of~(\ref{sumWo}) and~(\ref{sumWt}), and thus concludes the proof of Theorems~\ref{theoo}, \ref{theot} and~\ref{theoth} as well as the Main Theorem.\par
\seqskip

\section{Conclusion}
\label{concsec}
\indent We considered a tight-binding model of bilayer graphene describing spin-less fermions hopping on two hexagonal layers in Bernal stacking, in the presence of short range interactions. We assumed that only three hopping parameters are different from zero (those usually called $\gamma_0,\gamma_1$ and $\gamma_3$ in the literature), in which case the Fermi surface at half-filling degenerates to a collection of 8 Fermi points.  Under a smallness assumption on the interaction strength $U$ and on the transverse hopping $\epsilon$, we proved by rigorous RG methods that the specific ground state energy and correlation functions in the thermodynamic limit are analytic in $U$, uniformly in $\epsilon$. Our proof requires a detailed analysis of the crossover regimes from one in which the two layers are effectively decoupled, to one where the effective dispersion relation is approximately parabolic around the central Fermi points (and the inter-particle interaction is effectively marginal), to the deep infrared one, where the effective dispersion relation is approximately conical around each Fermi points (and the inter-particle interaction is effectively irrelevant).  Such an analysis, in which the influence of the flow of the effective constants in one regime has crucial repercussions in lower regimes, is, to our knowledge, novel.\par
\bigskip

\indent We expect our proof to be adaptable without substantial efforts to the case where $\gamma_4$ and $\Delta$ are different from zero, as in \eqref{relge}, the intra-layer next-to-nearest neighbor hopping $\gamma_0'$ is $O(\epsilon)$, the chemical potential is $O(\epsilon^3)$, and the temperature is larger than (const.)$\epsilon^4$. At smaller scales, the Fermi set becomes effectively one-dimensional, which thoroughly changes the scaling properties. In particular, the effective inter-particle interaction becomes marginal, again, and its flow tends to grow logarithmically. Perturbative analysis thus breaks down at exponentially small temperatures in $\epsilon$ and in $U$, and it should be possible to rigorously control the system down to such temperatures.  Such an analysis could prove difficult, because it requires fine control on the geometry of the Fermi surface, as in \cite{benZS} and in \cite{fktZFa,fktZFb,fktZFc}, where the Fermi liquid behavior of a system of interacting electrons was proved, respectively down to exponentially small and zero temperatures, under different physical conditions. We hope to come back to this issue in the future.\par
\bigskip

\indent Another possible extension would be the study of crossover effects on other physical observables, such as the conductivity, in the spirit of \cite{masOO}. In addition, it would be interesting to study the case of three-dimensional Coulomb interactions, which is physically interesting in describing {\it clean} bilayer graphene samples, i.e.  where screening effects are supposedly negligible. It may be possible to draw inspiration from the analysis of \cite{gmpOZ, gmpOOt} to construct the ground state, order by order in renormalized perturbation theory.  The construction of the theory in the second and third regimes would pave the way to understanding the universality of the conductivity in the deep infrared, beyond the regime studied in \cite{masOO}.\par
\bigskip

{\bf Acknowledgments} We acknowledge financial support from the ERC Starting Grant CoMBoS (grant agreement No. 239694)
and the PRIN National Grant {\it Geometric and analytic theory of Hamiltonian systems in finite and infinite dimensions}. 

\pagebreak
\appendix

\section{Computation of the Fermi points}
\label{fermiapp}
\indent In this appendix, we prove~(\ref{fermdef}).\par
\bigskip

\Theo{Proposition}\label{fermprop}
Given
$$\Omega(k):=1+2e^{-\frac32ik_x}\cos\left(\frac{\sqrt3}{2}k_y\right),$$
the solutions for $k\in\hat\Lambda_\infty$ (see~(\ref{lae}) and following lines for the definition of $\hat\Lambda$ and $\hat \Lambda_\infty$) of
\begin{equation}
\Omega^2(k)-\gamma_1\gamma_3\Omega^*(k)e^{-3ik_x}=0
\label{eqfermpts}\end{equation}
with
$$0<\gamma_1\gamma_3<2$$
are
\begin{equation}\left\{\begin{array}l
p_{F,0}^\omega:=\left(\frac{2\pi}{3},\omega\frac{2\pi}{3\sqrt3}\right)\\[0.5cm]
p_{F,1}^\omega:=\left(\frac{2\pi}{3},\omega\frac{2}{\sqrt3}\arccos\left(\frac{1-\gamma_1\gamma_3}{2}\right)\right)\\[0.5cm]
p_{F,2}^\omega:=\left(\frac{2\pi}{3}+\frac{2}{3}\arccos\left(\frac{\sqrt{1+\gamma_1\gamma_3}(2-\gamma_1\gamma_3)}{2}\right),\omega\frac{2}{\sqrt3}\arccos\left(\frac{1+\gamma_1\gamma_3}{2}\right)\right)\\[0.5cm]
p_{F,3}^\omega:=\left(\frac{2\pi}{3}-\frac{2}{3}\arccos\left(\frac{\sqrt{1+\gamma_1\gamma_3}(2-\gamma_1\gamma_3)}{2}\right),\omega\frac{2}{\sqrt3}\arccos\left(\frac{1+\gamma_1\gamma_3}{2}\right)\right)
\end{array}\right.\label{fermia}\end{equation}
for $\omega\in\{+,-\}$.\par
\endtheo
\bigskip

\indent\underline{Proof}: We define
$$
C:=\cos\left(\frac{3}{2}k_x\right),\quad
S:=\sin\left(\frac{3}{2}k_x\right),\quad
Y:=\cos\left(\frac{\sqrt3}{2}k_y\right),\quad
G:=\gamma_1\gamma_3
$$
in terms of which~(\ref{eqfermpts}) becomes
\begin{equation} 
\left\{\begin{array}l
4(2C^2-1)Y^2+2C(2-G)Y+1-G(2C^2-1)=0\\[0.2cm]
-2S(C(4Y^2-G)+Y(2-G))=0.
\end{array}\right.\label{eq:a3}\end{equation}
\bigskip

\pointt If $S=\sin((3/2)k_x)=0$, then $k_x\in\{0,2\pi/3\}$. Furthermore, since $k\in\hat\Lambda_\infty$, if $k_x=0$ then $k_y=0$, which is not a solution of~(\ref{eqfermpts}) as long as $G<3$. Therefore $k_x=2\pi/3$, so that $C=-1$, and $Y$ solves
$$4Y^2-2(2-G)Y+1-G=0$$
so that
$$Y=\frac{2-G\pm G}{4}$$
and therefore
$$
k_y=\pm\frac{2\pi}{3\sqrt3}
\quad\mathrm{or}\quad 
k_y=\pm\frac{2}{\sqrt3}\arccos\left(\frac{1-G}{2}\right)
.$$
\bigskip

\pointt If $S\neq0$, then
$$C(4Y^2-G)=-Y(2-G)$$
so that the first of \eqref{eq:a3} becomes $4Y^2=1+G$, which implies
$$Y=\pm\frac{\sqrt{1+G}}{2},\quad C=\mp\frac{\sqrt{1+G}(2-G)}{2}$$
so that
$$
k_x=\frac{2\pi}{3}+\frac{2}{3}\arccos\left(\frac{\sqrt{1+G}(2-G)}{2}\right),\quad
k_y=\pm\frac{2}{\sqrt3}\arccos\left(\frac{\sqrt{1+G}}{2}\right)
$$
or
$$
k_x=\frac{2\pi}{3}-\frac{2}{3}\arccos\left(\frac{\sqrt{1+G}(2-G)}{2}\right),\quad
k_y=\pm\frac{2}{\sqrt3}\arccos\left(\frac{\sqrt{1+G}}{2}\right).
$$
\penalty10000\hfill\penalty10000$\square$
\seqskip

\section{$4\times4$ matrix inversions}
\label{inversapp}
\indent In this appendix, we give the explicit expression of the determinant and the inverse of matrices that have the form of the inverse free propagator.
The result is collected in the following proposition and corollary, whose proofs are straightforward, brute force, computations. 
\par
\bigskip

\Theo{Proposition}\label{matinvprop}
Given a matrix
\begin{equation}
A=\left(\begin{array}{*{4}{c}}
i\mathfrak x&\mathfrak a^*&0&\mathfrak b^*\\
\mathfrak a&i\mathfrak x&\mathfrak b&0\\
0&\mathfrak b^*&i\mathfrak z&\mathfrak c\\
\mathfrak b&0&\mathfrak c^*&i\mathfrak z\end{array}\right)
\label{formmatrixinva}\end{equation}
with $(\mathfrak x,\mathfrak z)\in\mathbb{R}^2$ and $(\mathfrak a,\mathfrak b,\mathfrak c)\in\mathbb C^3$. We have
\begin{equation}
\det A=
(|\mathfrak b|^2+\mathfrak z\mathfrak x)^2+|\mathfrak a|^2\mathfrak z^2+|\mathfrak c|^2(\mathfrak x^2+|\mathfrak a|^2)-2\mathcal Re(\mathfrak a^*\mathfrak b^2\mathfrak c)
\label{detAexpr}\end{equation}
and
$$A^{-1}=
\frac{1}{\det A}\left(\begin{array}{*{4}{c}}\mathfrak g_{a,a}&\mathfrak g_{a,\tilde b}&\mathfrak g_{a,\tilde a}&\mathfrak g_{a,b}\\
\mathfrak g_{a,\tilde b}^+&\mathfrak g_{a,a}&\mathfrak g_{a,b}^+&\mathfrak g_{a,\tilde a}^+\\
\mathfrak g_{a,\tilde a}^+&\mathfrak g_{a,b}&\mathfrak g_{\tilde a,\tilde a}&\mathfrak g_{\tilde a,b}\\
\mathfrak g_{a,b}^+&\mathfrak g_{a,\tilde a}&\mathfrak g_{\tilde a,b}^+&\mathfrak g_{\tilde a,\tilde a}\end{array}\right)
$$
with
\begin{equation}\left\{\begin{array}l
\mathfrak g_{a,a}=-i\mathfrak z|\mathfrak b|^2-i\mathfrak x(\mathfrak z^2+|\mathfrak c|^2)\\[0.5cm]
\mathfrak g_{a,\tilde b}=\mathfrak z^2\mathfrak a^*-\mathfrak c^*((\mathfrak b^*)^2-\mathfrak a^*\mathfrak c)\\[0.5cm]
\mathfrak g_{a,\tilde a}=i\mathfrak z\mathfrak a^*\mathfrak b+i\mathfrak x\mathfrak b^*\mathfrak c^*\\[0.5cm]
\mathfrak g_{a,b}=\mathfrak b((\mathfrak b^*)^2-\mathfrak a^*\mathfrak c)+\mathfrak z\mathfrak x\mathfrak b^*\\[0.5cm]
\mathfrak g_{\tilde a,b}=-\mathfrak a((\mathfrak b^*)^2-\mathfrak a^*\mathfrak c)+\mathfrak x^2\mathfrak c\\[0.5cm]
\mathfrak g_{\tilde a,\tilde a}=-i\mathfrak z|\mathfrak a|^2-i\mathfrak x(\mathfrak x\mathfrak z+|\mathfrak b|^2).
\end{array}\right.\label{invAexpr}\end{equation}
and given a function $\mathfrak g(\mathfrak a,\mathfrak b,\mathfrak c,\mathfrak x,\mathfrak z)$,
$$
\mathfrak g^+(\mathfrak a,\mathfrak b,\mathfrak c,\mathfrak x,\mathfrak z):=\mathfrak g^*(\mathfrak a,\mathfrak b,\mathfrak c,-\mathfrak x,-\mathfrak z).
\nopagebreakaftereq$$
\endtheo
\restorepagebreakaftereq
\subseqskip

\Theo{Corollary}\label{Anokzprop}
If $\mathfrak z=\mathfrak x=0$, then
\begin{equation}
\det A=\left|\mathfrak b^2-\mathfrak a\mathfrak c^*\right|^2\geqslant0.
\label{detnokz}\end{equation}
In particular, $A$ is invertible if and only if $\mathfrak b^2\neq\mathfrak a\mathfrak c^*$.\par
\endtheo
\seqskip

\section{Block diagonalization}
\label{diagapp}
\indent In this appendix, we give the  formula for  block-diagonalizing $4\times4$ matrices, which is useful to separate the massive block from the massless one.
The result is collected in the following proposition, whose proof is straightforward. 
\par
\bigskip

\Theo{Proposition}\label{blockdiagprop}
Given a $4\times4$ complex matrix $B$, which can be written in block-form as
\begin{equation}
B=\left(\begin{array}{*{2}{c}}B^{\xi\xi}&B^{\xi\phi}\\B^{\xi\phi}&B^{\phi\phi}\end{array}\right)
\label{blockB}\end{equation}
in which $B^{\xi\xi}$, $B^{\xi\phi}$ and $B^{\phi\phi}$ are $2\times2$ complex matrices and $B^{\xi\xi}$ and $B^{\phi\phi}$ are invertible, we have
\begin{equation}
\left(\begin{array}{*{2}{c}}\mathds1&0\\-B^{\xi\phi}(B^{\xi\xi})^{-1}&\mathds1\end{array}\right)
B
\left(\begin{array}{*{2}{c}}\mathds1&-(B^{\xi\xi})^{-1}B^{\xi\phi}\\0&\mathds1\end{array}\right)
=\left(\begin{array}{*{2}{c}}B^{\xi\xi}&0\\0&B^{\phi\phi}-B^{\xi\phi}(B^{\xi\xi})^{-1}B^{\xi\phi}\end{array}\right).
\label{diagdB}\end{equation}
If $B^{\phi\phi}-B^{\xi\phi}(B^{\xi\xi})^{-1}B^{\xi\phi}$ is invertible then
$$(B^{\phi\phi}-B^{\xi\phi}(B^{\xi\xi})^{-1}B^{\xi\phi})^{-1}$$
is the lower-right block of $B^{-1}$.
\endtheo
\seqskip

\section{Bound of the propagator in the II-III intermediate regime}
\label{boundproptthapp}
\indent In this appendix, we prove the assertion between~(\ref{detatth}) and~(\ref{boundfreepropztth}), that is that the determinant of the inverse propagator is bounded below by $(\mathrm{const}.)\ \epsilon^8$ in the intermediate regime between the second and third regimes. Using the symmetry under $k_x\mapsto-k_x$ and under $2\pi/3$ rotations, we restrict our discussion to $\omega=+$ and $k_y-p_{F,0,y}^+>0$. In a coordinate frame centered at $p_{F,0}^+$, we denote with some abuse of notation 
${\bf k}'_{+,0}=(k_0,k_x,k_y)$ and $p_{F,1}^+=(0,D\bar\epsilon^2)$, where $\bar\epsilon\frac32\gamma_3$ 
and $D=\frac{8}{27}\frac{\gamma_1}{\gamma_3}(1+O(\epsilon^2))$ (see \eqref{eq:3.2bis}). Note that 
$D>0$ is uniformly bounded away from 0 for $\bar\epsilon$ small (recall that $\gamma_1=\epsilon$ and $\gamma_3=0.33 \epsilon$). In these coordinates, we restrict to $k_y>0$, and  
the first and third conditions in \eqref{inttth} read
\begin{equation} \sqrt{k_0^2+\bar\epsilon^2(k_x^2+k_y^2)}\geqslant \bar\kappa\bar\epsilon^3, \qquad \sqrt{k_0^2+\bar\epsilon^2(9k_x^2+(k_y-D\bar\epsilon^2)^2)}\geqslant \bar\kappa\bar\epsilon^3, 
\end{equation}
where $\bar\kappa\bar\kappa_2(\frac{2\epsilon}{3\gamma_3})^3$. The second condition in \eqref{inttth} implies that $(k_x^2+k_y^2)\leqslant (\mathrm{const}.)\ \epsilon^2$, in which case the 
desired bound (that is, $|\det \hat A|\geqslant (\mathrm{const}.)\  \epsilon^8$, with $\det\hat A$ as in \eqref{detatth}) reads
\begin{equation} \epsilon^2 k_0^2 +\frac{81}{16}\big|(ik_x+k_y)^2-D\bar \epsilon^2(-ik_x+k_y)\big|^2\geqslant (\mathrm{const}.)\ \epsilon^8.\end{equation}
Therefore, the desired estimate follows from the following Proposition, which is proved below. 

\par
\bigskip

\Theo{Proposition}\label{bounddetintprop}
For all $D,\epsilon>0$, if $(k_0,k_x,k_y)\in\mathbb{R}^3$ satisfies
$$
k_y>0,\quad
\sqrt{k_0^2+\bar\epsilon^2(k_x^2+k_y^2)}>\bar\kappa\bar\epsilon^3,\quad
\sqrt{k_0^2+\bar\epsilon^2(9k_x^2+(k_y-D\bar\epsilon^2)^2)}>\bar\kappa\bar\epsilon^3
$$
for some constant $\bar\kappa>0$, then, for all $\alpha>0$, we have
\begin{equation}
\bar\epsilon^2k_0^2+\alpha\left|(ik_x+k_y)^2-D\bar\epsilon^2(-ik_x+k_y)\right|^2>C\bar\epsilon^8,
\label{ineqdeta}\end{equation}
where
\par
$$C:=\min\left(1,\frac{\alpha D^2}{12},\frac{\alpha(473-3\sqrt{105})\bar\kappa^2}{288}\right)\frac{\bar\kappa^2}{4}.
\nopagebreakaftereq$$
\endtheo
\restorepagebreakaftereq
\bigskip

\indent\underline{Proof}: We rewrite the left side of~(\ref{ineqdeta}) as
$$
l:=\bar\epsilon^2k_0^2+\alpha\left(-k_x^2+k_y^2-D\bar\epsilon^2k_y\right)^2+\alpha k_x^2\left(2k_y+D\bar\epsilon^2\right)^2.
$$
If $|k_0|>\bar\kappa\bar\epsilon^3/2$, then $l>\bar\kappa^2\bar\epsilon^8/4$ from which~(\ref{ineqdeta}) follows. If $|k_0|\leqslant\bar\kappa\bar\epsilon^3/2$, then
$$k_x^2+k_y^2>\frac{3}{4}\bar\kappa^2\bar\epsilon^4,\quad 9k_x^2+(k_y-D\bar\epsilon^2)^2>\frac{3}{4}\bar\kappa^2\bar\epsilon^4.$$
If $|k_x|>(1/4\sqrt3)\bar\kappa\bar\epsilon^2$, then, using the fact that $k_y>0$, $l>\alpha(1/48)D^2\bar\kappa^2\bar\epsilon^8$ from which~(\ref{ineqdeta}) follows. If $|k_x|\leqslant(1/4\sqrt3)\bar\kappa\bar\epsilon^2$, then
$$
k_y>\sqrt{\frac{35}{48}}\bar\kappa\bar\epsilon^2,\quad
|k_y-D\bar\epsilon^2|>\frac{3}{4}\bar\kappa\bar\epsilon^2
$$
so that
$$\left|k_y(k_y-D\bar\epsilon^2)\right|-k_x^2>\frac{3\sqrt{105}-1}{48}\bar\kappa^2\bar\epsilon^4$$
and $l>\alpha((3\sqrt{105}-1)^2/2304)\bar\kappa^4\bar\epsilon^8$ from which~(\ref{ineqdeta}) follows.\penalty10000\hfill\penalty10000$\square$\par
\seqskip

\section{Symmetries}
\label{symapp}
\indent In this appendix, we prove that the symmetries listed in~(\ref{aglobaluo}) through~(\ref{ainversiont}) leave $h_0$ and $\mathcal V$ invariant. We first recall
\begin{equation}
h_0=-\frac{1}{\chi_0(2^{-M}|k_0|)\beta|\Lambda|}\sum_{\mathbf k\in\mathcal B^*_{\beta,L}}\left(\begin{array}{*{2}{c}}\hat\xi_{\mathbf k}^+&\hat\phi_{\mathbf k}^+\end{array}\right)\left(\begin{array}{*{2}{c}}A^{\xi\xi}(\mathbf k)&A^{\xi\phi}(\mathbf k)\\A^{\phi\xi}(\mathbf k)&A^{\phi\phi}(\mathbf k)\end{array}\right)\left(\begin{array}{c}\hat\xi^-_{\mathbf k}\\\hat\phi^-_{\mathbf k}\end{array}\right)
\label{ahzexplicit}\end{equation}
with
$$\begin{array}c
A^{\xi\xi}(\mathbf k):=\left(\begin{array}{*{2}{c}}ik_0&\gamma_1\\\gamma_1&ik_0\end{array}\right),\quad
A^{\xi\phi}(\mathbf k)\equiv A^{\phi\xi}(\mathbf k):=\left(\begin{array}{*{2}{c}}0&\Omega^*(k)\\\Omega(k)&0\end{array}\right),\\[1cm]
A^{\phi\phi}(\mathbf k):=\left(\begin{array}{*{2}{c}}ik_0&\gamma_3\Omega(k)e^{3ik_x}\\\gamma_3\Omega^*(k)e^{-3ik_x}&ik_0\end{array}\right)
\end{array}$$
and
\begin{equation}
\mathcal V(\psi)=\frac{U}{(\beta|\Lambda|)^3}\sum_{(\alpha,\alpha')}\sum_{\mathbf k_1,\mathbf k_2,\mathbf k_3}\hat v_{\alpha,\alpha'}(k_1-k_2)\hat\psi^+_{\mathbf k_1,\alpha}\hat\psi^-_{\mathbf k_2,\alpha}\hat\psi^+_{\mathbf k_3,\alpha'}\hat\psi^-_{\mathbf k_1-\mathbf k_2+\mathbf k_3,\alpha'}
\label{aVexplicit}\end{equation}
where
$$\hat v_{\alpha,\alpha'}(k):=\sum_{x\in\Lambda}\ e^{ik\cdot x}v(x+d_\alpha-d_{\alpha'}).$$
\subseqskip

\point{Global $U(1)$} Follows immediately from the fact that there are as many $\psi^+$ as $\psi^-$ in $h_0$ and $\mathcal V$.
\penalty10000\hfill\penalty10000$\square$\par
\subseqskip

\point{ $2\pi/3$ rotation} We have 
$$\Omega(e^{i\frac{2\pi}3\sigma_2}k)=e^{il_2 \cdot k}\Omega(k),\quad e^{3i(e^{i\frac{2\pi}3\sigma_2}k)|_x}=e^{-3il_2\cdot k}e^{3ik_x}$$
so that $\mathcal T_{\mathbf k}^\dagger A^{\phi\phi}(T^{-1}\mathbf k)\mathcal T_{\mathbf k}=A^{\phi\phi}(\mathbf k)$ and
$A^{\xi\phi}(T^{-1}\mathbf k)\mathcal T_{\mathbf k}=A^{\xi\phi}(\mathbf k)$. This, together with $A^{\xi\xi}(T^{-1}\mathbf k)=A^{\xi\xi}(\mathbf k)$,
implies that $h_0$ is invariant under~(\ref{arotation}).\par
\smallskip
\indent Furthermore, interpreting $e^{-i\frac{2\pi}3\sigma_2}$ as a rotation in $\mathbb R^3$ around the $z$ axis, 
$$e^{-i\frac{2\pi}3\sigma_2}d_a=d_a,\quad
e^{-i\frac{2\pi}3\sigma_2}d_{\tilde b}=d_{\tilde b}, \quad
e^{-i\frac{2\pi}3\sigma_2}d_{\tilde a}=l_2+d_{\tilde a},\quad 
e^{-i\frac{2\pi}3\sigma_2}d_b=-l_2+d_b,$$
which implies, denoting by $\hat v(k)$ the matrix with elements $\hat v_{\alpha,\alpha'}(k)$,
$$\hat v(e^{i\frac{2\pi}3\sigma_2}k)=
\left(\begin{array}{*{4}{c}}
\hat v_{a,a}(k)&\hat v_{a,\tilde b}(k)&e^{ik\cdot l_2}\hat v_{a,\tilde a}(k)&e^{-ik\cdot l_2}\hat v_{a,b}(k)\\
\hat v_{\tilde b,a}(k)&\hat v_{\tilde b,\tilde b}(k)&e^{ik\cdot l_2}\hat v_{\tilde b,\tilde a}(k)&e^{-ik\cdot l_2}\hat v_{\tilde b,b}(k)\\
e^{-ik\cdot l_2}\hat v_{\tilde a,a}(k)&e^{-ik\cdot l_2}\hat v_{\tilde a,\tilde b}(k)&\hat v_{\tilde a,\tilde a}&e^{-2ik\cdot l_2}\hat v_{\tilde a,b}(k)\\
e^{ik\cdot l_2}\hat v_{b,a}(k)&e^{ik\cdot l_2}\hat v_{b,\tilde b}(k)&e^{2ik\cdot l_2}\hat v_{b,\tilde a}(k)&\hat v_{b,b}(k)
\end{array}\right)$$
furthermore
$$\left(\begin{array}{c}
\hat\xi^+_{\mathbf k_1,a}\hat\xi^-_{\mathbf k_2,a}\\[0.2cm]
\hat\xi^+_{\mathbf k_1,\tilde b}\hat\xi^-_{\mathbf k_2,\tilde b}\\[0.2cm]
(\hat\phi^+_{\mathbf k_1}\mathcal T_{\mathbf k_1}^\dagger)_{\tilde a}(\mathcal T_{\mathbf k_2}\hat\phi^-_{\mathbf k_2})_{\tilde a}\\[0.2cm]
(\hat\phi^+_{\mathbf k_1}\mathcal T_{\mathbf k_1}^\dagger)_b(\mathcal T_{\mathbf k_2}\hat\phi^-_{\mathbf k_2})_b
\end{array}\right)
=\left(\begin{array}{c}\hat\psi_{\mathbf k_1,a}^+\hat\psi_{\mathbf k_1,a}^-\\[0.2cm]
\hat\psi_{\mathbf k_1,\tilde b}^+\hat\psi_{\mathbf k_1,\tilde b}^-\\[0.2cm]
e^{il_2(k_1-k_2)}\hat\psi_{\mathbf k_1,\tilde a}^+\hat\psi_{\mathbf k_1,\tilde a}^-\\[0.2cm]
e^{-il_2(k_1-k_2)}\hat\psi_{\mathbf k_1,b}^+\hat\psi_{\mathbf k_1,b}^-
\end{array}\right)$$
from which one easily concludes that $\mathcal V$ is invariant under~(\ref{arotation}).\penalty10000\hfill\penalty10000$\square$\par
\subseqskip

\point{Complex conjugation} Follows immediately from $\Omega(-k)=\Omega^*(k)$ and $v(-k)=v^*(k)$.\penalty10000\hfill\penalty10000$\square$\par
\subseqskip

\point{ Vertical reflection} Follows immediately from $\Omega(R_vk)=\Omega(k)$ and $v(R_vk)=v(k)$ (since the second component of $d_\alpha$ is 0).\penalty10000\hfill\penalty10000$\square$\par
\subseqskip

\point{ Horizontal reflection} We have $\Omega(R_hk)=\Omega^*(k)$, $\sigma_1A^{\xi\xi}({\bf k})\sigma_1=A^{\xi\xi}({\bf k})$, 
$$\sigma_1A^{\xi\phi}({\bf k})\sigma_1=\left(\begin{array}{*{2}{c}}0&\Omega(k)\\\Omega^*(k)&0\end{array}\right)
,\quad \sigma_1A^{\phi\phi}({\bf k})\sigma_1=\left(\begin{array}{*{2}{c}}ik_0&\gamma_3\Omega^*(k)e^{-3ik_x}\\\gamma_3\Omega(k)e^{3ik_x}&ik_0\end{array}\right)
$$
from which the invariance of $h_0$ follows immediately. Furthermore
$$v_{\alpha,\alpha'}(R_hk)=v_{\pi_h(\alpha),\pi_h(\alpha')}(k)$$
where $\pi_h$ is the permutation that exchanges $a$ with $\tilde b$ and $\tilde a$ with $b$, from which the invariance of $\mathcal V$ follows immediately.\penalty10000\hfill\penalty10000$\square$\par
\subseqskip

\point{ Parity} We have $\Omega(Pk)=\Omega^*(k)$ so that $\big[A^{\xi\phi }(P\mathbf k)\big]^T=A^{\xi\phi}(\mathbf k)$, $\big[A^{\phi\phi}(P\mathbf k)\big]^T=A^{\phi\phi}(\mathbf k)$, $\big[A^{\xi\xi}(P\mathbf k)\big]^T=A^{\xi\xi}(\mathbf k)$. Therefore $h_0$ is mapped to
$$
h_0\longmapsto-\frac{1}{\chi_0(2^{-M}|k_0|)\beta|\Lambda|}\sum_{\mathbf k\in\mathcal B^*_{\beta,L}}\left(\begin{array}{*{2}{c}}\hat\xi_{\mathbf k}^-&\hat\phi_{\mathbf k}^-\end{array}\right)\left(\begin{array}{*{2}{c}}A^{\xi\xi}(\mathbf k)&A^{\xi\phi}(\mathbf k)\\A^{\phi\xi}(\mathbf k)&A^{\phi\phi}(\mathbf k)\end{array}\right)^T\left(\begin{array}{c}\hat\xi^+_{\mathbf k}\\\hat\phi^+_{\mathbf k}\end{array}\right)
$$
which is equal to $h_0$ since exchanging $\hat\psi^-$ and $\hat\psi^+$ adds a minus sign. The invariance of $\mathcal V$ follows from the remark that under parity 
$\hat\psi^+_{{\bf k}_1,\alpha}\hat\psi^-_{{\bf k}_2,\alpha}\mapsto \hat\psi^+_{P{\bf k}_2,\alpha}\hat\psi^-_{P{\bf k}_1,\alpha}$, and $\hat v(k_1-k_2)=\hat v(P(k_2-k_1))$.
\penalty10000\hfill\penalty10000$\square$\par
\bigskip

\point{ Time inversion} We have 
$$\begin{array}c
\sigma_3A^{\xi\xi}(I\mathbf k)\sigma_3=-A^{\xi\xi}(\mathbf k),\quad
\sigma_3A^{\xi\phi}(I\mathbf k)\sigma_3=-A^{\xi\phi}(\mathbf k),\\[0.2cm]
\sigma_3A^{\phi\phi}(I\mathbf k)\sigma_3=-A^{\phi\phi}(\mathbf k)
\end{array}$$
from which the invariance of $h_0$ follows immediately. The invariance of $\mathcal V$ is trivial.\penalty10000\hfill\penalty10000$\square$\par
\seqskip

\section{Constraints due to the symmetries}
\label{constWtapp}

\indent In this appendix we discuss some of the consequences of the symmetries listed in section~\ref{symsec} on $\hat W_2^{(h)}(\mathbf k)$ and its derivatives.\par
\bigskip

\indent We recall the definitions of the symmetry transformations from section~\ref{symsec}:
\begin{equation}\begin{array}c
T\mathbf k:=(k_0,e^{-i\frac{2\pi}3\sigma_2}k),\quad
R_v\mathbf k:=(k_0,k_1,-k_2),\quad
R_h\mathbf k:=(k_0,-k_1,k_2),\\[0.2cm]
P\mathbf k:=(k_0,-k_1,-k_2),\quad
I\mathbf k:=(-k_0,k_1,k_2).
\end{array}\label{symdefs}\end{equation}
Furthermore, given a $4\times4$ matrix $\mathbf M$ whose components are indexed by $\{a,\tilde b,\tilde a,b\}$, we denote the sub-matrix with components in $\{a,\tilde b\}^2$ by $\mathbf M^{\xi\xi}$, that with $\{\tilde a,b\}^2$ by $\mathbf M^{\phi\phi}$, with $\{a,\tilde b\}\times\{\tilde a,b\}$ by $\mathbf M^{\xi\phi}$ and with $\{\tilde a,b\}\times\{a,\tilde b\}$ by $\mathbf M^{\phi\xi}$. In addition, given a complex matrix $M$, we denote its component-wise complex conjugate by $M^*$ (which is not to be confused with its adjoint $M^\dagger$).\par
\bigskip 
\Theo{Proposition}\label{symprop}
Given a $2\times2$ complex matrix $M(\mathbf k)$ parametrized by $\mathbf k\in\mathcal B_\infty$ (we recall that $\mathcal B_\infty$ was defined above the statement of the Main Theorem in section~\ref{maintheosec}) and a pair of points $(\mathbf p_{F}^+,\mathbf p_{F}^-)\in\mathcal B_\infty^2$, 
if $\forall\mathbf k\in\mathcal B_\infty$
\begin{equation}
M(\mathbf k)= M(-\mathbf k)^*= M(R_v\mathbf k)=\sigma_1 M(R_h\mathbf k)\sigma_1=-\sigma_3 M(I\mathbf k)\sigma_3
\label{invsymsM}\end{equation}
and
\begin{equation}
\mathbf p_{F}^\omega=-\mathbf p_{F}^{-\omega}=R_v\mathbf p_{F}^{-\omega}=R_h\mathbf p_{F}^\omega=I\mathbf p_{F}^{\omega}
\label{invsymspf}\end{equation}
for $\omega\in\{-,+\}$, then $\exists(\mu,\zeta,\nu,\varpi)\in\mathbb{R}^4$ such that
\begin{equation}\begin{array}c
M(\mathbf p_{F}^\omega)=\mu\sigma_1,\quad
\partial_{k_0} M(\mathbf p_{F}^\omega)=i\zeta\mathds1,\\[0.5cm]
\partial_{k_1} M(\mathbf p_{F}^\omega)=\nu\sigma_2,\quad
\partial_{k_2} M(\mathbf p_{F}^\omega)=\omega \varpi\sigma_1.
\end{array}\label{Apfzt}\end{equation}
\endtheo
\bigskip

\indent\underline{Proof}:\par\penalty10000
\medskip\penalty10000
\pointt We first prove that $M(\mathbf p_{F}^\omega)=\mu\sigma_1$. We write
$$ M(\mathbf p_{F}^\omega)=:t\mathds1+x\sigma_1+y\sigma_2+z\sigma_3$$
where $(t,x,y,z)\in\mathbb C^4$.
We have
$$
 M(\mathbf p_{F}^\omega)=M(\mathbf p_{F}^{-\omega})^*=M(\mathbf p_{F}^{-\omega})=\sigma_1 M(\mathbf p_{F}^\omega)\sigma_1=-\sigma_3 M(\mathbf p_{F}^\omega)\sigma_3.
$$
Therefore $(t,x,y,z)$ are independent of $\omega$, $t=y=z=0$ and $x\in\mathbb{R}$.\par
\bigskip

\pointt We now study $\partial_{k_0} M$ which we write as
$$\partial_{k_0} M(\mathbf p_{F}^\omega)=:t_0\mathds1+x_0\sigma_1+y_0\sigma_2+z_0\sigma_3.$$
We have
$$
\partial_{k_0} M(\mathbf p_{F}^\omega)=-(\partial_{k_0} M(\mathbf p_{F}^{-\omega}))^*=\partial_{k_0} M(\mathbf p_{F}^{-\omega})=\sigma_1\partial_{k_0} M(\mathbf p_{F}^\omega)\sigma_1=\sigma_3\partial_{k_0} M(\mathbf p_{F}^\omega)\sigma_3.
$$
Therefore $(t_0,x_0,y_0,z_0)$ are independent of $\omega$, $x_0=y_0=z_0=0$ and $t_0\in i\mathbb{R}$.\par
\bigskip

\pointt We now turn our attention to $\partial_{k_1}M$:
$$\partial_{k_1} M(\mathbf p_{F}^\omega)=:t_1\mathds1+x_1\sigma_1+y_1\sigma_2+z_1\sigma_3.$$
We have
$$
\partial_{k_1} M(\mathbf p_{F}^\omega)=-(\partial_{k_1} M(\mathbf p_{F}^{-\omega}))^*=\partial_{k_1} M(\mathbf p_{F}^{-\omega})=-\sigma_1\partial_{k_1} M(\mathbf p_{F}^\omega)\sigma_1
=-\sigma_3\partial_{k_1} M(\mathbf p_{F}^{\omega})\sigma_3.
$$
Therefore $(t_1,x_1,y_1,z_1)$ are independent of $\omega$, $t_1=x_1=z_1=0$ and $y_1\in\mathbb{R}$.\par
\bigskip

\pointt Finally, we consider $\partial_{k_y}M$:
$$\partial_{k_2} M(\mathbf p_{F}^\omega)=:t_2^{(\omega)}\mathds1+x_2^{(\omega)}\sigma_1+y_2^{(\omega)}\sigma_2+z_2^{(\omega)}\sigma_3.$$
We have
$$
\partial_{k_2} M(\mathbf p_{F}^\omega)=-(\partial_{k_2} M(\mathbf p_{F}^{-\omega}))^*=-\partial_{k_2} M(\mathbf p_{F}^{-\omega})=\sigma_1\partial_{k_2} M(\mathbf p_{F}^\omega)\sigma_1=-\sigma_3\partial_{k_2} M(\mathbf p_{F}^\omega)\sigma_3.
$$
Therefore $t_2^{(\omega)}=y_2^{(\omega)}=z_2^{(\omega)}=0$, $x_2^{(\omega)}=-x_2^{(-\omega)}\in\mathbb{R}$.\penalty10000\hfill\penalty10000$\square$\par
\subseqskip

\Theo{Proposition}\label{symfourprop}
Given a $4\times4$ complex matrix $\mathbf M(\mathbf k)$ parametrized by $\mathbf k\in\mathcal B_\infty$ and two points $(\mathbf p_{F}^+,\mathbf p_{F}^-)\in\mathcal B_\infty^2$, if $\forall(f,f')\in\{\xi,\phi\}^2$ and $\forall\omega\in\{-,+\}$,
\begin{equation}\begin{array}c
\mathbf M^{ff'}(\mathbf p_{F}^\omega)=\mu^{ff'}\sigma_1,\quad
\partial_{k_0}\mathbf M^{ff'}(\mathbf p_{F}^\omega)=i\zeta^{ff'}\mathds1,\\[0.5cm]
\partial_{k_1}\mathbf M^{ff'}(\mathbf p_{F}^\omega)=\nu^{ff'}\sigma_2,\quad
\partial_{k_2}\mathbf M^{ff'}(\mathbf p_{F}^\omega)=\omega\varpi^{ff'}\sigma_1
\end{array}\label{Msymhyprot}\end{equation}
with $(\mu^{ff'}, \zeta^{ff'}, \nu^{ff'}, \varpi^{ff'})\in\mathbb R^{4}$ independent of $\omega$, 
and $\forall\mathbf k\in\mathcal B_\infty$
\begin{equation}
\mathbf M(\mathbf k)=\mathbf M^T(P\mathbf k)
\label{invPM}\end{equation}
and
\begin{equation}
\mathbf p_{F}^\omega=P\mathbf p_{F}^{-\omega}
\label{invPpf}\end{equation}
then
\begin{equation}
\mu^{\phi\xi}=\mu^{\xi\phi},\quad
\zeta^{\phi\xi}=\zeta^{\xi\phi},\quad
\nu^{\phi\xi}=\nu^{\xi\phi},\quad
\varpi^{\phi\xi}=\varpi^{\xi\phi}.
\label{eqrotsymo}\end{equation}
Furthermore, if $\mathbf p_{F}^\omega=(0,\frac{2\pi}3,\omega\frac{2\pi}{3\sqrt3})$ and (recalling that $\mathcal T_{\mathbf k}=e^{-i(l_2\cdot k)\sigma_3}$, with $l_2=(3/2,-\sqrt3/2)$)
\begin{equation}
\mathbf M(\mathbf k)=\left(\begin{array}{*{2}{c}}\mathds1&0\\0&\mathcal T_{\mathbf k}^\dagger\end{array}\right)\mathbf M(T^{-1}\mathbf k)\left(\begin{array}{*{2}{c}}\mathds1&0\\0&\mathcal T_{\mathbf k}\end{array}\right)
\label{invTM}\end{equation}
then\par
\begin{equation}\begin{array}c
\nu^{\phi\phi}=-\varpi^{\phi\phi},\quad
\nu^{\xi\phi}=\varpi^{\xi\phi},\quad
\nu^{\phi\xi}=\varpi^{\phi\xi},\quad
\nu^{\xi\xi}=\varpi^{\xi\xi}=0,\\[0.2cm]
\mu^{\phi\phi}=\mu^{\xi\phi}=\mu^{\phi\xi}=0, \quad \zeta^{\phi\xi}=\zeta^{\xi\phi}=0.
\nopagebreakaftereq\end{array}\label{eqrotsym}\end{equation}
\endtheo
\restorepagebreakaftereq
\resetpointcounter
\bigskip

\indent\underline{Proof}: (\ref{eqrotsymo}) is straightforward, so we immediately turn to the proof of~(\ref{eqrotsym}).\par
\bigskip

\pointt We first focus on $\mathbf M^{\phi\phi}$ which satisfies
\begin{equation}
\mathbf M^{\phi\phi}(\mathbf k)=\mathcal T_{\mathbf k}^\dagger\mathbf M^{\phi\phi}(T^{-1}\mathbf k)\mathcal T_{\mathbf k}.\label{eq:mff}
\end{equation}
Evaluating this formula at ${\bf k}=\mathbf p_F^\omega$, recalling that $\mathbf M^{\phi\phi}(\mathbf p_{F}^\omega)=\mu^{\phi\phi}\sigma_1$, 
 and noting that $\mathcal T_{\mathbf p_{F}^\omega}=-\frac12\mathds1-i\omega\frac{\sqrt3}2\sigma_3$, we obtain $\mu^{\phi\phi}=0$.
 Therefore, deriving \eqref{eq:mff} with respect to $k_i$, $i=1,2$, and evaluating at $\mathbf p_F^\omega$, we get:
 $$\partial_{k_i}\mathbf M^{\phi\phi}(\mathbf p_{F}^\omega)=\sum_{j=1}^2T_{i,j}\mathcal T_{\mathbf p_{F}^\omega}^\dagger\partial_{k_j}\mathbf M^{\phi\phi}(\mathbf p_{F}^\omega)\mathcal T_{\mathbf p_{F}^\omega}$$
with
$$T=\frac{1}{2}\left(\begin{array}{*{2}{c}}-1&-\sqrt3\\[0.2cm]\sqrt3&-1\end{array}\right).$$
Furthermore, recalling that $\partial_{k_1}\mathbf M^{\phi\phi}=\nu^{\phi\phi}\sigma_2$ and $\partial_{k_2}\mathbf M^{\phi\phi}=\omega\varpi^{\phi\phi}\sigma_1$,
$$\mathcal T_{\mathbf p_{F}^\omega}^\dagger\partial_{k_1}\mathbf M^{\phi\phi}\mathcal T_{\mathbf p_{F}^\omega}=\nu^{\phi\phi}\Big(-\frac12\sigma_2-\omega\frac{\sqrt3}{2}\sigma_1\Big), \quad 
\mathcal T_{\mathbf p_{F}^\omega}^\dagger\partial_{k_2}\mathbf M^{\phi\phi}\mathcal T_{\mathbf p_{F}^\omega}=\omega\varpi^{\phi\phi}\Big(-\frac12\sigma_1+\omega\frac{\sqrt3}{2}\sigma_2\Big),$$
which implies
$$\left(\begin{array}{c}\nu^{\phi\phi}\sigma_2\\[0.2cm] \omega \varpi^{\phi\phi}\sigma_1\end{array}\right)=\frac{1}{4}\left(\begin{array}{*{2}{c}} \nu^{\phi\phi}-3 \varpi^{\phi\phi}&\omega\sqrt3(\nu^{\phi\phi}+\varpi^{\phi\phi})\\[0.2cm]-\sqrt3(\nu^{\phi\phi}+\varpi^{\phi\phi})&\omega(\varpi^{\phi\phi}-3\nu^{\phi\phi})\end{array}\right)\left(\begin{array}{c}\sigma_2\\[0.2cm]\sigma_1\end{array}\right)$$
so $\nu^{\phi\phi}=-\varpi^{\phi\phi}$.
\bigskip

\pointt We now study $\mathbf M^{\phi\xi}$ which satisfies
$$
\mathbf M^{\phi\xi}(\mathbf k)=\mathcal T_{\mathbf k}^\dagger\mathbf M^{\phi\xi}(T^{-1}\mathbf k).
$$
Evaluating this formula and its derivative with respect to $k_0$ at ${\bf k}=\mathbf p_F^\omega$, we obtain 
$\mu^{\phi\xi}=\zeta^{\phi\xi}=0$.
Evaluating the derivative of this formula with respect to $k_i$ at ${\bf k}=\mathbf p_F^\omega$, we obtain
$$\partial_{k_i}\mathbf M^{\phi\xi}(\mathbf p_{F}^\omega)=\sum_{j=1}^2T_{i,j}\mathcal T_{\mathbf p_{F}^\omega}^\dagger\partial_{k_j}\mathbf M^{\phi\xi}(\mathbf p_{F}^\omega).$$
Furthermore, 
$$\mathcal T_{\mathbf p_{F}^\omega}^\dagger\partial_{k_1}\mathbf M^{\phi\xi}=\nu^{\phi\xi}\Big(-\frac12\sigma_2+\omega\frac{\sqrt3}{2}\sigma_1\Big), \quad 
\mathcal T_{\mathbf p_{F}^\omega}^\dagger\partial_{k_2}\mathbf M^{\phi\xi}=\omega\varpi^{\phi\xi}\Big(-\frac12\sigma_1-\omega\frac{\sqrt3}{2}\sigma_2\Big),$$
which implies
$$\left(\begin{array}{c}\nu^{\phi\xi}\sigma_2\\[0.2cm] \omega \varpi^{\phi\xi}\sigma_1\end{array}\right)=\frac{1}{4}\left(\begin{array}{*{2}{c}} \nu^{\phi\xi}+3 \varpi^{\phi\xi}&-\omega\sqrt3(\nu^{\phi\xi}-\varpi^{\phi\xi})\\[0.2cm]-\sqrt3(\nu^{\phi\xi}-\varpi^{\phi\xi})&\omega(\varpi^{\phi\xi}+3\nu^{\phi\xi})\end{array}\right)\left(\begin{array}{c}\sigma_2\\[0.2cm]\sigma_1\end{array}\right)$$
so that $\nu^{\phi\xi}_h=\varpi^{\phi\xi}_h$. 
The case of $\mathbf M^{\xi\phi}$ is completely analogous and gives $\mu^{\xi\phi}=\zeta^{\xi\phi}=0$ and $\nu^{\xi\phi}_h=\varpi^{\xi\phi}_h$.
\par
\bigskip

\pointt We finally turn to $\mathbf M^{\xi\xi}$, which satisfies
$$
\mathbf M^{\xi\xi}(\mathbf k)=\mathbf M^{\xi\xi}(T^{-1}\mathbf k).
$$
Therefore for $i\in\{1,2\}$,
$$
\partial_{k_i}\mathbf M^{\xi\xi}(\mathbf p_{F}^\omega)=\sum_{j=1}^2T_{i,j}\partial_{k_j}\mathbf M^{\xi\xi}(\mathbf p_{F}^\omega)
$$
so that
$\partial_{k_i}\mathbf M^{\xi\xi}(\mathbf p_{F}^\omega)=0$, that is $\nu^{\xi\xi}=\varpi^{\xi\xi}=0$.\penalty10000\hfill\penalty10000$\square$\par

\pagebreak


\addcontentsline{toc}{section}{References}

\begin{thebibliography}{WWW99}
\bibitem[AR98]{arNE}
A.~Abdesselam, V.~Rivasseau - {\it Explicit Fermionic tree expansions}, Letters in Mathematical Physics, Vol.~44, n.~1, p.~77-88, 1998.

\bibitem[BF84]{batEF}
G.~Battle, P.~Federbush - {\it A note on cluster expansions, tree graph identities, extra $1/N!$ factors!!!}, Letters in Mathematical Physics, Vol.~8, p.~55-57, 1984.

\bibitem[BG90]{benNZ}
G.~Benfatto, G.~Gallavotti - {\it Perturbation theory of the Fermi surface in a quantum liquid - a general quasiparticle formalism and one-dimensional systems}, Journal of Statistical Physics, Vol.~59, n.~3-4, p.~541-664, 1990.

\bibitem[BG95]{benNFi}
G.~Benfatto, G.~Gallavotti - {\it Renormalization Group}, Princeton University Press, 1995.

\bibitem[BM02]{bmZT}
G.~Benfatto, V.~Mastropietro - {\it On the density-density critical indices in interacting Fermi systems}, Communications in Mathematical Physics, Vol.~231, n.~1, p.~97-134, 2002.

\bibitem[BGM06]{benZS}
G.~Benfatto, A.~Giuliani, V.~Mastropietro - {\it Fermi liquid behavior in the 2D Hubbard model}, Annales Henri Poincar\'e, Vol.~7, p.~809-898, 2006.

\bibitem[BF78]{brySeE}
D.~Brydges, P.~Federbush - {\it A new form of the Mayer expansion in classical statistical mechanics}, Journal of Mathematical Physics, Vol.~19, p.~2064, 1978.

\bibitem[BK87]{bkESe}
D.~Brydges, T.~Kennedy - {\it Mayer expansions and the Hamilton-Jacobi equation}, Journal of Statistical Physics, Vol.~48, n.~1-2, p.~19-49, 1987.

\bibitem[CTV12]{ctvOT}
V.~Cvetkovic, R.~Throckmorton, O.~Vafek - {\it Electronic multicriticality in bilayer graphene}, Physical Review B, Vol.~86, n.~075467, 2012.

\bibitem[DDe79]{doeSeN}
R.~Doezema, W.~Datars, H.~Schaber, A.~Van~Schyndel - {\it Far-infrared magnetospectroscopy of the Landau-level structure in graphite}, Physical Review B, Vol.~19, n.~8, p.~4224-4230, 1979.

\bibitem[DD02]{dreZT}
M.~Dresselhaus, G.~Dresselhaus - {\it Intercalation compounds of graphite}, Advances in Physics, Vol.~51, n.~1, p.~1-186, 2002.

\bibitem[FKT04a]{fktZFa}
J.~Feldman, H.~Kn\"orrer, E.~Trubowitz - {\it A two dimensional Fermi liquid. Part~1: Overview}, Communications in Mathematical Physics, Vol.~247, n.~1, p.~1-47, 2004.

\bibitem[FKT04b]{fktZFb}
J.~Feldman, H.~Kn\"orrer, E.~Trubowitz - {\it A two dimensional Fermi liquid. Part~2: Convergence}, Communications in Mathematical Physics, Vol.~247, n.~1, p.~49-111, 2004.

\bibitem[FKT04c]{fktZFc}
J.~Feldman, H.~Kn\"orrer, E.~Trubowitz - {\it A two dimensional Fermi liquid. Part~3: The Fermi surface}, Communications in Mathematical Physics, Vol.~247, n.~1, p.~113-177, 2004.

\bibitem[GN85]{galEFi}
G.~Gallavotti, F.~Nicol\`o - {\it Renormalization theory for four dimensional scalar fields}, Communications in Mathematical Physics, Vol.~100, p.~545-590 and Vol.~101, p.~247-282, 1985.

\bibitem[GN07]{geiZSe}
A.~Geim, K.~Novoselov - {\it The rise of graphene}, Nature Materials, Vol.~6, p.~183-191, 2007.

\bibitem[Ge10]{geOZ}
A.~Geim - {\it Random walk to graphene}, Nobel lecture, 2010.

\bibitem[GM01]{genZO}
G.~Gentile, V.~Mastropietro - {\it Renormalization group for one-dimensional fermions - a review on mathematical results}, Physics Reports, Vol.~352, p.~273-437, 2001.

\bibitem[GM10]{giuOZ}
A.~Giuliani, V.~Mastropietro - {\it The two-dimensional Hubbard model on the honeycomb lattice}, Communications in Mathematical Physics, Vol.~293, p.~301-364, 2010.

\bibitem[Gi10]{giuOZh}
A.~Giuliani - {\it The Ground State Construction of the Two-dimensional Hubbard Model on the Honeycomb Lattice}, Quantum Theory from Small to Large Scales, lecture notes of the Les Houches Summer School, Vol.~95, Oxford University Press, 2010.

\bibitem[GMP10]{gmpOZ}
A.~Giuliani, V.~Mastropietro, M.~Porta - {\it Lattice gauge theory model for graphene}, Physical Review B, Vol.~82, n.~121418(R), 2010.

\bibitem[GMP11]{gmpOO}
A.~Giuliani, V.~Mastropietro, M.~Porta - {\it Absence of interaction corrections in the optical conductivity of graphene}, Physical Review B, Vol.~83, n.~195401, 2011.

\bibitem[GMP11b]{gmpOOt}
A.~Giuliani, V.~Mastropietro, M.~Porta - {\it Lattice quantum electrodynamics for graphene}, Annals of Physics, Vol.~327, n.~2, p.~461-511, 2011.

\bibitem[GMP12]{gmpOT}
A.~Giuliani, V.~Mastropietro, M.~Porta - {\it Universality of conductivity in interacting graphene}, Communications in Mathematical Physics, Vol.~311, n.~2, p.~317-355, 2012.

\bibitem[Lu13]{luOTh}
L.~Lu - {\it Constructive analysis of two dimensional Fermi systems at finite temperature}, PhD thesis, Institute for Theoretical Physics, Heidelberg, \url{http://www.ub.uni-heidelberg.de/archiv/14947}, 2013.

\bibitem[MNe07]{malZSe}
L.~Malard, J.~Nilsson, D.~Elias, J.~Brant, F.~Plentz, E.~Alves, A.~Castro Neto, M.~Pimenta - {\it Probing the electronic structure of bilayer graphene by Raman scattering}, Physical Review B, Vol.~76, n.~201401, 2007.

\bibitem[Ma11]{masOO}
V.~Mastropietro - {\it Conductivity between Luttinger liquids: coupled chains and bilayer graphene}, Physical Review B, Vol.~84, n.~035109, 2011.

\bibitem[MF06]{mccZS}
E.~McCann, V.~Fal'ko - {\it Landau-level degeneracy and Quantum Hall Effect in a graphite bilayer}, Physical Review Letters, Vol.~86, 086805, 2006.

\bibitem[Mc57]{mccFiSe}
J.~McClure - {\it Band structure of graphite and de Haas-van Alphen effect}, Physical review, Vol.~108, p.~612-618, 1957.

\bibitem[MMD79]{misSeN}
A.~Misu, E.~Mendez, M.S.~Dresselhaus - {\it Near Infrared Reflectivity of Graphite under Hydrostatic Pressure}, Journal of the Physical Society of Japan, Vol.~47, n.~1, p.~199-207, 1979.

\bibitem[NGe04]{ngeZF}
K.~Novoselov, A.~Geim, S.~Morozov, D.~Jiang, Y.~Zhang, S.~Dubonos, I.~Grigorieva, A.~Firsov - {\it Electric field effect in atomically thin carbon films}, Science, vol.~306, p.~666-669, 2004.

\bibitem[NGe05]{ngeZFi}
K.~Novoselov, A.~Geim, S.~Morozov, D.~Jiang, M.~Katsnelson, I.~Grigorieva, S.~Dubonos, A.~Firsov - {\it Two-dimensional gas of massless Dirac fermions in graphene}, Nature, Vol.~438, n.~10, p.~197-200, 2005.

\bibitem[NMe06]{novZS}
K.~Novoselov, E.~McCann, S.~Morozov, V.~Fal'ko, M.~Katsnelson, U.~Zeitler, D.~Jiang, F.~Schedin, A.~Geim - {\it Unconventional quantum Hall effect and Berry's phase of $\pi$ in bilayer graphene}, Nature Physics, Vol.~2, p.~177-180, 2006.

\bibitem[PP06]{parZS}
B.~Partoens, F.~Peeters - {\it From graphene to graphite: electronic structure around the $K$ point}, Physical Review B, Vol.~74, n.~075404, 2006.

\bibitem[PS08]{psZE}
W.~Pedra, M.~Salmhofer - {\it Determinant bounds and the Matsubara UV problem of many-fermion systems}, Communications in Mathematical Physics, Vol.~282, n.~3, p.~797-818, 2008.

\bibitem[Sal13]{salOTh}
M.~Salmhofer - {\it Renormalization: an introduction}, Springer Science \& Business Media, 2013.

\bibitem[SW58]{sloFiE}
J.~Slonczewski, P.~Weiss - {\it Band structure of graphite}, Physical Review, Vol.~109, p.~272-279, 1958.

\bibitem[TV12]{tvOT}
R.~Throckmorton, O.~Vafek - {\it Fermions on bilayer graphene: symmetry breaking for $B=0$ and $\nu=0$}, Physical Review B, Vol.~86, 115447, 2012.

\bibitem[TDD77]{toySeSe}
W.~Toy, M.~Dresselhaus, G.~Dresselhaus - {\it Minority carriers in graphite and the H-point magnetoreflection spectra}, Physical Review B, Vol.~15, p.~4077-4090, 1977.

\bibitem[TMe92]{triNT}
S.~Trickey, F.~M\"uller-Plathe, G.~Diercksen, J.~Boettger - {\it Interplanar binding and lattice relaxation in a graphite dilayer}, Physical Review B, Vol.~45, p.~4460-4468, 1992.

\bibitem[Va10]{vafOZ}
O.~Vafek - {\it Interacting Fermions on the honeycomb bilayer: from weak to strong coupling}, Physical Review B, Vol.~82, 205106, 2010.

\bibitem[VY10]{vayOZ}
O.~Vafek, K.~Yang - {\it Many-body instability of Coulomb interacting bilayer graphene: renormalization group approach}, Physical Review B, Vol.~81, 041401, 2010.

\bibitem[Wa47]{walFSe}
P.~Wallace - {\it The band theory of graphite}, Physical Review, Vol.~71, n.~9, p.~622-634, 1947.

\bibitem[ZTe05]{zteZFi}
Y.~Zhang, Y.W.~Tan, H.~Stormer, P.~Kim - {\it Experimental observation of the quantum Hall effect and Berry's phase in graphene}, Nature, Vol.~438, n.~10, p.~201-204, 2005.

\bibitem[ZLe08]{zhaZE}
L.~Zhang, Z.~Li, D.~Basov, M.~Fogler, Z.~Hao, M.~Martin - {\it Determination of the electronic structure of bilayer graphene from infrared spectroscopy}, Physical Review B, Vol.~78, n.~235408, 2008.


\end{thebibliography}
\end{document}